\documentclass[twocolumn,numberedappendix,iop]{openjournal}
\usepackage{graphicx,amsmath,amssymb,amstext}
\usepackage{amsbsy,amsfonts,amsthm,color}
\usepackage[colorlinks,linkcolor=blue,citecolor=blue,urlcolor=blue ]{hyperref}
\usepackage[utf8]{inputenc}
\usepackage{float}
\usepackage{placeins} 
\usepackage{xcolor}
\usepackage{ulem}
\usepackage[T1]{fontenc}
\usepackage[title]{appendix}
\usepackage{booktabs}
\usepackage{array}
\usepackage{multirow}
\usepackage{comment}

\usepackage{subcaption}
\usepackage{caption}
\makeatletter
\newenvironment{rcases}
  {\left.\begin{aligned}}
  {\end{aligned}\right\rbrace}
\makeatother
\let\longtable*\relax

\makeatletter
\let\longtable\relax

\expandafter\let\csname longtable*\endcsname\relax
\expandafter\let\csname endlongtable*\endcsname\relax
\makeatother

\begin{document}

\title{Joint cosmological fits to DESI-DR1 full-shape clustering \\ and weak gravitational lensing in configuration space\vspace{-4em}}
\shorttitle{DESI-DR1 Shear $\times$ RSD Analysis}

\author{
A.~Semenaite,$^{1,*}$
C.~Blake,$^{1}$
A.~Porredon,$^{2,3,4,5}$
J.~Aguilar,$^{6}$
S.~Ahlen,$^{7}$
D.~Bianchi,$^{8,9}$
D.~Brooks,$^{10}$
F.~J.~Castander,$^{11,12}$
T.~Claybaugh,$^{6}$
A.~Cuceu,$^{6}$
K.~S.~Dawson,$^{13}$
A.~de la Macorra,$^{14}$
Biprateep~Dey,$^{15,16}$
P.~Doel,$^{10}$
A.~Eggemeier,$^{17}$
A.~Elliott,$^{18,5}$
N.~Emas,$^{1}$
S.~Ferraro,$^{6,19}$
A.~Font-Ribera,$^{20}$
J.~E.~Forero-Romero,$^{21,22}$
C.~Garcia-Quintero,$^{23,56}$
E.~Gaztañaga,$^{11,24,12}$
S.~Gontcho A Gontcho,$^{6,25}$
G.~Gutierrez,$^{26}$
J.~Guy,$^{6}$
B.~Hadzhiyska,$^{27,19}$
H.~K.~Herrera-Alcantar,$^{28,29}$
S.~Heydenreich,$^{30}$
K.~Honscheid,$^{31,18,5}$
C.~Howlett,$^{32}$
D.~Huterer,$^{33,34}$
M.~Ishak,$^{35}$
S.~Joudaki,$^{2}$
R.~Joyce,$^{36}$
E.~Jullo,$^{37}$
D.~Kirkby,$^{38}$
A.~Kremin,$^{6}$
A.~Krolewski,$^{39,40,41}$
O.~Lahav,$^{10}$
C.~Lamman,$^{5}$
M.~Landriau,$^{6}$
J.~U.~Lange,$^{42}$
L.~Le~Guillou,$^{43}$
A.~Leauthaud,$^{30,44}$
M.~E.~Levi,$^{6}$
M.~Manera,$^{45,20}$
A.~Meisner,$^{36}$
R.~Miquel,$^{46,20}$
J.~Moustakas,$^{47}$
S.~Nadathur,$^{24}$
J.~ A.~Newman,$^{16}$
N.~Palanque-Delabrouille,$^{29,6}$
W.~J.~Percival,$^{39,40,41}$
A.~Pezzotta,$^{9}$
C.~Poppett,$^{6,48,19}$
F.~Prada,$^{49}$
I.~P\'erez-R\`afols,$^{50}$
A.~Robertson,$^{36}$
G.~Rossi,$^{51}$
R.~Ruggeri,$^{52}$
A.~Sanchez,$^{53}$
E.~Sanchez,$^{2}$
C.~Saulder,$^{53}$
D.~Schlegel,$^{6}$
M.~Schubnell,$^{33,34}$
H.~Seo,$^{54}$
J.~Silber,$^{6}$
D.~Sprayberry,$^{36}$
G.~Tarl\'{e},$^{34}$
B.~A.~Weaver,$^{36}$
P.~Zarrouk,$^{43}$
R.~Zhou,$^{6}$
and H.~Zou$^{55}$
\\
{\it (Affiliations can be found after the references)}
}
\thanks{$^*$E-mail: asemenaite@swin.edu.au}
\shortauthors{A.~Semenaite et al.}

\begin{abstract}
We present a joint $3\times2$-pt cosmological analysis of auto- and cross-correlations between the Dark Energy Spectroscopic Instrument Data Release 1 (DESI-DR1) Bright Galaxy Survey (BGS) and Luminous Red Galaxy (LRG) samples and overlapping shear measurements from the KiDS-1000, DES-Y3 and HSC-Y3 weak lensing surveys.  We perform our analysis in configuration space and, in addition to the cosmic shear correlation functions for each weak lensing dataset, we fit the tangential shear of the weak lensing source galaxies around DESI lens galaxies. Finally, we make use of the anisotropic BGS and LRG clustering information by fitting the full shape of the two-point correlation function multipoles measured over the full DESI-DR1 footprint, presenting the first full-shape analysis of DESI measurements in configuration space.  We find that the addition of weak lensing information serves to improve, with respect to the clustering-only case, the measurements of the power spectrum amplitude parameters $\ln(10^{10}A_{\rm{s}})$ and $\sigma_{12}$ by $15\%$ and $36\%$, respectively. It also improves measurements of the linear bias of the lens galaxies by $15-20\%$, depending on the tracer. Our results show excellent consistency, regardless of the weak lensing survey considered, and are furthermore consistent with a companion analysis that fits $3\times2$-pt correlations including DESI projected clustering measurements, as well as the results published by the weak lensing collaborations themselves. Our measured values for weak lensing amplitude are $S_{8}^{\mathrm{DESI\times HSC}}=0.787\pm0.020$, $S_{8}^{\mathrm{DESI\times DES}}=0.791\pm0.016$, $S_{8}^{\mathrm{DESI\times KiDS}}=0.771\pm0.017$, which are $1.9\sigma-2.9\sigma$ below the $S_8$ value preferred by \textit{Planck}. Finally, our clustering-only results are in good agreement with the Fourier space full-shape analysis of all DESI tracers, although we see some indications of the presence of projection effects.  This work paves the way for future `same-sky' analyses of cross-correlations between the upcoming DESI data releases and overlapping shear datasets.
\end{abstract}

\maketitle

\section{Introduction}

\begin{figure*}
\centering
\includegraphics[width=0.9\textwidth]{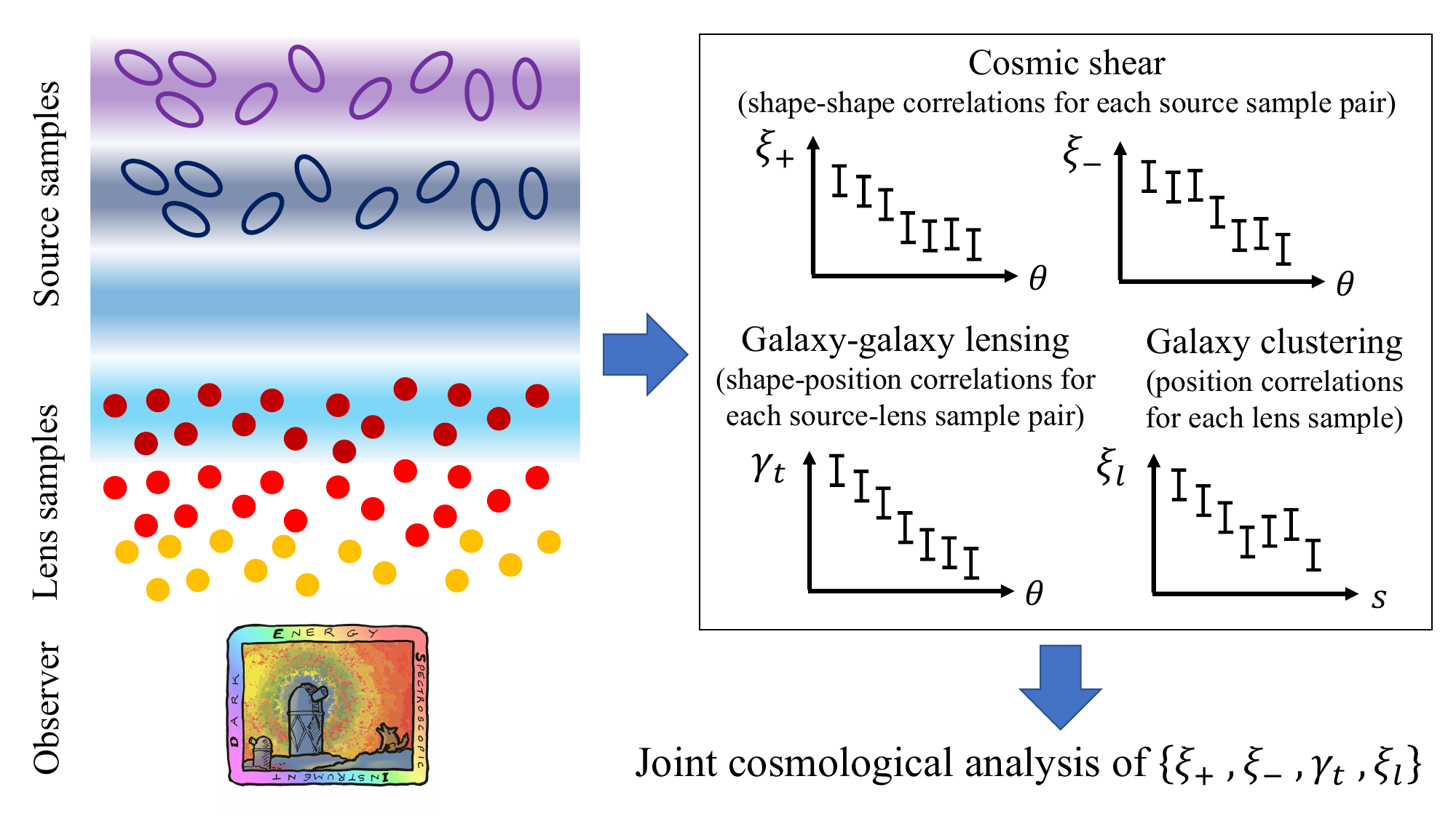}
\caption{A cartoon representation of the different source and lens correlation statistics combined by this paper.  We use datasets of the shapes of distant source galaxies in photometric redshift bins, obtained from deep imaging surveys, where these shapes are correlated by weak gravitational lensing.  DESI maps out a sample of foreground lens galaxies with spectroscopic redshifts.  Our analysis then combines correlations between the shapes of the source galaxies (cosmic shear, $\xi_+$ and $\xi_-$), correlations between the shapes of the source galaxies and positions of the lens galaxies (average tangential shear due to galaxy-galaxy lensing, $\gamma_t$), and the auto-correlation between the redshift-space positions of the lens galaxies (clustering multipoles, $\xi_\ell$).  Whilst conventional ``$3 \times 2$-pt'' analyses use projected clustering statistics of the lenses, our study also includes the redshift-space distortion information present in the monopole and quadrupole, $\ell = \{ 0,2 \}$.}
\label{fig:cartoon}
\end{figure*}

The completion and release of the first year of observations performed by the Dark Energy Spectroscopic Instrument (DESI) marked the start of the so-called `Stage IV' era of modern cosmological surveys, delivering orders of magnitude larger datasets than their predecessors. Much of this work is motivated by the wish to understand the nature of dark energy, which makes up the majority of the energy budget of the Universe today. Although significant progress has been made since the first observations confirming the dark energy-driven acceleration of the Universe \citep{riess1998, perlmutter1999}, none of the Stage-III analyses have been able to measure a deviation from the standard $\Lambda$CDM model which predicts constant dark energy density with the equation of state parameter of $w=-1$ \citep{alam2021, troster021, abbott2023}. Whilst a number of cosmological observables prefer $\Lambda$CDM as their best-fit model, some discrepancies, or `tensions', in the measured values of its parameters have emerged between different probes \citep{DiValentino2021, 2021A&A...646A.140H, freedman2021, riess2022, abdalla2022, 2025JCAP...02..021A, 2025PhRvD.112h3514A, 2025PhRvD.112h3515A}.  These results potentially point us towards scenarios beyond $\Lambda$CDM \citep{Schoneberg_2022, Herold_2023, 2025PhRvD.112h3515A, tang2025, ishak2025}, while simultaneously motivating us to improve our understanding of the interplay between different observables, with the aim of building a coherent cosmological picture. 

Galaxy clustering and weak gravitational lensing are naturally complementary probes of the large-scale structure of the Universe: whilst galaxy clustering allows us to obtain high signal-to-noise measurements of the visible part of the cosmic web, gravitational lensing, despite being a much weaker effect, provides a powerful probe of the total matter distribution. The combination of these observables has been used to explore the consistency between these phenomena as well as improve our understanding of the galaxy-halo connection on small scales \citep[for example,][]{leauthaud2017, zacharegkas2021, Miyatake2022II, amon2023, Li2025}, and to perform different versions of combined large-scale cosmological fits in the form of the so-called `$3\times2$-pt' analyses \citep{2021A&A...646A.140H, abbott2023, sugiyama2023}. The latter refers to a set of three two-point correlation functions between the source and lens galaxies: cosmic shear, which is the auto-correlation of the observed source galaxy shapes that are distorted by the gravitational potential of the large-scale structure, galaxy-galaxy lensing, which refers to the cross-correlation of source galaxy shapes with lens galaxy positions, and galaxy clustering, which is measured through auto-correlation of lens galaxy positions.  A visualisation of the statistics we employ in this study is shown in Figure \ref{fig:cartoon}.

Whilst joint galaxy clustering and weak lensing analyses can, in principle, provide a comprehensive description of the cosmic web, by breaking cosmological parameter degeneracies and calibrating systematic effects \citep{hu2004, bernstein2009, joachimi2010, Miyatake2022I}, performing a combined analysis is non-trivial. Gravitational lensing is a relatively low signal-to-noise effect that requires datasets an order of magnitude larger than galaxy clustering, which can only be achieved by photometric surveys. Lensing additionally involves mixing of scales due to projection along the line of sight, meaning that, in order to accurately describe our measurements, we need to carefully model the complex physics on small scales or implement broad scale cuts, sacrificing some of the information \citep{martinelli2021,amon2022, pranjal2025, 2025arXiv251005539E, derose2025}. Galaxy clustering has the advantage of needing smaller galaxy samples, but high-quality redshifts are of extreme importance. Modelling the clustering signal involves complex bias models on top of perturbative theories that can account for the redshift space distortions (RSD) induced by galaxy peculiar velocities, both on virial scales within a dark matter halo, as well as on the larger scales where coherent infall velocities trace the growth of the cosmic web \citep{senatore2014, chen2020, eggemeier2025}.

Joint analyses of these probes have been approached in one of two main ways: either by using a uniform power spectrum model for the full $3\times2$-pt data vector, but fitting projected clustering, which loses the information carried by the RSD \citep[as done in, for example,][]{abbott2023, sugiyama2023}, or by retaining the redshift information of galaxy clustering measurements, but using a combination of different modelling prescriptions for shear and galaxy clustering observables \citep[for example,][]{2021A&A...646A.140H}\footnote{It is, however, worth noting that there have been successful examples of consistent joint modelling of CMB lensing and galaxy clustering, mainly due to the small-scales of CMB lensing data vector being noise-dominated \citep[see, for example, ][]{maus2025cmb,chen2022}.}. The first approach has the advantage of allowing us to extract information from both spectroscopic and photometric samples without relying on perturbative models, allowing for a uniform analysis, but loses some information on the structure growth, as well as the information carried by the BAO peak in the two-point correlation function. 

DESI overlaps with significant areas of three major publicly-available weak lensing datasets: the Kilo-Degree Survey \citep[KiDS-1000,][]{2021A&A...645A.105G}, the Dark Energy Survey \citep[DES-Y3,][]{2021MNRAS.504.4312G, 2022PhRvD.105b3514A, 2022PhRvD.105b3515S}, and the Hyper Suprime-Cam survey \citep[HSC-Y3,][]{2022PASJ...74..421L}.  This overlap offers a unique opportunity for joint galaxy clustering and weak lensing analyses with a high-quality redshift lens sample for galaxy-galaxy lensing and anisotropic clustering measurements. The unprecedented spatial extent of DESI allows us not only to obtain precise measurements of cosmological parameters, by making use of the complementarity between weak lensing and galaxy clustering, but also to carry out consistency checks by comparing the results between the combinations with different surveys, and (although we do not incorporate it in this study) improve calibration of source redshift distributions \citep{2025arXiv251025419L, RossanaDESI, DianaDESI}. 

In this work we hence present a galaxy clustering and weak lensing cosmological analysis of the shear measurements by three major weak lensing surveys, KiDS-1000, DES-Y3 and HSC-Y3, and DESI data corresponding to its first year of observations \citep[Data Release 1, ][]{2025arXiv250314745D}. We use a unified pipeline to fit DESI galaxy clustering together with the shear and galaxy-galaxy lensing measurements in the overlapping areas between DESI and the weak lensing datasets. We make use of the full clustering information by fitting the full shape of DESI galaxy two-point correlation function multipoles, whereas our companion paper \cite{AnnaDESI}, performs a $3\times2$-pt analysis fitting DESI projected clustering measurements. Our study also presents the first DESI full-shape galaxy clustering analysis in configuration space. We limit this analysis to $\Lambda$CDM fits, noting that exploring dark energy extensions, at the level of precision of DESI-DR1, requires additional combination with external probes \citep{desi_fs} and is beyond the scope of this work.

Our study is similar in spirit to unified weak lensing analyses such as \cite{deskids}, however, both this and our companion papers are the first to consider the full set of three two-point correlation functions for a number of weak lensing surveys. We, therefore, present this analysis as an important consistency test for the `same-sky' observations available today. We also consider our study as a benchmark analysis exploring the constraining power that can be obtained from joint weak lensing and galaxy clustering analyses, while retaining distinct power spectrum prescriptions for the two types of probe. This sacrifices some potential gains in direct nuisance and astrophysical parameter calibration, but allows us to make use of the full range of scales accessible for each probe. Our companion paper \cite{Joeheft} will instead present a Fourier-space joint analysis that includes a uniform shear and RSD model based on \cite{chen2024} and \cite{derose2025}.

We begin by describing our galaxy clustering and weak lensing datasets in Sec.~\ref{sec:datasets}. We proceed with a description of our methodology in Sec.~\ref{sec:methodology}, including the models we adopt for galaxy clustering, galaxy-galaxy lensing and shear data vectors, as well as our covariance and priors. The model validation is presented in Sec.~\ref{sec:validation}, where we also briefly describe the lightcone mocks used to validate our pipeline. Finally, we present our results in Sec.~\ref{sec:results} and conclude in Sec.~\ref{sec:conclusions}.

\section{Datasets}
\label{sec:datasets}

\subsection{DESI galaxies}

We use the Dark Energy Spectroscopic Instrument Data Release 1 \citep[DESI-DR1,][]{2025arXiv250314745D} as the dataset for our galaxy clustering and galaxy-galaxy lensing measurements.  Main survey operations for DESI commenced in May 2021, and over its 8-year program it will collect more than 60 million spectra of galaxies and quasars across $17{,}000$\,deg$^2$ \citep{2016arXiv161100036D, 2016arXiv161100037D, 2024AJ....167...62D}.  The DESI observing strategy is fully described by \cite{2023AJ....166..259S}.  In brief, the survey obtains spectra for four principal target classes: the Bright Galaxy Survey (BGS), Luminous Red Galaxy Survey (LRG), Emission Line Galaxy Survey (ELG), and Quasar Survey (QSO), which are all photometrically-selected from the DESI Legacy optical imaging surveys \citep{2019AJ....157..168D}.  The instrument design is described by \cite{2022AJ....164..207D}: a robotic positioner \citep{2023AJ....165....9S} is used to assign 5000 fibres to targets in the focal plane \citep{2024AJ....168..245P} aided by an optical corrector \citep{2024AJ....168...95M}, and the resulting data are processed by the DESI spectroscopic pipeline \citep{2023AJ....165..144G}.

DESI-DR1 consists of all data acquired during the first 13 months of the DESI main survey up to June 2022, including high-confidence redshifts for 13.1M galaxies.  In our analysis we use DESI-DR1 Key Project catalogues \citep{2025JCAP...07..017A} from the Bright Galaxy Survey \citep{2023AJ....165..253H} and Luminous Red Galaxy survey \citep{2023AJ....165...58Z}.  Bright Galaxies span redshift range $0.1 < z < 0.4$, and LRGs occupy $0.4 < z < 1.1$, divided into three redshift bins $(0.4-0.6, 0.6-0.8, 0.8-1.1)$.  Following the Key Project analysis, we apply an absolute magnitude cut $M_r < -21.5$ to the k-corrected $r$-band to define the BGS sample, with a redshift-correction due to linear evolution, such that we can utilise the DESI-DR1 clustering measurements \citep{2025JCAP...07..017A}.  In this respect the BGS samples studied in our paper differ from those analysed for galaxy-galaxy lensing and $3 \times 2$-pt correlations by \cite{2025arXiv250621677H} and \cite{AnnaDESI}, who utilise fainter BGS samples in a series of narrower redshift bins \citep{2024MNRAS.533..589Y}.

\subsection{Shear surveys}

In addition to our galaxy clustering dataset, we also utilise weak gravitational lensing shear measurements from the Kilo-Degree Survey \citep[KiDS-1000,][]{2021A&A...645A.105G}, the Dark Energy Survey Year 3 sample \citep[DES-Y3,][]{2021MNRAS.504.4312G, 2022PhRvD.105b3514A, 2022PhRvD.105b3515S}, and the Hyper Suprime-Cam  survey Year 3 catalogue \citep[HSC-Y3,][]{2022PASJ...74..421L}, which all share significant overlap with DESI-DR1. We additionally make use of the HSC Year 1 measurements for our light cone mock construction \citep[HSC-Y1,][]{hscy1}, as discussed in Sec.~\ref{sec:validation}. The KiDS-1000 catalogue spans 1006 deg$^2$ with an effective source number density of $6.2$ arcmin$^{-2}$, split into five tomographic source bins; the DES-Y3 dataset covers 4143 deg$^2$ with a source density of 5.6 arcmin$^{-2}$, divided into four tomographic sub-samples; finally, the HSC-Y3 catalogue is the deepest of the samples, with a source density of 19.9 arcmin$^{-2}$ over 433 deg$^2$, again consisting of four tomographic sub-samples.  Measurements of cosmological parameters from these datasets have been presented by each of these lensing collaborations \cite[e.g.,][]{2021A&A...646A.140H, 2022PhRvD.105b3520A, 2023OJAp....6E..36D, 2023PhRvD.108l3518L, 2023PhRvD.108l3519D}.  Whilst our work was in progress, cosmological analysis of the extended KiDS-Legacy survey was presented by that collaboration \citep{2025arXiv250319441W}.

\subsection{Correlation measurements}

We analyze a combination of configuration-space correlation measurements: the cosmic shear correlation functions of the weak lensing surveys, $\xi_{\pm}(\theta)$, as a function of angular separation $\theta$; the average tangential shear of the weak lensing source galaxies around the DESI lens galaxies, $\gamma_t(\theta)$; and the DESI galaxy correlation function multipoles, $\xi^\ell(s)$, as a function of comoving separation $s$, where $\ell = \{ 0, 2, 4 \}$.  In this section we briefly describe how we obtain each measurement; our covariance model will be discussed in Sec.~\ref{sec:cov}.

For the cosmic shear correlation functions between the tomographic source samples of each weak lensing survey, we use the publicly-released measurements of the KiDS-1000, DES-Y3 and HSC-Y3 collaborations.  KiDS-1000 has presented $\xi_{\pm}(\theta)$ measurements in 9 logarithmically-spaced separation bins between limits $0.5 < \theta < 300'$ \citep{2021A&A...645A.104A}; DES-Y3 used 20 separation bins in the range $2.5 < \theta < 250'$ \citep{2022PhRvD.105b3520A}; and HSC-Y3 used 7 bins in the range $7.1 < \theta < 56.6'$  for $\xi_+$, and 7 bins in the range $31.2 < \theta < 248'$ for $\xi_-$ \citep{2023PhRvD.108l3518L}.  These are the same cosmic shear datasets as used in the $3 \times 2$-pt cosmological analysis presented by \cite{AnnaDESI}.

For the galaxy correlation function multipoles, we employ the DESI collaboration measurements derived from the DR1 Key Project catalogues \citep{2025JCAP...07..017A}, including one set of multipoles for the BGS sample ($0.1 < z < 0.4$) together with additional measurements for three independent LRG redshift ranges $(0.4-0.6, 0.6-0.8, 0.8-1.1)$.  The correlation function measurements span 45 linearly-spaced separation bins between limits $20 < s < 200 \, h^{-1}$ Mpc, with bin width $\Delta s = 4 \, h^{-1}$ Mpc.  Our correlation function measurements exclude small-scale PIP (pairwise inverse probability) weights \citep{2025JCAP...04..074B}, because their effect is negligible at the separations we are considering, although we include the ``theta-cut'' corrections \citep{2025JCAP...07..017A}.  The correlation function measurements were performed by \cite{2025JCAP...07..017A} using {\sc pycorr}\footnote{\url{https://github.com/cosmodesi/pycorr}}, which wraps a version of the {\sc corrfunc} package\footnote{\url{https://corrfunc.readthedocs.io/en/master}} \citep{2020MNRAS.491.3022S}.

We measure the average tangential shear of the tomographic source catalogues around each DESI galaxy sample using {\sc treecorr}{\footnote{\url{https://rmjarvis.github.io/TreeCorr}}} \citep{2004MNRAS.352..338J}, employing 15 logarithmic angular separation bins in the range $0.3 < \theta < 300$ arcmin.  Our $\gamma_t$ measurements for the LRG samples match the DR1 galaxy-galaxy lensing data presented by \cite{2025arXiv250621677H}; our BGS measurements differ because we are using the Key Project catalogue including the absolute magnitude cut $M_r < -21.5$.  Before performing the galaxy-galaxy lensing measurement, we cut the source and lens samples to overlapping sky areas \citep{2025arXiv250621677H} corresponding to  447, 851 and 441 deg$^2$ for KiDS-1000, DES-Y3 and HSC-Y3, respectively.  When measuring the average tangential shear, we subtract a signal around randomly-distributed lenses with the same sky distribution as the data, and we do not perform a boost correction to account for source-lens clustering on small scales \citep{sheldon2004}, as this is negligible on the separations considered in this analysis \citep{lange2024}.

\section{Methodology}
\label{sec:methodology}

In this work, we are interested in modeling shear and galaxy clustering using the approaches that have traditionally been used to describe the respective scales, i.e., we are using a halo model-based approach for weak lensing tracers and a perturbation theory-based approach for galaxy clustering. 

Perturbative approaches model the density field by considering small perturbations around its mean value, and performing higher-order expansion to describe contributions from nonlinear evolution. The tracer population is then described by stating that galaxies trace the dark matter field in a biased manner and assigning bias parameters to the perturbative expansion terms to describe the amplitude of each contribution. While this approach has the advantage of being less dependent on a specific small-scale dynamics description, perturbative models tend to break down at small scales where the amplitude of density perturbations, relative to the mean value of the field, exceeds unity. The standard one-loop galaxy bias expansion in redshift space has been validated to be accurate up to $k_{\mathrm{max}} \sim 0.2-0.3 \, h \, \mathrm{Mpc^{-1}}$ for most tracers \cite[e.g.,][]{maus2025, pezzotta2021}.

The halo model assumes that all the matter in the Universe is contained in dark matter halos, and considers the contributions of halos from different masses as well as the clustering between different halos \cite[for an overview, see][]{asgari2023}. Combining this with a Halo Occupation Distribution (HOD) prescription yields a description for biased tracers. These models allow for a more accurate description of the non-linear regime, but are heavily dependent on the assumed halo properties and accurate model of different galaxy formation effects \cite[as shown in, for example][]{chavesmontero2023}. Furthermore, models formulated within the halo-model formalism generally fail to account for the damping of the BAO peak and show inaccuracies in the region of transition between clustering within a single halo (1-halo term) and inter-halo clustering (2-halo term) \cite[see, for example,][]{mead2015}, although \cite{2025arXiv250810902B} present a recent counter-example.

In order to circumvent these limitations and extend the validity of these approaches, a number of hybrid models have been developed, complementing perturbative approaches with information from numerical simulations, such as the response function based model \citep{nishimichi2017} or Hybrid Effective Field Theory (HEFT) models \citep{modi2020}. The latter class of approaches has been particularly successful for joint modelling of galaxy clustering and galaxy galaxy lensing observables \citep[][]{hadzhiyska2021, chen2024}. Furthermore, a recent attempt to use EFT to model projected weak lensing quantities was presented by \cite{damico2025} who showed that this approach results in $3\times2$pt constraints that are consistent within $1.3\sigma-1.5\sigma$ with the fiducial analysis for DES Y3. 

In this work, however, we are interested in the constraints achieved without a fully consistent model for galaxy clustering and weak lensing, but by instead employing the modelling that allows us to access the range of scales that are traditionally employed in flagship cosmological analyses of the respective probes. For a fully-consistent HEFT based analysis of DESI galaxy clustering and weak lensing observables, see \cite{Joeheft}, whereas for a simulation-based analysis of DESI galaxy clustering and galaxy-galaxy lensing between DESI and weak lensing surveys see \cite{Langedesi}. Our companion paper, \cite{AnnaDESI}, presents a $3\times$2pt analysis that fits to projected galaxy clustering and is, therefore, using a halo model - based prescription to model the power spectra for all three correlation functions. 

\subsection{Galaxy clustering model}

We model our galaxy clustering two-point correlation function by directly emulating the full shape of the multipoles in redshift space. Our emulator is based on the Fourier-space power spectrum emulator {\sc comet}\footnote{\url{https://comet-emu.readthedocs.io/en/latest/}} \citep{eggemeier2023}, which is trained to predict the component spectra under the perturbative framework of the Effective Field Theory \cite[EFT][]{baumann2012}. In particular, we focus on a variation of EFT, which, instead of performing a full perturbative expansion, treats the redshift space distortion effect arising from the small scale virialised galaxy velocities by considering the velocity difference generating function \cite[$\mathrm{VDG}_{\infty}$,][]{eggemeier2025}. We Fourier transform the modified {\sc comet}'s training set and build a new emulator using the transformed spectra in configuration space.  We describe the theory used to model the clustering multipoles in Sec.~\ref{sec:vdg} and provide the details of the configuration-space emulator setup in Sec.~\ref{sec:cometxi}.

\subsubsection{Galaxy bias}

In order to obtain a tracer power spectrum, $P_{\rm{gg}}$, we need a relation to connect the matter density field fluctuations $\delta$ to the resulting galaxy density field $\delta_{\rm{g}}$. Under the perturbative framework, this can be written down as a sum of operators that are functions of the gravitational potential $\Phi$ and velocity potential $\Phi_{\rm{v}}$. Following the notation in \cite{eggemeier2019} and expanding to include contributions up to one-loop, we obtain:
\begin{equation}
        \delta_{\rm{g}} = b_1\delta+\frac{b_2}{2}\delta^2+\gamma_2\mathcal{G}_2(\Phi_{\rm{v}})+\gamma_{21}\mathcal{G}_2(\varphi_1, \varphi_2).
\label{eq:bias}
\end{equation}
Here $b_1$ and $b_2$ are, respectively, linear and quadratic bias terms, which describe galaxy bias as a function of the local matter density contrast, whereas $\gamma_{2}$ and $\gamma_{21}$ are tidal bias parameters, which account for non-local effects that produce anisotropies in the gravitational collapse. These tidal effects are described in terms of the Galilean invariant operator $\mathcal{G}_2$ and are given by:
\begin{align}
    \mathcal{G}_2(\Phi_{\rm{v}})&=(\nabla_{ij}\Phi_{\rm{v}})^2 - (\nabla^2\Phi_{\rm{v}})^2,\\
    \mathcal{G}_2(\varphi_1, \varphi_2)&=\nabla_{ij}\varphi_2\nabla_{ij}\varphi_1 - \nabla^2\varphi_{2}\nabla^2\varphi_{1},
\label{eq:tidalbias}
\end{align}
such that the first term represents the tidal stress tensor generated by the velocity potential and the second term includes a higher-order correction obtained by expressing the non-linear velocity potential up to the second order  ($\Phi_{\rm{v}}=\Phi_{\rm{v}}^{1}+\Phi_{\rm{v}}^{2}$ and $\varphi_{1}=-\Phi_{\rm{v}}^{1}$, $\varphi_{2}=-\Phi_{\rm{v}}^{2}$). The full one-loop bias expression in Eq.\ref{eq:bias} additionally includes a higher-derivative bias contribution, which scales as $\nabla^{2}\delta$, but this is identical to the scaling of the first of the counterterms and is, therefore, absorbed by it.  

The tidal bias parameters $\gamma_{2}$ and $\gamma_{21}$ both enter the bias expansion at a higher order and are degenerate with each other. Following \cite{semenaite2022}, we fix both of these parameters in terms of linear bias $b_1$, such that:
\begin{equation}
\gamma_{2} = 0.524-0.547b_1+0.046b_1^2,
\label{eq:gamma2}
\end{equation}
\begin{equation}
\gamma_{21} = \frac{2}{21}(b_1-1)+\frac{6}{7}\gamma_2.
\label{eq:gamma21}
\end{equation}
The first of these relations, Eq.\ref{eq:gamma2}, describes an excursion set result by \cite{sheth1996} and has been shown by \cite{eggemeier2020} to be more accurate than the relation obtained by assuming local bias in Lagrangian space for tracers with $b_1\gtrsim1.3$. The second relation, Eq.\ref{eq:gamma21}, is derived assuming conserved evolution of galaxies after their formation \citep{fry1996, catelan1998, chan2012}. The coevolution relation for $\gamma_{21}$ was thoroughly tested on mocks and data for BOSS galaxies, demonstrating excellent agreement \citep{eggemeier2021, semenaite2022}. We additionally perform tests on the {\sc Abacus} lightcone mocks introduced in Sec.~\ref{sec:abacusmocks} below, allowing both $\gamma_{2}$ and $\gamma_{21}$ to vary freely, and confirm that, also for these mocks, fixing the tidal bias parameters to the coevolution relations from Eq.\ref{eq:gamma2} and \ref{eq:gamma21} improves the accuracy with which the input cosmology is recovered. Finally, \cite{maus2025} confirmed that similar coevolution (or ``minimal bias'') schemes, where non-local bias parameters are fixed to a relationship derived by assuming zero tidal bias at the time of galaxy formation, provide consistent and accurate cosmology constraints for DESI-like galaxies. 

The one-loop bias expansion has been shown to recover unbiased constraints when implemented in different perturbation theory codes \citep{maus2025, pezzotta2021} and matches the approach that was used in DESI's key full-shape analysis \citep{desi_fs}. While our chosen bias parametrisation differs from the Eulerian perturbation implementation in {\sc velocileptors} \citep{chen2020} used by \cite{desi_fs}, it can be directly mapped onto alternative bias bases, such as the ones used by \cite{assassi2014} and \cite{ivanov2020} or \cite{damico2020}, via simple linear relations \cite[see, for example, Eq.18 and Eq.19 in ][]{eggemeier2023}. The agreement between the different bias prescriptions was further investigated and confirmed by \cite{maus2025}. Finally, \cite{ramirez2025} confirmed, on {\sc Abacus} simulation box mocks, that one-loop Lagrangian bias expansion with tidal bias parameter fixed to a coevolution relation recovers input cosmology for two-point correlation function multipoles for DESI LRG, ELG and QSO-like samples. Their full-shape two-point correlation function fits are modeled using the Gaussian streaming model. 

\subsubsection{Redshift space distortions}
\label{sec:rsd}

In order to model the \textit{observed} power spectrum, we need to include the effect of the redshift space distortions -- the contribution to the observed redshift that arises due to galaxies' peculiar velocities along the line of sight. The relationship between density fields in real and redshift space is derived by assuming mass conservation. The anisotropic galaxy power spectrum in redshift space, $P^{s}_{\rm{gg}}$, under the distant-observer approximation can then be written down as \citep{scoccimarro1999}:
\begin{equation}
    P^{s}_{\rm{gg}}(k,\mu) = \int \frac{d\pmb{r}}{(2\pi)^3}e^{i\pmb{k}\cdot\pmb{r}}\Big\langle e^{\lambda\Delta u_{\parallel}}\times D_s(\pmb{x})D_s(\pmb{x'})\Big\rangle
    \label{eq:prealtoz}
\end{equation}
and 
\begin{equation}
    D_s(\pmb{x})\equiv \delta_{\rm{g}}(\pmb{x})+f\nabla_{\parallel}u_{\parallel}(\pmb{x}).
\end{equation}
Here $\pmb{r}\equiv \pmb{x}-\pmb{x'}$, $\pmb{u}\equiv-\pmb{v}/[faH(a)]$ is the normalised velocity field with the line of sight velocity component difference $\Delta u_{\parallel}\equiv u_{\parallel}(\pmb{x})-u_{\parallel}(\pmb{x'})$, $\lambda=-ifk\mu$, and we note that the subscript ``$s$'' denoting redshift-space quantities is dropped for conciseness in the following equations.  We evaluate the density and velocity fields at the scale factor $a$, with $H(a)$ denoting the Hubble parameter, and $f$ the logarithmic derivative of the linear growth factor $D(a)$, such that $f(a)\equiv d\log D(a)/d\log a$.

\subsubsection{EFT framework}
\label{sec:vdg}

The general description of the galaxy power spectrum under EFT, $P_{\rm{gg}, \mathrm{EFT
}}$, can be written down as a sum of four terms:
\begin{equation}
\begin{split}
    P_{\rm{gg}, EFT}(\pmb{k}) &= P^{\,\rm{tree}}_{\rm{gg, SPT}}(\pmb{k}) + P^{\,\rm{1-loop}}_{\rm{gg, SPT}}(\pmb{k}) + P^{\,\rm{stoch}}_{\rm{gg}}(\pmb{k}) \\ &+ P^{\,\rm{ctr}}_{\rm{gg}}(\pmb{k}).
\end{split}
\label{eq:peft}
\end{equation}
Here the first two terms correspond to the standard perturbation theory (SPT) contributions at the leading (tree-level) and next-to-leading (1-loop) order, obtained from expansion of the real- to redshift-space mapping (Eq.\ref{eq:prealtoz}), while taking into account the non-linear evolution of matter and velocity fields and applying a bias prescription to map from dark matter to galaxy field. When including galaxy bias in Fourier space, we need to additionally take into account the stochastic contribution from highly non-linear scales, $P^{\,\rm{stoch}}_{\rm{gg}}(\pmb{k})$. This term corresponds to a constant noise which, in configuration space, would contribute to separations close to zero and is therefore not included in our two-point correlation function model. Finally, any further effects coming from the highly non-linear modes that cannot be modelled using perturbation theory are accounted for by a series of free amplitude parameters referred to as \textit{counterterms}, whose contribution is represented by the fourth term in the expansion, $ P^{\,\rm{ctr}}_{\rm{gg}}$.

To the leading order (LO), the counterterm contribution can be written down as:
\begin{equation}
P_{\rm{gg}}^{\,\rm{ctr, LO}}(k, \mu) = -2 \sum^{2}_{n=0}c_{2n}L_{2n}(\mu)k^2P_{\rm{lin}}(k).
\label{eq:pctrlo}
\end{equation}
This form reflects the fact that the leading-order counterterms for the galaxy power spectrum in redshift space scale with the linear power spectrum, $P_{\rm{lin}}$, as  $\sim\mu^{2n}k^2P_{\rm{lin}(k)} $ with $n=0,1,2$ \cite[as shown in e.g.][]{senatore2014,desjacques18}. In the expression above, instead of scaling with the cosine of the angle between the separation vector and the line of sight, $\mu$, we can use a linear transformation of the counterterm parameters to define three free parameters, $c_{0}, c_{2}$ and $c_{4}$, and the corresponding Legendre polynomials of order $2n$, $L_{2n}$. Each of these free amplitude parameters then mainly contributes to a single multipole only.

The linear-order counterterm contribution, as defined in Eq.\ref{eq:pctrlo}, accounts for such nonlinear effects as a breakdown of the perfect fluid approximation for the matter field and velocity bias \citep{desjacques18, carrasco2012, baumann2012, pueblas2009} but does not properly capture the non-linear redshift space distortions (RSD), arising due to galaxy virial motions, also known as the fingers-of-God (FoG) effect. While it is possible to treat this effect purely perturbatively by considering a power series expansion of the pairwise velocity difference generating function, the associated lengthscale (the inverse linear velocity dispersion, $\sigma_{v}^{-1}$) can be significantly smaller than the non-linearity scale of the matter field. This implies that terms beyond the one-loop expansion cannot be neglected. In order to account for this contribution, \citet{ivanov2020} suggested including the next-to-leading order term, $P^{\,\rm{ctr, NLO}}_{\rm{gg}}$, with an associated free amplitude parameter, $c_{\rm{nlo}}$, such that:
\begin{equation}
    P^{\,\rm{ctr, NLO}}_{\rm{gg}}(k, \mu) = c_{\rm{nlo}}(\mu k f)^{4}P^{\,\rm{tree}}_{\rm{gg},SPT}(\pmb{k}).
\label{eq:ctrnlo}
\end{equation}
EFT models that introduce counterterms to account for the redshift space distortions sourced by the small-scale virialised galaxy velocities have been successfully used in a number of galaxy clustering analyses, including, for example, \citet{philcox2022, ivanov2020, damico2020}. 

Nonetheless, the next-to-leading order counterterm contribution poses a challenge for obtaining its configuration space equivalent. As evident from Eq.\ref{eq:ctrnlo}, this term is proportional to $k^4$, quickly diverging with $k$. As a result, $P^{\,\rm{ctr, NLO}}_{\rm{gg}}$ contributes increasingly significantly to the smallest scales, raising a challenge when attempting to perform a Fourier transform. The standard fully perturbative approach is, therefore, less suitable for our set-up and we opt to emulate the $\mathrm{VDG}_{\infty}$-based two-point correlation function instead.  \cite{2022JCAP...02..036Z} present an alternative treatment of these issues using Pade approximants to damp the high-$k$ power.

\subsubsection{Overview of $\mathrm{VDG}_{\infty}$}

Given that the FoG effect is sourced by highly non-linear scales, it is more appropriate, instead of performing a power series expansion of the velocity difference generating function, to model its effect on the galaxy power spectrum by defining an effective damping function, as done in the $\mathrm{VDG}_{\infty}$ approach. The functional form of this damping function is derived by considering the large scale limit of the pairwise velocity distribution generating function and is given by \citep{sanchez2017}:
\begin{equation}
W_{\infty}(\lambda)=\frac{1}{\sqrt{1-\lambda^2a^2_{\rm{vir}}}}\,\exp\left(\frac{\lambda^2\sigma^2_v}{1-\lambda^2\mathnormal{a}^2_{\rm{vir}}}\right),
\label{eq:winfty}
\end{equation}
where $\lambda$ is defined in terms of the wavevector $k$, the angle $\mu$ and the growth rate $f$, such that $\lambda=-ifk\mu$, $\sigma_{v}$ is the linear velocity dispersion and $a_{\rm{vir}}$ is a free parameter characterising the kurtosis of the small-scale velocity distribution, such that $a_{\rm{vir}}=0$ corresponds to a Gaussian distribution. This is a crucial part of the functional form, as it has been shown that the pairwise velocity probability distribution function has significant exponential tails even in the large scale limit \citep{sheth1996, juszkiewicz1998, scoccimarro2004,cuestalazaro2020}.

This form of FoG modelling has been used in previous full-shape galaxy clustering analyses, such as \cite{sanchez2017, hou2021, semenaite2022, semenaite2023}. Furthermore, \cite{eggemeier2025} showed that the $\mathrm{VDG}_{\infty}$ treatment of RSD extends EFT validity range for the one loop power spectrum from approximately $0.2 \, h \, \rm{Mpc}^{-1}$ to $0.3 \, h \, \rm{Mpc}^{-1}$. Crucially, this approach allows us circumvent the issue of trying to perform a Fourier transform of the diverging $P^{\,\rm{ctr, NLO}}_{\rm{gg}}$ term.

The final power spectrum model in $\mathrm{VDG}_{\infty}$ can then be obtained from its EFT expression following:
\begin{equation}
    P_{\rm{gg, VDG_{\infty}}}(\pmb{k}) = W_{\infty}(\pmb{k})\big[P_{\rm{gg}, EFT}(\pmb{k})-\Delta P(\pmb{k})-P_{\rm{stoch}}\big]+P_{\rm{stoch}},
\label{eq:pggvdg}
\end{equation}
where $\Delta P(\pmb{k})$ is a correction term that accounts for the fact that the mapping from real to redshift space is not expanded perturbatively \cite[for a full expression for $\Delta P(\pmb{k})$ see Eq.12 in][]{eggemeier2023} and the stochastic noise term, $P_{\rm{stoch}}$, is unaffected by the damping. 

\subsubsection{Alcock-Paczynski distortions}
\label{sec:ap}

When measuring galaxy clustering we need to assume a fiducial cosmology to perform the conversion between observed redshifts and the corresponding distances. A mis-match between the fiducial and true cosmology will introduce anisotropic distortions known as Alcock-Paczynski (AP) distortions \citep{alcock1979} and modelled by introducing the geometric distortion factors:
\begin{align}
q_{\bot} &=D_{\rm{M}}(z_{\rm eff})/D_{\rm{M}}'(z_{\rm eff}),\\
q_{\parallel} &=H'(z_{\rm eff})/H(z_{\rm eff}). 
\end{align}
Here primed quantities mark fiducial cosmology values, $D_{\rm{M}}(z)$ is the comoving angular diameter distance, $H(z)$ is the Hubble parameter and $z_{\rm eff}$ is the effective redshift of the galaxy sample. 

These distortion factors can be directly applied to scale the separations of galaxy pairs, $s$ and the angles $\mu$, as given by:
\begin{align}
s &=s'\left( q_{\parallel}^2\mu'^2+q^2_{\bot}(1-\mu'^2)\right),\\
\mu &=\mu'\frac{q_\parallel}{\sqrt{q_{\parallel}^2\mu'^2+q^2_{\bot}(1-\mu'^2)}}.
\label{eq:smuap}
\end{align}
Our fiducial cosmology matches the cosmology of the fiducial output of the {\sc Abacus} simulations ($\omega_{\rm{c}} = 0.120$, $\omega_{\rm{b}} = 0.02237$, $h = 0.6736$), consistent with the key DESI full-shape analysis.

\subsection{Configuration-space emulator}
\label{sec:cometxi}

We build our configuration-space emulator based on the Fourier-space Gaussian Process emulator {\sc comet}. In particular, we build on the $\mathrm{VDG}_{\infty}$ implementation and we do not include massive neutrinos. We do, however, include the updated discrete sine transform IR-resummation scheme that is implemented in the current version of {\sc comet} \cite[for a detailed description, see Sec.~C4 in][]{pezzotta2025}. This presents a more accurate way to account for the large-scale bulk flows that smear the BAO wiggles.

While we could perform a Fourier transform on the output of the original {\sc comet}, training the emulator on its Fourier-transformed training set and emulating directly in configuration space allows us to achieve a speed-up in model evaluation as well as better control over the accuracy of the Fourier transform.  A detailed description of the emulator, including validation, is provided in \cite{agneemu}, here we provide a summary of the emulator's main features and structure. 

\subsubsection{Evolution mapping}
\label{sec:evomapping}

{\sc comet} is able to provide accurate theory predictions for arbitrary fiducial cosmologies, a wide range of redshifts and a number of cosmologies beyond the standard $\Lambda$CDM\footnote{The evolution mapping framework, as presented here, does not account for extensions that include scale-dependent growth, like modified gravity cosmologies or cosmologies with massive neutrinos. However, additional corrections can be made to map scale-dependent effects, for a detailed description on including neutrinos in evolution mapping framework see \cite{pezzotta2025}.}. This is achieved by utilising the \textit{evolution mapping} approach \citep{sanchez2022}, which allows us to obtain predictions for a range of cosmologies and redshifts, without having to include them in our training set. 

Evolution mapping divides cosmological parameters into ``shape'' and ``evolution'' parameters based on their effect on the power spectrum. A crucial part of this approach is that both shape and evolution parameters need to be defined in $h$-independent units, i.e., we want to use the \textit{physical} densities for all energy species $i$, $\omega_{i}=\Omega_{i}h^2$ (as $h$ enters $\Omega_{i}$ through the critical density) and we use $\sigma_{12}$ (linear density field variance defined on the scale of $12 \, \rm{Mpc}$) to describe the amplitude of the power spectrum today \citep{sanchez2020}. This is the equivalent to $\sigma_{8}$ with the relevant scale in Mpc chosen such that $\sigma_{12}$ recovers the same value as $\sigma_{8}$ in a \textit{Planck} $\Lambda$CDM cosmology. 

Evolution mapping framework is then based on the observation that the shape parameters ($\omega_{\rm{m}}$, $\omega_{\rm{b}}$, $n_{\rm{s}}$...) define the shape of the power spectrum and, for a fixed set of shape parameter values, evolution parameters ($A_{\rm{s}}$, $w$, $w_{a}$, ...) only scale the amplitude of the power spectrum. As $h$ represents a sum of shape parameters (matter density $\omega_{\rm{m}}$ and baryon density $\omega_{\rm{b}}$) and evolution parameters (dark energy density $\omega_{\rm{DE}}$), it cannot be assigned to either group, which is why we need to ensure that our parameter space is not defined via $h$\footnote{The model prediction can then be scaled to $\mathrm{Mpc}~h^{-1}$ to fit the data vector.}. 

The evolution mapping framework implies that, in order to map between different evolution parameter combinations, we can just scale the power spectrum obtained for the same shape parameters to the correct amplitude, as described by $\sigma_{12}$. This is exact for the linear power spectrum, and is true to within 8\% at $k = 1.5 \, \rm{Mpc}^{-1}$ for non-linear power spectrum for the most extreme mappings, with the deviation occuring due to different recent structure formation histories. While the deviation can be corrected for to produce better than 1\% agreement, as described in \cite{sanchez2022}, for perturbative theories which are based on integrals over the linear power spectrum, the mapping is exact. This approach can be furthermore extended to also model the non-linear velocity field \citep{esposito2024} and the halo mass function \citep{fiorilli2025}.

By implementing this mapping, {\sc comet} is able to provide exact (within emulation error) predictions for a number of $\Lambda$CDM extensions that it is not explicitly trained on, including time-varying dark energy equation of state cosmologies and cosmologies with non-zero curvature (both dark energy density and curvature energy density are evolution parameters). It can also easily and exactly map between different redshifts, as redshift is an evolution parameter as well. 

The neutrino mass sum cannot be classified as either a shape or evolution parameter due to its scale-dependent effect. As a result, it was not included in the initial versions of {\sc comet}. The most recent update of {\sc comet} includes a prescription that extends evolution mapping for models with non-zero neutrino masses \citep{pezzotta2025}, however, that requires an additional emulator. For simplicity, we do not include this option in our configuration-space emulator (i.e. our theory predictions assume massless neutrinos), as galaxy clustering does not provide strong constraints on the neutrino mass sum when considered without including CMB information.

\subsubsection{Emulator structure}

Following \cite{eggemeier2023}, we emulate the two-point correlation function components $\xi_{\mathcal{B}}$ arising from the galaxy density field model defined in Eq.\ref{eq:bias}.  These components correspond to Fourier-transformed spectra of the original {\sc comet} with one main difference -- each of the component spectra is damped, as prescribed by the $\mathrm{VDG}_{\infty}$ framework (Eq.\ref{eq:winfty} and \ref{eq:pggvdg}), before the Fourier transform is performed. This is due to the fact that the Fourier-space multiplication with $W_{\infty}$ would correspond to a convolution operation in configuration space. In order to maintain the speed of the emulator we choose to perform the damping in Fourier space. As a result, the configuration-space emulator has to be explicitly trained on an additional parameter, $a_{\rm{vir}}$.

The evolution mapping statement for the two-point correlation function components, $\xi_{\mathcal{B}}$, as a function of the redshift space separation $\pmb{s}$, redshift $z$ and a set of shape and evolution parameters, $\Theta_{\rm{s}}$ and $\Theta_{\rm{e}}$, can be written as:
\begin{equation}
\begin{split}
&\xi_{\mathcal{B}}(\pmb{s}|z, a_{\rm{vir}}, \Theta_{\rm{s}}, \Theta_{\rm{e}}) = \\& \xi_{\mathcal{B}}\{\pmb{s}| a_{\rm{vir}}, f(z, \Theta_{\rm{s}}, \Theta_{\rm{e}}), \xi_{\rm{lin}}(r|\Theta_{\rm{s}}, \sigma_{12}(z, \Theta_{\rm{s}}, \Theta_{\rm{e}}))\} .
\end{split}
\label{eq:xievo}
\end{equation} 
Here, $f$ is the linear growth rate required for the real to redshift space mapping and $\xi_{\rm{lin}}$ is the real-space linear two-point correlation function, which is calculated for the value of $\sigma_{12}$ that corresponds to the same shape parameters and the desired combination of evolution parameters (including $z$).

In order to correctly capture the anisotropic information in $\xi_{\mathcal{B}}(\pmb{s})$, we project it into Legendre multipoles and separately emulate the components for monopole, quadrupole and hexadecapole. The full $\xi_{\mathcal{B}}(\pmb{s})$ can then be reconstructed using the Legendre expansion. Such reconstruction is, however, not exact and, following \cite{eggemeier2023}, we correct for the resulting error by additionally including the $\ell=6$ multipole $\xi_{\mathcal{B},6}$ at a fixed redshift $z=1$ and fixed cosmology matching \textit{Planck} $\Lambda$CDM. Analogously to the other multipoles, the $\ell=6$ term depends on $f$ and $\sigma_{12}$, whose variation with shape and evolution parameters is correctly accounted for and can be factored out. As in the Fourier-space version of {\sc comet}, to reduce the error in the fixed cosmology approximation of $\xi_{\mathcal{B},6}$, these terms are split into contributions with different powers of $f$ and their amplitudes are scaled to the required value of $\sigma_{12}$. Finally, in order to account for the variation with $a_{\rm{vir}}$, we create a grid of theory predictions for the FoG-damped $\xi_{\mathcal{B},6}|_{\rm{Planck}}$ that covers our $a_{\rm{vir}}$ training space, and interpolate to obtain $\xi_{\mathcal{B},6}(a_{\rm{vir}})|_{\rm{Planck}}$ for an arbitrary value of $a_{\rm{vir}}$. 

The final expression for the anisotropic two-point correlation function components $\xi_{\mathcal{B}}(\pmb{s})$ is then:
\begin{equation}
\begin{split}
    &\xi_{\mathcal{B}}(\pmb{s}|z, a_{\rm{vir}},\Theta_{\rm{s}}, \Theta_{\rm{e}}) \approx  \sum^{2}_{\ell=0}\xi_{\mathcal{B},2\ell}(s|z, a_{\rm{vir}}, \Theta_{\rm{s}}, \Theta_{\rm{e}})L_{2\ell}(\mu) + \\& \Big(\frac{\sigma_{12}(z, \Theta_{\rm{s}}, \Theta_{\rm{e}})}{\sigma_{12}(z=1, \Theta_{\rm{s}}^{\rm{Planck}}, \Theta_{\rm{e}}^{\rm{Planck}})}\Big)^{2L}\xi_{\mathcal{B},6}(a_{\rm{vir}})|_{\rm{Planck}}L_{6}(\mu),
\end{split}
\label{eq:fullxis}
\end{equation}
where $L=1$ for the linear terms and $L=2$ for all the one-loop terms. The sum in Eq.\ref{eq:fullxis} goes over all the linear and one-loop contributions that make up the bias expansion (Eq.\ref{eq:bias}), resulting in 14 different terms.\footnote{We emulate the linear and one-loop terms for $\mathcal{B}=b_1$ and $\mathcal{B}=1$ together.  For a full list of terms see Table 1 in \cite{eggemeier2023}, excluding the terms involving $c_{\rm{NLO}}$.} 

Finally, the AP distortions (Sec.~\ref{sec:ap}) are applied analytically, as $q_{\perp}$ and $q_{\parallel}$ mix both shape and evolution parameters. We emulate $\xi_{\mathcal{B}}(\pmb{s})$ without including the AP distortions and then rescale the separations and angles as given by Eq.\ref{eq:smuap}.\\

\subsubsection{Configuration-space emulator summary}

Following the evolution mapping description in Eq.\ref{eq:xievo}, we emulate the components of the two-point correlation function multipoles at different combinations of shape parameters, $f$, $a_{\rm{vir}}$ and $\sigma_{12}$. Emulating the components allows us to obtain the full biased two-point correlation function theory model for arbitrary bias values without having to include them in training. 

We additionally emulate the value for $\sigma_{12}$ for fixed evolution parameters (including redshift) and different combinations of shape parameters. In total, we have two emulators:
\begin{enumerate}
    \item $\xi_{\mathcal{B}, \ell}(s|\Theta_{\rm{s}}, \sigma_{12}, f, a_{\rm{vir}})$ for $\ell=0,2,4$ and for all $\mathcal{B}$ corresponding to the one-loop bias expansion (Eq.\ref{eq:bias}).
    \item $\sigma_{12} (z=1, \Theta_{\rm{s}} \Theta_{\rm{e}}^{\rm{fixed}})$. 
\end{enumerate}
The particular values for the fixed cosmology, $\Theta_{\rm{e}}^{\rm{fixed}}$, are not significant here, as we only require a reference amplitude that is then scaled to obtain the correct $\sigma_{12}$ for the desired evolution parameters. Our $\sigma_{12} (z=1, \Theta_{\rm{s}}, \Theta_{\rm{e}}^{\rm{fixed}})$ emulator is identical to the one used in the Fourier-space version of {\sc comet} \footnote{We fix our evolution cosmology values such that $w=-1$, $z=1$, $h=0.695$ and $A_{\rm{s}}=2.2078559 \times 10^{-9}$, with all other evolution parameters set to 0.}.

\subsubsection{Parameter space and training}

\begin{table}[]
    \centering
    \begin{tabular}{c|c|c }
        \hline
        Parameter & Min. emulator range & Max. emulator range\\
        \hline
        $\omega_{\rm{b}}$ & 0.0205 & 0.02415\\
        $\omega_{\rm{c}}$ & 0.085 & 0.155\\
        $n_{\rm{s}}$ & 0.92 & 1.01\\
        \hline
        $\sigma_{12}$ & 0.2 & 1.0\\
        $f$ & 0.5 & 1.05\\
        $a_{\rm{vir}} [\rm{Mpc}]$ & 0.0 & 15.0\\
        \hline
    \end{tabular}
    \caption{The ranges of validity for the configuration-space {\sc comet} extension.}
    \label{tab:emu_paramranges}
\end{table}

In total, our training parameter space consists of six parameters: the growth rate $f$, the amplitude of the power spectrum $\sigma_{12}$, the kurtosis parameter of the small-scale velocity distribution $a_{\rm{vir}}$, and three shape parameters: matter and baryon densities, $\omega_{\rm{m}}$ and $\omega_{\rm{b}}$, and the spectral index, $n_{\rm{s}}$. We provide the minimum and maximum values for all the parameters in Table \ref{tab:emu_paramranges}. 

The training parameter space for the shape parameters, $f$ and $\sigma_{12}$ is identical to the one used for the Fourier-space {\sc comet} \cite[version 1,][]{eggemeier2023}. The range of supported redshifts is limited by the training interval for $\sigma_{12}$ for the maximum redshift and shape parameter values for the minimum redshift. These correspond to a redshift validity range of, approximately, $0.1\lesssim z \lesssim 3$.  The minimum and maximum values $a_{\rm{vir}}$ are chosen such that we can replicate a Gaussian velocity distribution ($a_{\rm{vir}} = 0$ Mpc) and cover the standard prior used in clustering analyses \citep[see, for example, ][]{hou2021, semenaite2022}.

In order to create our configuration-space training set, we need to perform a Fourier transform of the component spectra.  We do so by making use of the {\sc hankl}\footnote{\url{https://hankl.readthedocs.io/en/latest/}} package \citep{karamanis2021}. In order to suppress the ringing effects introduced by the Fourier transform, we extend the component power spectrum theory predictions up to $k_{\rm{max}} = 15 \, {\rm{Mpc}}^{-1}$ (and extrapolate beyond this point) and apply an additional suppression factor $\exp(-k/k_{\rm{cut}})^{2}$ with $k_{\rm{cut}} = 2.5 \, {\rm{Mpc}}^{-1}$. The resulting two-point correlation function components are evaluated at 200 separation $s$ bins that cover the range $10 < s < 250 \, {\rm{Mpc}}$.

The Gaussian Process (GP) emulation is implemented using the publicly available package {\sc GPy}\footnote{\url{https://gpy.readthedocs.io/en}} and we use a sum of the squared exponential and Matérn kernels as our kernel function, which represents the covariance between different points of the training set. The process of training the emulator  involves finding the hyperparameters of the kernel that maximise the log-likelihood of the GP models with respect to the training data. We pre-process the training spectra by subtracting their mean value and dividing by the variance across the training set. This is a standard practice that allows us to reduce the dynamical range of the training set for more accurate emulation. 

We validate the resulting emulator by comparing it with 1500 theory predictions that were not used in training, and find that we are able to accurately (with the maximum difference of $1\sigma$) recover the theory multipoles for volumes corresponding to more than a factor of ten times the volume of the DESI LRG sample and for scales $20 < s < 200 \, {\rm{Mpc}}$. We find the greatest relative deviations at small scales due to the fact that the constraining power at these separations is highest, and so even a small emulation error will be significant. 

Finally, as a sanity check, we also perform a cosmological analysis of {\sc Abacus} lightcone clustering measurements using the original Fourier-space {\sc comet}, and after Fourier transforming its output. The resulting contours are in perfect agreement, confirming that the configuration-space emulator is working as expected.

\subsection{Weak lensing model}

Our model for the weak lensing observables, cosmic shear and galaxy-galaxy lensing, follows the choices of \cite{AnnaDESI}. The model and its {\sc cosmosis} implementation was validated using the {\sc Buzzard} simulations \citep{derose2019, 2022PhRvD.105l3520D} and presented in \cite{2025OJAp....8E..24B} and \cite{2025arXiv251005539E}. 

\subsubsection{Cosmic shear}

The configuration-space model of the shear auto-correlation function for tomographic bins $i$ and $j$ can be obtained from the corresponding angular convergence power spectrum $C^{ij}_{\kappa\kappa}$ by assuming the flat-sky approximation and performing the transform:
\begin{equation}
    \xi^{ij}_{\pm}(\theta)=\int \frac{d\ell \, \ell}{2\pi}J_{0/4}(\ell\theta)C^{ij}_{\kappa\kappa}(\ell),
\label{eq:xipmwl}
\end{equation}
where $J_{0}$ and $J_{4}$ are the 0th (for $\xi_{+}$) and 4th (for $\xi_{-}$) order Bessel functions of the first kind.

The convergence power spectrum can be further connected to the matter power spectrum, $P_{\rm{mm}}$ through the lensing efficiency kernel $W$, by assuming the Limber approximation and performing an integral over the comoving radial distance (in a flat Universe) $\chi$ \citep{guzik2001, hu2004, joachimi2010}:
\begin{equation}
    C^{ij}_{\kappa\kappa} = \int d\chi \frac{W^{i}_{\kappa}(\chi)W^{j}_{\kappa}(\chi)}{\chi^{2}}P_{\rm{mm}}\Bigg(\frac{\ell+1/2}{\chi}, z(\chi)\Bigg),
    \label{eq:ckappakappa}
\end{equation}
with the lensing efficiency kernel for bin $i$ defined as:
\begin{equation}
    W^{i}_{\kappa,s}(\chi) = \frac{3}{2}\frac{\Omega_{\mathrm{m}}H^{2}_{0}}{c^{2}}\frac{\chi}{a(\chi)}\int^{\chi_{h}}_{\chi}\mathrm{d}\chi'n_{s}^{i}(\chi')\frac{(\chi'-\chi)}{\chi'}. 
    \label{eq:wkappa}
\end{equation}
Here, $H_{0}$ is the Hubble constant today, $c$ is the speed of light, and $a(\chi)$ is the scale factor. The lensing efficiency kernel for a tomographic bin $i$ is determined by the source galaxy redshift distribution for that bin, $n^{i}_{s}$, which is integrated up to the comoving distance of the horizon $\chi_{h}$. 

The nonlinear matter power spectrum, $P_{\rm{mm}}$, for shear observables is calculated using {\sc camb} \citep{lewis2000} using the {\sc hmcode2020} model that includes baryon feedback \citep{mead2021}. We do not include baryon feedback when performing validation tests on mocks, as this systematic is not present in the simulations. 

In addition to the contribution from weak lensing shear, the measured shape of the source galaxies has some correlation with the large-scale structure not due to the bending of the light by weak lensing, but due to physical alignment sourced by the gravitational potential of the structure in the environment. This is referred to as the Intrinsic Alignment (IA) effect \citep[for a review see, for example, ][]{lamman2024} and can be modeled by adding two additional contributions to $\xi_{\pm}$, such that the observed $\xi_{\pm\rm{obs}}$ is the sum \citep{hirata2004}:
\begin{equation}
    \xi_{\pm \rm{obs}}=\xi_{\pm} + \xi_{\pm}^{\rm{II}}+\xi_{\pm}^{\rm{GI}},
    \label{eq:xipmtot}
\end{equation}
where $\xi_{\pm}^{\rm{II}}$ is the contribution due to the correlations between the intrinsic ellipticities of the neighbouring galaxies, and $\xi_{\pm}^{\rm{GI}}$ measures the correlation between the intrinsic ellipticity of a foreground galaxy and the shear experienced by a source galaxy. While there are several IA models available, we follow the modelling choices of \cite{AnnaDESI} and adopt one of the most commonly used prescriptions, the Non-Linear Alignment (NLA) model \citep{bridle2007}. Under this model, the intrinsic alignment power spectra are related to the matter power spectrum via:
\begin{align}
    P_{\rm{II}}(k,z)&=a_{1}^{2}(z)P_{\rm{mm}}(k,z),\\
    P_{\rm{GI}}(k,z)&=a_{1}(z)P_{\rm{mm}}(k,z),
\end{align}
with the amplitude of the IA effect set by:
\begin{equation}
    a_{1}(z) = -A_{\rm{IA}}C_{1}\rho_{\rm{crit}}\frac{\Omega_{\rm{m}}}{D_{+}(z)}\Bigg(\frac{1+z}{1+z_0}\Bigg)^{\eta_{1}},
\end{equation}
where $D_{+}(z)$ is the linear growth factor, $C_{1}$ is a normalisation constant with value $C_{1}=5\times 10^{-14}h^{-2}M_{\odot}^{-1}\rm{Mpc}^3$, $\rho_{\rm{crit}}$ is the critical density, $z_0=0.62$ is the pivot redshift, and we have neglected the luminosity dependence of $a_{1}$. The IA power spectra can then be converted to the angular power spectrum contributions following:
\begin{equation}
C^{ij}_{\rm{II}}(\ell) = \int d\chi \frac{n^{i}_{\rm{s}}(\chi)n^{i}_{\rm{s}}(\chi)}{\chi^{2}}P_{\rm{II}}\Bigg(\frac{\ell+1/2}{\chi}, z(\chi)\Bigg)
\end{equation}
and
\begin{equation}
\begin{split}
C^{ij}_{\rm{GI}}(\ell) = \int d\chi &\Bigg( \frac{W^{i}_{\kappa}(\chi)n^{j}_{\rm{s}}(\chi)+W^{j}_{\kappa}(\chi)n^{i}_{\rm{s}}(\chi)}{\chi^{2}} \Bigg) \\ & P_{\rm{II}}\Bigg(\frac{\ell+1/2}{\chi}, z(\chi)\Bigg).
\end{split}
\end{equation}
These expressions are then transformed, equivalently to $C^{ij}_{\kappa\kappa}$ in Eq.\ref{eq:xipmwl}, to obtain the corresponding contributions to the observed shear signal, as given in  Eq.\ref{eq:xipmtot}.  We note that our mocks do not include the effect of IA and, therefore, we neglect the IA contributions when performing validation tests on the mock lightcone measurements. 

\subsubsection{Galaxy-galaxy lensing}

The cross-correlation between lens galaxy positions in tomographic lens galaxy bin $i$, and source galaxy shapes in tomographic source galaxy bin $j$, can be represented by the average tangential shear, $\gamma_{\rm{t}}^{ij}$. Analogously to the shear Eq.\ref{eq:xipm}, under the flat-sky approximation we can write:
\begin{equation}
    \gamma_{\rm{t}}^{ij}(\theta)=\int \frac{d\ell \, \ell}{2\pi}J_{2}(\ell\theta)C^{ij}_{g\kappa}(\ell),
\label{eq:xigt}
\end{equation}
where $J_{2}$ is the 2nd order Bessel function of the first kind and $C^{ij}_{\rm{g}\kappa}(\ell)$ is the galaxy-convergence cross-power spectrum.

As we did for shear, we can connect the configuration-space signal to $P_{\rm{mm}}$ by applying the Limber approximation to obtain:
\begin{equation}
    C^{ij}_{\rm{g}\kappa}(\ell) = \int d\chi \frac{W^{i}_{\rm{g}}(\chi)W^{j}_{\kappa}(\chi)}{\chi^{2}}P_{\rm{mm}}\Bigg(\frac{\ell+1/2}{\chi}, z(\chi)\Bigg)
\end{equation}
where $W^{i}_{\rm{g}}$ is the galaxy density kernel for lens redshift bin $i$, given by:
\begin{equation}
    W^{i}_{\rm{g}}(\chi) = b_{1}^{i}n^{i}_{l}(z)\frac{dz}{d\chi}.
\label{eq:wg}
\end{equation}
Here $b_{1}$ is the linear galaxy bias that corresponds to the lens redshift bin $i$, and $n^{i}_{l}(z)$ is its lens redshift distribution. In our set-up, the linear bias parameter $b_{1}$ is the same as the linear bias in the one-loop bias expansion in Eq.\ref{eq:bias} used for the galaxy clustering two-point correlation function.

Similarly to $\xi_{\pm}$, the observed tangential shear includes an intrinsic alignment contribution, $\gamma^{ij}_{\rm{t}}(\theta)_{\rm{gI}}$, which is obtained by applying the transform in Eq.\ref{eq:xigt} to the corresponding angular power spectrum given by:
\begin{equation}
    C^{ij}_{\rm{gI}}(\ell) = \int d\chi \frac{W^{i}_{\rm{g}}(\chi)n^{j}_{s}(\chi)}{\chi^{2}}P_{\rm{GI}}\Bigg(\frac{\ell+1/2}{\chi}, z(\chi)\Bigg).
    \label{eq:cgi}
\end{equation}
In addition to shape distortion, weak lensing introduces a magnification effect that needs to be taken into account when computing tangential shear. Following \cite{poole2023}, the magnification contribution to the angular power spectrum, $C^{ij}_{g\kappa,\rm{mag}}$, can be written as:
\begin{equation}
\begin{split}
    C^{ij}_{g\kappa,\rm{mag}}(\ell)=2(\alpha^{i}-1)\int &d\chi \frac{{}W^{i}_{\kappa,l}(\chi)W^{j}_{\kappa,s}}{\chi^{2}} \\ &P_{\rm{mm}}\Bigg(\frac{\ell+1/2}{\chi}, z(\chi)\Bigg),
    \label{eq:cgkappamag}
\end{split}
\end{equation}
where the lensing efficiencies are as defined in Eq.\ref{eq:wkappa} but $W^{i}_{\kappa,l}$ is calculated over the \textit{lens} redshift distribution, $n^{i}_{l}$, as opposed to the source redshift distribution that is used to obtain  $W^{j}_{\kappa,s}$. 

The final observed galaxy-galaxy lensing signal is then:
\begin{equation}
    \gamma^{ij}_{\rm{t}}(\theta)_{\rm{obs}} = \gamma^{ij}_{\rm{t}}(\theta)+\gamma^{ij}_{\rm{t}}(\theta)_{\rm{gI}}+\gamma^{ij}_{\rm{t}}(\theta)_{\rm{mag}},
\end{equation}
which includes the cross-correlation between source shapes and lens positions introduced by weak lensing, $\gamma^{ij}_{\rm{t}}(\theta)$ (obtained from $C^{ij}_{\rm{g}\kappa}$ in Eq.\ref{eq:ckappakappa}), an additional IA cross-correlation term due to the physical alignment of galaxies within their environments, $\gamma^{ij}_{\rm{t}}(\theta)_{\rm{gI}}$ (obtained from $C^{ij}_{\rm{gI}}$ in Eq.\ref{eq:cgi}), and the magnification contribution, $\gamma^{ij}_{\rm{t}}(\theta)_{\rm{mag}}$ (obtained from $C^{ij}_{\rm{g\kappa, mag}}$ in Eq.\ref{eq:cgkappamag}).

\subsection{Covariance}
\label{sec:cov}

In this section we summarise the covariance model we adopt for the set of configuration-space correlation functions spanning the cosmic shear and redshift-space density fields, $\{ \xi_\pm(\theta), \gamma_t(\theta), \xi^\ell(s) \}$.  The foundation of the model is a Gaussian analytical covariance including sample variance, noise and mixed terms.  Our implementation of these terms for the cosmic shear and galaxy-galaxy lensing correlations is presented by \cite{2025OJAp....8E..24B}, where our computations follow the methods of \cite{2017MNRAS.470.2100K} and \cite{2021A&A...646A.129J}.  For the cosmic shear covariance we include super-sample variance \citep{2013PhRvD..87l3504T}, and we also apply a small-scale noise correction based on the number of observed pairs within the survey geometry \citep{2018MNRAS.479.4998T}.  The galaxy-galaxy lensing covariance is corrected for boundary effects using an ensemble of fast mocks matched to the footprint of the lensing surveys, as discussed by \cite{2024MNRAS.533..589Y}.  We do not include a non-Gaussian contribution to the cosmic shear and galaxy-galaxy lensing covariance, which is currently negligible for the cases of interest \citep{2021A&A...646A.129J}.

In this paper we extend the Gaussian analytical covariance to include the correlation function multipoles $\xi^\ell(s)$, and their cross-covariance with $\xi_\pm(\theta)$ and $\gamma_t(\theta)$.  We believe that these analytical cross-covariances in configuration space are new to the literature and we present them in Appendix \ref{sec:covxil}.  Analytical auto-covariances for $\xi^\ell(s)$ have been previously discussed by \cite{2016MNRAS.457.1577G}, and the cross-covariance in Fourier space was studied by \cite{2022PhRvD.106f3536T}, who concluded that it was negligible for cosmological studies.  In our study we repeated the tests for the importance of these cross-covariance terms, as presented in Appendix \ref{sec:crosscovariance}. We confirm that the contribution of these terms is also negligible in configuration space for this analysis, however, we keep the full cross-covariance for the sake of completeness.

The Gaussian analytical covariance is less accurate for the auto-covariances of $\xi^\ell(s)$ than for the lensing correlation functions, because $\xi^\ell(s)$ is dominated by sample variance rather than noise, rendering the ``boundary effects'' of the sample geometry more significant in this case.  For this reason, in our fiducial covariance we replaced the Gaussian analytical $\xi^\ell(s)$ auto-covariance with a more accurate semi-analytical determination using the {\sc RascalC}\footnote{\url{https://rascalc.readthedocs.io/en/latest/}} code \citep{2025JCAP...01..145R}, which includes the effects of survey geometry and non-Gaussianity, and is the covariance model provided by the DESI-DR1 Key Project analysis \citep{2025JCAP...07..017A}.

Our shear and clustering datasets only partially overlap on the sky.  We correct our covariance model for this effect by assuming that measurements in non-overlapping regions are uncorrelated \citep[which was found to be a safe assumption by][]{2023OJAp....6E..36D}.  If the areal coverage of two datasets is $\Omega_1$ and $\Omega_2$, and $\Omega_O$ is the overlap area common to both, then the covariance of the correlation functions of these datasets is scaled by $\Omega_O^2/\Omega_1 \Omega_2$ \citep{2025OJAp....8E..24B}.

\subsection{Cosmological inference}
\label{sec:inference}

\begin{table}
	\centering
	\caption{Priors used in our clustering-alone and joint weak lensing and clustering analyses. $U$ indicates a flat uniform prior within the specified range. $\mathcal{N}(\mu, \sigma)$ indicates a Gaussian prior with mean $\mu$ and standard deviation $\sigma$. The nuisance parameters are varied separately for each DESI lens redshift bin $i$ with identical priors. Note that any weak lensing contours that do not include galaxy clustering are obtained using extended priors for $A_{\rm{s}} \times 10^9$ of $U(0.5, 5.0)$, and for $\Omega_{c}h^{2}$ of $U(0.03, 0.75)$.} 
	\label{tab:priors}
	\begin{tabular}{cc} 
		\hline
		Parameter & Prior \\
		\hline
		\hline
		\multicolumn{2}{c}{Cosmological parameters} \\
		\hline
		$\Omega_{\rm{b}}h^2$ & $U(0.0205, 0.02415)$ \\
		$\Omega_{\rm{c}}h^2$ & $U(0.085, 0.155)$ \\
		$h$ & $U(0.5, 0.9)$ \\
		$A_{\rm{s}} \times 10^9$ & $U(1.0, 3.5)$ \\
		$\rm{n}_s$ & $U(0.92, 1.01)$ \\
		\hline
		\multicolumn{2}{c}{Clustering nuisance parameters} \\
		\hline
        $b_1^{i}$ & $U(0.5, 9.0)$ \\
        $b_2^{i}$ & $U(-4.0, 8.0)$ \\
        $a_{\text{vir}}^{i}[\mathrm{Mpc}]$ & $U(0.0, 12.0)$ \\
        $c_0^{i}[\mathrm{Mpc^{2}}]$ & $U(-1000, 1000)$\\
        $c_2^{i}[\mathrm{Mpc^{2}}]$ & $U(-1000, 1000)$\\
        \hline
        \multicolumn{2}{c}{Weak lensing nuisance parameters}\\
        \hline
        $A_{\rm{IA}}$ & $U(-5.0, 5.0)$\\
        $\eta_{1}$ & $U(-5.0, 5.0)$\\
        $\log T_{\rm{AGN}}$ & $U(7.3, 8.3)$\\
        $\alpha^{1}$ & $U(-2, 6.0)$\\
        $\alpha^{2}$ & $\mathcal{N}(2.54, 0.036)$\\
        $\alpha^{3}$ & $\mathcal{N}(2.49, 0.12)$\\
        $\alpha^{4}$ & $\mathcal{N}(2.58, 0.48)$\\

        \multicolumn{2}{c}{\bf{KiDS-1000}}\\
        $\Delta z_{1}$  & $\mathcal{N}(0.0, 1.0)$\\
        $\Delta z_{2}$  & $\mathcal{N}(-0.181, 1.0)$\\
        $\Delta z_{3}$  & $\mathcal{N}(-1.110.0, 1.0)$\\
        $\Delta z_{4}$  & $\mathcal{N}(-1.395, 1.0)$\\
        $\Delta z_{5}$  & $\mathcal{N}(1.265, 1.0)$\\
        $m_{1}$ & $\mathcal{N}(-0.009, 0.019)$\\
        $m_{2}$ & $\mathcal{N}(-0.011, 0.020)$\\
        $m_{3}$ & $\mathcal{N}(-0.015, 0.017)$\\
        $m_{4}$ & $\mathcal{N}(0.002, 0.012)$\\
        $m_{5}$ & $\mathcal{N}(0.007, 0.010)$\\
        \multicolumn{2}{c}{\bf{DES-Y3}}\\
        $\Delta z$ & $\mathcal{N}(0, [0.018,0.015,0.011,0.017])$\\
        $m_{1}$ & $\mathcal{N}(-0.0063,0.0091)$\\
        $m_{2}$ & $\mathcal{N}(-0.0198,0.0078)$\\
        $m_{3}$ & $\mathcal{N}(-0.0241,0.0076)$\\
        $m_{4}$ & $\mathcal{N}(-0.0369,0.0076)$\\
        \multicolumn{2}{c}{\bf{HSC-Y1}}\\
        $\Delta z$ & $\mathcal{N}(0, [0.0374,0.0124,0.0326,0.0343])$\\
        $m$ & $\mathcal{N}(0, [0.01,0.01,0.01,0.01])$\\
        \multicolumn{2}{c}{\bf{HSC-Y3}}\\
        $\Delta z$ & $\mathcal{N}(0, [0.024, 0.022])$\\
        $m$ & $\mathcal{N}(0, [0.01,0.01,0.01,0.01])$\\
        \hline

	\end{tabular}
\end{table}

We perform our Bayesian inference using {\sc cosmosis}\footnote{\url{https://cosmosis.readthedocs.io/en/latest/index.html}} platform \citep{zuntz2015}, in particular, we make use of the {\sc nautilus} sampler\footnote{\url{https://nautilus-sampler.readthedocs.io/en/latest/}} \citep{lange2023}, which uses neural networks for accurate and fast nested sampling.

Our joint shear, galaxy-galaxy lensing and galaxy clustering two-point correlation function likelihood is built such that it obtains the theory predictions and scale cuts for shear and galaxy-galaxy lensing modelling from the same module that was used for the DESI large scale $3\times2$-pt analysis by \cite{AnnaDESI} and we add our {\sc comet}-based clustering likelihood as a separate module that provides theory predictions and scale cuts for galaxy clustering two-point correlation function multipoles. We use an analytical covariance that includes cross-correlation between weak gravitational lensing and galaxy clustering, but we replace the clustering autocorrelation part of the covariance with predictions from {\sc RascalC}, as described in Sec.~\ref{sec:cov}. 

Throughout this work, we are interested in the $\textit{physical}$ parameter space -- i.e., we focus on the parameters that are not defined through $h$, as discussed in Sec.~\ref{sec:evomapping}. This means that we present our constraints on the physical matter and dark energy densities, $\omega_{\rm{m}}$ and $\omega_{\rm{DE}}$ as well as linear density field variance defined on the scale of 12 Mpc, $\sigma_{12}$. Where appropriate, we will also include the constraints on $S_{8}=\sigma_{8}\sqrt{\Omega_{\rm{m}}/0.3}$, as this parameter corresponds to the amplitude of the weak lensing signal and may be of interest for validation purposes or when comparing with other weak lensing studies. 

Using the physical parameter space in cosmological analyses allows for an easier and more consistent interpretation, as these parameters describe the features of the power spectrum directly, without marginalising over the posterior for $h$ recovered by a particular probe, as discussed in \cite{sanchez2020, semenaite2022} and \cite{semenaite2023} \citep[and further explored in, for example,][]{forconi2025}. This choice is more impactful in the cases where the Hubble parameter is less well constrained (such as when analysing shear measurements or when investigating cosmologies beyond $\Lambda$CDM) or in the cases where the aim of the analysis is to provide accurate comparisons of different datasets which might have different posteriors for $h$.

Our cosmology priors for clustering only and joint clustering and weak lensing analyses are defined such that we do not sample outside of the emulator validity range with the priors for $\Omega_{\rm{c}}h^2$ and $A_{s}$ somewhat narrower than what is commonly used in clustering analyses. Nonetheless, our mock validation tests presented in Sec.~\ref{sec:mockfits} show that our clustering-alone and joint constraints are well within the chosen prior ranges. We use flat priors for all the cosmological parameters with narrow informative priors for $\Omega_{\rm{b}}h^2$ and $n_{\rm{s}}$, which are not constrained by clustering. The latter priors correspond to the ranges approximately 12$\sigma$ around \textit{Planck} 2018 best-fit value for $\Omega_{\rm{b}}h^2$ and 11$\sigma$ for $n_{\rm{s}}$.

In addition to our galaxy clustering-only and joint galaxy clustering, galaxy-galaxy lensing and shear fits, we also display the constraints from shear and galaxy-galaxy lensing measurements without the addition of clustering autocorrelation measurements. For this set up, we use wider priors for the cosmological parameters that are otherwise constrained by clustering, $A_s$ and $\Omega_{c}h^{2}$. We choose these priors to match the set up used in \cite{AnnaDESI}, resulting in the uniform prior for $A_{\rm{s}} \times 10^9$ of $U(0.5, 5.0)$, and the uniform prior for $\Omega_{c}h^{2}$ of $U(0.03, 0.75)$.

Our priors for shear and galaxy-galaxy lensing modelling parameters follow those of \cite{AnnaDESI} with the exception of the linear bias parameters, which are defined by the priors used for the galaxy clustering two-point correlation function fits (the clustering setup uses a slightly wider uniform prior but the particular choice is not highly significant, as this parameter is well constrained by the clustering multipoles). The priors for magnification bias parameters ($\alpha^{i}$) are derived from the measurements presented in \cite{2025arXiv250621677H}, which provides the measurements and their corresponding errors for the three LRG lens bins. We do not have a corresponding measurement for our BGS bin, therefore, we choose to use a flat prior for this case.

\subsection{Scale cuts}

For shear fits, we choose the same scales as \cite{AnnaDESI} and as validated by \cite{2025arXiv251005539E}. These correspond to matching the fiducial shear scale cuts, as determined by each of the weak lensing surveys themselves, with the exception of KiDS-1000, where \cite{AnnaDESI} change the $\xi_{-}$ scale cuts from $4'$ to $6.06'$ For galaxy-galaxy lensing we use a somewhat larger minimum scale of $12 \, h^{-1}\rm{Mpc}$ (instead of $6 \, h^{-1}\rm{Mpc}$ used by \cite{AnnaDESI}). This is because we do not perform point-mass marginalisation, which analytically marginalises over the galaxy-galaxy lensing signal originating from the mass enclosed at smaller scales and allows for a fit to an extended range of scales \citep{2020MNRAS.491.5498M}. We confirm that this choice does not have a significant effect on the final constraints, and we also note that we do not expect significant deviations from the NLA intrinsic alignment model on these scales \citep{2023MNRAS.524.2195S}.

Our galaxy clustering observable differs significantly from that of \cite{AnnaDESI}, as we are interested in the anisotropic large-scale clustering information, as measured by the two-point correlation function multipoles (instead of the projected clustering two-point correlation function that extends to more non-linear scales). For galaxy clustering multipoles, we set the scale cuts independently of the lensing observables and fit the minimum and maximum scales $s$ of $22 < s < 160 \, h^{-1}\rm{Mpc}$. This is both within the limits of validity of our theory emulator and is consistent with what was used for BOSS configuration-space analyses \citep[for example, ][]{semenaite2022, hou2021}. Furthermore, \cite{ramirez2025} confirmed that, for DESI LRG-like tracers and when using perturbative models to fit the clustering two-point correlation function multipoles, the minimum separation of $s_{\rm{min}}=22 \, h^{-1}\rm{Mpc}$ results in the tightest constraints while still remaining unbiased. 

\section{Model validation}
\label{sec:validation}

\subsection{Abacus mocks}
\label{sec:abacusmocks}

We tested our cosmological analysis pipeline using mock light-cone catalogues matching the properties of the DESI and weak lensing surveys, derived from the {\sc AbacusSummit} suite of $N$-body simulations \citep{2021MNRAS.508..575G, 2021MNRAS.508.4017M}, which consist of 25 realisations in the DESI fiducial cosmology.  These mock catalogues have already been presented and validated by \cite{2023MNRAS.525.4367H}, and in this section we provide a brief summary of their properties.  Representative DESI BGS and LRG populations were sampled from the mocks using a baseline Halo Occupation Distribution model \citep{2022MNRAS.509.2194H, 2024MNRAS.530..947Y, 2024MNRAS.533..589Y} consistent with the projected clustering of early DESI observations \citep{2024AJ....167...62D}, sub-sampled to the observed redshift distributions of each survey component.  We created random catalogues matching these galaxy datasets, by drawing angular coordinates from the light cone mask at each redshift, and sampling random redshifts from the set of mock galaxy redshifts.  We also sampled tomographic weak lensing source populations from the dark matter halo catalogues, matching the statistical properties of the KiDS, DES and HSC datasets, including the source redshift distributions, weights, shape noise, photometric redshift error and shear calibration bias, as described by \cite{2023MNRAS.525.4367H}.  We note that these mock lensing catalogues do not currently contain intrinsic alignments.  Also, during this analysis and after the production of HSC-Y1 mocks, HSC-Y3 data became available, so we note that we could only explicitly validate our pipeline at the level of precision of HSC-Y1.

We now describe the construction of the mock data vectors representing the overlapping and non-overlapping areas between the different weak lensing surveys and DESI-DR1.  First, we divided each {\sc Abacus} simulation light cone into closely-packed regions approximating the overlap areas of DESI-DR1 and KiDS-1000, DES-Y3 and HSC-Y1, using a {\sc healpix}\footnote{\url{https://healpy.readthedocs.io/en/latest/}} $n_{\rm side} = 8$ pixelisation.  In this way, from each individual light cone we formed 4 DESI-KiDS regions, 2 DESI-DES regions and 12 DESI-HSC regions, in which we performed $\gamma_t$ measurements, where these regions are comprised of 9, 16 and 3 pixels, respectively.  We then combined these regions to form wider catalogues as inputs for the correlation function multipole and cosmic shear measurements.  Our mock DESI BGS and LRG datasets spanned the complete light cone (5157 deg$^2$ for BGS and 3921 deg$^2$ for LRG), and the mock KiDS, DES and HSC lensing surveys covered 967, 3867 and 161 deg$^2$, respectively, of which 483, 966 and 161 deg$^2$ overlapped with the mock DESI catalogue.  We used this approach to create a total of 10 mock realisations for each combination of DESI and weak lensing survey.  We measured the cosmic shear and average tangential shear correlation functions of these mock datasets using {\sc treecorr}, using the same tomographic samples and separation bins as the real survey data, and we obtained the clustering multipoles using {\sc corrfunc}.  Finally, we determined the Gaussian analytical and {\sc RascalC} covariances for each mock configuration, following the methods outlined in Sec.~\ref{sec:cov}

\subsection{Fitting results}
\label{sec:mockfits}

We fit the mean measurements of shear, galaxy-galaxy lensing and galaxy clustering two-point correlation function multipoles (monopole and quadrupole), as measured on {\sc Abacus} simulations. While our measurements correspond to the mean across the available {\sc Abacus} realisations, we do not scale the covariance, but use the analytical covariance calculated for a single realisation volume. We do so to be able to consider realistic projection effects in our tests. 

Our priors are largely the same as used for the data analysis and listed in Table \ref{tab:priors}, except for the lensing nuisance parameters. As discussed above, our mocks do not include intrinsic alignment effects, therefore, we do not model any intrinsic alignment during the validation process. We also do not model baryon feedback, as this information is not included in the mocks, and our scale cuts are chosen so as to remove the scales significantly impacted by baryonic effects. Furthermore, the priors in any remaining weak-lensing modelling parameters are set to be centred around their no-contamination values -- i.e., we centre all photometric redshift, $\Delta z$, and all shear calibration, $m$, priors around 0, and we centre all magnification, $\alpha$, priors around 1.  

When assessing the agreement between our fits and the fiducial cosmology, we quote the significance of any deviations from the {\sc Abacus} cosmology value $\theta_{\rm{true}}$ assuming a Gaussian marginalised posterior and, for each cosmological parameter of interest, considering its mean $\bar\theta$ and standard deviation $\sigma_{\theta}$, such that $\Delta\theta(\sigma)= (\theta_{\rm{true}}-\bar\theta)/\sigma_{\theta}$. 

\subsubsection{Fiducial mock fits}
\label{sec:mockfits}

\begin{figure}
\includegraphics[width=0.99\columnwidth]{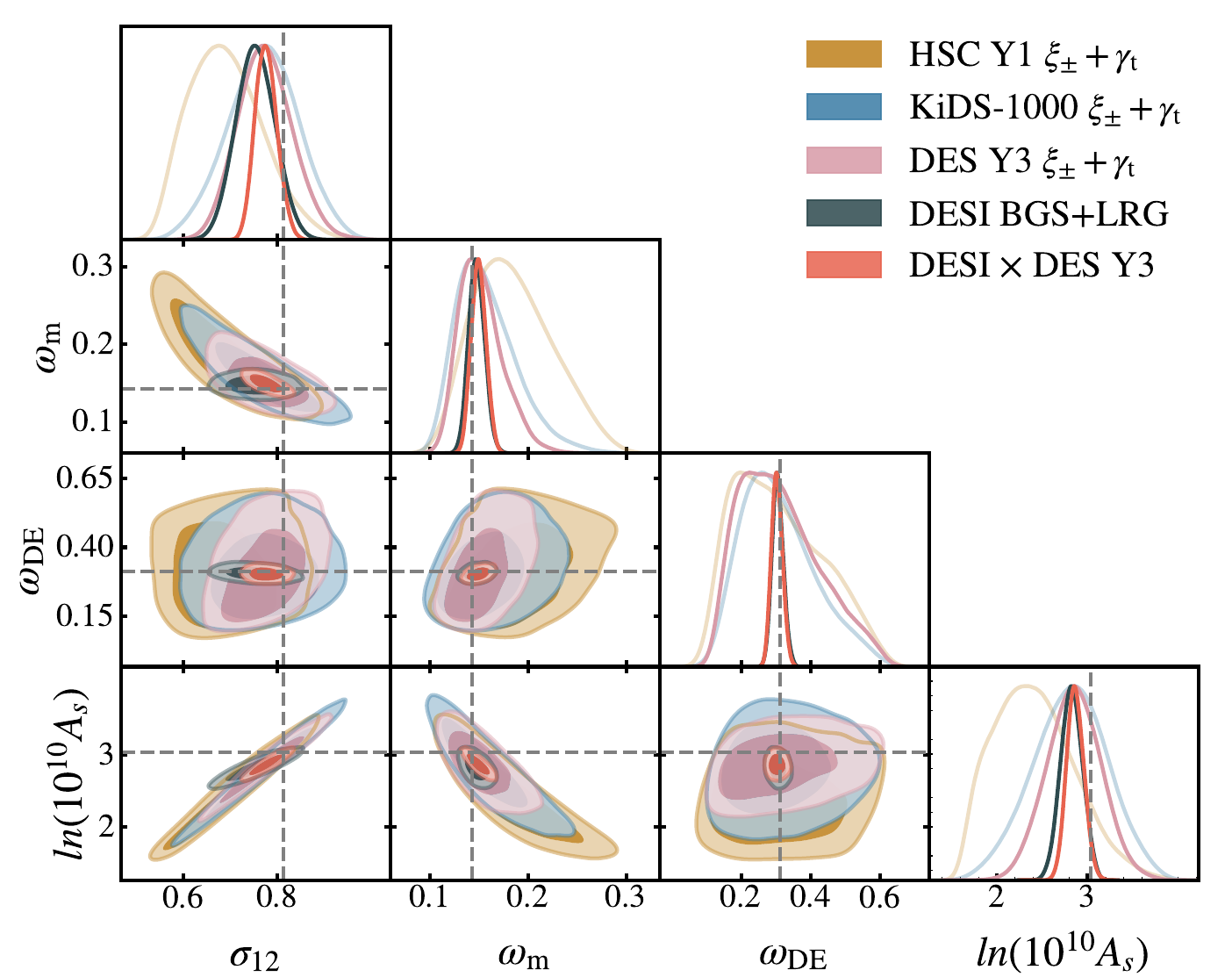}
\caption{Flat $\Lambda$CDM parameter constraints derived from fits to mock-mean measurements on {\sc Abacus} lightcones. We present the constraints derived from fitting the DESI-like BGS and LRG clustering multipoles, $\xi_{0,2}$, as well as joint fits to $\xi_{\pm} + \gamma_{\rm{t}}$ measurements corresponding to auto- and cross-correlations between HSC-Y1, KiDS-1000, DES-Y3, and DESI mock data. We also show the constraints obtained from fits to mock DESI$\times$DES-Y3 measurements, $\xi_{0,2}+\xi_{\pm}+\gamma_{\rm{t}}$ (shear$\times$RSD).}
\label{fig:abacusdes3x2pt}
\end{figure}

\begin{figure}
\includegraphics[width=0.99\columnwidth]{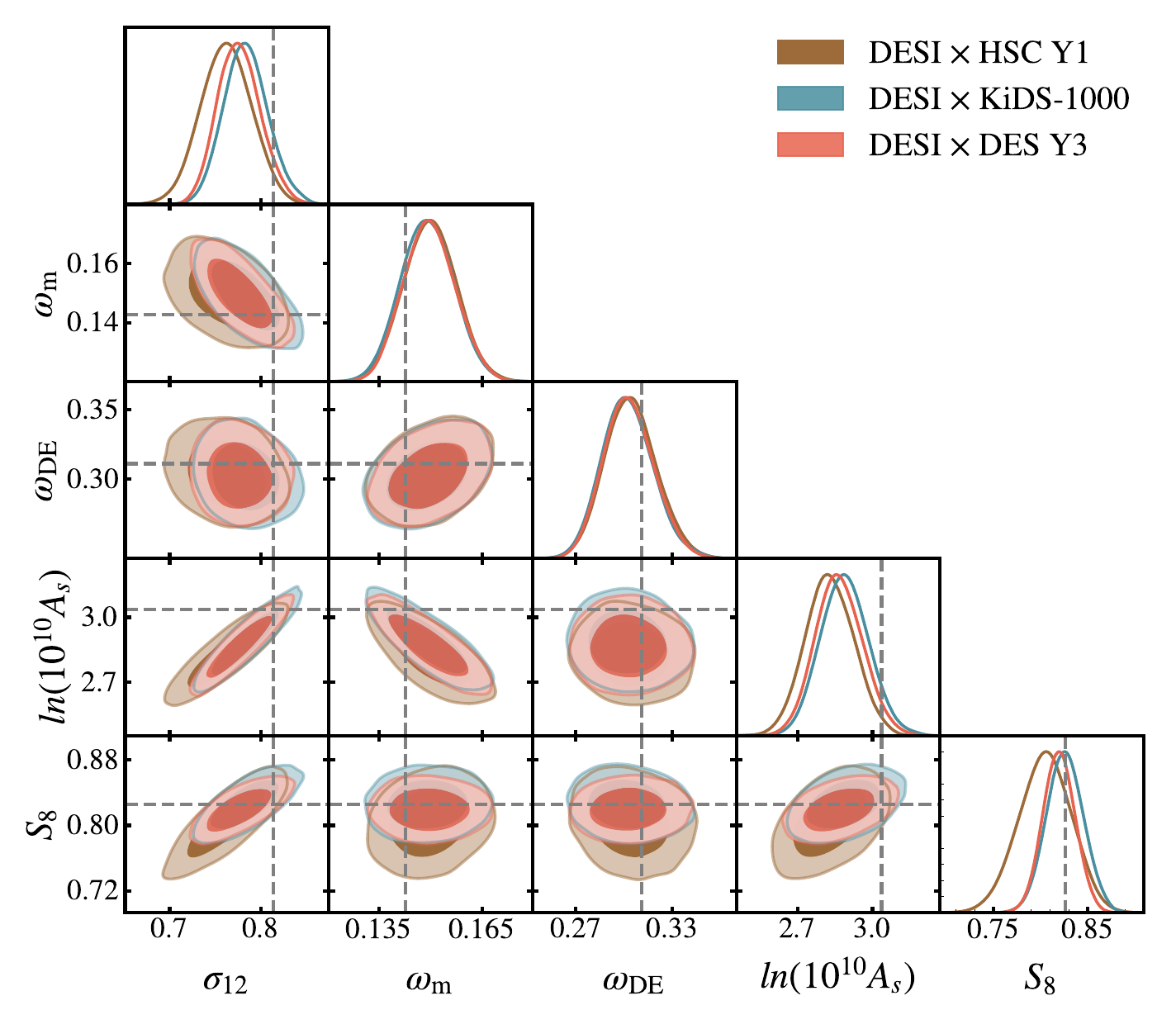}
\caption{Flat $\Lambda$CDM constraints derived from joint clustering and weak lensing fits to mean measurements on {\sc Abacus} lightcones. Here we compare the constraints obtained using the combination of DESI with different weak lensing surveys: HSC-Y1 (in brown), KiDS-1000 (in blue) and DES (in red).}
\label{fig:abacus_all3x2pt}
\end{figure}

The results of our fiducial mock fits for the combination of DESI-like lenses and DES-like sources are displayed in Figure \ref{fig:abacusdes3x2pt}, where we compare the fits to weak-lensing-only measurements (shear and galaxy-galaxy lensing), galaxy clustering two-point correlation function multipoles, and the joint fits to the combined data vector. We compare the joint clustering and weak lensing fits for all the weak lensing surveys in Figure \ref{fig:abacus_all3x2pt}.

As expected, the joint constraints are primarily dominated by the clustering information; however, the addition of weak lensing, which traces the matter field directly, improves constraints on the matter power spectrum amplitude parameters $\sigma_{12}$ and $\ln(10^{10}A_{\rm{s}})$. Galaxy clustering fits provide strong constraints on the matter density and dark energy density parameters $\omega_{\rm{m}}$ and $\omega_{\rm{DE}}$ through the shape of the two-point correlation function and the constraint on the Hubble parameter, $h^{2}=\sum\omega_{i}$, originating from the scale of the BAO peak.  We recover a slightly high $\omega_{\rm{m}}$ value (at around $\sim 0.6\sigma$ for clustering alone and $\sim 0.9\sigma$ for joint analysis), which is expected following our massless neutrino assumption.

\begin{figure}
\includegraphics[width=1.0\columnwidth]{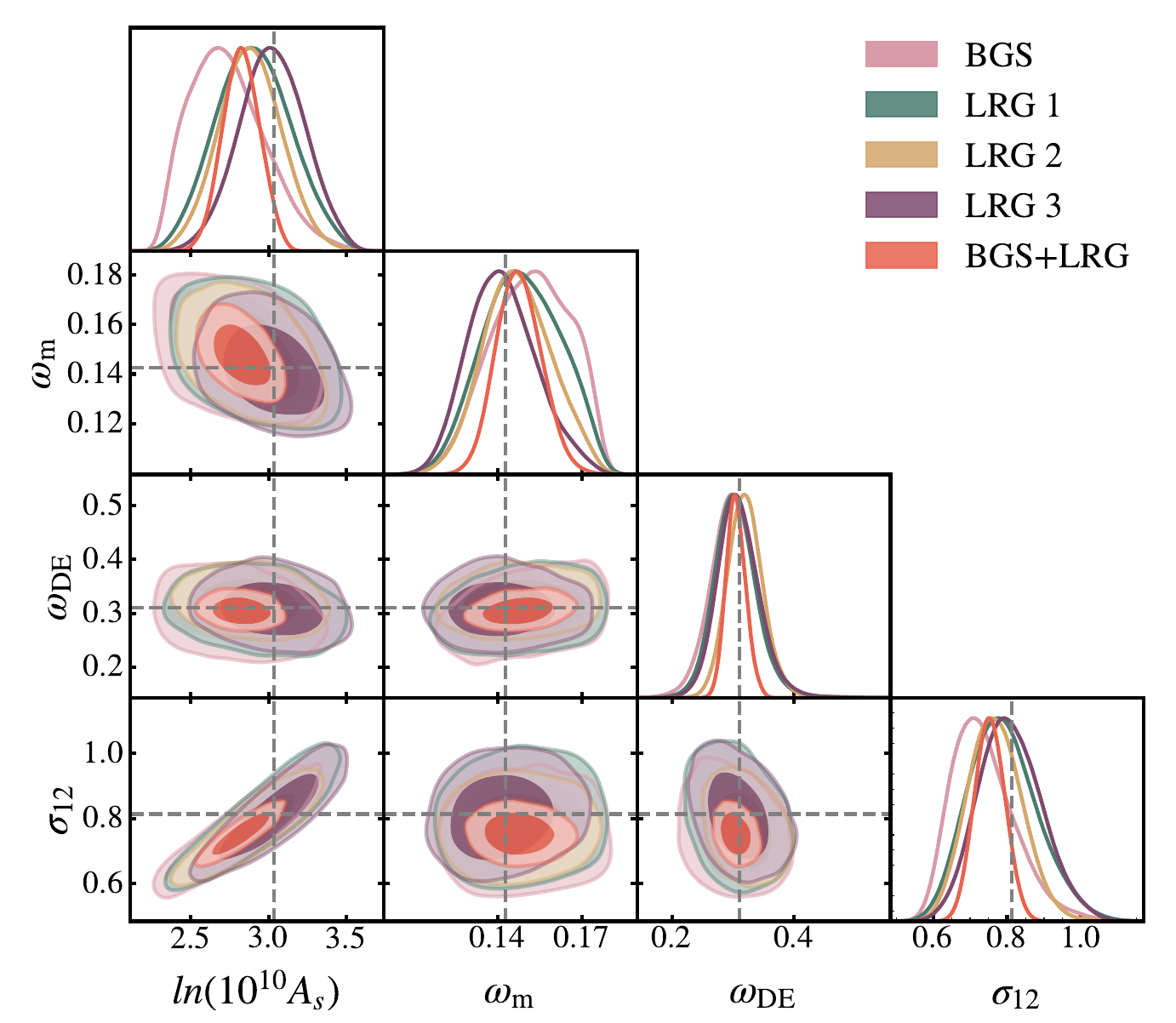}
\caption{Flat $\Lambda$CDM parameter constraints derived from galaxy clustering two-point correlation function multipole measurements on {\sc Abacus} lightcones, for different DESI-like tracers.}
\label{fig:abacus_xi02_singlebins}
\end{figure}

Most significantly, the clustering-only and joint constraints show offsets in the amplitude parameters $\sigma_{12}$ and $\ln(10^{10}A_{\rm{s}})$, which are strongly degenerate parameters.  The shifts in mean values compared to the input fiducial cosmology are slightly more significant for $\ln(10^{10}A_{\rm{s}})$ ($\sim 1.78\sigma$ for both clustering-only and joint cases) than $\sigma_{12}$ ($\sim1.46\sigma$ for clustering-only and $\sim1.6\sigma$ for joint case). Both of these parameters are known to be affected by projection effects in the multi-dimensional parameter space, when the mean marginalised posterior is shifted away from the maximum likelihood value \citep{carrilho2023, simon2022, tsedrik2025b}. These effects are more prominent in the presence of strong degeneracies and poor constraining power and are, therefore, especially significant when the dark energy equation of state parameter $w$ is allowed to vary. However, even in $\Lambda$CDM cosmologies, we expect to observe a shift, as power spectrum amplitude parameters are degenerate with poorly constrained bias parameters and, in the weak lensing case (for real data), the intrinsic alignment amplitude. Any additional effects that change the amplitude of the power spectrum, such as the RSD effects or the Alcock-Paczynski distortions will further worsen the projection effects.

\begin{figure}
\includegraphics[width=0.9\columnwidth]{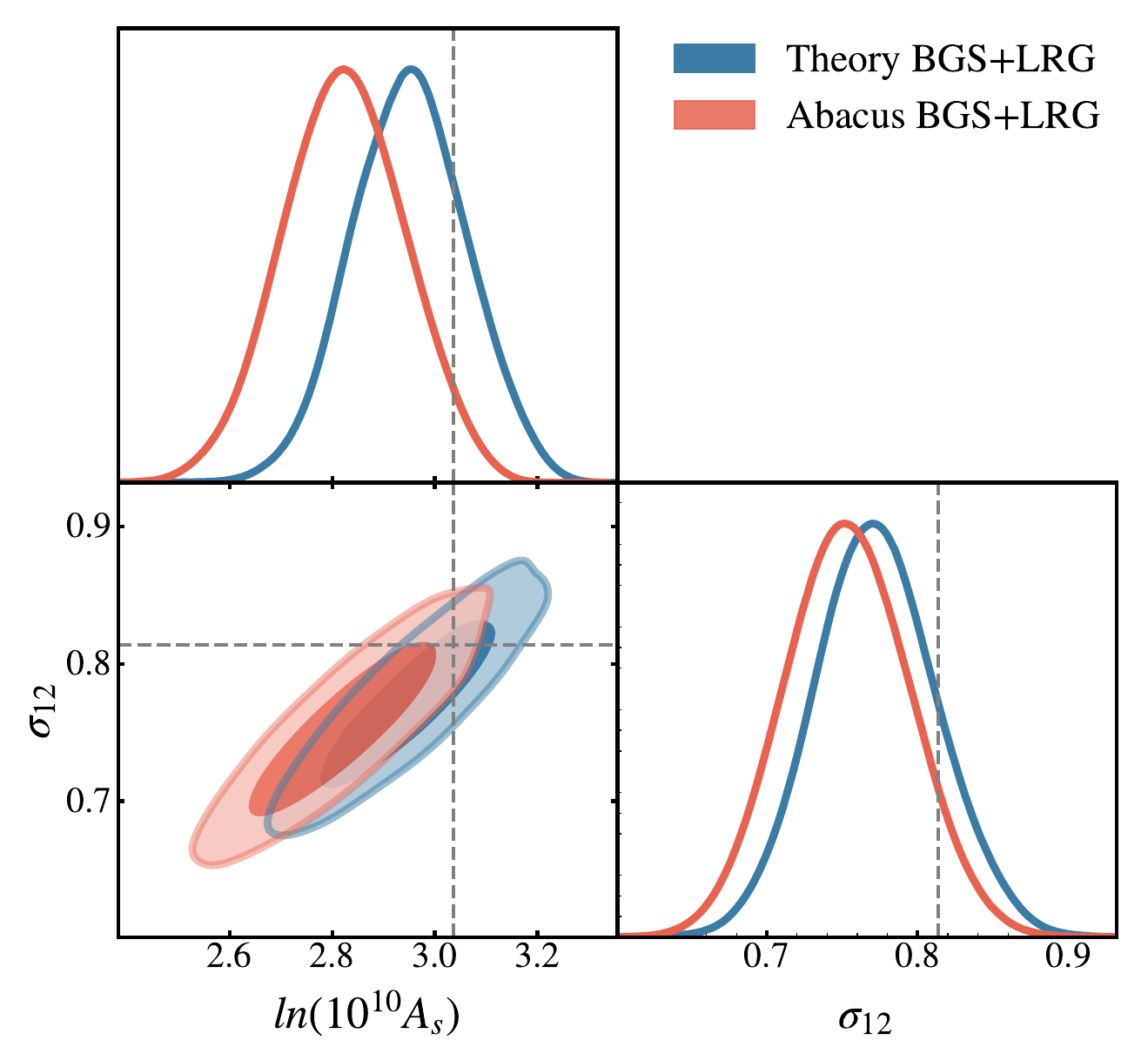}
\caption{Comparison of marginalised posterior contours for the amplitude parameters $\sigma_{12}$ and $\ln(10^{10}A_{\rm{s}})$ obtained by fitting a theory vector that corresponds to {\sc Abacus} clustering and the actual mock clustering measurements.}
\label{fig:projection}
\end{figure}

In order to estimate the projection effects' contribution to the parameter offsets, we fit a fiducial theory vector using the same covariance we use for the mock analysis.  In the absence of projection effects, we would expect a perfect recovery of the input cosmology. As the offsets are driven by clustering information, we performed this test using the two-point correlation function multipoles and we show the result of this fit in Figure \ref{fig:projection}.  We find that the expected offsets related to projection effects are around $1\sigma$ for $\sigma_{12}$ and around $0.76\sigma$ for $\ln(10^{10}A_{\rm{s}})$. Our mock clustering fits match the theory result for $\sigma_{12}$ well, but show a more significant offset in $\ln(10^{10}A_{\rm{s}})$. We note that the theory vector is created for a simplified case of zero counter-terms and the resulting contour for DESI-like clustering may shift slightly depending on the particular nuisance parameter values.  In summary, we experience significant projection effects even in $\Lambda$CDM, although the size of this effect for $\ln(10^{10}A_{\rm{s}})$ seems to be more significant than indicated by the theory vector fits. While there is a number of methods commonly used to alleviate projection effects, such as sampling in parameter combinations that reduce parameter degeneracies, and including additional informed priors \citep[see, for example, ][]{desi_fs, tsedrik2025b, paradiso2025, chen2024, donald2023}, we leave further exploration of such methods for the upcoming DESI-DR2 analyses.

We furthermore want to test whether any individual tracer shows deviations from the true cosmology. We present clustering constraints for each tracer in Figure \ref{fig:abacus_xi02_singlebins}, which shows that higher-redshift tracers exhibit better agreement with the fiducial cosmology values.  While all LRG bins recover both $\sigma_{12}$ and $\ln(10^{10}A_{\rm{s}})$ to well within $1\sigma$ from their fiducial values, the BGS fit shows a stronger bias, pulling the joint constraints away from the fiducial cosmology. 

While the BGS sample exhibits less constraining power than the LRG tracers, and will hence experience stronger projection effects, another potential explanation for the more significant offset for BGS-like tracer fits could be the limitations of the HOD model used to generate this sample. Unlike the LRG tracers, which were created using the {\sc AbacusHOD} model\footnote{\url{https://abacusutils.readthedocs.io/en/latest/hod.html}} \citep{yuan2022} and tuned to match the 2D correlation function of the DESI one-percent survey, the BGS sample was modeled by fitting the \textit{projected} two-point correlation function \citep{2024MNRAS.530..947Y}.  The {\sc AbacusHOD} fits were performed on DESI one-percent survey BGS clustering measurements in four redshift bins in order to capture the redshift evolution of the sample, as described in \cite{2024MNRAS.530..947Y}. In our study, the best fit model values for an absolute magnitude cut of $M_{R}<-20$ were used to create the mock {\sc Abacus} BGS galaxies. As our BGS clustering measurement is obtained in one continuous redshift bin, and we are fitting clustering multipoles, it is possible that the mocks exhibit some mis-match compared to the real data and that our model is sensitive to this difference.

Considering that our individual-tracer clustering fits provide an accurate recovery of the fiducial cosmological parameters, we note the biases observed in the combined BGS+LRG fits and proceed with the analysis. Given the limitations of the current lightcone mocks, we leave further testing for the DR2 work on the new generation of more realistic mocks.  We next validate the DESI combinations with the remaining two weak lensing surveys, KiDS and HSC.  We show the comparison of the combined cosmological constraints for all three surveys in Figure \ref{fig:abacus_all3x2pt} and find them to be highly consistent with each other.

We test the effect of several further analysis options for fitting the clustering multipoles, and find no significant changes to the resulting posteriors.  First, we test the effect of including the hexadecapole in the set of clustering statistics, and confirm that the addition of this multipole does not significantly improve the constraints, as also shown in the {\sc Abacus} galaxy clustering two-point correlation function fitting study by \cite{ramirez2025}. We furthermore test that removing the first three separation bins, such that the minimum scale fitted is either 37 $h^{-1}\rm{Mpc}$, 32 $h^{-1}\rm{Mpc}$, or 27 $h^{-1}\rm{Mpc}$ results in looser constraints with no significant improvement in biases, confirming that we do not suffer from significant modeling failures related to small-scale physics but these data points instead help in reducing projection effects. We also test varying the tidal parameters $\gamma_{2}$ and $\gamma_{21}$ freely, instead of fixing them to the coevolution relations, and find that this similarly results in a slight increase in biases in the amplitude parameters $\ln(10^{10}A_{\rm{s}})$ and $\sigma_{12}$. These tests are further presented in Appendix \ref{sec:validation_extra}.

\section{Results}
\label{sec:results}

\begin{figure}
\includegraphics[width=0.99\columnwidth]{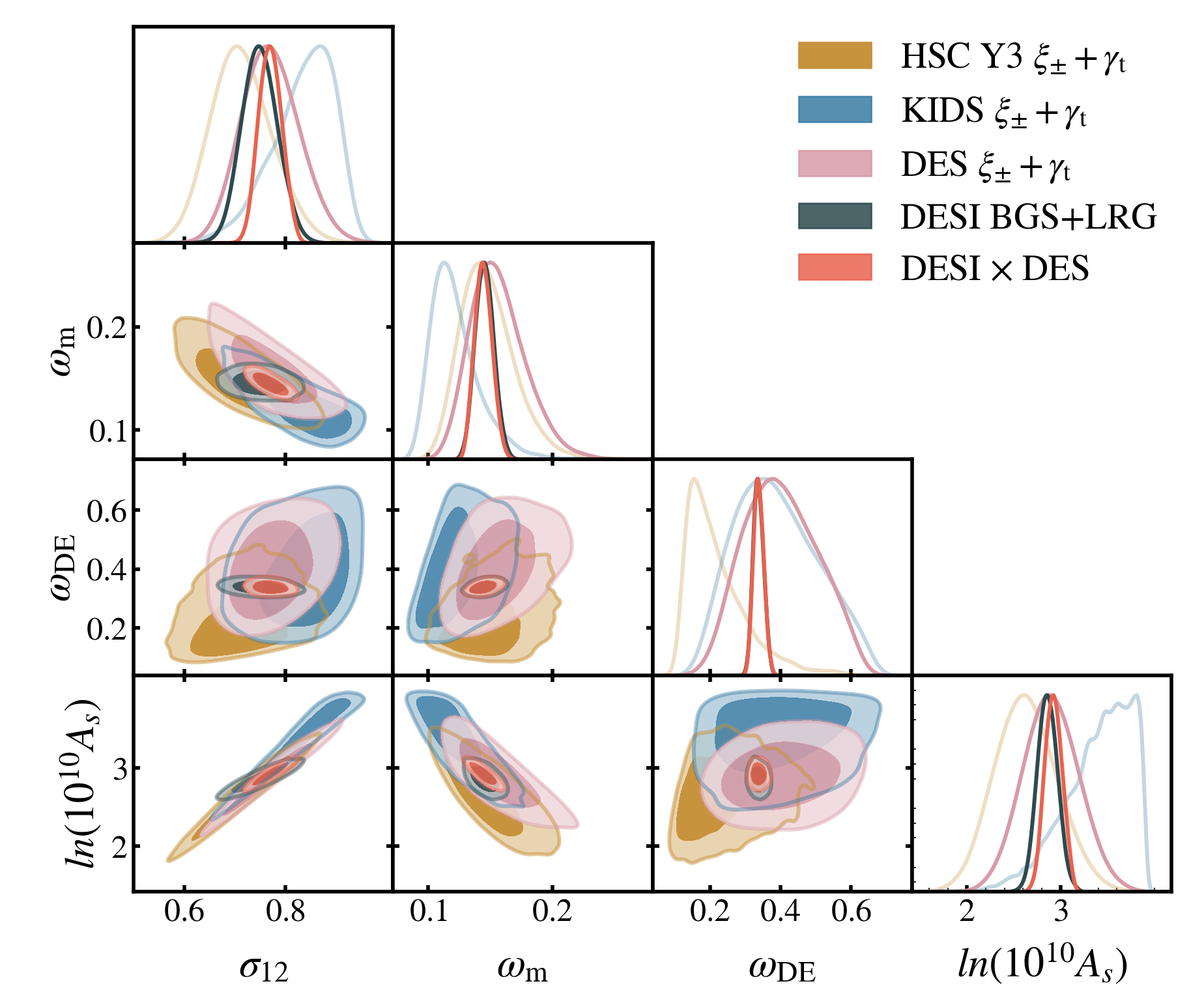}
\caption{Flat $\Lambda$CDM parameter constraints derived from fits to DESI clustering multipoles, $\xi_{0,2}$, as well as joint fits to $\xi_{\pm} + \gamma_{\rm{t}}$ measurements corresponding to auto- and cross-correlations between HSC-Y3, KiDS-1000, DES-Y3, and DESI. We also show the constraints obtained from fits to DESI$\times$DES-Y3 measurements, $\xi_{0,2}+\xi_{\pm}+\gamma_{\rm{t}}$ (shear$\times$RSD).}
\label{fig:desixdes}
\end{figure}

\begin{figure}
\includegraphics[width=0.99\columnwidth]{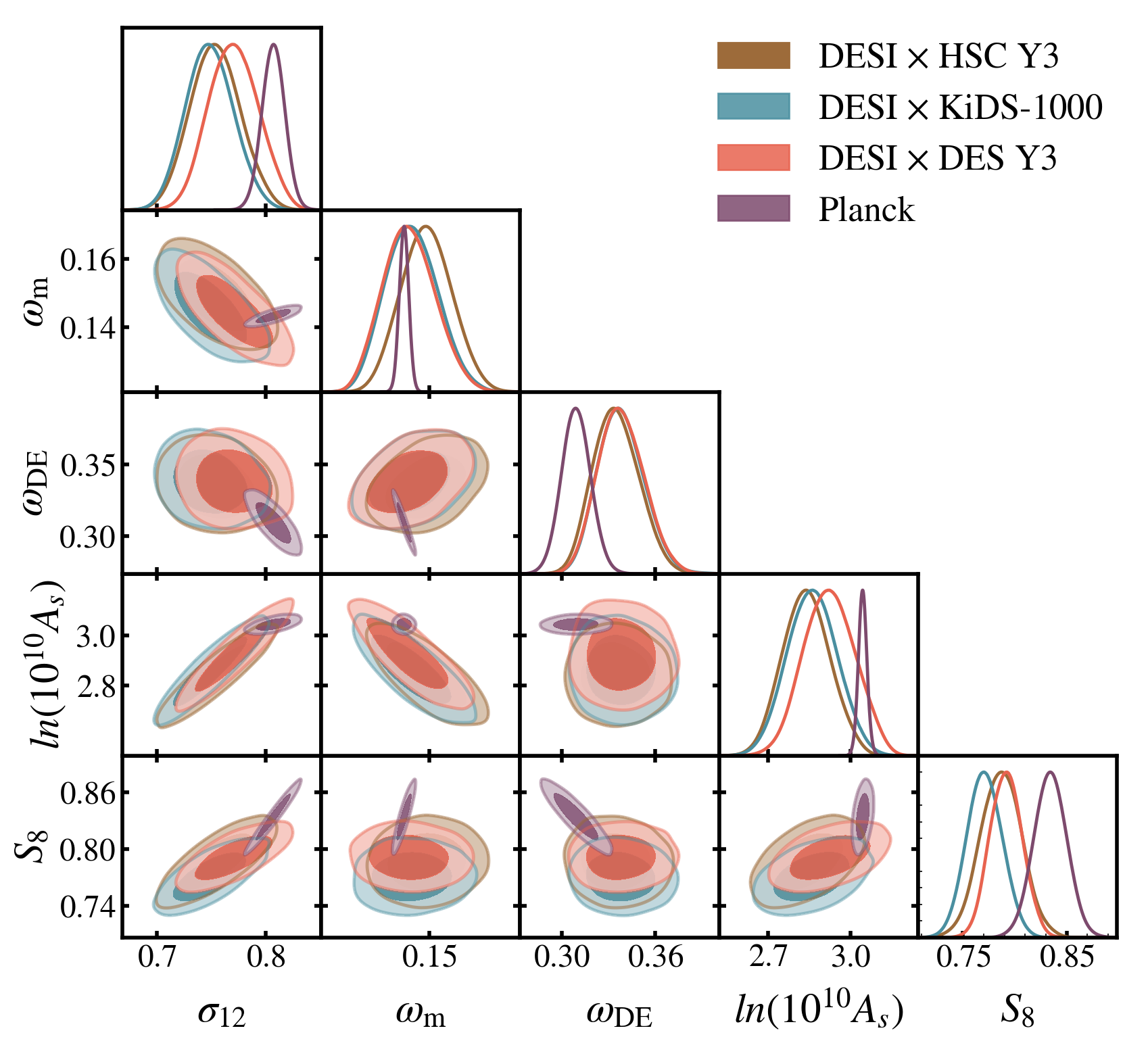}
\caption{Flat $\Lambda$CDM parameter constraints derived from shear$\times$RSD fits to DESI clustering multipoles and auto and cross-correlations between HSC-Y3, KiDS-1000, DES-Y3, and DESI. We also include constraints from fits to \textit{Planck} CMB temperature and polarization power spectra, for comparison.}
\label{fig:desixwl}
\end{figure}

\begin{table*}

	\centering
	\caption{Marginalised posterior constraints (mean values with 68 per-cent confidence interval) derived from the full-shape analysis of DESI BGS+LRG galaxy two-point correlation function multipoles (clustering-only) as well as the constraints obtained from shear and galaxy-galaxy lensing measurements between DESI lenses and HSC-Y3, KiDS-1000 and DES-Y3 sources, and the `shear$\times$RSD' fits to the combined set of two-point correlation functions of DESI clustering and its cross-correlations with weak lensing surveys.}
	\begin{tabular}{ccccccccc} 
		\toprule
		\multicolumn{1}{c|}{Parameter} & \multicolumn{1}{c|}{Clustering-only ($\xi_{0,2}$)} & \multicolumn{3}{c|}{Shear + g-g lensing ($\xi_{\pm}+ \gamma_{\rm{t}}$)} & \multicolumn{3}{c}{Shear$\times$RSD $(\xi_{0,2} + \xi_{\pm}+ \gamma_{\rm{t}})$}  \\ 
		\midrule
		 & DESI & DES-Y3 & KiDS-1000 & HSC-Y3 & DES-Y3 & KiDS-1000 & HSC-Y3 \\
         & BGS+LRG & $\times$ DESI & $\times$ DESI & $\times$ DESI & $\times$ DESI & $\times$ DESI & $\times$ DESI \\ 
		\midrule
		$\sigma_{12}$ & $0.750\pm 0.036$ & $0.771^{+0.053}_{-0.061}$ & $0.835^{+0.078}_{-0.041}$ & $0.714^{+0.055}_{-0.066}$ & $0.771\pm 0.023$ & $0.749\pm 0.022$ & $0.754\pm 0.023$ \\ [+0.1cm]
		$\omega_{\rm{m}}$ & $0.1455\pm 0.0075$ & $0.156^{+0.018}_{-0.026}$ & $0.122^{+0.011}_{-0.023}$ & $0.148^{+0.017}_{-0.025}$ & $0.1446^{+0.0067}_{-0.0075}$ & $0.1453^{+0.0067}_{-0.0075}$ & $0.1493\pm 0.0068$\\[+0.1cm]
        $\rm{ln}10^{10}\rm{A}_{\rm{s}}$ & $2.86\pm 0.11$ & $2.90\pm 0.30$ & $3.40^{+0.48}_{-0.16}$ & $2.63^{+0.32}_{-0.36}$ & $2.923\pm 0.093$ & $2.861\pm 0.090$ & $2.840\pm 0.086$\\[+0.1cm]
		$\omega_{\rm{DE}}$ & $0.338^{+0.014}_{-0.016}$ & $0.402^{+0.096}_{-0.12}$ & $0.390^{+0.098}_{-0.15}$ & $0.221^{+0.033}_{-0.10}$ & $0.338^{+0.014}_{-0.016}$ & $0.338^{+0.013}_{-0.016}$ & $0.335^{+0.014}_{-0.016}$\\ [+0.1cm]
		
		\midrule
        $S_8$ & $0.776\pm 0.035$ & $0.805\pm 0.024$ & $0.777^{+0.020}_{-0.022}$ & $0.771\pm 0.030$ & $0.791\pm 0.016$ & $0.771\pm 0.017$ & $0.787\pm 0.020$ \\ [+0.1cm]
        $\Omega_{\mathrm{m}}$ & $0.301\pm 0.011$ & $0.288^{+0.041}_{-0.059}$ & $0.249^{+0.034}_{-0.068}$ & $0.418\pm 0.082$ & $0.300^{+0.010}_{-0.011}$ & $0.301\pm 0.011$ & $0.308\pm 0.011$ \\ [+0.1cm]
        \bottomrule
        
	\end{tabular}
\label{tab:constraints}
\end{table*}

In this section, we present the results of the data fits. This work represents the first constraints from DESI full-shape configuration-space galaxy clustering measurements, as well as the first results from cross-correlations between these clustering measurements and shear measurements from HSC-Y3, KiDS-1000 and DES-Y3. We will consider three cases: clustering-only ($\xi_{0,2}$), joint shear and galaxy-galaxy lensing ($\xi_{\pm}+\gamma_{\rm{t}}$) and shear$\times$RSD ($\xi_{0,2} + \xi_{\pm}+ \gamma_{\rm{t}}$) fits. As described in Section \ref{sec:validation}, we caution that we were not able to explicitly validate the pipeline used to fit HSC-Y3 dataset but rather performed the validation tests on HSC-Y1-like mocks.   

When presenting our results, we focus on the physical cosmological parameters (i.e., not defined through the Hubble parameter $h$, see sections Sec.~\ref{sec:evomapping} and \ref{sec:inference}) and present a summary of the physical parameter space constraints in Table \ref{tab:constraints}. Where relevant throughout this section, we will also include constraints on non-physical parameters of interest: the weak lensing signal amplitude $S_8$, the relative matter density $\Omega_{\rm{m}}$ and the Hubble parameter $h_0$. 

We note that for results that do not include galaxy two-point correlation function multipoles (the combination of shear and galaxy-galaxy lensing fits, $\xi_{\pm}+\gamma_{\rm{t}}$), we use wider priors on the cosmological parameters that are otherwise constrained by clustering, $\ln(10^{10}A_{\rm{s}})$ and $\omega_{\rm{c}}$, as described in Sec.~\ref{sec:inference}.

\subsection{Cosmological constraints from shear$\times$RSD fits}

We begin by considering DESI and its cross-correlations with DES-Y3, which is the weak lensing survey with the greatest overlap. The combined shear and galaxy-galaxy lensing constraints are in excellent agreement with the clustering-only results, as seen in Figure \ref{fig:desixdes}. When considering the combination in shear$\times$RSD fits, we see that the resulting posteriors are dominated by the clustering multipole constraints, particularly for matter and dark energy densities. Nonetheless, the addition of DES-Y3 shear measurements and its cross-correlations improves the matter power spectrum amplitude parameters, $\sigma_{12}$ and $\ln(10^{10}A_{\rm{s}})$. This is expected, as weak lensing is not as sensitive to the shape cosmological parameters, but the amplitude of the weak lensing signal is directly sensitive to the amplitude of density fluctuations in the matter field, which helps break the degeneracy between the amplitude parameters and galaxy bias. 

Notably, the improvement is greater for $\sigma_{12}$ (improvement of $\sim36\%$) than for $\ln(10^{10}A_{\rm{s}})$ (improvement of $\sim15\%$). The origin of this difference lies in the degeneracy directions between $\sigma_{12}$ and $\ln(10^{10}A_{\rm{s}})$ and the matter density $\omega_{\rm{m}}$: while the DES-Y3 $\xi_{\pm}+\gamma_{\rm{t}}$ contours show convincing anti-correlation between the power spectrum amplitude parameters and matter density, galaxy clustering is only aligned in that same direction for $\ln(10^{10}A_{\rm{s}})$. In contrast, in $\omega_{\rm{m}}$-$\sigma_{12}$ parameter subspace, DESI clustering multipoles are able to constrain $\omega_{\rm{m}}$ relatively independently of $\sigma_{12}$ (importantly, the extent of this posterior is limited by the informative prior on $n_{\rm{s}}$, which is degenerate with $\omega_{\rm{m}}$). This results in a slightly rotated contour with respect to the direction preferred by the combination of shear and galaxy-galaxy lensing, and a greater constraining power gain for $\sigma_{12}$ when the two sets of measurements are combined in the shear$\times$RSD fit.

We then consider cross-correlations with KiDS-1000 and HSC-Y3. Here as well, we find that clustering measurements are largely consistent with shear and galaxy-galaxy lensing. We note that KiDS-1000 combined shear and galaxy-galaxy lensing measurements prefer higher values of $\sigma_{12}$ and $\ln(10^{10}A_{\rm{s}})$, with the resulting posterior limited by the upper limit of the imposed prior. This is largely consistent with the combined shear and galaxy-galaxy lensing result obtained by the KiDS collaboration \citep[see Table C.1 of ][]{2021A&A...646A.140H}. The fits to lensing auto- and cross-correlations using HSC-Y3 sources similarly lack the constraining power on $\omega_{\rm{DE}}$, with most of the posterior volume concentrated in the region of low values of $\omega_{\rm{DE}}$ (and, consequently, low values of $h$). As noted before, the overlap area of DESI with KiDS-1000 and HSC-Y3 is more limited, and so it is not surprising that the constraining power is deteriorated compared to DES-Y3 fits. 

Finally, we can compare the shear$\times$RSD fits for DESI cross-correlations with all three weak lensing surveys. As shown in Figure \ref{fig:desixwl}, the combined constraints are very consistent with each other with the choice of a particular weak lensing survey most relevant for $\sigma_{12}$ and, to a lesser extent, $\ln(10^{10}A_{\rm{s}})$ posteriors. The gain in constraining power for the amplitude parameters is similar for all weak lensing surveys considered, and the rest of the parameter space is dominated by DESI clustering information. 

\subsection {Linear bias constraints}

\begin{figure*}
\centering
\hspace*{-0.1\columnwidth}%
\makebox[\textwidth][c]{
\includegraphics[width=2.2\columnwidth]{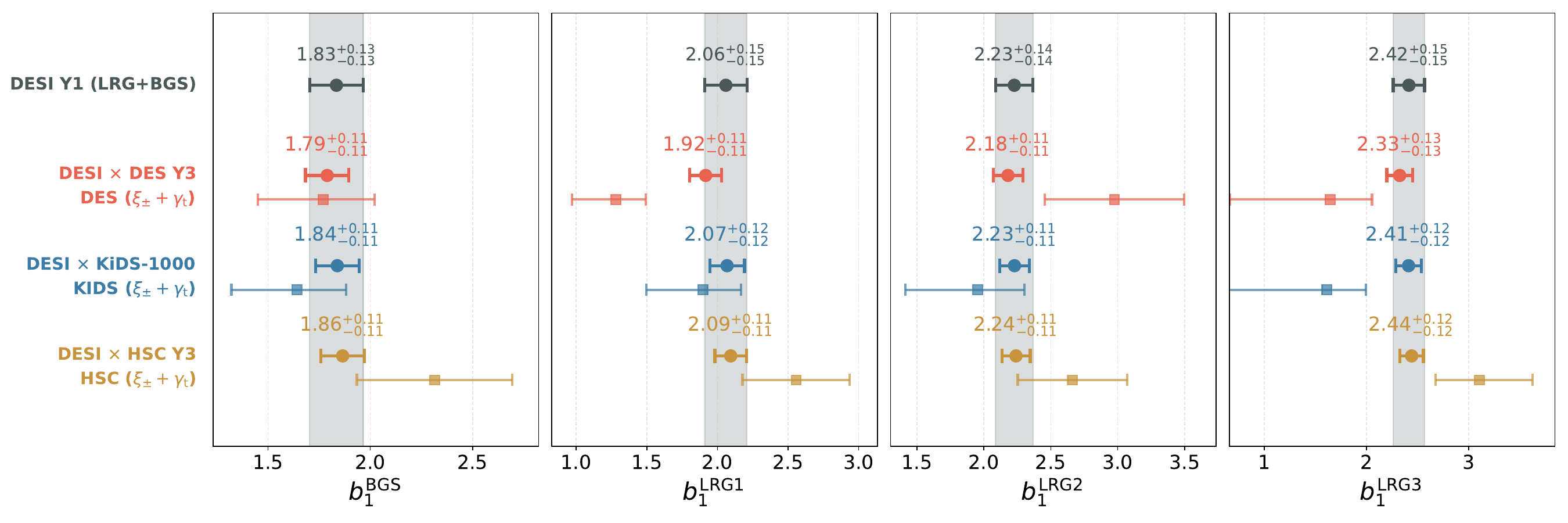}}
\caption{Linear bias constraints for DESI lens bins. We compare the clustering-only fits to DESI-DR1 galaxy clustering multipoles (joint fits to BGS and LRG and clustering multipoles, shaded area) with our shear$\times$RSD fits to two-point auto- and cross-correlations between DESI and DES-Y3, KiDS-1000 and HSC-Y3. We additionally present no-clustering constraints from shear and galaxy-galaxy lensing fits (square marker). All values correspond to mean with 68$\%$ confidence interval.}
\label{fig:desibias}
\end{figure*}

As the improvements in the measured amplitude parameters are directly related to the fact that weak lensing measurements allow us to break the degeneracy with galaxy bias, it is interesting to look at the bias constraints directly. Our galaxy-galaxy lensing and clustering models, while based on different prescriptions, share linear bias, so we can further investigate whether the clustering-only and combined shear and galaxy-galaxy lensing measurements agree with each other and how much information we gain by combining the two sets of measurements together. 

We present our linear bias fits for all DESI lens bins in Figure \ref{fig:desibias}. Regardless of which survey we consider, adding weak lensing information allows us to improve our constraints on $b_1$ by $15-20\%$, depending on the tracer. All linear bias values recovered are in an excellent agreement with the mock fits, except for the BGS bin, where {\sc Abacus} measurements prefer a lower bias value of $b_1=1.58\pm 0.13$, further suggesting that the HOD model used to create this sample may not represent real data accurately, as discussed in Section \ref{sec:mockfits}. 

Furthermore, even with the relatively limited overlap area, we are able to obtain some constraints using combined shear and galaxy-galaxy measurements on their own (with the exception of LRG3, which, due to its high redshift, can only be constrained by the HSC-Y3 data within the prior boundary). When comparing these measurements with shear$\times$RSD fits, we see that KiDS-1000 provides the most consistent constraints with clustering-only case (the mean values are consistent at the level of well within 1$\sigma$ for all tracers, except for LRG3), while HSC-Y3 prefers slightly higher bias values (although, again, mostly consistent with clustering only case). Out of the three weak lensing surveys considered, only DES-Y3 shows some more significant discrepancies, most notably for LRG1, where the $b_{1}$ value preferred by the joint shear and galaxy-galaxy lensing fit is over $2\sigma$ lower than the clustering-only constraint.

Any discrepancy in bias measurements must be considered bearing in mind that the clustering-only and joint shear and galaxy-galaxy lensing measurements are not performed on exactly the same set of lenses. While our clustering multipoles are obtained from tracers covering the full footprint of DESI-DR1, we can only measure galaxy-galaxy lensing on the areas overlapping with the weak lensing surveys, which means that each of these measurements corresponds to a different and limited part of the full clustering sample. As a result, we may expect some shifts in preferred bias values corresponding to cosmic variance or varying completeness in that particular patch of the sky.  The latter effect will be significantly mitigated in the DESI-DR2 sample. 

Ultimately, any potential discrepancies between clustering-only and joint shear and galaxy-galaxy lensing fits have limited effect on the final shear$\times$RSD cosmology. Depending on which shear survey we consider, the final bias and, therefore, $\sigma_{12}$ constraint shifts slightly, all of these shifts are below $1\sigma$, with the DESI$\times$DES-Y3 measurement experiencing the greatest shift towards slightly higher values of $\sigma_{12}$.

\subsection{Comparison with other analyses and surveys}

\begin{figure}
\includegraphics[width=0.99\columnwidth]{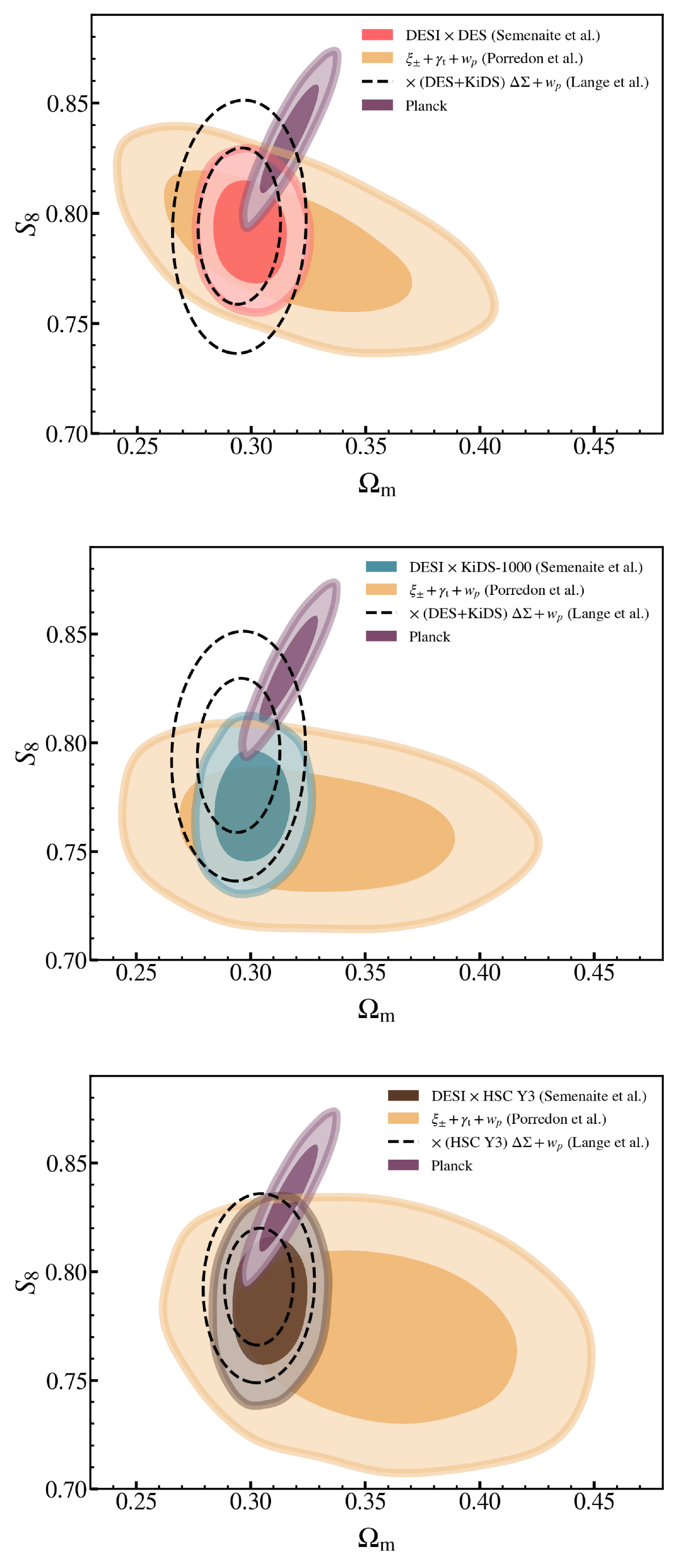}
\caption{Comparison of the marginalised $S_8-\Omega_{\mathrm{m}}$ posteriors derived from fits to DESI cross-correlations with DES Y3 (top), KiDS-1000 (middle) and HSC Y3 (bottom). In each case we display the results from fits to configuration space measurements including the RSD effects presented in this work together with the results of our companion $3\times2$-pt analysis that fits projected DESI clustering ($w_{p}$) instead of clustering multipoles, as presented in \cite{AnnaDESI}. We also present, in dashed lines, the results from the simulation-based modelling analysis by \cite{Langedesi}, which considers the cross correlations between DESI and the combination of DES and KiDS (top and middle panels), and cross correlations between DESI and HSC Y3 (bottom panel). The simulation-based analysis includes the galaxy-galaxy lensing effect by fitting the excess surface density statistic, $\Delta\Sigma$.}
\label{fig:s8omegam}
\end{figure}

\begin{figure}
\includegraphics[width=0.99\columnwidth]{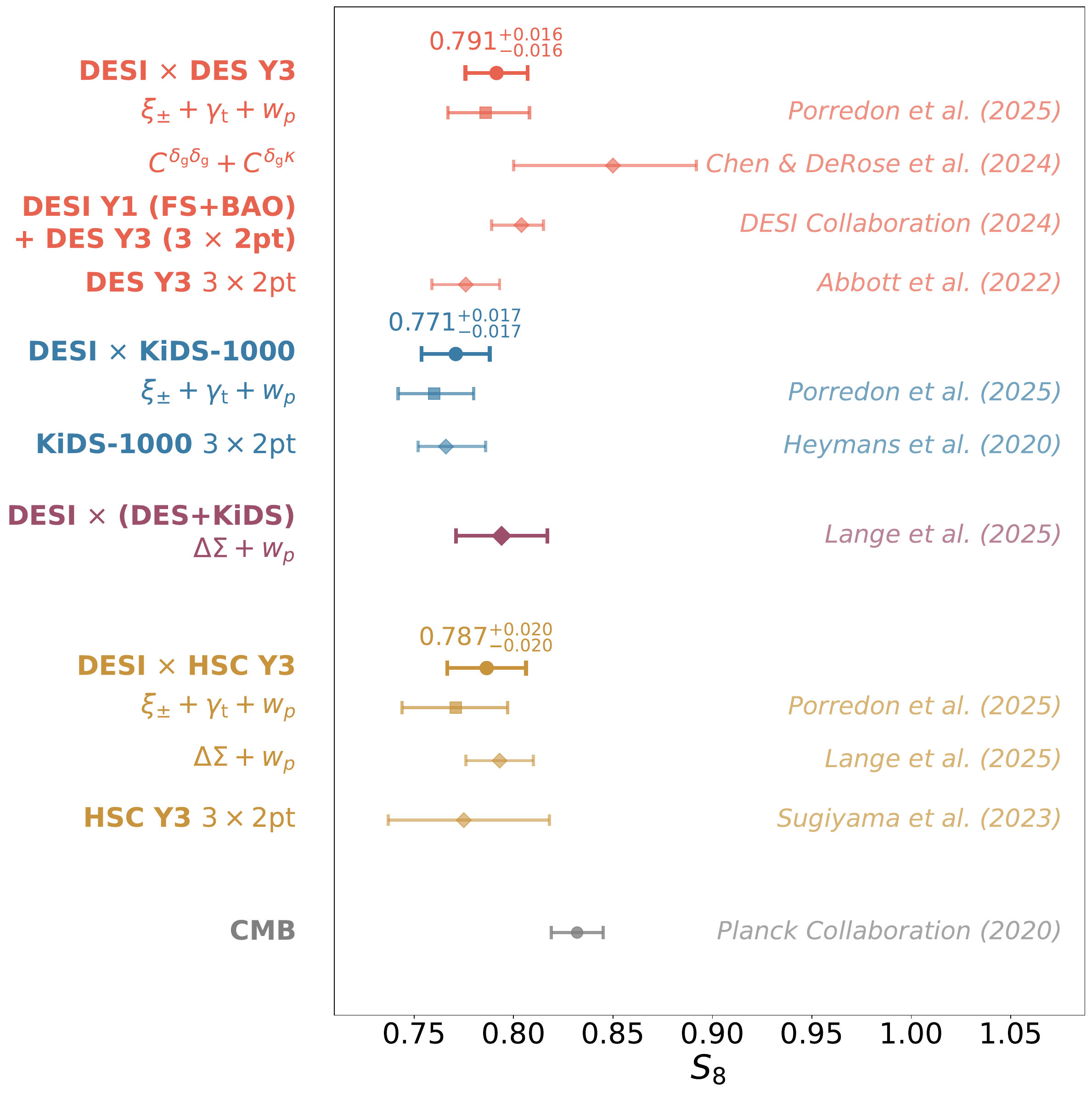
}
\caption{Comparison of the measurements of the lensing amplitude parameter $S_8=\sigma_{8}\sqrt{\Omega_{\mathrm{m}}/0.3}$ derived in this work, and similar analyses for each weak lensing survey. In each case, the top constraint (round marker) corresponds to our shear$\times$RSD fits for (top to bottom) DESI$\times$DES-Y3, DESI$\times$KiDS-1000 and DESI$\times$HSC-Y3. We also also present the results from a companion $3\times2$-pt analysis that fits projected DESI clustering ($w_{p}$) instead of clustering multipoles \citep[square marker, $w_p+\xi_{\pm}+\gamma_{\rm{t}}$,][]{AnnaDESI}. The constraints obtained by each weak lensing survey's fiducial  analyses are marked as $3\times2$-pt. Finally, for DES-Y3, we additionally provide the results from fitting cross-correlations with DESI in Fourier space using the HEFT modelling approach \citep{chen2024}, and the result obtained from the joint DESI full-shape clustering and DES-Y3 $3\times2$-pt analysis that does not consider cross-correlations between the two surveys \citep[rhombus markers,][]{desi_fs}. Finally, we also show the results from the simulation-based analysis by \cite{Langedesi} who fit the projected clustering and excess surface mass density, $\Delta\Sigma$, measurements between DESI and the combination of DES and KiDS (displayed separately, in purple), as well as the cross-correlations between DESI and HSC. }
\label{fig:s8}
\end{figure}

We now place our results in the context of previous weak lensing and clustering analyses, as well as CMB measurements from \textit{Planck}. In order to do so, it useful to, in addition to our physical constraints, consider the Hubble parameter, $h_0$, which simply corresponds to the sum of the physical densities of all energy species (and at $z=0$ is dominated by $\omega_{\rm{m}}$ and $\omega_{\rm{DE}}$), and the relative matter density $\Omega_{\rm{m}}$.  From DESI clustering we find,
\begin{equation}
\begin{rcases}
h_{0}=0.695^{+0.013}_{-0.014} \\
\Omega_{\mathrm{m}} = 0.301 \pm 0.012
\end{rcases}
\quad \text{DESI BGS+LRG $\xi_{0,2}$} \\[12pt]
\end{equation}
Combining galaxy clustering with shear and galaxy-galaxy lensing measurements does not have any significant effect on these constraints. We show our marginalised $S_{8}-\Omega_{\mathrm{m}}$ posterior for all weak lensing survey combinations in Figure \ref{fig:s8omegam} and we present our constraints on the lensing amplitude parameter $S_8$ in Figure \ref{fig:s8}.

First, we are interested in how our results compare to DESI's fiducial full-shape galaxy clustering-only analysis in Fourier space presented by \cite{desi_fs} and \cite{2025JCAP...09..008A}. When considering the combined tracer clustering results and comparing them with DESI key full-shape cosmology constraints, we first note that, in addition to considering Fourier-space measurements, the fiducial analysis makes use of all DESI tracers (i.e., in addition to BGS and LRG, they fit emission line galaxy, ELG, and quasar, QSO, measurements, as well as making use of Ly$\alpha$ forest) and includes reconstructed BAO signal. The fiducial analysis additionally varies the neutrino mass sum and utilises a Big Bang Nucleosynthesis (BBN) prior and a somewhat narrower Gaussian prior on $n_{\rm{s}}$. As expected, the fiducial analysis is, therefore, able to provide higher precision measurements for both $\Omega_{\rm{m}}$ and $h_0$: $\Omega_{\rm{m}} = 0.2962\pm0.0095$ and $h_0=0.6856\pm0.0075$, with both values nonetheless extremely consistent with our configuration-space constraints. 

We see a greater discrepancy in the initial power spectrum amplitude constraint, with the fiducial DESI analysis \citep{desi_fs} obtaining $\ln(10^{10}A_{\rm{s}})=3.117\pm0.092$, which is $\sim1.8\sigma$ higher than our measurement of $\ln(10^{10}A_{\rm{s}})=2.86\pm0.11$. Although the two constraints are still consistent with each other, this difference is likely to at least partially be the consequence of the projection effects.  While \cite{desi_fs} implement a sampling technique that mitigates against projection effects, we saw that, without such mitigation, we might expect shifts to low $\ln(10^{10}A_{\rm{s}})$ values, as discussed in Sec.~\ref{sec:mockfits}. 

Considering individual tracer clustering and comparing with DESI key Fourier space fits presented by \cite{2025JCAP...09..008A}, we find a good agreement between the two sets of posteriors with an excellent agreement in $h_0$ but somewhat lower values of $\ln(10^{10}A_{\rm{s}})$ for all bins but LRG2. We note that the individual tracer posteriors are affected by our limited $\ln(10^{10}A_{\rm{s}})$ prior and, therefore, we do not quote these constraints but present the individual tracer comparison in Appendix \ref{sec:tracers}. 

We can furthermore compare our DESI$\times$DES-Y3 constraints with the joint DESI+DES-Y3 (3$\times$2-pt) analysis performed in \cite{desi_fs}, where the two datasets are treated independently. Our DESI$\times$DES-Y3 measurement of $\sigma_{8}=0.793\pm 0.022$ agrees to well within $1\sigma$ with the DESI+DES-Y3 result of $\sigma_{8}=0.807^{+0.016}_{-0.020}$. Finally, the KiDS collaboration has presented $\sigma_{12}$ constraints for their 3$\times$2-pt cosmology fits \citep[see][]{2021A&A...646A.140H}, so we can compare our DESI$\times$KiDS-1000 constraint of $\sigma_{12}=0.749\pm0.022$ with the KiDS-1000$\times$BOSS+eBOSS QSO constraint of  $\sigma_{12}= 0.743_{-0.026}^{+0.03}$ and find the two to be in excellent agreement.

Our companion analysis, \cite{AnnaDESI}, uses the same shear and galaxy-galaxy model as our study, but, instead of fitting galaxy clustering multipoles, utilises the projected clustering of DESI tracers in different lens bins -- in their analysis, the BGS sample is split into three separate tomographic bins with different magnitude cuts, and the LRG3 bin is not used.  Projected clustering degrades the constraining power within the physical parameter space, and the loss of the BAO feature prevents projected clustering from tracing the standard ruler information to constrain $h_0$. Nonetheless, the analysis provides powerful constraints on the parameter combination $S_8=\sigma_{8}\sqrt{\Omega_{\mathrm{m}}/0.3}$, which describes the amplitude of the weak lensing signal. As can be seen in Figure \ref{fig:s8omegam} and Figure \ref{fig:s8}, our results are in an excellent agreement with those of \cite{AnnaDESI} and measure $S_8$ with comparable precision. Our measurements are additionally consistent with the DESI+DES-Y3 result from \cite{desi_fs} and the DESI$\times$DES-Y3 galaxy-galaxy lensing and clustering Fourier-space fits by \cite{chen2024}. The latter measurement represents a significantly distinct setup compared to this analysis: they define a different DESI lens sample, including the photometric sample from DESI Legacy, and implement an effective field theory prescription for both the clustering and galaxy-galaxy lensing model, including a fully perturbative model for IA effects \citep[for more details see][]{derose2025}.

Finally, in Figure \ref{fig:s8omegam} we additionally compare our results with the simulation-based analysis of \cite{Langedesi} who fit projected galaxy clustering and excess surface mass density measurements between DESI and the combination of DES Y3 and KiDS-1000, as well as the cross-correlations between DESI and HSC Y3. Once again, this analysis differs significantly from our approach: it relies on simulations and uses a complex HOD model which includes assembly bias in order to extract information from highly non-linear scales. The fits are performed on measurements that use different lens sample selection (excluding LRG3 and using two BGS bins) with varied $S_8$ and $\Omega_{\mathrm{m}}$, while fixing the remaining cosmology parameters to the values preferred by the CMB measurements. Despite these differences, the resulting constraints are in an excellent agreement with our fits.

Our final results are, therefore, consistent with the fiducial DESI full-shape galaxy clustering power spectra fits as well as our companion $3\times2$-pt analysis that considers the projected galaxy clustering two-point correlation function. In the figures presented in this section we also include the high-$z$ constraints obtained by fits to the \textit{Planck} CMB temperature and polarization power spectra \citep{Planck2018}. We find that, overall, our results are consistent with the values preferred by these measurements: our marginalised $\omega_{\rm{m}}$ posterior is in excellent agreement with the CMB constraint, whereas our dark energy density $\omega_{\rm{DE}}$ measurement is in good agreement, with a slightly lower value than that preferred by \textit{Planck}, resulting in a consistent $h_0$ constraint. When combining with weak lensing surveys, we measure an $S_8$ value slightly lower than that predicted by \textit{Planck}, but consistent with the fiducial cosmological analyses by the weak lensing surveys considered in our study. Our best $\sigma_{8}$ measurement $\sigma_{8}=0.793\pm 0.022$ (DESI $\times$DES-Y3) is also in agreement, within 1$\sigma$, with the measurement coming from the CMB survey combination SPT-3G+ACT DR6+\textit{Planck} \citep{camphuis2025} of  $\sigma_{8}=0.8137\pm 0.0038$. 

Finally, our results are also consistent with \cite{maus2025cmb}, who cross-correlated DESI galaxies with Planck PR4 and ACT DR6 CMB lensing maps. The resulting constraints, $\sigma_8=0.803\pm0.017$, $\Omega_{\rm{m}}=0.3037\pm0.0069$ and $S_8=0.808\pm0.01$ are consistent with our measurements within $1\sigma$ (with the exception of the $S_{8}$ measurement from DESI$\times$KiDS, which differs by $\sim1.5\sigma$), but yield higher values for all three parameters.

\section{Conclusions}
\label{sec:conclusions}

In this work we present cosmological constraints derived from configuration-space shear$\times$RSD fits to DESI BGS+LRG two-point correlation function multipoles, and shear and galaxy-galaxy lensing measurements, that correspond to the auto- and cross-correlations of DES-Y3, KiDS-1000 and HSC-Y3 sources and DESI-DR1 lenses. We make use of each observable's traditional modelling approach, using a halo-model based power spectrum prescription to model shear and galaxy-galaxy lensing data vectors, and a perturbative EFT-based model for clustering multipoles. We use an analytical covariance that accounts for cross-correlations between weak lensing and galaxy clustering observables, although we find that excluding these terms does not significantly affect the final contours (as presented in Appendix \ref{sec:crosscovariance}). This analysis serves as an exploration of the information that can be extracted from a joint LSS analysis, making use of overlapping photometric and spectroscopic observations, whilst retaining distinct modelling approaches that allow us to fit the full range of scales considered in the respective key cosmological analyses for each observable. 

We validate our approach on {\sc Abacus} lightcone mocks, which are constructed to match the properties of DESI and weak lensing surveys. We find that, whilst we are able to accurately recover the fiducial cosmology for shear and galaxy-galaxy lensing-only fits as well as the clustering multipole fits to individual tracers, when performing a combined analysis, the marginalised posterior mean values for the amplitude parameters $\sigma_{12}$ and $\ln(10^{10}A_{\rm{s}})$ are low by $1.5-1.8\sigma$. These parameters are known to be sensitive to projection effects, so we investigate the potential contribution of these effects to the observed shifts by fitting a theory data vector, which suggests expected projection effects of around $0.76\sigma$-$1\sigma$. When considering individual bin fits, we note that the clustering fits for the low-redshift BGS bin experience a more significant bias than the LRG clustering fits, which may potentially be attributed to a different HOD model used to create this tracer population. We leave more extensive validation work to the upcoming DESI-DR2 analysis effort, which will include production of an updated set of lightcone mocks with improved HOD modelling of DESI tracers.

We present the first configuration-space constraints from DESI full-shape two-point correlation function fits, and find them to be consistent with the fiducial Fourier-space analysis, with the greatest discrepancy observed in the preferred power spectrum amplitude. We measure the amplitude parameters $\sigma_{12}=0.750\pm0.036$ and $\ln(10^{10}A_{\rm{s}}) = 2.86\pm0.11$, which correspond to a $\sim 1.8\sigma$ lower amplitude compared to the $\ln(10^{10}A_{\rm{s}})$ value obtained by the power spectrum fits to the full DESI data set. Nonetheless, we find that the maximum posterior point corresponds to a $\sim1.2\sigma$ higher value of $\ln(10^{10}A_{\rm{s}})$ than the marginalised posterior mean, indicating presence of some projection effects.  

We furthermore find that our clustering-only constraints are consistent with shear and galaxy-galaxy lensing measurements from all three weak lensing surveys considered in this work. We find that adding cross-correlations with weak lensing improves the constraints for $\ln(10^{10}A_{\rm{s}})$ and $\sigma_{12}$ by $15\%$ and $36\%$, respectively. We also find that shear information improves the measurement of linear bias with respect to the clustering-only case by $15-20\%$. We confirm that the clustering-only bias measurements are consistent with the results obtained from joint shear and galaxy-galaxy lensing fits, although we see some variation with the lensing survey considered, suggesting cosmic variance effects or potential sensitivity to varying completeness across the DESI-DR1 footprint. 

Our final shear$\times$RSD constraints are consistent across all weak lensing surveys, and are in excellent agreement with the cosmological analyses published by each lensing collaboration, as well as with the results of our companion $3\times$2-pt analysis by \cite{AnnaDESI}, with comparable precision. Our measured DESI$\times$DES amplitude, $\sigma_{8}=0.793\pm0.22$, lies well within 1$\sigma$ of the result obtained from DESI+DES fits performed as part of the DESI-DR1 fiducial full-shape analysis \citep{desi_fs}.  Our preferred value for $\sigma_{12}$ for DESI$\times$KiDS-1000 is similarly in good agreement with the KiDS$\times$BOSS+eBOSS measurement \citep{2021A&A...646A.140H}.

In this work, in order to establish the base validation of our pipelines, we limit ourselves to $\Lambda$CDM fits to LSS observables, and do not consider any extensions beyond $\Lambda$CDM or combinations with CMB or supernovae measurements. Whilst we were able to provide the first validation for our setup, further tests are needed to understand the origin of the deviations in the power spectrum amplitude parameters seen on the {\sc Abacus} mocks. In addition to improvements in mock HOD realism to ensure accurate representation of the clustering samples, a key missing ingredient in our current validation setup is the effect of intrinsic alignments. As IA affects the amplitude of the shear signal, modelling this effect is important for accurate recovery of $S_8$. Furthermore, the amplitude of the alignment can be a significant source of projection effects, which requires further investigation.

Whilst we do not anticipate very significant projection effects within $\Lambda$CDM model fits, we nonetheless found offsets of around $1\sigma$ for clustering-only fits in our current setup. In order to perform extended model analyses, it is, therefore, imperative to explore potential mitigating techniques, such as informative priors or sampling schemes using parametrisations that avoid sampling in degenerate directions. Combining galaxy clustering with additional probes is another way to limit projection effects, and weak lensing has been explored as a potential option \citep{tsedrik2025a}. The improvement in the power spectrum amplitude and bias constraints that we found in this work demonstrates that weak lensing is a promising probe, that could potentially allow us to mitigate against projection effects using complementary measurements for the redshift ranges probed by spectroscopic surveys.  We leave a more detailed exploration of projection effects in shear$\times$RSD analyses for future work.

DESI-DR2 BAO results \citep{2025PhRvD.112h3514A, 2025PhRvD.112h3515A} provided hints of a preference for evolving dark energy models, when DESI observations are combined with measurements from other cosmological probes, underlying the importance of multi-probe analyses. The combination of BAO and DES-Y3 $3\times$2-pt measurements alone shows a preference for the $w_0w_a\mathrm{CDM}$ model over the standard $\Lambda$CDM, at the level of around 2.2$\sigma$. While this is only an initial result, it showcases the power of large-scale structure measurements to provide reliable constraints without requiring external probes. The shear$\times$RSD analysis presented in this work shows that joint galaxy clustering and weak lensing configuration space fits provide precise and consistent cosmological constraints across the cosmological parameter space. It, therefore, serves as an important validation of the key cosmological results, as well as a powerful tool to measure the properties of large-scale structure. 

\section*{Acknowledgements}

We thank Samuel Brieden, Joe DeRose, Alejandro Pérez Fernández and Martin White for helpful discussions and comments.

AS and CB acknowledge financial support from Australian Research Council Discovery Project DP220101609. AP acknowledges financial support from the European Union's Marie Skłodowska-Curie grant agreement 101068581, and from the \textit{César Nombela} Research Talent Attraction grant from the Community of Madrid (Ref. 2023-T1/TEC-29011).

This material is based upon work supported by the U.S. Department of Energy (DOE), Office of Science, Office of High-Energy Physics, under Contract No. DE–AC02–05CH11231, and by the National Energy Research Scientific Computing Center, a DOE Office of Science User Facility under the same contract. Additional support for DESI was provided by the U.S. National Science Foundation (NSF), Division of Astronomical Sciences under Contract No. AST-0950945 to the NSF’s National Optical-Infrared Astronomy Research Laboratory; the Science and Technology Facilities Council of the United Kingdom; the Gordon and Betty Moore Foundation; the Heising-Simons Foundation; the French Alternative Energies and Atomic Energy Commission (CEA); the National Council of Humanities, Science and Technology of Mexico (CONAHCYT); the Ministry of Science, Innovation and Universities of Spain (MICIU/AEI/10.13039/501100011033), and by the DESI Member Institutions: \url{https://www.desi.lbl.gov/collaborating-institutions}. Any opinions, findings, and conclusions or recommendations expressed in this material are those of the author(s) and do not necessarily reflect the views of the U. S. National Science Foundation, the U. S. Department of Energy, or any of the listed funding agencies.

The authors are honored to be permitted to conduct scientific research on I'oligam Du'ag (Kitt Peak), a mountain with particular significance to the Tohono O’odham Nation.

This research made use of the following python packages in addition to those already cited in the manuscript: {\sc astropy} \citep{astropy}, {\sc numpy} \citep{numpy}, {\sc scipy} \citep{scipy} and {\sc matplotlib} \citep{matplotlib}.

\section*{Data Availability}

Data points for all the figures are available at \url{https://doi.org/10.5281/zenodo.17934989}

\bibliographystyle{mnras}
\bibliography{example}

@ARTICLE{2015MNRAS.447..234W,
       author = {{White}, Martin and {Reid}, Beth and {Chuang}, Chia-Hsun and {Tinker}, Jeremy L. and {McBride}, Cameron K. and {Prada}, Francisco and {Samushia}, Lado},
        title = "{Tests of redshift-space distortions models in configuration space for the analysis of the BOSS final data release}",
      journal = {\mnras},
         year = 2015,
        month = feb,
       volume = {447},
       number = {1},
        pages = {234-245},
          doi = {10.1093/mnras/stu2460},
archivePrefix = {arXiv},
       eprint = {1408.5435},
 primaryClass = {astro-ph.CO},
       adsurl = {https://ui.adsabs.harvard.edu/abs/2015MNRAS.447..234W}
}

@ARTICLE{2016MNRAS.457.1577G,
       author = {{Grieb}, Jan Niklas and {S{\'a}nchez}, Ariel G. and {Salazar-Albornoz}, Salvador and {Dalla Vecchia}, Claudio},
        title = "{Gaussian covariance matrices for anisotropic galaxy clustering measurements}",
      journal = {\mnras},
         year = 2016,
        month = apr,
       volume = {457},
       number = {2},
        pages = {1577-1592},
          doi = {10.1093/mnras/stw065},
archivePrefix = {arXiv},
       eprint = {1509.04293},
 primaryClass = {astro-ph.CO},
       adsurl = {https://ui.adsabs.harvard.edu/abs/2016MNRAS.457.1577G}
}

@ARTICLE{2024MNRAS.533..589Y,
       author = {{Yuan}, Sihan and {Blake}, Chris and {Krolewski}, Alex and {Lange}, Johannes and {Elvin-Poole}, Jack and {Leauthaud}, Alexie and {DeRose}, Joseph and {Aguilar}, Jessica Nicole and {Ahlen}, Steven and {Beltz-Mohrmann}, Gillian and {Brooks}, David and {Claybaugh}, Todd and {de la Macorra}, Axel and {Doel}, Peter and {Emas}, Ni Putu Audita Placida and {Ferraro}, Simone and {Forero-Romero}, Jaime E. and {Garcia-Quintero}, Cristhian and {Gazta{\~n}aga}, Enrique and {Gontcho}, Satya Gontcho A. and {Hadzhiyska}, Boryana and {Heydenreich}, Sven and {Honscheid}, Klaus and {Ishak}, Mustapha and {Joudaki}, Shahab and {Jullo}, Eric and {Kisner}, Theodore and {Kremin}, Anthony and {Lambert}, Andrew and {Landriau}, Martin and {Manera}, Marc and {Meisner}, Aaron and {Miquel}, Ramon and {Nie}, Jundan and {Palanque-Delabrouille}, Nathalie and {Poppett}, Claire and {Porredon}, Anna and {Rezaie}, Mehdi and {Ross}, Ashley J. and {Rossi}, Graziano and {Ruggeri}, Rossana and {Sanchez}, Eusebio and {Saulder}, Christoph and {Seo}, Hee-Jong and {Silber}, Joseph Harry and {Tarl{\'n}}, Gregory and {Vargas-Maga{\~n}a}, Mariana and {Weaver}, Benjamin Alan and {Xhakaj}, Enia and {Zhou}, Zhimin and {Zou}, Hu},
        title = "{Redshift evolution and covariances for joint lensing and clustering studies with DESI Y1}",
      journal = {\mnras},
         year = 2024,
        month = sep,
       volume = {533},
       number = {1},
        pages = {589-607},
          doi = {10.1093/mnras/stae1792},
archivePrefix = {arXiv},
       eprint = {2403.00915},
 primaryClass = {astro-ph.CO},
       adsurl = {https://ui.adsabs.harvard.edu/abs/2024MNRAS.533..589Y}
}

@ARTICLE{2025OJAp....8E..24B,
       author = {{Blake}, Chris and {Garcia-Quintero}, C. and {Ahlen}, S. and {Bianchi}, D. and {Brooks}, D. and {Claybaugh}, T. and {de la Macorra}, A. and {DeRose}, J. and {Dey}, A. and {Doel}, P. and {Emas}, N. and {Ferraro}, S. and {Forero-Romero}, J.~E. and {Gutierrez}, G. and {Heydenreich}, S. and {Honscheid}, K. and {Howlett}, C. and {Ishak}, M. and {Joudaki}, S. and {Jullo}, E. and {Kehoe}, R. and {Kirkby}, D. and {Kremin}, A. and {Krolewski}, A. and {Landriau}, M. and {Lange}, J.~U. and {Leauthaud}, A. and {Levi}, M.~E. and {Manera}, M. and {Miquel}, R. and {Moustakas}, J. and {Niz}, G. and {Percival}, W.~J. and {P{\'e}rez-R{\`a}fols}, I. and {Porredon}, A. and {Rossi}, G. and {Ruggeri}, R. and {Sanchez}, E. and {Saulder}, C. and {Schlegel}, D. and {Sprayberry}, D. and {Sun}, Z. and {Tarl{\'e}}, G. and {Weaver}, B.~A.},
        title = "{The DESI-Lensing Mock Challenge: large-scale cosmological analysis of 3x2-pt statistics}",
      journal = {The Open Journal of Astrophysics},
         year = 2025,
        month = mar,
       volume = {8},
          eid = {24},
        pages = {24},
          doi = {10.33232/001c.131903},
archivePrefix = {arXiv},
       eprint = {2412.12548},
 primaryClass = {astro-ph.CO},
       adsurl = {https://ui.adsabs.harvard.edu/abs/2025OJAp....8E..24B}
}

@ARTICLE{2022PhRvD.106f3536T,
       author = {{Taylor}, Peter L. and {Markovi{\v{c}}}, Katarina},
        title = "{Covariance of photometric and spectroscopic two-point statistics: Implications for cosmological parameter inference}",
      journal = {\prd},
         year = 2022,
        month = sep,
       volume = {106},
       number = {6},
          eid = {063536},
        pages = {063536},
          doi = {10.1103/PhysRevD.106.063536},
archivePrefix = {arXiv},
       eprint = {2205.14167},
 primaryClass = {astro-ph.CO},
       adsurl = {https://ui.adsabs.harvard.edu/abs/2022PhRvD.106f3536T}
}

@ARTICLE{2024AJ....167...62D,
       author = {{DESI Collaboration} and {Adame}, A.~G. and {Aguilar}, J. and {Ahlen}, S. and {Alam}, S. and {Aldering}, G. and {Alexander}, D.~M. and {Alfarsy}, R. and {Allende Prieto}, C. and {Alvarez}, M. and {Alves}, O. and {Anand}, A. and {Andrade-Oliveira}, F. and {Armengaud}, E. and {Asorey}, J. and {Avila}, S. and {Aviles}, A. and {Bailey}, S. and {Balaguera-Antol{\'\i}nez}, A. and {Ballester}, O. and {Baltay}, C. and {Bault}, A. and {Bautista}, J. and {Behera}, J. and {Beltran}, S.~F. and {BenZvi}, S. and {Beraldo e Silva}, L. and {Bermejo-Climent}, J.~R. and {Berti}, A. and {Besuner}, R. and {Beutler}, F. and {Bianchi}, D. and {Blake}, C. and {Blum}, R. and {Bolton}, A.~S. and {Brieden}, S. and {Brodzeller}, A. and {Brooks}, D. and {Brown}, Z. and {Buckley-Geer}, E. and {Burtin}, E. and {Cabayol-Garcia}, L. and {Cai}, Z. and {Canning}, R. and {Cardiel-Sas}, L. and {Carnero Rosell}, A. and {Castander}, F.~J. and {Cervantes-Cota}, J.~L. and {Chabanier}, S. and {Chaussidon}, E. and {Chaves-Montero}, J. and {Chen}, S. and {Chen}, X. and {Chuang}, C. and {Claybaugh}, T. and {Cole}, S. and {Cooper}, A.~P. and {Cuceu}, A. and {Davis}, T.~M. and {Dawson}, K. and {de Belsunce}, R. and {de la Cruz}, R. and {de la Macorra}, A. and {de Mattia}, A. and {Demina}, R. and {Demirbozan}, U. and {DeRose}, J. and {Dey}, A. and {Dey}, B. and {Dhungana}, G. and {Ding}, J. and {Ding}, Z. and {Doel}, P. and {Doshi}, R. and {Douglass}, K. and {Edge}, A. and {Eftekharzadeh}, S. and {Eisenstein}, D.~J. and {Elliott}, A. and {Escoffier}, S. and {Fagrelius}, P. and {Fan}, X. and {Fanning}, K. and {Fawcett}, V.~A. and {Ferraro}, S. and {Ereza}, J. and {Flaugher}, B. and {Font-Ribera}, A. and {Forero-S{\'a}nchez}, D. and {Forero-Romero}, J.~E. and {Frenk}, C.~S. and {G{\"a}nsicke}, B.~T. and {Garc{\'\i}a}, L. {\'A}. and {Garc{\'\i}a-Bellido}, J. and {Garcia-Quintero}, C. and {Garrison}, L.~H. and {Gil-Mar{\'\i}n}, H. and {Golden-Marx}, J. and {Gontcho A Gontcho}, S. and {Gonzalez-Morales}, A.~X. and {Gonzalez-Perez}, V. and {Gordon}, C. and {Graur}, O. and {Green}, D. and {Gruen}, D. and {Guy}, J. and {Hadzhiyska}, B. and {Hahn}, C. and {Han}, J.~J. and {Hanif}, M.~M.~S. and {Herrera-Alcantar}, H.~K. and {Honscheid}, K. and {Hou}, J. and {Howlett}, C. and {Huterer}, D. and {Ir{\v{s}}i{\v{c}}}, V. and {Ishak}, M. and {Jana}, A. and {Jiang}, L. and {Jimenez}, J. and {Jing}, Y.~P. and {Joudaki}, S. and {Jullo}, E. and {Joyce}, R. and {Juneau}, S. and {Kizhuprakkat}, N. and {Kara{\c{c}}ayl{\i}}, N.~G. and {Karim}, T. and {Kehoe}, R. and {Kent}, S. and {Khederlarian}, A. and {Kim}, S. and {Kirkby}, D. and {Kisner}, T. and {Kitaura}, F. and {Kneib}, J. and {Koposov}, S.~E. and {Kov{\'a}cs}, A. and {Kremin}, A. and {Krolewski}, A. and {L'Huillier}, B. and {Lahav}, O. and {Lambert}, A. and {Lamman}, C. and {Lan}, T. -W. and {Landriau}, M. and {Lang}, D. and {Lange}, J.~U. and {Lasker}, J. and {Le Guillou}, L. and {Leauthaud}, A. and {Levi}, M.~E. and {Li}, T.~S. and {Linder}, E. and {Lyons}, A. and {Magneville}, C. and {Manera}, M. and {Manser}, C.~J. and {Margala}, D. and {Martini}, P. and {McDonald}, P. and {Medina}, G.~E. and {Medina-Varela}, L. and {Meisner}, A. and {Mena-Fern{\'a}ndez}, J. and {Meneses-Rizo}, J. and {Mezcua}, M. and {Miquel}, R. and {Montero-Camacho}, P. and {Moon}, J. and {Moore}, S. and {Moustakas}, J. and {Mueller}, E. and {Mundet}, J. and {Mu{\~n}oz-Guti{\'e}rrez}, A. and {Myers}, A.~D. and {Nadathur}, S. and {Napolitano}, L. and {Neveux}, R. and {Newman}, J.~A. and {Nie}, J. and {Niz}, G. and {Norberg}, P. and {Noriega}, H.~E. and {Paillas}, E. and {Palanque-Delabrouille}, N. and {Palmese}, A. and {Zhiwei}, P. and {Parkinson}, D. and {Penmetsa}, S. and {Percival}, W.~J. and {P{\'e}rez-Fern{\'a}ndez}, A. and {P{\'e}rez-R{\`a}fols}, I. and {Pieri}, M. and {Poppett}, C. and {Porredon}, A. and {Prada}, F. and {Pucha}, R. and {Raichoor}, A. and {Ram{\'\i}rez-P{\'e}rez}, C.},
        title = "{Validation of the Scientific Program for the Dark Energy Spectroscopic Instrument}",
      journal = {\aj},
         year = 2024,
        month = feb,
       volume = {167},
       number = {2},
          eid = {62},
        pages = {62},
          doi = {10.3847/1538-3881/ad0b08},
archivePrefix = {arXiv},
       eprint = {2306.06307},
 primaryClass = {astro-ph.CO},
       adsurl = {https://ui.adsabs.harvard.edu/abs/2024AJ....167...62D}
}

@ARTICLE{2024AJ....168..245P,
       author = {{Poppett}, Claire and {Tyas}, Luke and {Aguilar}, J. and {Bebek}, Christopher and {Bramall}, D. and {Claybaugh}, T. and {Edelstein}, J. and {Fagrelius}, P. and {Heetderks}, H. and {Jelinsky}, P. and {Jelinsky}, S. and {Lafever}, Robin and {Lambert}, A. and {Lampton}, M. and {Levi}, Michael E. and {Martini}, P. and {Rockosi}, C. and {Schmoll}, J. and {Sharples}, Ray M. and {Sirk}, Martin and {Wishnow}, Edward and {Yu}, Jiaxi and {Ahlen}, S. and {Bault}, A. and {BenZvi}, S. and {Brooks}, D. and {Cole}, S. and {de la Macorra}, A. and {Dey}, Arjun and {Doel}, P. and {Fanning}, K. and {Font-Ribera}, A. and {Forero-Romero}, J.~E. and {Gazta{\~n}aga}, E. and {Gontcho A Gontcho}, S. and {Gonzalez-Morales}, A.~X. and {Hahn}, C. and {Honscheid}, K. and {Jimenez}, J. and {Juneau}, S. and {Kirkby}, D. and {Kremin}, A. and {Landriau}, M. and {Le Guillou}, L. and {Manera}, M. and {Meisner}, A. and {Miquel}, R. and {Moustakas}, J. and {Mueller}, E. and {Mu{\~n}oz-Guti{\'e}rrez}, A. and {Myers}, A.~D. and {Nie}, J. and {Niz}, G. and {Palanque-Delabrouille}, N. and {Percival}, W.~J. and {Prada}, F. and {Rabinowitz}, D. and {Rezaie}, M. and {Rossi}, G. and {Sanchez}, E. and {Schlafly}, Edward F. and {Schlegel}, D. and {Schubnell}, M. and {Seo}, H. and {Sprayberry}, D. and {Tarl{\'e}}, G. and {Vargas-Maga{\~n}a}, M. and {Weaver}, B.~A. and {Zhou}, R.},
        title = "{Overview of the Fiber System for the Dark Energy Spectroscopic Instrument}",
      journal = {\aj},
         year = 2024,
        month = dec,
       volume = {168},
       number = {6},
          eid = {245},
        pages = {245},
          doi = {10.3847/1538-3881/ad76a4},
       adsurl = {https://ui.adsabs.harvard.edu/abs/2024AJ....168..245P}
}

@ARTICLE{2023AJ....165....9S,
       author = {{Silber}, Joseph Harry and {Fagrelius}, Parker and {Fanning}, Kevin and {Schubnell}, Michael and {Aguilar}, Jessica Nicole and {Ahlen}, Steven and {Ameel}, Jon and {Ballester}, Otger and {Baltay}, Charles and {Bebek}, Chris and {Benton Beard}, Dominic and {Besuner}, Robert and {Cardiel-Sas}, Laia and {Casas}, Ricard and {Castander}, Francisco Javier and {Claybaugh}, Todd and {Dobson}, Carl and {Duan}, Yutong and {Dunlop}, Patrick and {Edelstein}, Jerry and {Emmet}, William T. and {Elliott}, Ann and {Evatt}, Matthew and {Gershkovich}, Irena and {Guy}, Julien and {Harris}, Stu and {Heetderks}, Henry and {Heetderks}, Ian and {Honscheid}, Klaus and {Illa}, Jose Maria and {Jelinsky}, Patrick and {Jelinsky}, Sharon R. and {Jimenez}, Jorge and {Karcher}, Armin and {Kent}, Stephen and {Kirkby}, David and {Kneib}, Jean-Paul and {Lambert}, Andrew and {Lampton}, Mike and {Leitner}, Daniela and {Levi}, Michael and {McCauley}, Jeremy and {Meisner}, Aaron and {Miller}, Timothy N. and {Miquel}, Ramon and {Mundet}, Juli{\'a} and {Poppett}, Claire and {Rabinowitz}, David and {Reil}, Kevin and {Roman}, David and {Schlegel}, David and {Serrano}, Santiago and {Van Shourt}, William and {Sprayberry}, David and {Tarl{\'e}}, Gregory and {Tie}, Suk Sien and {Weaverdyck}, Curtis and {Zhang}, Kai and {Azzaro}, Marco and {Bailey}, Stephen and {Becerril}, Santiago and {Blackwell}, Tami and {Bouri}, Mohamed and {Brooks}, David and {Buckley-Geer}, Elizabeth and {Castro}, Jose Pe{\~n}ate and {Derwent}, Mark and {Dey}, Arjun and {Dhungana}, Govinda and {Doel}, Peter and {Eisenstein}, Daniel J. and {Fahim}, Nasib and {Garcia-Bellido}, Juan and {Gazta{\~n}aga}, Enrique and {A Gontcho}, Satya Gontcho and {Gutierrez}, Gaston and {H{\"o}rler}, Philipp and {Kehoe}, Robert and {Kisner}, Theodore and {Kremin}, Anthony and {Kronig}, Luzius and {Landriau}, Martin and {Le Guillou}, Laurent and {Martini}, Paul and {Moustakas}, John and {Palanque-Delabrouille}, Nathalie and {Peng}, Xiyan and {Percival}, Will and {Prada}, Francisco and {Allende Prieto}, Carlos and {de Rivera}, Guillermo Gonzalez and {Sanchez}, Eusebio and {Sanchez}, Justo and {Sharples}, Ray and {Soares-Santos}, Marcelle and {Schlafly}, Edward and {Weaver}, Benjamin Alan and {Zhou}, Zhimin and {Zhu}, Yaling and {Zou}, Hu and {DESI Collaboration}},
        title = "{The Robotic Multiobject Focal Plane System of the Dark Energy Spectroscopic Instrument (DESI)}",
      journal = {\aj},
         year = 2023,
        month = jan,
       volume = {165},
       number = {1},
          eid = {9},
        pages = {9},
          doi = {10.3847/1538-3881/ac9ab1},
archivePrefix = {arXiv},
       eprint = {2205.09014},
 primaryClass = {astro-ph.IM},
       adsurl = {https://ui.adsabs.harvard.edu/abs/2023AJ....165....9S}
}

@ARTICLE{2023AJ....165..144G,
       author = {{Guy}, J. and {Bailey}, S. and {Kremin}, A. and {Alam}, Shadab and {Alexander}, D.~M. and {Allende Prieto}, C. and {BenZvi}, S. and {Bolton}, A.~S. and {Brooks}, D. and {Chaussidon}, E. and {Cooper}, A.~P. and {Dawson}, K. and {de la Macorra}, A. and {Dey}, A. and {Dey}, Biprateep and {Dhungana}, G. and {Eisenstein}, D.~J. and {Font-Ribera}, A. and {Forero-Romero}, J.~E. and {Gazta{\~n}aga}, E. and {Gontcho A Gontcho}, S. and {Green}, D. and {Honscheid}, K. and {Ishak}, M. and {Kehoe}, R. and {Kirkby}, D. and {Kisner}, T. and {Koposov}, Sergey E. and {Lan}, Ting-Wen and {Landriau}, M. and {Le Guillou}, L. and {Levi}, Michael E. and {Magneville}, C. and {Manser}, Christopher J. and {Martini}, P. and {Meisner}, Aaron M. and {Miquel}, R. and {Moustakas}, J. and {Myers}, Adam D. and {Newman}, Jeffrey A. and {Nie}, Jundan and {Palanque-Delabrouille}, N. and {Percival}, W.~J. and {Poppett}, C. and {Prada}, F. and {Raichoor}, A. and {Ravoux}, C. and {Ross}, A.~J. and {Schlafly}, E.~F. and {Schlegel}, D. and {Schubnell}, M. and {Sharples}, Ray M. and {Tarl{\'e}}, Gregory and {Weaver}, B.~A. and {Y{\'e}che}, Christophe and {Zhou}, Rongpu and {Zhou}, Zhimin and {Zou}, H.},
        title = "{The Spectroscopic Data Processing Pipeline for the Dark Energy Spectroscopic Instrument}",
      journal = {\aj},
         year = 2023,
        month = apr,
       volume = {165},
       number = {4},
          eid = {144},
        pages = {144},
          doi = {10.3847/1538-3881/acb212},
archivePrefix = {arXiv},
       eprint = {2209.14482},
 primaryClass = {astro-ph.IM},
       adsurl = {https://ui.adsabs.harvard.edu/abs/2023AJ....165..144G}
}

@ARTICLE{2023AJ....166..259S,
       author = {{Schlafly}, Edward F. and {Kirkby}, David and {Schlegel}, David J. and {Myers}, Adam D. and {Raichoor}, Anand and {Dawson}, Kyle and {Aguilar}, Jessica and {Allende Prieto}, Carlos and {Bailey}, Stephen and {BenZvi}, Segev and {Bermejo-Climent}, Jose and {Brooks}, David and {de la Macorra}, Axel and {Dey}, Arjun and {Doel}, Peter and {Fanning}, Kevin and {Font-Ribera}, Andreu and {Forero-Romero}, Jaime E. and {Garc{\'\i}a-Bellido}, Juan and {Gontcho A Gontcho}, Satya and {Guy}, Julien and {Hahn}, ChangHoon and {Honscheid}, Klaus and {Ishak}, Mustapha and {Juneau}, St{\'e}phanie and {Kehoe}, Robert and {Kisner}, Theodore and {Kremin}, Anthony and {Landriau}, Martin and {Lang}, Dustin A. and {Lasker}, James and {Levi}, Michael E. and {Magneville}, Christophe and {Manser}, Christopher J. and {Martini}, Paul and {Meisner}, Aaron M. and {Miquel}, Ramon and {Moustakas}, John and {Newman}, Jeffrey A. and {Nie}, Jundan and {Palanque-Delabrouille}, Nathalie. and {Percival}, Will J. and {Poppett}, Claire and {Rockosi}, Constance and {Ross}, Ashley J. and {Rossi}, Graziano and {Tarl{\'e}}, Gregory and {Weaver}, Benjamin A. and {Y{\`e}che}, Christophe and {Zhou}, Rongpu and {DESI Collaboration}},
        title = "{Survey Operations for the Dark Energy Spectroscopic Instrument}",
      journal = {\aj},
         year = 2023,
        month = dec,
       volume = {166},
       number = {6},
          eid = {259},
        pages = {259},
          doi = {10.3847/1538-3881/ad0832},
archivePrefix = {arXiv},
       eprint = {2306.06309},
 primaryClass = {astro-ph.CO},
       adsurl = {https://ui.adsabs.harvard.edu/abs/2023AJ....166..259S}
}

@ARTICLE{2023AJ....165..253H,
       author = {{Hahn}, ChangHoon and {Wilson}, Michael J. and {Ruiz-Macias}, Omar and {Cole}, Shaun and {Weinberg}, David H. and {Moustakas}, John and {Kremin}, Anthony and {Tinker}, Jeremy L. and {Smith}, Alex and {Wechsler}, Risa H. and {Ahlen}, Steven and {Alam}, Shadab and {Bailey}, Stephen and {Brooks}, David and {Cooper}, Andrew P. and {Davis}, Tamara M. and {Dawson}, Kyle and {Dey}, Arjun and {Dey}, Biprateep and {Eftekharzadeh}, Sarah and {Eisenstein}, Daniel J. and {Fanning}, Kevin and {Forero-Romero}, Jaime E. and {Frenk}, Carlos S. and {Gazta{\~n}aga}, Enrique and {A Gontcho}, Satya Gontcho and {Guy}, Julien and {Honscheid}, Klaus and {Ishak}, Mustapha and {Juneau}, St{\'e}phanie and {Kehoe}, Robert and {Kisner}, Theodore and {Lan}, Ting-Wen and {Landriau}, Martin and {Le Guillou}, Laurent and {Levi}, Michael E. and {Magneville}, Christophe and {Martini}, Paul and {Meisner}, Aaron and {Myers}, Adam D. and {Nie}, Jundan and {Norberg}, Peder and {Palanque-Delabrouille}, Nathalie and {Percival}, Will J. and {Poppett}, Claire and {Prada}, Francisco and {Raichoor}, Anand and {Ross}, Ashley J. and {Gaines}, Sasha and {Saulder}, Christoph and {Schlafly}, Eddie and {Schlegel}, David and {Sierra-Porta}, David and {Tarle}, Gregory and {Weaver}, Benjamin A. and {Y{\`e}che}, Christophe and {Zarrouk}, Pauline and {Zhou}, Rongpu and {Zhou}, Zhimin and {Zou}, Hu},
        title = "{The DESI Bright Galaxy Survey: Final Target Selection, Design, and Validation}",
      journal = {\aj},
         year = 2023,
        month = jun,
       volume = {165},
       number = {6},
          eid = {253},
        pages = {253},
          doi = {10.3847/1538-3881/accff8},
archivePrefix = {arXiv},
       eprint = {2208.08512},
 primaryClass = {astro-ph.CO},
       adsurl = {https://ui.adsabs.harvard.edu/abs/2023AJ....165..253H}
}

@ARTICLE{2023AJ....165...58Z,
       author = {{Zhou}, Rongpu and {Dey}, Biprateep and {Newman}, Jeffrey A. and {Eisenstein}, Daniel J. and {Dawson}, K. and {Bailey}, S. and {Berti}, A. and {Guy}, J. and {Lan}, Ting-Wen and {Zou}, H. and {Aguilar}, J. and {Ahlen}, S. and {Alam}, Shadab and {Brooks}, D. and {de la Macorra}, A. and {Dey}, A. and {Dhungana}, G. and {Fanning}, K. and {Font-Ribera}, A. and {Gontcho}, S. Gontcho A. and {Honscheid}, K. and {Ishak}, Mustapha and {Kisner}, T. and {Kov{\'a}cs}, A. and {Kremin}, A. and {Landriau}, M. and {Levi}, Michael E. and {Magneville}, C. and {Manera}, Marc and {Martini}, P. and {Meisner}, Aaron M. and {Miquel}, R. and {Moustakas}, J. and {Myers}, Adam D. and {Nie}, Jundan and {Palanque-Delabrouille}, N. and {Percival}, W.~J. and {Poppett}, C. and {Prada}, F. and {Raichoor}, A. and {Ross}, A.~J. and {Schlafly}, E. and {Schlegel}, D. and {Schubnell}, M. and {Tarl{\'e}}, Gregory and {Weaver}, B.~A. and {Wechsler}, R.~H. and {Y{\'e}che}, Christophe and {Zhou}, Zhimin},
        title = "{Target Selection and Validation of DESI Luminous Red Galaxies}",
      journal = {\aj},
         year = 2023,
        month = feb,
       volume = {165},
       number = {2},
          eid = {58},
        pages = {58},
          doi = {10.3847/1538-3881/aca5fb},
archivePrefix = {arXiv},
       eprint = {2208.08515},
 primaryClass = {astro-ph.CO},
       adsurl = {https://ui.adsabs.harvard.edu/abs/2023AJ....165...58Z}
}

@ARTICLE{2022AJ....164..207D,
       author = {{DESI Collaboration} and {Abareshi}, B. and {Aguilar}, J. and {Ahlen}, S. and {Alam}, Shadab and {Alexander}, David M. and {Alfarsy}, R. and {Allen}, L. and {Allende Prieto}, C. and {Alves}, O. and {Ameel}, J. and {Armengaud}, E. and {Asorey}, J. and {Aviles}, Alejandro and {Bailey}, S. and {Balaguera-Antol{\'\i}nez}, A. and {Ballester}, O. and {Baltay}, C. and {Bault}, A. and {Beltran}, S.~F. and {Benavides}, B. and {BenZvi}, S. and {Berti}, A. and {Besuner}, R. and {Beutler}, Florian and {Bianchi}, D. and {Blake}, C. and {Blanc}, P. and {Blum}, R. and {Bolton}, A. and {Bose}, S. and {Bramall}, D. and {Brieden}, S. and {Brodzeller}, A. and {Brooks}, D. and {Brownewell}, C. and {Buckley-Geer}, E. and {Cahn}, R.~N. and {Cai}, Z. and {Canning}, R. and {Capasso}, R. and {Carnero Rosell}, A. and {Carton}, P. and {Casas}, R. and {Castander}, F.~J. and {Cervantes-Cota}, J.~L. and {Chabanier}, S. and {Chaussidon}, E. and {Chuang}, C. and {Circosta}, C. and {Cole}, S. and {Cooper}, A.~P. and {da Costa}, L. and {Cousinou}, M. -C. and {Cuceu}, A. and {Davis}, T.~M. and {Dawson}, K. and {de la Cruz-Noriega}, R. and {de la Macorra}, A. and {de Mattia}, A. and {Della Costa}, J. and {Demmer}, P. and {Derwent}, M. and {Dey}, A. and {Dey}, B. and {Dhungana}, G. and {Ding}, Z. and {Dobson}, C. and {Doel}, P. and {Donald-McCann}, J. and {Donaldson}, J. and {Douglass}, K. and {Duan}, Y. and {Dunlop}, P. and {Edelstein}, J. and {Eftekharzadeh}, S. and {Eisenstein}, D.~J. and {Enriquez-Vargas}, M. and {Escoffier}, S. and {Evatt}, M. and {Fagrelius}, P. and {Fan}, X. and {Fanning}, K. and {Fawcett}, V.~A. and {Ferraro}, S. and {Ereza}, J. and {Flaugher}, B. and {Font-Ribera}, A. and {Forero-Romero}, J.~E. and {Frenk}, C.~S. and {Fromenteau}, S. and {G{\"a}nsicke}, B.~T. and {Garcia-Quintero}, C. and {Garrison}, L. and {Gazta{\~n}aga}, E. and {Gerardi}, F. and {Gil-Mar{\'\i}n}, H. and {Gontcho A Gontcho}, S. and {Gonzalez-Morales}, Alma X. and {Gonzalez-de-Rivera}, G. and {Gonzalez-Perez}, V. and {Gordon}, C. and {Graur}, O. and {Green}, D. and {Grove}, C. and {Gruen}, D. and {Gutierrez}, G. and {Guy}, J. and {Hahn}, C. and {Harris}, S. and {Herrera}, D. and {Herrera-Alcantar}, Hiram K. and {Honscheid}, K. and {Howlett}, C. and {Huterer}, D. and {Ir{\v{s}}i{\v{c}}}, V. and {Ishak}, M. and {Jelinsky}, P. and {Jiang}, L. and {Jimenez}, J. and {Jing}, Y.~P. and {Joyce}, R. and {Jullo}, E. and {Juneau}, S. and {Kara{\c{c}}ayl{\i}}, N.~G. and {Karamanis}, M. and {Karcher}, A. and {Karim}, T. and {Kehoe}, R. and {Kent}, S. and {Kirkby}, D. and {Kisner}, T. and {Kitaura}, F. and {Koposov}, S.~E. and {Kov{\'a}cs}, A. and {Kremin}, A. and {Krolewski}, Alex and {L'Huillier}, B. and {Lahav}, O. and {Lambert}, A. and {Lamman}, C. and {Lan}, Ting-Wen and {Landriau}, M. and {Lane}, S. and {Lang}, D. and {Lange}, J.~U. and {Lasker}, J. and {Le Guillou}, L. and {Leauthaud}, A. and {Le Van Suu}, A. and {Levi}, Michael E. and {Li}, T.~S. and {Magneville}, C. and {Manera}, M. and {Manser}, Christopher J. and {Marshall}, B. and {Martini}, Paul and {McCollam}, W. and {McDonald}, P. and {Meisner}, Aaron M. and {Mena-Fern{\'a}ndez}, J. and {Meneses-Rizo}, J. and {Mezcua}, M. and {Miller}, T. and {Miquel}, R. and {Montero-Camacho}, P. and {Moon}, J. and {Moustakas}, J. and {Mueller}, E. and {Mu{\~n}oz-Guti{\'e}rrez}, Andrea and {Myers}, Adam D. and {Nadathur}, S. and {Najita}, J. and {Napolitano}, L. and {Neilsen}, E. and {Newman}, Jeffrey A. and {Nie}, J.~D. and {Ning}, Y. and {Niz}, G. and {Norberg}, P. and {Noriega}, Hern{\'a}n E. and {O'Brien}, T. and {Obuljen}, A. and {Palanque-Delabrouille}, N. and {Palmese}, A. and {Zhiwei}, P. and {Pappalardo}, D. and {PENG}, X. and {Percival}, W.~J. and {Perruchot}, S. and {Pogge}, R. and {Poppett}, C. and {Porredon}, A. and {Prada}, F. and {Prochaska}, J. and {Pucha}, R. and {P{\'e}rez-Fern{\'a}ndez}, A. and {P{\'e}rez-R{\`a}fols}, I. and {Rabinowitz}, D. and {Raichoor}, A.},
        title = "{Overview of the Instrumentation for the Dark Energy Spectroscopic Instrument}",
      journal = {\aj},
         year = 2022,
        month = nov,
       volume = {164},
       number = {5},
          eid = {207},
        pages = {207},
          doi = {10.3847/1538-3881/ac882b},
archivePrefix = {arXiv},
       eprint = {2205.10939},
 primaryClass = {astro-ph.IM},
       adsurl = {https://ui.adsabs.harvard.edu/abs/2022AJ....164..207D}
}

@ARTICLE{2025arXiv250314745D,
       author = {{DESI Collaboration} and {Abdul-Karim}, M. and {Adame}, A.~G. and {Aguado}, D. and {Aguilar}, J. and {Ahlen}, S. and {Alam}, S. and {Aldering}, G. and {Alexander}, D.~M. and {Alfarsy}, R. and {Allen}, L. and {Allende Prieto}, C. and {Alves}, O. and {Anand}, A. and {Andrade}, U. and {Armengaud}, E. and {Avila}, S. and {Aviles}, A. and {Awan}, H. and {Bailey}, S. and {Baleato Lizancos}, A. and {Ballester}, O. and {Bault}, A. and {Bautista}, J. and {BenZvi}, S. and {Beraldo e Silva}, L. and {Bermejo-Climent}, J.~R. and {Beutler}, F. and {Bianchi}, D. and {Blake}, C. and {Blum}, R. and {Bolton}, A.~S. and {Bonici}, M. and {Brieden}, S. and {Brodzeller}, A. and {Brooks}, D. and {Buckley-Geer}, E. and {Burtin}, E. and {Canning}, R. and {Carnero Rosell}, A. and {Carr}, A. and {Carrilho}, P. and {Casas}, L. and {Castander}, F.~J. and {Cereskaite}, R. and {Cervantes-Cota}, J.~L. and {Chaussidon}, E. and {Chaves-Montero}, J. and {Chen}, S. and {Chen}, X. and {Claybaugh}, T. and {Cole}, S. and {Cooper}, A.~P. and {Cousinou}, M. -C. and {Cuceu}, A. and {Davis}, T.~M. and {Dawson}, K.~S. and {de Belsunce}, R. and {de la Cruz}, R. and {de la Macorra}, A. and {de Mattia}, A. and {Deiosso}, N. and {Della Costa}, J. and {Demina}, R. and {Demirbozan}, U. and {DeRose}, J. and {Dey}, A. and {Dey}, B. and {Ding}, J. and {Ding}, Z. and {Doel}, P. and {Douglass}, K. and {Dowicz}, M. and {Ebina}, H. and {Edelstein}, J. and {Eisenstein}, D.~J. and {Elbers}, W. and {Emas}, N. and {Escoffier}, S. and {Fagrelius}, P. and {Fan}, X. and {Fanning}, K. and {Fawcett}, V.~A. and {Fern\textbackslash'andez-Garc\textbackslash'ia}, E. and {Ferraro}, S. and {Findlay}, N. and {Font-Ribera}, A. and {Forero-Romero}, J.~E. and {Forero-S\textbackslash'anchez}, D. and {Frenk}, C.~S. and {G\textbackslash''ansicke}, B.~T. and {Galbany}, L. and {Garc\textbackslash'ia-Bellido}, J. and {Garcia-Quintero}, C. and {Garrison}, L.~H. and {Gazta\textbackslash\raisebox{-0.5ex}\textasciitildenaga}, E. and {Gil-Mar\textbackslash'in}, H. and {Gnedin}, O.~Y. and {Gontcho}, S. Gontcho A and {Gonzalez-Morales}, A.~X. and {Gonzalez-Perez}, V. and {Gordon}, C. and {Graur}, O. and {Green}, D. and {Gruen}, D. and {Gsponer}, R. and {Guandalin}, C. and {Gutierrez}, G. and {Guy}, J. and {Hahn}, C. and {Han}, J.~J. and {Han}, J. and {He}, S. and {Herrera-Alcantar}, H.~K. and {Honscheid}, K. and {Hou}, J. and {Howlett}, C. and {Huterer}, D. and {Ir\textbackslashv\{s\}i\textbackslashv\{c\}}, V. and {Ishak}, M. and {Jacques}, A. and {Jimenez}, J. and {Jing}, Y.~P. and {Joachimi}, B. and {Joudaki}, S. and {Joyce}, R. and {Jullo}, E. and {Juneau}, S. and {Kara\textbackslashc\{c\}ayl\{\textbackslashi\}}, N.~G. and {Karim}, T. and {Kehoe}, R. and {Kent}, S. and {Khederlarian}, A. and {Kirkby}, D. and {Kisner}, T. and {Kitaura}, F. -S. and {Kizhuprakkat}, N. and {Kong}, H. and {Koposov}, S.~E. and {Kremin}, A. and {Krolewski}, A. and {Lahav}, O. and {Lai}, Y. and {Lamman}, C. and {Lan}, T. -W. and {Landriau}, M. and {Lang}, D. and {Lange}, J.~U. and {Lasker}, J. and {Le Goff}, J.~M. and {Le Guillou}, L. and {Leauthaud}, A. and {Levi}, M.~E. and {Li}, S. and {Li}, T.~S. and {Lodha}, K. and {Lokken}, M. and {Luo}, Y. and {Magneville}, C. and {Manera}, M. and {Manser}, C.~J. and {Margala}, D. and {Martini}, P. and {Maus}, M. and {McCullough}, J. and {McDonald}, P. and {Medina}, G.~E. and {Medina-Varela}, L. and {Meisner}, A. and {Mena-Fern\textbackslash'andez}, J. and {Menegas}, A. and {Mezcua}, M. and {Miquel}, R. and {Montero-Camacho}, P. and {Moon}, J. and {Moustakas}, J. and {Mu\textbackslash\raisebox{-0.5ex}\textasciitildenoz-Guti\textbackslash'errez}, A. and {Mu\textbackslash\raisebox{-0.5ex}\textasciitildenoz-Santos}, D. and {Myers}, A.~D. and {Myles}, J. and {Nadathur}, S. and {Najita}, J. and {Napolitano}, L. and {Newman}, J.~A. and {Nikakhtar}, F. and {Nikutta}, R. and {Niz}, G. and {Noriega}, H.~E. and {Padmanabhan}, N. and {Paillas}, E. and {Palanque-Delabrouille}, N. and {Palmese}, A. and {Pan}, J. and {Pan}, Z. and {Parkinson}, D. and {Peacock}, J. and {Percival}, W.~J. and {P\textbackslash'erez-Fern\textbackslash'andez}, A. and {P\textbackslash'erez-R\textbackslash`afols}, I. and {Peterson}, P.},
        title = "{Data Release 1 of the Dark Energy Spectroscopic Instrument}",
      journal = {arXiv e-prints},
         year = 2025,
        month = mar,
          eid = {arXiv:2503.14745},
        pages = {arXiv:2503.14745},
          doi = {10.48550/arXiv.2503.14745},
archivePrefix = {arXiv},
       eprint = {2503.14745},
 primaryClass = {astro-ph.CO},
       adsurl = {https://ui.adsabs.harvard.edu/abs/2025arXiv250314745D}
}

@ARTICLE{2019AJ....157..168D,
       author = {{Dey}, Arjun and {Schlegel}, David J. and {Lang}, Dustin and {Blum}, Robert and {Burleigh}, Kaylan and {Fan}, Xiaohui and {Findlay}, Joseph R. and {Finkbeiner}, Doug and {Herrera}, David and {Juneau}, St{\'e}phanie and {Landriau}, Martin and {Levi}, Michael and {McGreer}, Ian and {Meisner}, Aaron and {Myers}, Adam D. and {Moustakas}, John and {Nugent}, Peter and {Patej}, Anna and {Schlafly}, Edward F. and {Walker}, Alistair R. and {Valdes}, Francisco and {Weaver}, Benjamin A. and {Y{\`e}che}, Christophe and {Zou}, Hu and {Zhou}, Xu and {Abareshi}, Behzad and {Abbott}, T.~M.~C. and {Abolfathi}, Bela and {Aguilera}, C. and {Alam}, Shadab and {Allen}, Lori and {Alvarez}, A. and {Annis}, James and {Ansarinejad}, Behzad and {Aubert}, Marie and {Beechert}, Jacqueline and {Bell}, Eric F. and {BenZvi}, Segev Y. and {Beutler}, Florian and {Bielby}, Richard M. and {Bolton}, Adam S. and {Brice{\~n}o}, C{\'e}sar and {Buckley-Geer}, Elizabeth J. and {Butler}, Karen and {Calamida}, Annalisa and {Carlberg}, Raymond G. and {Carter}, Paul and {Casas}, Ricard and {Castander}, Francisco J. and {Choi}, Yumi and {Comparat}, Johan and {Cukanovaite}, Elena and {Delubac}, Timoth{\'e}e and {DeVries}, Kaitlin and {Dey}, Sharmila and {Dhungana}, Govinda and {Dickinson}, Mark and {Ding}, Zhejie and {Donaldson}, John B. and {Duan}, Yutong and {Duckworth}, Christopher J. and {Eftekharzadeh}, Sarah and {Eisenstein}, Daniel J. and {Etourneau}, Thomas and {Fagrelius}, Parker A. and {Farihi}, Jay and {Fitzpatrick}, Mike and {Font-Ribera}, Andreu and {Fulmer}, Leah and {G{\"a}nsicke}, Boris T. and {Gaztanaga}, Enrique and {George}, Koshy and {Gerdes}, David W. and {Gontcho}, Satya Gontcho A. and {Gorgoni}, Claudio and {Green}, Gregory and {Guy}, Julien and {Harmer}, Diane and {Hernandez}, M. and {Honscheid}, Klaus and {Huang}, Lijuan Wendy and {James}, David J. and {Jannuzi}, Buell T. and {Jiang}, Linhua and {Joyce}, Richard and {Karcher}, Armin and {Karkar}, Sonia and {Kehoe}, Robert and {Kneib}, Jean-Paul and {Kueter-Young}, Andrea and {Lan}, Ting-Wen and {Lauer}, Tod R. and {Le Guillou}, Laurent and {Le Van Suu}, Auguste and {Lee}, Jae Hyeon and {Lesser}, Michael and {Perreault Levasseur}, Laurence and {Li}, Ting S. and {Mann}, Justin L. and {Marshall}, Robert and {Mart{\'\i}nez-V{\'a}zquez}, C.~E. and {Martini}, Paul and {du Mas des Bourboux}, H{\'e}lion and {McManus}, Sean and {Meier}, Tobias Gabriel and {M{\'e}nard}, Brice and {Metcalfe}, Nigel and {Mu{\~n}oz-Guti{\'e}rrez}, Andrea and {Najita}, Joan and {Napier}, Kevin and {Narayan}, Gautham and {Newman}, Jeffrey A. and {Nie}, Jundan and {Nord}, Brian and {Norman}, Dara J. and {Olsen}, Knut A.~G. and {Paat}, Anthony and {Palanque-Delabrouille}, Nathalie and {Peng}, Xiyan and {Poppett}, Claire L. and {Poremba}, Megan R. and {Prakash}, Abhishek and {Rabinowitz}, David and {Raichoor}, Anand and {Rezaie}, Mehdi and {Robertson}, A.~N. and {Roe}, Natalie A. and {Ross}, Ashley J. and {Ross}, Nicholas P. and {Rudnick}, Gregory and {Safonova}, Sasha and {Saha}, Abhijit and {S{\'a}nchez}, F. Javier and {Savary}, Elodie and {Schweiker}, Heidi and {Scott}, Adam and {Seo}, Hee-Jong and {Shan}, Huanyuan and {Silva}, David R. and {Slepian}, Zachary and {Soto}, Christian and {Sprayberry}, David and {Staten}, Ryan and {Stillman}, Coley M. and {Stupak}, Robert J. and {Summers}, David L. and {Sien Tie}, Suk and {Tirado}, H. and {Vargas-Maga{\~n}a}, Mariana and {Vivas}, A. Katherina and {Wechsler}, Risa H. and {Williams}, Doug and {Yang}, Jinyi and {Yang}, Qian and {Yapici}, Tolga and {Zaritsky}, Dennis and {Zenteno}, A. and {Zhang}, Kai and {Zhang}, Tianmeng and {Zhou}, Rongpu and {Zhou}, Zhimin},
        title = "{Overview of the DESI Legacy Imaging Surveys}",
      journal = {\aj},
         year = 2019,
        month = may,
       volume = {157},
       number = {5},
          eid = {168},
        pages = {168},
          doi = {10.3847/1538-3881/ab089d},
archivePrefix = {arXiv},
       eprint = {1804.08657},
 primaryClass = {astro-ph.IM},
       adsurl = {https://ui.adsabs.harvard.edu/abs/2019AJ....157..168D}
}

@ARTICLE{2025JCAP...07..017A,
       author = {{DESI Collaboration} and {Adame}, A.~G. and {Aguilar}, J. and {Ahlen}, S. and {Alam}, S. and {Alexander}, D.~M. and {Alvarez}, M. and {Alves}, O. and {Anand}, A. and {Andrade}, U. and {Armengaud}, E. and {Avila}, S. and {Aviles}, A. and {Awan}, H. and {Bailey}, S. and {Baltay}, C. and {Bault}, A. and {Behera}, J. and {BenZvi}, S. and {Beutler}, F. and {Bianchi}, D. and {Blake}, C. and {Blum}, R. and {Brieden}, S. and {Brodzeller}, A. and {Brooks}, D. and {Brown}, Z. and {Buckley-Geer}, E. and {Burtin}, E. and {Calderon}, R. and {Canning}, R. and {Carnero Rosell}, A. and {Cereskaite}, R. and {Cervantes-Cota}, J.~L. and {Chabanier}, S. and {Chaussidon}, E. and {Chaves-Montero}, J. and {Chen}, S. and {Chen}, X. and {Claybaugh}, T. and {Cole}, S. and {Cuceu}, A. and {Davis}, T.~M. and {Dawson}, K. and {de la Macorra}, A. and {de Mattia}, A. and {Deiosso}, N. and {Demina}, R. and {Dey}, A. and {Dey}, B. and {Ding}, Z. and {Doel}, P. and {Edelstein}, J. and {Eftekharzadeh}, S. and {Eisenstein}, D.~J. and {Elliott}, A. and {Fagrelius}, P. and {Fanning}, K. and {Ferraro}, S. and {Ereza}, J. and {Findlay}, N. and {Flaugher}, B. and {Font-Ribera}, A. and {Forero-S{\'a}nchez}, D. and {Forero-Romero}, J.~E. and {Frenk}, C.~S. and {Garcia-Quintero}, C. and {Gazta{\~n}aga}, E. and {Gil-Mar{\'\i}n}, H. and {Gontcho}, S. Gontcho A. and {Gonzalez-Morales}, A.~X. and {Gonzalez-Perez}, V. and {Gordon}, C. and {Green}, D. and {Gruen}, D. and {Gsponer}, R. and {Gutierrez}, G. and {Guy}, J. and {Hadzhiyska}, B. and {Hahn}, C. and {Hanif}, M.~M.~S. and {Herrera-Alcantar}, H.~K. and {Honscheid}, K. and {Hou}, J. and {Howlett}, C. and {Huterer}, D. and {Ir{\v{s}}i{\v{c}}}, V. and {Ishak}, M. and {Juneau}, S. and {Kara{\c{c}}ayl{\i}}, N.~G. and {Kehoe}, R. and {Kent}, S. and {Kirkby}, D. and {Kitaura}, F. -S. and {Kong}, H. and {Kremin}, A. and {Krolewski}, A. and {Lai}, Y. and {Lan}, T. -W. and {Landriau}, M. and {Lang}, D. and {Lasker}, J. and {Le Goff}, J.~M. and {Le Guillou}, L. and {Leauthaud}, A. and {Levi}, M.~E. and {Li}, T.~S. and {Lodha}, K. and {Magneville}, C. and {Manera}, M. and {Margala}, D. and {Martini}, P. and {Maus}, M. and {McDonald}, P. and {Medina-Varela}, L. and {Meisner}, A. and {Mena-Fern{\'a}ndez}, J. and {Miquel}, R. and {Moon}, J. and {Moore}, S. and {Moustakas}, J. and {Mudur}, N. and {Mueller}, E. and {Mu{\~n}oz-Guti{\'e}rrez}, A. and {Myers}, A.~D. and {Nadathur}, S. and {Napolitano}, L. and {Neveux}, R. and {Newman}, J.~A. and {Nguyen}, N.~M. and {Nie}, J. and {Niz}, G. and {Noriega}, H.~E. and {Padmanabhan}, N. and {Paillas}, E. and {Palanque-Delabrouille}, N. and {Pan}, J. and {Penmetsa}, S. and {Percival}, W.~J. and {Pieri}, M.~M. and {Pinon}, M. and {Poppett}, C. and {Porredon}, A. and {Prada}, F. and {P{\'e}rez-Fern{\'a}ndez}, A. and {P{\'e}rez-R{\`a}fols}, I. and {Rabinowitz}, D. and {Raichoor}, A. and {Ram{\'\i}rez-P{\'e}rez}, C. and {Ramirez-Solano}, S. and {Rashkovetskyi}, M. and {Ravoux}, C. and {Rezaie}, M. and {Rich}, J. and {Rocher}, A. and {Rockosi}, C. and {Roe}, N.~A. and {Rosado-Marin}, A. and {Ross}, A.~J. and {Rossi}, G. and {Ruggeri}, R. and {Ruhlmann-Kleider}, V. and {Samushia}, L. and {Sanchez}, E. and {Saulder}, C. and {Schlafly}, E.~F. and {Schlegel}, D. and {Scholte}, D. and {Schubnell}, M. and {Seo}, H. and {Sharples}, R. and {Silber}, J. and {Slosar}, A. and {Smith}, A. and {Sprayberry}, D. and {Tan}, T. and {Tarl{\'e}}, G. and {Trusov}, S. and {Vaisakh}, R. and {Valcin}, D. and {Valdes}, F. and {Vargas-Maga{\~n}a}, M. and {Verde}, L. and {Walther}, M. and {Wang}, B. and {Wang}, M.~S. and {Weaver}, B.~A. and {Weaverdyck}, N. and {Wechsler}, R.~H. and {Weinberg}, D.~H. and {White}, M. and {Wilson}, M.~J. and {Yu}, J. and {Yu}, Y. and {Yuan}, S. and {Y{\`e}che}, C. and {Zaborowski}, E.~A. and {Zarrouk}, P. and {Zhang}, H. and {Zhao}, C. and {Zhao}, R.},
        title = "{DESI 2024 II: sample definitions, characteristics, and two-point clustering statistics}",
      journal = {\jcap},
         year = 2025,
        month = jul,
       volume = {2025},
       number = {7},
          eid = {017},
        pages = {017},
          doi = {10.1088/1475-7516/2025/07/017},
archivePrefix = {arXiv},
       eprint = {2411.12020},
 primaryClass = {astro-ph.CO},
       adsurl = {https://ui.adsabs.harvard.edu/abs/2025JCAP...07..017A}
}

@ARTICLE{2025JCAP...02..021A,
       author = {{DESI Collaboration} and {Adame}, A.~G. and {Aguilar}, J. and {Ahlen}, S. and {Alam}, S. and {Alexander}, D.~M. and {Alvarez}, M. and {Alves}, O. and {Anand}, A. and {Andrade}, U. and {Armengaud}, E. and {Avila}, S. and {Aviles}, A. and {Awan}, H. and {Bahr-Kalus}, B. and {Bailey}, S. and {Baltay}, C. and {Bault}, A. and {Behera}, J. and {BenZvi}, S. and {Bera}, A. and {Beutler}, F. and {Bianchi}, D. and {Blake}, C. and {Blum}, R. and {Brieden}, S. and {Brodzeller}, A. and {Brooks}, D. and {Buckley-Geer}, E. and {Burtin}, E. and {Calderon}, R. and {Canning}, R. and {Carnero Rosell}, A. and {Cereskaite}, R. and {Cervantes-Cota}, J.~L. and {Chabanier}, S. and {Chaussidon}, E. and {Chaves-Montero}, J. and {Chen}, S. and {Chen}, X. and {Claybaugh}, T. and {Cole}, S. and {Cuceu}, A. and {Davis}, T.~M. and {Dawson}, K. and {de la Macorra}, A. and {de Mattia}, A. and {Deiosso}, N. and {Dey}, A. and {Dey}, B. and {Ding}, Z. and {Doel}, P. and {Edelstein}, J. and {Eftekharzadeh}, S. and {Eisenstein}, D.~J. and {Elliott}, A. and {Fagrelius}, P. and {Fanning}, K. and {Ferraro}, S. and {Ereza}, J. and {Findlay}, N. and {Flaugher}, B. and {Font-Ribera}, A. and {Forero-S{\'a}nchez}, D. and {Forero-Romero}, J.~E. and {Frenk}, C.~S. and {Garcia-Quintero}, C. and {Gazta{\~n}aga}, E. and {Gil-Mar{\'\i}n}, H. and {Gontcho a Gontcho}, S. and {Gonzalez-Morales}, A.~X. and {Gonzalez-Perez}, V. and {Gordon}, C. and {Green}, D. and {Gruen}, D. and {Gsponer}, R. and {Gutierrez}, G. and {Guy}, J. and {Hadzhiyska}, B. and {Hahn}, C. and {Hanif}, M.~M.~S. and {Herrera-Alcantar}, H.~K. and {Honscheid}, K. and {Howlett}, C. and {Huterer}, D. and {Ir{\v{s}}i{\v{c}}}, V. and {Ishak}, M. and {Juneau}, S. and {Kara{\c{c}}ayl{\i}}, N.~G. and {Kehoe}, R. and {Kent}, S. and {Kirkby}, D. and {Kremin}, A. and {Krolewski}, A. and {Lai}, Y. and {Lan}, T.-W. and {Landriau}, M. and {Lang}, D. and {Lasker}, J. and {Le Goff}, J.~M. and {Le Guillou}, L. and {Leauthaud}, A. and {Levi}, M.~E. and {Li}, T.~S. and {Linder}, E. and {Lodha}, K. and {Magneville}, C. and {Manera}, M. and {Margala}, D. and {Martini}, P. and {Maus}, M. and {McDonald}, P. and {Medina-Varela}, L. and {Meisner}, A. and {Mena-Fern{\'a}ndez}, J. and {Miquel}, R. and {Moon}, J. and {Moore}, S. and {Moustakas}, J. and {Mueller}, E. and {Mu{\~n}oz-Guti{\'e}rrez}, A. and {Myers}, A.~D. and {Nadathur}, S. and {Napolitano}, L. and {Neveux}, R. and {Newman}, J.~A. and {Nguyen}, N.~M. and {Nie}, J. and {Niz}, G. and {Noriega}, H.~E. and {Padmanabhan}, N. and {Paillas}, E. and {Palanque-Delabrouille}, N. and {Pan}, J. and {Penmetsa}, S. and {Percival}, W.~J. and {Pieri}, M.~M. and {Pinon}, M. and {Poppett}, C. and {Porredon}, A. and {Prada}, F. and {P{\'e}rez-Fern{\'a}ndez}, A. and {P{\'e}rez-R{\`a}fols}, I. and {Rabinowitz}, D. and {Raichoor}, A. and {Ram{\'\i}rez-P{\'e}rez}, C. and {Ramirez-Solano}, S. and {Rashkovetskyi}, M. and {Ravoux}, C. and {Rezaie}, M. and {Rich}, J. and {Rocher}, A. and {Rockosi}, C. and {Roe}, N.~A. and {Rosado-Marin}, A. and {Ross}, A.~J. and {Rossi}, G. and {Ruggeri}, R. and {Ruhlmann-Kleider}, V. and {Samushia}, L. and {Sanchez}, E. and {Saulder}, C. and {Schlafly}, E.~F. and {Schlegel}, D. and {Schubnell}, M. and {Seo}, H. and {Shafieloo}, A. and {Sharples}, R. and {Silber}, J. and {Slosar}, A. and {Smith}, A. and {Sprayberry}, D. and {Tan}, T. and {Tarl{\'e}}, G. and {Taylor}, P. and {Trusov}, S. and {Ure{\~n}a-L{\'o}pez}, L.~A. and {Vaisakh}, R. and {Valcin}, D. and {Valdes}, F. and {Vargas-Maga{\~n}a}, M. and {Verde}, L. and {Walther}, M. and {Wang}, B. and {Wang}, M.~S. and {Weaver}, B.~A. and {Weaverdyck}, N. and {Wechsler}, R.~H. and {Weinberg}, D.~H. and {White}, M. and {Yu}, J. and {Yu}, Y. and {Yuan}, S. and {Y{\`e}che}, C. and {Zaborowski}, E.~A. and {Zarrouk}, P. and {Zhang}, H. and {Zhao}, C. and {Zhao}, R. and {Zhou}, R. and {Zhuang}, T.},
        title = "{DESI 2024 VI: cosmological constraints from the measurements of baryon acoustic oscillations}",
      journal = {\jcap},
         year = 2025,
        month = feb,
       volume = {2025},
       number = {2},
          eid = {021},
        pages = {021},
          doi = {10.1088/1475-7516/2025/02/021},
archivePrefix = {arXiv},
       eprint = {2404.03002},
 primaryClass = {astro-ph.CO},
       adsurl = {https://ui.adsabs.harvard.edu/abs/2025JCAP...02..021A}
}

@ARTICLE{2025arXiv250621677H,
       author = {{Heydenreich}, S. and {Leauthaud}, A. and {Blake}, C. and {Sun}, Z. and {Lange}, J.~U. and {Zhang}, T. and {DeMartino}, M. and {Ross}, A.~J. and {Aguilar}, J. and {Ahlen}, S. and {Bianchi}, D. and {Brooks}, D. and {Castander}, F.~J. and {Claybaugh}, T. and {Cuceu}, A. and {de la Macorra}, A. and {DeRose}, J. and {Dey}, Arjun and {Dey}, Biprateep and {Doel}, P. and {Emas}, N. and {Ferraro}, S. and {Font-Ribera}, A. and {Forero-Romero}, J.~E. and {Garcia-Quintero}, C. and {Gazta{\~n}aga}, E. and {Gontcho}, S. Gontcho A and {Gutierrez}, G. and {Hadzhiyska}, B. and {Honscheid}, K. and {Huterer}, D. and {Ishak}, M. and {Jeffrey}, N. and {Joudaki}, S. and {Jullo}, E. and {Juneau}, S. and {Kirkby}, D. and {Kisner}, T. and {Kremin}, A. and {Krolewski}, A. and {Lahav}, O. and {Lamman}, C. and {Landriau}, M. and {Le Guillou}, L. and {Manera}, M. and {Meisner}, A. and {Miquel}, R. and {Nadathur}, S. and {Palanque-Delabrouille}, N. and {Percival}, W.~J. and {Porredon}, A. and {Prada}, F. and {P{\'e}rez-R{\`a}fols}, I. and {Rossi}, G. and {Ruggeri}, R. and {Sanchez}, E. and {Saulder}, C. and {Schlegel}, D. and {Semenaite}, A. and {Silber}, J. and {Sprayberry}, D. and {Tarl{\'e}}, G. and {Weaver}, B.~A. and {Yuan}, S. and {Zarrouk}, P. and {Zhou}, R. and {Zou}, H.},
        title = "{Lensing Without Borders: Measurements of galaxy-galaxy lensing and projected galaxy clustering in DESI DR1}",
      journal = {arXiv e-prints},
         year = 2025,
        month = jun,
          eid = {arXiv:2506.21677},
        pages = {arXiv:2506.21677},
          doi = {10.48550/arXiv.2506.21677},
archivePrefix = {arXiv},
       eprint = {2506.21677},
 primaryClass = {astro-ph.CO},
       adsurl = {https://ui.adsabs.harvard.edu/abs/2025arXiv250621677H}
}

@ARTICLE{2016arXiv161100036D,
       author = {{DESI Collaboration} and {Aghamousa}, Amir and {Aguilar}, Jessica and {Ahlen}, Steve and {Alam}, Shadab and {Allen}, Lori E. and {Allende Prieto}, Carlos and {Annis}, James and {Bailey}, Stephen and {Balland}, Christophe and {Ballester}, Otger and {Baltay}, Charles and {Beaufore}, Lucas and {Bebek}, Chris and {Beers}, Timothy C. and {Bell}, Eric F. and {Bernal}, Jos{\'e} Luis and {Besuner}, Robert and {Beutler}, Florian and {Blake}, Chris and {Bleuler}, Hannes and {Blomqvist}, Michael and {Blum}, Robert and {Bolton}, Adam S. and {Briceno}, Cesar and {Brooks}, David and {Brownstein}, Joel R. and {Buckley-Geer}, Elizabeth and {Burden}, Angela and {Burtin}, Etienne and {Busca}, Nicolas G. and {Cahn}, Robert N. and {Cai}, Yan-Chuan and {Cardiel-Sas}, Laia and {Carlberg}, Raymond G. and {Carton}, Pierre-Henri and {Casas}, Ricard and {Castander}, Francisco J. and {Cervantes-Cota}, Jorge L. and {Claybaugh}, Todd M. and {Close}, Madeline and {Coker}, Carl T. and {Cole}, Shaun and {Comparat}, Johan and {Cooper}, Andrew P. and {Cousinou}, M. -C. and {Crocce}, Martin and {Cuby}, Jean-Gabriel and {Cunningham}, Daniel P. and {Davis}, Tamara M. and {Dawson}, Kyle S. and {de la Macorra}, Axel and {De Vicente}, Juan and {Delubac}, Timoth{\'e}e and {Derwent}, Mark and {Dey}, Arjun and {Dhungana}, Govinda and {Ding}, Zhejie and {Doel}, Peter and {Duan}, Yutong T. and {Ealet}, Anne and {Edelstein}, Jerry and {Eftekharzadeh}, Sarah and {Eisenstein}, Daniel J. and {Elliott}, Ann and {Escoffier}, St{\'e}phanie and {Evatt}, Matthew and {Fagrelius}, Parker and {Fan}, Xiaohui and {Fanning}, Kevin and {Farahi}, Arya and {Farihi}, Jay and {Favole}, Ginevra and {Feng}, Yu and {Fernandez}, Enrique and {Findlay}, Joseph R. and {Finkbeiner}, Douglas P. and {Fitzpatrick}, Michael J. and {Flaugher}, Brenna and {Flender}, Samuel and {Font-Ribera}, Andreu and {Forero-Romero}, Jaime E. and {Fosalba}, Pablo and {Frenk}, Carlos S. and {Fumagalli}, Michele and {Gaensicke}, Boris T. and {Gallo}, Giuseppe and {Garcia-Bellido}, Juan and {Gaztanaga}, Enrique and {Pietro Gentile Fusillo}, Nicola and {Gerard}, Terry and {Gershkovich}, Irena and {Giannantonio}, Tommaso and {Gillet}, Denis and {Gonzalez-de-Rivera}, Guillermo and {Gonzalez-Perez}, Violeta and {Gott}, Shelby and {Graur}, Or and {Gutierrez}, Gaston and {Guy}, Julien and {Habib}, Salman and {Heetderks}, Henry and {Heetderks}, Ian and {Heitmann}, Katrin and {Hellwing}, Wojciech A. and {Herrera}, David A. and {Ho}, Shirley and {Holland}, Stephen and {Honscheid}, Klaus and {Huff}, Eric and {Hutchinson}, Timothy A. and {Huterer}, Dragan and {Hwang}, Ho Seong and {Illa Laguna}, Joseph Maria and {Ishikawa}, Yuzo and {Jacobs}, Dianna and {Jeffrey}, Niall and {Jelinsky}, Patrick and {Jennings}, Elise and {Jiang}, Linhua and {Jimenez}, Jorge and {Johnson}, Jennifer and {Joyce}, Richard and {Jullo}, Eric and {Juneau}, St{\'e}phanie and {Kama}, Sami and {Karcher}, Armin and {Karkar}, Sonia and {Kehoe}, Robert and {Kennamer}, Noble and {Kent}, Stephen and {Kilbinger}, Martin and {Kim}, Alex G. and {Kirkby}, David and {Kisner}, Theodore and {Kitanidis}, Ellie and {Kneib}, Jean-Paul and {Koposov}, Sergey and {Kovacs}, Eve and {Koyama}, Kazuya and {Kremin}, Anthony and {Kron}, Richard and {Kronig}, Luzius and {Kueter-Young}, Andrea and {Lacey}, Cedric G. and {Lafever}, Robin and {Lahav}, Ofer and {Lambert}, Andrew and {Lampton}, Michael and {Landriau}, Martin and {Lang}, Dustin and {Lauer}, Tod R. and {Le Goff}, Jean-Marc and {Le Guillou}, Laurent and {Le Van Suu}, Auguste and {Lee}, Jae Hyeon and {Lee}, Su-Jeong and {Leitner}, Daniela and {Lesser}, Michael and {Levi}, Michael E. and {L'Huillier}, Benjamin and {Li}, Baojiu and {Liang}, Ming and {Lin}, Huan and {Linder}, Eric and {Loebman}, Sarah R. and {Luki{\'c}}, Zarija and {Ma}, Jun and {MacCrann}, Niall and {Magneville}, Christophe and {Makarem}, Laleh and {Manera}, Marc and {Manser}, Christopher J. and {Marshall}, Robert and {Martini}, Paul and {Massey}, Richard and {Matheson}, Thomas and {McCauley}, Jeremy and {McDonald}, Patrick and {McGreer}, Ian D. and {Meisner}, Aaron and {Metcalfe}, Nigel and {Miller}, Timothy N. and {Miquel}, Ramon and {Moustakas}, John and {Myers}, Adam and {Naik}, Milind and {Newman}, Jeffrey A. and {Nichol}, Robert C. and {Nicola}, Andrina and {Nicolati da Costa}, Luiz and {Nie}, Jundan and {Niz}, Gustavo and {Norberg}, Peder and {Nord}, Brian and {Norman}, Dara and {Nugent}, Peter and {O'Brien}, Thomas and {Oh}, Minji and {Olsen}, Knut A.~G.},
        title = "{The DESI Experiment Part I: Science,Targeting, and Survey Design}",
      journal = {arXiv e-prints},
         year = 2016,
        month = oct,
          eid = {arXiv:1611.00036},
        pages = {arXiv:1611.00036},
          doi = {10.48550/arXiv.1611.00036},
archivePrefix = {arXiv},
       eprint = {1611.00036},
 primaryClass = {astro-ph.IM},
       adsurl = {https://ui.adsabs.harvard.edu/abs/2016arXiv161100036D}
}

@ARTICLE{2016arXiv161100037D,
       author = {{DESI Collaboration} and {Aghamousa}, Amir and {Aguilar}, Jessica and {Ahlen}, Steve and {Alam}, Shadab and {Allen}, Lori E. and {Allende Prieto}, Carlos and {Annis}, James and {Bailey}, Stephen and {Balland}, Christophe and {Ballester}, Otger and {Baltay}, Charles and {Beaufore}, Lucas and {Bebek}, Chris and {Beers}, Timothy C. and {Bell}, Eric F. and {Bernal}, Jos{\'e} Luis and {Besuner}, Robert and {Beutler}, Florian and {Blake}, Chris and {Bleuler}, Hannes and {Blomqvist}, Michael and {Blum}, Robert and {Bolton}, Adam S. and {Briceno}, Cesar and {Brooks}, David and {Brownstein}, Joel R. and {Buckley-Geer}, Elizabeth and {Burden}, Angela and {Burtin}, Etienne and {Busca}, Nicolas G. and {Cahn}, Robert N. and {Cai}, Yan-Chuan and {Cardiel-Sas}, Laia and {Carlberg}, Raymond G. and {Carton}, Pierre-Henri and {Casas}, Ricard and {Castander}, Francisco J. and {Cervantes-Cota}, Jorge L. and {Claybaugh}, Todd M. and {Close}, Madeline and {Coker}, Carl T. and {Cole}, Shaun and {Comparat}, Johan and {Cooper}, Andrew P. and {Cousinou}, M. -C. and {Crocce}, Martin and {Cuby}, Jean-Gabriel and {Cunningham}, Daniel P. and {Davis}, Tamara M. and {Dawson}, Kyle S. and {de la Macorra}, Axel and {De Vicente}, Juan and {Delubac}, Timoth{\'e}e and {Derwent}, Mark and {Dey}, Arjun and {Dhungana}, Govinda and {Ding}, Zhejie and {Doel}, Peter and {Duan}, Yutong T. and {Ealet}, Anne and {Edelstein}, Jerry and {Eftekharzadeh}, Sarah and {Eisenstein}, Daniel J. and {Elliott}, Ann and {Escoffier}, St{\'e}phanie and {Evatt}, Matthew and {Fagrelius}, Parker and {Fan}, Xiaohui and {Fanning}, Kevin and {Farahi}, Arya and {Farihi}, Jay and {Favole}, Ginevra and {Feng}, Yu and {Fernandez}, Enrique and {Findlay}, Joseph R. and {Finkbeiner}, Douglas P. and {Fitzpatrick}, Michael J. and {Flaugher}, Brenna and {Flender}, Samuel and {Font-Ribera}, Andreu and {Forero-Romero}, Jaime E. and {Fosalba}, Pablo and {Frenk}, Carlos S. and {Fumagalli}, Michele and {Gaensicke}, Boris T. and {Gallo}, Giuseppe and {Garcia-Bellido}, Juan and {Gaztanaga}, Enrique and {Pietro Gentile Fusillo}, Nicola and {Gerard}, Terry and {Gershkovich}, Irena and {Giannantonio}, Tommaso and {Gillet}, Denis and {Gonzalez-de-Rivera}, Guillermo and {Gonzalez-Perez}, Violeta and {Gott}, Shelby and {Graur}, Or and {Gutierrez}, Gaston and {Guy}, Julien and {Habib}, Salman and {Heetderks}, Henry and {Heetderks}, Ian and {Heitmann}, Katrin and {Hellwing}, Wojciech A. and {Herrera}, David A. and {Ho}, Shirley and {Holland}, Stephen and {Honscheid}, Klaus and {Huff}, Eric and {Hutchinson}, Timothy A. and {Huterer}, Dragan and {Hwang}, Ho Seong and {Illa Laguna}, Joseph Maria and {Ishikawa}, Yuzo and {Jacobs}, Dianna and {Jeffrey}, Niall and {Jelinsky}, Patrick and {Jennings}, Elise and {Jiang}, Linhua and {Jimenez}, Jorge and {Johnson}, Jennifer and {Joyce}, Richard and {Jullo}, Eric and {Juneau}, St{\'e}phanie and {Kama}, Sami and {Karcher}, Armin and {Karkar}, Sonia and {Kehoe}, Robert and {Kennamer}, Noble and {Kent}, Stephen and {Kilbinger}, Martin and {Kim}, Alex G. and {Kirkby}, David and {Kisner}, Theodore and {Kitanidis}, Ellie and {Kneib}, Jean-Paul and {Koposov}, Sergey and {Kovacs}, Eve and {Koyama}, Kazuya and {Kremin}, Anthony and {Kron}, Richard and {Kronig}, Luzius and {Kueter-Young}, Andrea and {Lacey}, Cedric G. and {Lafever}, Robin and {Lahav}, Ofer and {Lambert}, Andrew and {Lampton}, Michael and {Landriau}, Martin and {Lang}, Dustin and {Lauer}, Tod R. and {Le Goff}, Jean-Marc and {Le Guillou}, Laurent and {Le Van Suu}, Auguste and {Lee}, Jae Hyeon and {Lee}, Su-Jeong and {Leitner}, Daniela and {Lesser}, Michael and {Levi}, Michael E. and {L'Huillier}, Benjamin and {Li}, Baojiu and {Liang}, Ming and {Lin}, Huan and {Linder}, Eric and {Loebman}, Sarah R. and {Luki{\'c}}, Zarija and {Ma}, Jun and {MacCrann}, Niall and {Magneville}, Christophe and {Makarem}, Laleh and {Manera}, Marc and {Manser}, Christopher J. and {Marshall}, Robert and {Martini}, Paul and {Massey}, Richard and {Matheson}, Thomas and {McCauley}, Jeremy and {McDonald}, Patrick and {McGreer}, Ian D. and {Meisner}, Aaron and {Metcalfe}, Nigel and {Miller}, Timothy N. and {Miquel}, Ramon and {Moustakas}, John and {Myers}, Adam and {Naik}, Milind and {Newman}, Jeffrey A. and {Nichol}, Robert C. and {Nicola}, Andrina and {Nicolati da Costa}, Luiz and {Nie}, Jundan and {Niz}, Gustavo and {Norberg}, Peder and {Nord}, Brian and {Norman}, Dara and {Nugent}, Peter and {O'Brien}, Thomas and {Oh}, Minji and {Olsen}, Knut A.~G.},
        title = "{The DESI Experiment Part II: Instrument Design}",
      journal = {arXiv e-prints},
         year = 2016,
        month = oct,
          eid = {arXiv:1611.00037},
        pages = {arXiv:1611.00037},
          doi = {10.48550/arXiv.1611.00037},
archivePrefix = {arXiv},
       eprint = {1611.00037},
 primaryClass = {astro-ph.IM},
       adsurl = {https://ui.adsabs.harvard.edu/abs/2016arXiv161100037D}
}

@ARTICLE{AnnaDESI,
       author = {{Porredon}, A. and {Blake}, C. and {Lange}, J.~U. and {Emas}, N. and {Aguilar}, J. and {Ahlen}, S. and {Bera}, A. and {Bianchi}, D. and {Brooks}, D. and {Castander}, F.~J. and {Claybaugh}, T. and {Coloma Nadal}, J. and {Cuceu}, A. and {Dawson}, K.~S. and {de la Macorra}, A. and {Dey}, Biprateep and {Doel}, P. and {Elliott}, A. and {Ferraro}, S. and {Font-Ribera}, A. and {Forero-Romero}, J.~E. and {Garcia-Quintero}, C. and {Gazta{\~n}aga}, E. and {Gontcho}, S. Gontcho A and {Gutierrez}, G. and {Guy}, J. and {Hadzhiyska}, B. and {Herrera-Alcantar}, H.~K. and {Heydenreich}, S. and {Honscheid}, K. and {Howlett}, C. and {Huterer}, D. and {Ishak}, M. and {Joudaki}, S. and {Joyce}, R. and {Kirkby}, D. and {Kremin}, A. and {Krolewski}, A. and {Lahav}, O. and {Lamman}, C. and {Landriau}, M. and {Le Guillou}, L. and {Leauthaud}, A. and {Levi}, M.~E. and {Manera}, M. and {Meisner}, A. and {Miquel}, R. and {Nadathur}, S. and {Newman}, J.~A. and {Niz}, G. and {Palanque-Delabrouille}, N. and {Percival}, W.~J. and {Poppett}, C. and {Prada}, F. and {P{\'e}rez-R{\`a}fols}, I. and {Robertson}, A. and {Rossi}, G. and {Ruggeri}, R. and {Sanchez}, E. and {Saulder}, C. and {Schlegel}, D. and {Schubnell}, M. and {Semenaite}, A. and {Seo}, H. and {Silber}, J. and {Souki}, A. and {Sprayberry}, D. and {Tarl{\'e}}, G. and {Vargas-Maga{\~n}a}, M. and {Weaver}, B.~A. and {Zhou}, C. and {Zhou}, R. and {Zou}, H.},
        title = "{DESI-DR1 $3 \times 2$-pt analysis: consistent cosmology across weak lensing surveys}",
      journal = {arXiv e-prints},
         year = 2025,
        month = dec,
          eid = {arXiv:2512.15960},
        pages = {arXiv:2512.15960},
          doi = {10.48550/arXiv.2512.15960},
archivePrefix = {arXiv},
       eprint = {2512.15960},
 primaryClass = {astro-ph.CO},
       adsurl = {https://ui.adsabs.harvard.edu/abs/2025arXiv251215960P}
}

@ARTICLE{Langedesi,
       author = {{Lange}, Johannes U. and {Wells}, Alexandra and {Hearin}, Andrew and {Beltz-Mohrmann}, Gillian and {Leauthaud}, Alexie and {Heydenreich}, Sven and {Blake}, Chris and {Aguilar}, Jessica Nicole and {Ahlen}, Steven and {Anand}, Abhijeet and {Bianchi}, Davide and {Brooks}, David and {Castander}, Francisco Javier and {Claybaugh}, Todd and {Cole}, Shaun and {Cuceu}, Andrei and {Dawson}, Kyle and {de la Macorra}, Axel and {Dey}, Biprateep and {Doel}, Peter and {Elliott}, Ann and {Putu Audita Placida Emas}, Ni and {Ferraro}, Simone and {Font-Ribera}, Andreu and {Forero-Romero}, Jaime E. and {Garcia-Quintero}, Cristhian and {Gazta{\~n}aga}, Enrique and {Gontcho}, Satya Gontcho A and {Gutierrez}, Gaston and {Guy}, Julien and {Honscheid}, Klaus and {Huterer}, Dragan and {Ishak}, Mustapha and {Joudaki}, Shahab and {Joyce}, Dick and {Kehoe}, Robert and {Kirkby}, David and {Kisner}, Theodore and {Kremin}, Anthony and {Krolewski}, Alex and {Lahav}, Ofer and {Lamman}, Claire and {Landriau}, Martin and {Le Guillou}, Laurent and {Levi}, Michael and {Manera}, Marc and {Martini}, Paul and {Meisner}, Aaron and {Miquel}, Ramon and {Moustakas}, John and {Mueller}, Eva-Maria and {Nadathur}, Seshadri and {Newman}, Jeffrey A. and {Niz}, Gustavo and {Palanque-Delabrouille}, Nathalie and {Percival}, Will and {Poppett}, Claire and {Porredon}, Anna and {Prada}, Francisco and {P{\'e}rez-R{\`a}fols}, Ignasi and {Robertson}, Amy and {Rossi}, Graziano and {Ruggeri}, Rossana and {Sanchez}, Eusebio and {Saulder}, Christoph and {Schlegel}, David and {Schubnell}, Michael and {Semenaite}, Agne and {Seo}, Hee-Jong and {Silber}, Joseph Harry and {Sprayberry}, David and {Sun}, Zechang and {Tarl{\'e}}, Gregory and {Vargas Magana}, Mariana and {Weaver}, Benjamin Alan and {Wechsler}, Risa and {Zarrouk}, Pauline and {Zhou}, Rongpu and {Zou}, Hu},
        title = "{Cosmological Constraints from Full-Scale Clustering and Galaxy-Galaxy Lensing with DESI DR1}",
      journal = {arXiv e-prints},
         year = 2025,
        month = dec,
          eid = {arXiv:2512.15962},
        pages = {arXiv:2512.15962},
          doi = {10.48550/arXiv.2512.15962},
archivePrefix = {arXiv},
       eprint = {2512.15962},
 primaryClass = {astro-ph.CO},
       adsurl = {https://ui.adsabs.harvard.edu/abs/2025arXiv251215962L}
}

@ARTICLE{RossanaDESI,
       author = {{Ruggeri}, R. and {Blake}, C. and {Aguilar}, J. and {Ahlen}, S. and {Bianchi}, D. and {Brooks}, D. and {Castander}, F.~J. and {Claybaugh}, T. and {Cuceu}, A. and {Dawson}, K.~S. and {de la Macorra}, A. and {Dey}, B. and {Doel}, P. and {Elliott}, A. and {Emas}, N. and {Ferraro}, S. and {Font-Ribera}, A. and {Forero-Romero}, J.~E. and {Garcia-Quintero}, C. and {Gazta{\~n}aga}, E. and {Gontcho}, S. Gontcho A and {Gutierrez}, G. and {Guy}, J. and {Hadzhiyska}, B. and {Herrera-Alcantar}, H.~K. and {Heydenreich}, S. and {Honscheid}, K. and {Howlett}, C. and {Huterer}, D. and {Ishak}, M. and {Joudaki}, S. and {Joyce}, R. and {Kirkby}, D. and {Krolewski}, A. and {Lahav}, O. and {Lamman}, C. and {Landriau}, M. and {Lange}, J.~U. and {Leauthaud}, A. and {Levi}, M.~E. and {Manera}, M. and {Meisner}, A. and {Miquel}, R. and {Moustakas}, J. and {Nadathur}, S. and {Newman}, J.~A. and {Percival}, W.~J. and {Poppett}, C. and {Porredon}, A. and {Prada}, F. and {P{\'e}rez-R{\`a}fols}, I. and {Robertson}, A. and {Rossi}, G. and {Sanchez}, E. and {Saulder}, C. and {Schlegel}, D. and {Schubnell}, M. and {Semenaite}, A. and {Seo}, H. and {Silber}, J. and {Sprayberry}, D. and {Tarl{\'e}}, G. and {Weaver}, B.~A. and {Zarrouk}, P. and {Zhou}, R. and {Zou}, H.},
        title = "{Clustering redshift distribution calibration of weak lensing surveys using the DESI-DR1 spectroscopic dataset}",
      journal = {arXiv e-prints},
         year = 2025,
        month = dec,
          eid = {arXiv:2512.15963},
        pages = {arXiv:2512.15963},
          doi = {10.48550/arXiv.2512.15963},
archivePrefix = {arXiv},
       eprint = {2512.15963},
 primaryClass = {astro-ph.CO},
       adsurl = {https://ui.adsabs.harvard.edu/abs/2025arXiv251215963R}
}

@ARTICLE{DianaDESI,
       author = {{Blanco}, Diana and {Leauthaud}, Alexie and {Ulf Lange}, Johannes and {Wright}, Angus H. and {Hildebrandt}, Hendrik and {Heydenreich}, Sven and {Ravulapalli}, Darshika and {Ratajczak}, Joshua and {Dawson}, Kyle S. and {McCullough}, Jamie and {Dey}, Biprateep and {Aguilar}, Jessica N. and {Ahlen}, Steven and {Anand}, Abhijeet and {Bianchi}, Davide and {Blake}, Chris and {Brooks}, David and {Castander}, Francisco J. and {Claybaugh}, Todd and {Cuceu}, Andrei and {de la Macorra}, Axel and {Della Costa}, John and {Dey}, Arjun and {Elliott}, Ann and {Putu Audita Placida Emas}, Ni and {Ferraro}, Simone and {Font-Ribera}, Andreu and {Forero-Romero}, Jaime E. and {Garcia-Quintero}, Cristhian and {Gazta{\~n}aga}, Enrique and {Gontcho}, Satya Gontcho A and {Gutierrez}, Gaston and {Huterer}, Dragan and {Ishak}, Mustapha and {Jimenez}, Jorge and {Joudaki}, Shahab and {Joyce}, Dick and {Juneau}, Stephanie and {Kirkby}, David and {Kremin}, Anthony and {Krolewski}, Alex and {Lamman}, Claire and {Landriau}, Martin and {Le Guillou}, Laurent and {Manera}, Marc and {Meisner}, Aaron and {Miquel}, Ramon and {Moustakas}, John and {Nadathur}, Seshadri and {Newman}, Jeffrey A. and {Percival}, Will and {Porredon}, Anna and {Prada}, Francisco and {P{\'e}rez-R{\`a}fols}, Ignasi and {Robertson}, Amy and {Rossi}, Graziano and {Sanchez}, Eusebio and {Saulder}, Christoph and {Semenaite}, Agne and {Schlegel}, David and {Seo}, Hee-Jong and {Silber}, Joseph H. and {Sprayberry}, David and {Tarl{\'e}}, Gregory and {Weaver}, Benjamin A. and {Zhou}, Rongpu and {Zou}, Hu},
        title = "{The Power of DESI for Photometric Redshift Calibration: A Case Study with KiDS-1000}",
      journal = {arXiv e-prints},
         year = 2025,
        month = dec,
          eid = {arXiv:2512.15964},
        pages = {arXiv:2512.15964},
          doi = {10.48550/arXiv.2512.15964},
archivePrefix = {arXiv},
       eprint = {2512.15964},
 primaryClass = {astro-ph.CO},
       adsurl = {https://ui.adsabs.harvard.edu/abs/2025arXiv251215964B}
}

@unpublished{Joeheft,
    author = {{DeRose et al.\ in prep.}},
    title = "{}",
    year = 2025,
    note = "{}"
}

@unpublished{agneemu,
    author = {{Semenaite et al.\ in prep.}},
    title = "{}",
    year = 2025,
    note = "{}"
}

@ARTICLE{2025arXiv251005539E,
       author = {{Emas}, N. and {Porredon}, A. and {Blake}, C. and {DeRose}, J. and {Aguilar}, J. and {Ahlen}, S. and {Bianchi}, D. and {Brooks}, D. and {Castander}, F.~J. and {Claybaugh}, T. and {Cuceu}, A. and {de la Macorra}, A. and {Dey}, A. and {Dey}, B. and {Doel}, P. and {Ferraro}, S. and {Forero-Romero}, J.~E. and {Garcia-Quintero}, C. and {Gazta{\~n}aga}, E. and {Gontcho}, S. Gontcho A and {Gutierrez}, G. and {Heydenreich}, S. and {Honscheid}, K. and {Huterer}, D. and {Ishak}, M. and {Joudaki}, S. and {Joyce}, R. and {Jullo}, E. and {Juneau}, S. and {Kehoe}, R. and {Kirkby}, D. and {Kisner}, T. and {Kremin}, A. and {Krolewski}, A. and {Lahav}, O. and {Landriau}, M. and {Lange}, J.~U. and {Le Guillou}, L. and {Leauthaud}, A. and {Manera}, M. and {Miquel}, R. and {Nadathur}, S. and {Percival}, W.~J. and {Prada}, F. and {Rossi}, G. and {Ruggeri}, R. and {Sanchez}, E. and {Saulder}, C. and {Semenaite}, A. and {Seo}, H. and {Silber}, J. and {Sprayberry}, D. and {Sun}, Z. and {Tarl{\'e}}, G. and {Weaver}, B.~A. and {Wechsler}, R.~H. and {Zhou}, R.},
        title = "{Validation of the DESI-DR1 3x2-pt analysis: scale cut and shear ratio tests}",
      journal = {arXiv e-prints},
         year = 2025,
        month = oct,
          eid = {arXiv:2510.05539},
        pages = {arXiv:2510.05539},
          doi = {10.48550/arXiv.2510.05539},
archivePrefix = {arXiv},
       eprint = {2510.05539},
 primaryClass = {astro-ph.CO},
       adsurl = {https://ui.adsabs.harvard.edu/abs/2025arXiv251005539E}
}

@ARTICLE{2021A&A...645A.105G,
       author = {{Giblin}, Benjamin and {Heymans}, Catherine and {Asgari}, Marika and {Hildebrandt}, Hendrik and {Hoekstra}, Henk and {Joachimi}, Benjamin and {Kannawadi}, Arun and {Kuijken}, Konrad and {Lin}, Chieh-An and {Miller}, Lance and {Tr{\"o}ster}, Tilman and {van den Busch}, Jan Luca and {Wright}, Angus H. and {Bilicki}, Maciej and {Blake}, Chris and {de Jong}, Jelte and {Dvornik}, Andrej and {Erben}, Thomas and {Getman}, Fedor and {Napolitano}, Nicola R. and {Schneider}, Peter and {Shan}, HuanYuan and {Valentijn}, Edwin},
        title = "{KiDS-1000 catalogue: Weak gravitational lensing shear measurements}",
      journal = {\aap},
         year = 2021,
        month = jan,
       volume = {645},
          eid = {A105},
        pages = {A105},
          doi = {10.1051/0004-6361/202038850},
archivePrefix = {arXiv},
       eprint = {2007.01845},
 primaryClass = {astro-ph.CO},
       adsurl = {https://ui.adsabs.harvard.edu/abs/2021A&A...645A.105G}
}

@ARTICLE{2021MNRAS.504.4312G,
       author = {{Gatti}, M. and {Sheldon}, E. and {Amon}, A. and {Becker}, M. and {Troxel}, M. and {Choi}, A. and {Doux}, C. and {MacCrann}, N. and {Navarro-Alsina}, A. and {Harrison}, I. and {Gruen}, D. and {Bernstein}, G. and {Jarvis}, M. and {Secco}, L.~F. and {Fert{\'e}}, A. and {Shin}, T. and {McCullough}, J. and {Rollins}, R.~P. and {Chen}, R. and {Chang}, C. and {Pandey}, S. and {Tutusaus}, I. and {Prat}, J. and {Elvin-Poole}, J. and {Sanchez}, C. and {Plazas}, A.~A. and {Roodman}, A. and {Zuntz}, J. and {Abbott}, T.~M.~C. and {Aguena}, M. and {Allam}, S. and {Annis}, J. and {Avila}, S. and {Bacon}, D. and {Bertin}, E. and {Bhargava}, S. and {Brooks}, D. and {Burke}, D.~L. and {Carnero Rosell}, A. and {Carrasco Kind}, M. and {Carretero}, J. and {Castander}, F.~J. and {Conselice}, C. and {Costanzi}, M. and {Crocce}, M. and {da Costa}, L.~N. and {Davis}, T.~M. and {De Vicente}, J. and {Desai}, S. and {Diehl}, H.~T. and {Dietrich}, J.~P. and {Doel}, P. and {Drlica-Wagner}, A. and {Eckert}, K. and {Everett}, S. and {Ferrero}, I. and {Frieman}, J. and {Garc{\'\i}a-Bellido}, J. and {Gerdes}, D.~W. and {Giannantonio}, T. and {Gruendl}, R.~A. and {Gschwend}, J. and {Gutierrez}, G. and {Hartley}, W.~G. and {Hinton}, S.~R. and {Hollowood}, D.~L. and {Honscheid}, K. and {Hoyle}, B. and {Huff}, E.~M. and {Huterer}, D. and {Jain}, B. and {James}, D.~J. and {Jeltema}, T. and {Krause}, E. and {Kron}, R. and {Kuropatkin}, N. and {Lima}, M. and {Maia}, M.~A.~G. and {Marshall}, J.~L. and {Miquel}, R. and {Morgan}, R. and {Myles}, J. and {Palmese}, A. and {Paz-Chinch{\'o}n}, F. and {Rykoff}, E.~S. and {Samuroff}, S. and {Sanchez}, E. and {Scarpine}, V. and {Schubnell}, M. and {Serrano}, S. and {Sevilla-Noarbe}, I. and {Smith}, M. and {Soares-Santos}, M. and {Suchyta}, E. and {Swanson}, M.~E.~C. and {Tarle}, G. and {Thomas}, D. and {To}, C. and {Tucker}, D.~L. and {Varga}, T.~N. and {Wechsler}, R.~H. and {Weller}, J. and {Wester}, W. and {Wilkinson}, R.~D.},
        title = "{Dark energy survey year 3 results: weak lensing shape catalogue}",
      journal = {\mnras},
         year = 2021,
        month = jul,
       volume = {504},
       number = {3},
        pages = {4312-4336},
          doi = {10.1093/mnras/stab918},
archivePrefix = {arXiv},
       eprint = {2011.03408},
 primaryClass = {astro-ph.CO},
       adsurl = {https://ui.adsabs.harvard.edu/abs/2021MNRAS.504.4312G}
}

@ARTICLE{2022PASJ...74..421L,
       author = {{Li}, Xiangchong and {Miyatake}, Hironao and {Luo}, Wentao and {More}, Surhud and {Oguri}, Masamune and {Hamana}, Takashi and {Mandelbaum}, Rachel and {Shirasaki}, Masato and {Takada}, Masahiro and {Armstrong}, Robert and {Kannawadi}, Arun and {Takita}, Satoshi and {Miyazaki}, Satoshi and {Nishizawa}, Atsushi J. and {Plazas Malagon}, Andres A. and {Strauss}, Michael A. and {Tanaka}, Masayuki and {Yoshida}, Naoki},
        title = "{The three-year shear catalog of the Subaru Hyper Suprime-Cam SSP Survey}",
      journal = {\pasj},
         year = 2022,
        month = apr,
       volume = {74},
       number = {2},
        pages = {421-459},
          doi = {10.1093/pasj/psac006},
archivePrefix = {arXiv},
       eprint = {2107.00136},
 primaryClass = {astro-ph.CO},
       adsurl = {https://ui.adsabs.harvard.edu/abs/2022PASJ...74..421L}
}

@ARTICLE{2021A&A...646A.140H,
       author = {{Heymans}, Catherine and {Tr{\"o}ster}, Tilman and {Asgari}, Marika and {Blake}, Chris and {Hildebrandt}, Hendrik and {Joachimi}, Benjamin and {Kuijken}, Konrad and {Lin}, Chieh-An and {S{\'a}nchez}, Ariel G. and {van den Busch}, Jan Luca and {Wright}, Angus H. and {Amon}, Alexandra and {Bilicki}, Maciej and {de Jong}, Jelte and {Crocce}, Martin and {Dvornik}, Andrej and {Erben}, Thomas and {Fortuna}, Maria Cristina and {Getman}, Fedor and {Giblin}, Benjamin and {Glazebrook}, Karl and {Hoekstra}, Henk and {Joudaki}, Shahab and {Kannawadi}, Arun and {K{\"o}hlinger}, Fabian and {Lidman}, Chris and {Miller}, Lance and {Napolitano}, Nicola R. and {Parkinson}, David and {Schneider}, Peter and {Shan}, HuanYuan and {Valentijn}, Edwin A. and {Verdoes Kleijn}, Gijs and {Wolf}, Christian},
        title = "{KiDS-1000 Cosmology: Multi-probe weak gravitational lensing and spectroscopic galaxy clustering constraints}",
      journal = {\aap},
         year = 2021,
        month = feb,
       volume = {646},
          eid = {A140},
        pages = {A140},
          doi = {10.1051/0004-6361/202039063},
archivePrefix = {arXiv},
       eprint = {2007.15632},
 primaryClass = {astro-ph.CO},
       adsurl = {https://ui.adsabs.harvard.edu/abs/2021A&A...646A.140H}
}

@ARTICLE{2022PhRvD.105b3520A,
       author = {{Abbott}, T.~M.~C. and {Aguena}, M. and {Alarcon}, A. and {Allam}, S. and {Alves}, O. and {Amon}, A. and {Andrade-Oliveira}, F. and {Annis}, J. and {Avila}, S. and {Bacon}, D. and {Baxter}, E. and {Bechtol}, K. and {Becker}, M.~R. and {Bernstein}, G.~M. and {Bhargava}, S. and {Birrer}, S. and {Blazek}, J. and {Brandao-Souza}, A. and {Bridle}, S.~L. and {Brooks}, D. and {Buckley-Geer}, E. and {Burke}, D.~L. and {Camacho}, H. and {Campos}, A. and {Carnero Rosell}, A. and {Carrasco Kind}, M. and {Carretero}, J. and {Castander}, F.~J. and {Cawthon}, R. and {Chang}, C. and {Chen}, A. and {Chen}, R. and {Choi}, A. and {Conselice}, C. and {Cordero}, J. and {Costanzi}, M. and {Crocce}, M. and {da Costa}, L.~N. and {da Silva Pereira}, M.~E. and {Davis}, C. and {Davis}, T.~M. and {De Vicente}, J. and {DeRose}, J. and {Desai}, S. and {Di Valentino}, E. and {Diehl}, H.~T. and {Dietrich}, J.~P. and {Dodelson}, S. and {Doel}, P. and {Doux}, C. and {Drlica-Wagner}, A. and {Eckert}, K. and {Eifler}, T.~F. and {Elsner}, F. and {Elvin-Poole}, J. and {Everett}, S. and {Evrard}, A.~E. and {Fang}, X. and {Farahi}, A. and {Fernandez}, E. and {Ferrero}, I. and {Fert{\'e}}, A. and {Fosalba}, P. and {Friedrich}, O. and {Frieman}, J. and {Garc{\'\i}a-Bellido}, J. and {Gatti}, M. and {Gaztanaga}, E. and {Gerdes}, D.~W. and {Giannantonio}, T. and {Giannini}, G. and {Gruen}, D. and {Gruendl}, R.~A. and {Gschwend}, J. and {Gutierrez}, G. and {Harrison}, I. and {Hartley}, W.~G. and {Herner}, K. and {Hinton}, S.~R. and {Hollowood}, D.~L. and {Honscheid}, K. and {Hoyle}, B. and {Huff}, E.~M. and {Huterer}, D. and {Jain}, B. and {James}, D.~J. and {Jarvis}, M. and {Jeffrey}, N. and {Jeltema}, T. and {Kovacs}, A. and {Krause}, E. and {Kron}, R. and {Kuehn}, K. and {Kuropatkin}, N. and {Lahav}, O. and {Leget}, P. -F. and {Lemos}, P. and {Liddle}, A.~R. and {Lidman}, C. and {Lima}, M. and {Lin}, H. and {MacCrann}, N. and {Maia}, M.~A.~G. and {Marshall}, J.~L. and {Martini}, P. and {McCullough}, J. and {Melchior}, P. and {Mena-Fern{\'a}ndez}, J. and {Menanteau}, F. and {Miquel}, R. and {Mohr}, J.~J. and {Morgan}, R. and {Muir}, J. and {Myles}, J. and {Nadathur}, S. and {Navarro-Alsina}, A. and {Nichol}, R.~C. and {Ogando}, R.~L.~C. and {Omori}, Y. and {Palmese}, A. and {Pandey}, S. and {Park}, Y. and {Paz-Chinch{\'o}n}, F. and {Petravick}, D. and {Pieres}, A. and {Plazas Malag{\'o}n}, A.~A. and {Porredon}, A. and {Prat}, J. and {Raveri}, M. and {Rodriguez-Monroy}, M. and {Rollins}, R.~P. and {Romer}, A.~K. and {Roodman}, A. and {Rosenfeld}, R. and {Ross}, A.~J. and {Rykoff}, E.~S. and {Samuroff}, S. and {S{\'a}nchez}, C. and {Sanchez}, E. and {Sanchez}, J. and {Sanchez Cid}, D. and {Scarpine}, V. and {Schubnell}, M. and {Scolnic}, D. and {Secco}, L.~F. and {Serrano}, S. and {Sevilla-Noarbe}, I. and {Sheldon}, E. and {Shin}, T. and {Smith}, M. and {Soares-Santos}, M. and {Suchyta}, E. and {Swanson}, M.~E.~C. and {Tabbutt}, M. and {Tarle}, G. and {Thomas}, D. and {To}, C. and {Troja}, A. and {Troxel}, M.~A. and {Tucker}, D.~L. and {Tutusaus}, I. and {Varga}, T.~N. and {Walker}, A.~R. and {Weaverdyck}, N. and {Wechsler}, R. and {Weller}, J. and {Yanny}, B. and {Yin}, B. and {Zhang}, Y. and {Zuntz}, J. and {DES Collaboration}},
        title = "{Dark Energy Survey Year 3 results: Cosmological constraints from galaxy clustering and weak lensing}",
      journal = {\prd},
         year = 2022,
        month = jan,
       volume = {105},
       number = {2},
          eid = {023520},
        pages = {023520},
          doi = {10.1103/PhysRevD.105.023520},
archivePrefix = {arXiv},
       eprint = {2105.13549},
 primaryClass = {astro-ph.CO},
       adsurl = {https://ui.adsabs.harvard.edu/abs/2022PhRvD.105b3520A}
}

@ARTICLE{2023PhRvD.108l3518L,
       author = {{Li}, Xiangchong and {Zhang}, Tianqing and {Sugiyama}, Sunao and {Dalal}, Roohi and {Terasawa}, Ryo and {Rau}, Markus M. and {Mandelbaum}, Rachel and {Takada}, Masahiro and {More}, Surhud and {Strauss}, Michael A. and {Miyatake}, Hironao and {Shirasaki}, Masato and {Hamana}, Takashi and {Oguri}, Masamune and {Luo}, Wentao and {Nishizawa}, Atsushi J. and {Takahashi}, Ryuichi and {Nicola}, Andrina and {Osato}, Ken and {Kannawadi}, Arun and {Sunayama}, Tomomi and {Armstrong}, Robert and {Bosch}, James and {Komiyama}, Yutaka and {Lupton}, Robert H. and {Lust}, Nate B. and {MacArthur}, Lauren A. and {Miyazaki}, Satoshi and {Murayama}, Hitoshi and {Nishimichi}, Takahiro and {Okura}, Yuki and {Price}, Paul A. and {Tait}, Philip J. and {Tanaka}, Masayuki and {Wang}, Shiang-Yu},
        title = "{Hyper Suprime-Cam Year 3 results: Cosmology from cosmic shear two-point correlation functions}",
      journal = {\prd},
         year = 2023,
        month = dec,
       volume = {108},
       number = {12},
          eid = {123518},
        pages = {123518},
          doi = {10.1103/PhysRevD.108.123518},
archivePrefix = {arXiv},
       eprint = {2304.00702},
 primaryClass = {astro-ph.CO},
       adsurl = {https://ui.adsabs.harvard.edu/abs/2023PhRvD.108l3518L}
}

@ARTICLE{2023OJAp....6E..36D,
       author = {{Dark Energy Survey and Kilo-Degree Survey Collaboration} and {Abbott}, T.~M.~C. and {Aguena}, M. and {Alarcon}, A. and {Alves}, O. and {Amon}, A. and {Andrade-Oliveira}, F. and {Asgari}, M. and {Avila}, S. and {Bacon}, D. and {Bechtol}, K. and {Becker}, M.~R. and {Bernstein}, G.~M. and {Bertin}, E. and {Bilicki}, M. and {Blazek}, J. and {Bocquet}, S. and {Brooks}, D. and {Burger}, P. and {Burke}, D.~L. and {Camacho}, H. and {Campos}, A. and {Carnero Rosell}, A. and {Carrasco Kind}, M. and {Carretero}, J. and {Castander}, F.~J. and {Cawthon}, R. and {Chang}, C. and {Chen}, R. and {Choi}, A. and {Conselice}, C. and {Cordero}, J. and {Crocce}, M. and {da Costa}, L.~N. and {da Silva Pereira}, M.~E. and {Dalal}, R. and {Davis}, C. and {de Jong}, J.~T.~A. and {DeRose}, J. and {Desai}, S. and {Diehl}, H.~T. and {Dodelson}, S. and {Doel}, P. and {Doux}, C. and {Drlica-Wagner}, A. and {Dvornik}, A. and {Eckert}, K. and {Eifler}, T.~F. and {Elvin-Poole}, J. and {Everett}, S. and {Fang}, X. and {Ferrero}, I. and {Fert{\'e}}, A. and {Flaugher}, B. and {Friedrich}, O. and {Frieman}, J. and {Garc{\'\i}a-Bellido}, J. and {Gatti}, M. and {Giannini}, G. and {Giblin}, B. and {Gruen}, D. and {Gruendl}, R.~A. and {Gutierrez}, G. and {Harrison}, I. and {Hartley}, W.~G. and {Herner}, K. and {Heymans}, C. and {Hildebrandt}, H. and {Hinton}, S.~R. and {Hoekstra}, H. and {Hollowood}, D.~L. and {Honscheid}, K. and {Huang}, H. and {Huff}, E.~M. and {Huterer}, D. and {James}, D.~J. and {Jarvis}, M. and {Jeffrey}, N. and {Jeltema}, T. and {Joachimi}, B. and {Joudaki}, S. and {Kannawadi}, A. and {Krause}, E. and {Kuehn}, K. and {Kuijken}, K. and {Kuropatkin}, N. and {Lahav}, O. and {Leget}, P. -F. and {Lemos}, P. and {Li}, S. -S. and {Li}, X. and {Liddle}, A.~R. and {Lima}, M. and {Lin}, C. -A. and {Lin}, H. and {MacCrann}, N. and {Mahony}, C. and {Marshall}, J.~L. and {McCullough}, J. and {Mena-Fern{\'a}ndez}, J. and {Menanteau}, F. and {Miquel}, R. and {Mohr}, J.~J. and {Muir}, J. and {Myles}, J. and {Napolitano}, N. and {Navarro-Alsina}, A. and {Ogando}, R.~L.~C. and {Palmese}, A. and {Pandey}, S. and {Park}, Y. and {Paterno}, M. and {Peacock}, J.~A. and {Petravick}, D. and {Pieres}, A. and {Plazas Malag{\'o}n}, A.~A. and {Porredon}, A. and {Prat}, J. and {Radovich}, M. and {Raveri}, M. and {Reischke}, R. and {Robertson}, N.~C. and {Rollins}, R.~P. and {Romer}, A.~K. and {Roodman}, A. and {Rykoff}, E.~S. and {Samuroff}, S. and {S{\'a}nchez}, C. and {Sanchez}, E. and {Sanchez}, J. and {Schneider}, P. and {Secco}, L.~F. and {Sevilla-Noarbe}, I. and {Shan}, H. -Y. and {Sheldon}, E. and {Shin}, T. and {Sif{\'o}n}, C. and {Smith}, M. and {Soares-Santos}, M. and {St{\"o}lzner}, B. and {Suchyta}, E. and {Swanson}, M.~E.~C. and {Tarle}, G. and {Thomas}, D. and {To}, C. and {Troxel}, M.~A. and {Tr{\"o}ster}, T. and {Tutusaus}, I. and {van den Busch}, J.~L. and {Varga}, T.~N. and {Walker}, A.~R. and {Weaverdyck}, N. and {Wechsler}, R.~H. and {Weller}, J. and {Wiseman}, P. and {Wright}, A.~H. and {Yanny}, B. and {Yin}, B. and {Yoon}, M. and {Zhang}, Y. and {Zuntz}, J.},
        title = "{DES Y3 + KiDS-1000: Consistent cosmology combining cosmic shear surveys}",
      journal = {The Open Journal of Astrophysics},
         year = 2023,
        month = oct,
       volume = {6},
          eid = {36},
        pages = {36},
          doi = {10.21105/astro.2305.17173},
archivePrefix = {arXiv},
       eprint = {2305.17173},
 primaryClass = {astro-ph.CO},
       adsurl = {https://ui.adsabs.harvard.edu/abs/2023OJAp....6E..36D}
}

@ARTICLE{2023PhRvD.108l3519D,
       author = {{Dalal}, Roohi and {Li}, Xiangchong and {Nicola}, Andrina and {Zuntz}, Joe and {Strauss}, Michael A. and {Sugiyama}, Sunao and {Zhang}, Tianqing and {Rau}, Markus M. and {Mandelbaum}, Rachel and {Takada}, Masahiro and {More}, Surhud and {Miyatake}, Hironao and {Kannawadi}, Arun and {Shirasaki}, Masato and {Taniguchi}, Takanori and {Takahashi}, Ryuichi and {Osato}, Ken and {Hamana}, Takashi and {Oguri}, Masamune and {Nishizawa}, Atsushi J. and {Malag{\'o}n}, Andr{\'e}s A. Plazas and {Sunayama}, Tomomi and {Alonso}, David and {Slosar}, An{\v{z}}e and {Luo}, Wentao and {Armstrong}, Robert and {Bosch}, James and {Hsieh}, Bau-Ching and {Komiyama}, Yutaka and {Lupton}, Robert H. and {Lust}, Nate B. and {MacArthur}, Lauren A. and {Miyazaki}, Satoshi and {Murayama}, Hitoshi and {Nishimichi}, Takahiro and {Okura}, Yuki and {Price}, Paul A. and {Tait}, Philip J. and {Tanaka}, Masayuki and {Wang}, Shiang-Yu},
        title = "{Hyper Suprime-Cam Year 3 results: Cosmology from cosmic shear power spectra}",
      journal = {\prd},
         year = 2023,
        month = dec,
       volume = {108},
       number = {12},
          eid = {123519},
        pages = {123519},
          doi = {10.1103/PhysRevD.108.123519},
archivePrefix = {arXiv},
       eprint = {2304.00701},
 primaryClass = {astro-ph.CO},
       adsurl = {https://ui.adsabs.harvard.edu/abs/2023PhRvD.108l3519D}
}

@ARTICLE{2025arXiv250319441W,
       author = {{Wright}, Angus H. and {St{\"o}lzner}, Benjamin and {Asgari}, Marika and {Bilicki}, Maciej and {Giblin}, Benjamin and {Heymans}, Catherine and {Hildebrandt}, Hendrik and {Hoekstra}, Henk and {Joachimi}, Benjamin and {Kuijken}, Konrad and {Li}, Shun-Sheng and {Reischke}, Robert and {von Wietersheim-Kramsta}, Maximilian and {Yoon}, Mijin and {Burger}, Pierre and {Chisari}, Nora Elisa and {de Jong}, Jelte and {Dvornik}, Andrej and {Georgiou}, Christos and {Harnois-D{\'e}raps}, Joachim and {Jalan}, Priyanka and {William}, Anjitha John and {Joudaki}, Shahab and {Lesci}, Giorgio Francesco and {Linke}, Laila and {Loureiro}, Arthur and {Mahony}, Constance and {Maturi}, Matteo and {Miller}, Lance and {Moscardini}, Lauro and {Napolitano}, Nicola R. and {Porth}, Lucas and {Radovich}, Mario and {Schneider}, Peter and {Tr{\"o}ster}, Tilman and {Wittje}, Anna and {Yan}, Ziang and {Zhang}, Yun-Hao},
        title = "{KiDS-Legacy: Cosmological constraints from cosmic shear with the complete Kilo-Degree Survey}",
      journal = {arXiv e-prints},
         year = 2025,
        month = mar,
          eid = {arXiv:2503.19441},
        pages = {arXiv:2503.19441},
          doi = {10.48550/arXiv.2503.19441},
archivePrefix = {arXiv},
       eprint = {2503.19441},
 primaryClass = {astro-ph.CO},
       adsurl = {https://ui.adsabs.harvard.edu/abs/2025arXiv250319441W}
}

@ARTICLE{2022PhRvD.105b3514A,
       author = {{Amon}, A. and {Gruen}, D. and {Troxel}, M.~A. and {MacCrann}, N. and {Dodelson}, S. and {Choi}, A. and {Doux}, C. and {Secco}, L.~F. and {Samuroff}, S. and {Krause}, E. and {Cordero}, J. and {Myles}, J. and {DeRose}, J. and {Wechsler}, R.~H. and {Gatti}, M. and {Navarro-Alsina}, A. and {Bernstein}, G.~M. and {Jain}, B. and {Blazek}, J. and {Alarcon}, A. and {Fert{\'e}}, A. and {Lemos}, P. and {Raveri}, M. and {Campos}, A. and {Prat}, J. and {S{\'a}nchez}, C. and {Jarvis}, M. and {Alves}, O. and {Andrade-Oliveira}, F. and {Baxter}, E. and {Bechtol}, K. and {Becker}, M.~R. and {Bridle}, S.~L. and {Camacho}, H. and {Carnero Rosell}, A. and {Carrasco Kind}, M. and {Cawthon}, R. and {Chang}, C. and {Chen}, R. and {Chintalapati}, P. and {Crocce}, M. and {Davis}, C. and {Diehl}, H.~T. and {Drlica-Wagner}, A. and {Eckert}, K. and {Eifler}, T.~F. and {Elvin-Poole}, J. and {Everett}, S. and {Fang}, X. and {Fosalba}, P. and {Friedrich}, O. and {Gaztanaga}, E. and {Giannini}, G. and {Gruendl}, R.~A. and {Harrison}, I. and {Hartley}, W.~G. and {Herner}, K. and {Huang}, H. and {Huff}, E.~M. and {Huterer}, D. and {Kuropatkin}, N. and {Leget}, P. and {Liddle}, A.~R. and {McCullough}, J. and {Muir}, J. and {Pandey}, S. and {Park}, Y. and {Porredon}, A. and {Refregier}, A. and {Rollins}, R.~P. and {Roodman}, A. and {Rosenfeld}, R. and {Ross}, A.~J. and {Rykoff}, E.~S. and {Sanchez}, J. and {Sevilla-Noarbe}, I. and {Sheldon}, E. and {Shin}, T. and {Troja}, A. and {Tutusaus}, I. and {Tutusaus}, I. and {Varga}, T.~N. and {Weaverdyck}, N. and {Yanny}, B. and {Yin}, B. and {Zhang}, Y. and {Zuntz}, J. and {Aguena}, M. and {Allam}, S. and {Annis}, J. and {Bacon}, D. and {Bertin}, E. and {Bhargava}, S. and {Brooks}, D. and {Buckley-Geer}, E. and {Burke}, D.~L. and {Carretero}, J. and {Costanzi}, M. and {da Costa}, L.~N. and {Pereira}, M.~E.~S. and {De Vicente}, J. and {Desai}, S. and {Dietrich}, J.~P. and {Doel}, P. and {Ferrero}, I. and {Flaugher}, B. and {Frieman}, J. and {Garc{\'\i}a-Bellido}, J. and {Gaztanaga}, E. and {Gerdes}, D.~W. and {Giannantonio}, T. and {Gschwend}, J. and {Gutierrez}, G. and {Hinton}, S.~R. and {Hollowood}, D.~L. and {Honscheid}, K. and {Hoyle}, B. and {James}, D.~J. and {Kron}, R. and {Kuehn}, K. and {Lahav}, O. and {Lima}, M. and {Lin}, H. and {Maia}, M.~A.~G. and {Marshall}, J.~L. and {Martini}, P. and {Melchior}, P. and {Menanteau}, F. and {Miquel}, R. and {Mohr}, J.~J. and {Morgan}, R. and {Ogando}, R.~L.~C. and {Palmese}, A. and {Paz-Chinch{\'o}n}, F. and {Petravick}, D. and {Pieres}, A. and {Romer}, A.~K. and {Sanchez}, E. and {Scarpine}, V. and {Schubnell}, M. and {Serrano}, S. and {Smith}, M. and {Soares-Santos}, M. and {Tarle}, G. and {Thomas}, D. and {To}, C. and {Weller}, J. and {DES Collaboration}},
        title = "{Dark Energy Survey Year 3 results: Cosmology from cosmic shear and robustness to data calibration}",
      journal = {\prd},
         year = 2022,
        month = jan,
       volume = {105},
       number = {2},
          eid = {023514},
        pages = {023514},
          doi = {10.1103/PhysRevD.105.023514},
archivePrefix = {arXiv},
       eprint = {2105.13543},
 primaryClass = {astro-ph.CO},
       adsurl = {https://ui.adsabs.harvard.edu/abs/2022PhRvD.105b3514A}
}

@ARTICLE{2022PhRvD.105b3515S,
       author = {{Secco}, L.~F. and {Samuroff}, S. and {Krause}, E. and {Jain}, B. and {Blazek}, J. and {Raveri}, M. and {Campos}, A. and {Amon}, A. and {Chen}, A. and {Doux}, C. and {Choi}, A. and {Gruen}, D. and {Bernstein}, G.~M. and {Chang}, C. and {DeRose}, J. and {Myles}, J. and {Fert{\'e}}, A. and {Lemos}, P. and {Huterer}, D. and {Prat}, J. and {Troxel}, M.~A. and {MacCrann}, N. and {Liddle}, A.~R. and {Kacprzak}, T. and {Fang}, X. and {S{\'a}nchez}, C. and {Pandey}, S. and {Dodelson}, S. and {Chintalapati}, P. and {Hoffmann}, K. and {Alarcon}, A. and {Alves}, O. and {Andrade-Oliveira}, F. and {Baxter}, E.~J. and {Bechtol}, K. and {Becker}, M.~R. and {Brandao-Souza}, A. and {Camacho}, H. and {Carnero Rosell}, A. and {Carrasco Kind}, M. and {Cawthon}, R. and {Cordero}, J.~P. and {Crocce}, M. and {Davis}, C. and {Di Valentino}, E. and {Drlica-Wagner}, A. and {Eckert}, K. and {Eifler}, T.~F. and {Elidaiana}, M. and {Elsner}, F. and {Elvin-Poole}, J. and {Everett}, S. and {Fosalba}, P. and {Friedrich}, O. and {Gatti}, M. and {Giannini}, G. and {Gruendl}, R.~A. and {Harrison}, I. and {Hartley}, W.~G. and {Herner}, K. and {Huang}, H. and {Huff}, E.~M. and {Jarvis}, M. and {Jeffrey}, N. and {Kuropatkin}, N. and {Leget}, P. -F. and {Muir}, J. and {Mccullough}, J. and {Navarro Alsina}, A. and {Omori}, Y. and {Park}, Y. and {Porredon}, A. and {Rollins}, R. and {Roodman}, A. and {Rosenfeld}, R. and {Ross}, A.~J. and {Rykoff}, E.~S. and {Sanchez}, J. and {Sevilla-Noarbe}, I. and {Sheldon}, E.~S. and {Shin}, T. and {Troja}, A. and {Tutusaus}, I. and {Varga}, T.~N. and {Weaverdyck}, N. and {Wechsler}, R.~H. and {Yanny}, B. and {Yin}, B. and {Zhang}, Y. and {Zuntz}, J. and {Abbott}, T.~M.~C. and {Aguena}, M. and {Allam}, S. and {Annis}, J. and {Bacon}, D. and {Bertin}, E. and {Bhargava}, S. and {Bridle}, S.~L. and {Brooks}, D. and {Buckley-Geer}, E. and {Burke}, D.~L. and {Carretero}, J. and {Costanzi}, M. and {da Costa}, L.~N. and {De Vicente}, J. and {Diehl}, H.~T. and {Dietrich}, J.~P. and {Doel}, P. and {Ferrero}, I. and {Flaugher}, B. and {Frieman}, J. and {Garc{\'\i}a-Bellido}, J. and {Gaztanaga}, E. and {Gerdes}, D.~W. and {Giannantonio}, T. and {Gschwend}, J. and {Gutierrez}, G. and {Hinton}, S.~R. and {Hollowood}, D.~L. and {Honscheid}, K. and {Hoyle}, B. and {James}, D.~J. and {Jeltema}, T. and {Kuehn}, K. and {Lahav}, O. and {Lima}, M. and {Lin}, H. and {Maia}, M.~A.~G. and {Marshall}, J.~L. and {Martini}, P. and {Melchior}, P. and {Menanteau}, F. and {Miquel}, R. and {Mohr}, J.~J. and {Morgan}, R. and {Ogando}, R.~L.~C. and {Palmese}, A. and {Paz-Chinch{\'o}n}, F. and {Petravick}, D. and {Pieres}, A. and {Plazas Malag{\'o}n}, A.~A. and {Rodriguez-Monroy}, M. and {Romer}, A.~K. and {Sanchez}, E. and {Scarpine}, V. and {Schubnell}, M. and {Scolnic}, D. and {Serrano}, S. and {Smith}, M. and {Soares-Santos}, M. and {Suchyta}, E. and {Swanson}, M.~E.~C. and {Tarle}, G. and {Thomas}, D. and {To}, C. and {DES Collaboration}},
        title = "{Dark Energy Survey Year 3 results: Cosmology from cosmic shear and robustness to modeling uncertainty}",
      journal = {\prd},
         year = 2022,
        month = jan,
       volume = {105},
       number = {2},
          eid = {023515},
        pages = {023515},
          doi = {10.1103/PhysRevD.105.023515},
archivePrefix = {arXiv},
       eprint = {2105.13544},
 primaryClass = {astro-ph.CO},
       adsurl = {https://ui.adsabs.harvard.edu/abs/2022PhRvD.105b3515S}
}

@ARTICLE{2021A&A...645A.104A,
       author = {{Asgari}, Marika and {Lin}, Chieh-An and {Joachimi}, Benjamin and {Giblin}, Benjamin and {Heymans}, Catherine and {Hildebrandt}, Hendrik and {Kannawadi}, Arun and {St{\"o}lzner}, Benjamin and {Tr{\"o}ster}, Tilman and {van den Busch}, Jan Luca and {Wright}, Angus H. and {Bilicki}, Maciej and {Blake}, Chris and {de Jong}, Jelte and {Dvornik}, Andrej and {Erben}, Thomas and {Getman}, Fedor and {Hoekstra}, Henk and {K{\"o}hlinger}, Fabian and {Kuijken}, Konrad and {Miller}, Lance and {Radovich}, Mario and {Schneider}, Peter and {Shan}, HuanYuan and {Valentijn}, Edwin},
        title = "{KiDS-1000 cosmology: Cosmic shear constraints and comparison between two point statistics}",
      journal = {\aap},
         year = 2021,
        month = jan,
       volume = {645},
          eid = {A104},
        pages = {A104},
          doi = {10.1051/0004-6361/202039070},
archivePrefix = {arXiv},
       eprint = {2007.15633},
 primaryClass = {astro-ph.CO},
       adsurl = {https://ui.adsabs.harvard.edu/abs/2021A&A...645A.104A}
}

@ARTICLE{2025JCAP...04..074B,
       author = {{Bianchi}, D. and {Hanif}, M.~M.~S. and {Carnero Rosell}, A. and {Lasker}, J. and {Ross}, A.~J. and {Pinon}, M. and {de Mattia}, A. and {White}, M. and {Ahlen}, S. and {Bailey}, S. and {Brooks}, D. and {Burtin}, E. and {Chaussidon}, E. and {Claybaugh}, T. and {Cole}, S. and {de la Macorra}, A. and {Ferraro}, S. and {Font-Ribera}, A. and {Forero-Romero}, J.~E. and {Gazta{\~n}aga}, E. and {Gontcho}, S. Gontcho A. and {Gutierrez}, G. and {Guy}, J. and {Hahn}, C. and {Honscheid}, K. and {Howlett}, C. and {Juneau}, S. and {Kirkby}, D. and {Kisner}, T. and {Kremin}, A. and {Landriau}, M. and {Le Guillou}, L. and {Levi}, M.~E. and {McDonald}, P. and {Meisner}, A. and {Miquel}, R. and {Moustakas}, J. and {Palanque-Delabrouille}, N. and {Percival}, W.~J. and {Prada}, F. and {P{\'e}rez-R{\`a}fols}, I. and {Raichoor}, A. and {Rossi}, G. and {Sanchez}, E. and {Schlegel}, D. and {Schubnell}, M. and {Sharples}, R. and {Silber}, J. and {Sprayberry}, D. and {Tarl{\'e}}, G. and {Vargas-Maga{\~n}a}, M. and {Weaver}, B.~A. and {Zarrouk}, P. and {Zhou}, R. and {Zou}, H.},
        title = "{Characterization of DESI fiber assignment incompleteness effect on 2-point clustering and mitigation methods for DR1 analysis}",
      journal = {\jcap},
         year = 2025,
        month = apr,
       volume = {2025},
       number = {4},
          eid = {074},
        pages = {074},
          doi = {10.1088/1475-7516/2025/04/074},
archivePrefix = {arXiv},
       eprint = {2411.12025},
 primaryClass = {astro-ph.CO},
       adsurl = {https://ui.adsabs.harvard.edu/abs/2025JCAP...04..074B}
}

@ARTICLE{2020MNRAS.491.3022S,
       author = {{Sinha}, Manodeep and {Garrison}, Lehman H.},
        title = "{CORRFUNC - a suite of blazing fast correlation functions on the CPU}",
      journal = {\mnras},
         year = 2020,
        month = jan,
       volume = {491},
       number = {2},
        pages = {3022-3041},
          doi = {10.1093/mnras/stz3157},
archivePrefix = {arXiv},
       eprint = {1911.03545},
 primaryClass = {astro-ph.CO},
       adsurl = {https://ui.adsabs.harvard.edu/abs/2020MNRAS.491.3022S}
}

@ARTICLE{2004MNRAS.352..338J,
       author = {{Jarvis}, M. and {Bernstein}, G. and {Jain}, B.},
        title = "{The skewness of the aperture mass statistic}",
      journal = {\mnras},
         year = 2004,
        month = jul,
       volume = {352},
       number = {1},
        pages = {338-352},
          doi = {10.1111/j.1365-2966.2004.07926.x},
archivePrefix = {arXiv},
       eprint = {astro-ph/0307393},
 primaryClass = {astro-ph},
       adsurl = {https://ui.adsabs.harvard.edu/abs/2004MNRAS.352..338J}
}

@ARTICLE{2017MNRAS.470.2100K,
       author = {{Krause}, Elisabeth and {Eifler}, Tim},
        title = "{cosmolike - cosmological likelihood analyses for photometric galaxy surveys}",
      journal = {\mnras},
         year = 2017,
        month = sep,
       volume = {470},
       number = {2},
        pages = {2100-2112},
          doi = {10.1093/mnras/stx1261},
archivePrefix = {arXiv},
       eprint = {1601.05779},
 primaryClass = {astro-ph.CO},
       adsurl = {https://ui.adsabs.harvard.edu/abs/2017MNRAS.470.2100K}
}

@ARTICLE{2021A&A...646A.129J,
       author = {{Joachimi}, B. and {Lin}, C. -A. and {Asgari}, M. and {Tr{\"o}ster}, T. and {Heymans}, C. and {Hildebrandt}, H. and {K{\"o}hlinger}, F. and {S{\'a}nchez}, A.~G. and {Wright}, A.~H. and {Bilicki}, M. and {Blake}, C. and {van den Busch}, J.~L. and {Crocce}, M. and {Dvornik}, A. and {Erben}, T. and {Getman}, F. and {Giblin}, B. and {Hoekstra}, H. and {Kannawadi}, A. and {Kuijken}, K. and {Napolitano}, N.~R. and {Schneider}, P. and {Scoccimarro}, R. and {Sellentin}, E. and {Shan}, H.~Y. and {von Wietersheim-Kramsta}, M. and {Zuntz}, J.},
        title = "{KiDS-1000 methodology: Modelling and inference for joint weak gravitational lensing and spectroscopic galaxy clustering analysis}",
      journal = {\aap},
         year = 2021,
        month = feb,
       volume = {646},
          eid = {A129},
        pages = {A129},
          doi = {10.1051/0004-6361/202038831},
archivePrefix = {arXiv},
       eprint = {2007.01844},
 primaryClass = {astro-ph.CO},
       adsurl = {https://ui.adsabs.harvard.edu/abs/2021A&A...646A.129J}
}

@ARTICLE{2018MNRAS.479.4998T,
       author = {{Troxel}, M.~A. and {Krause}, E. and {Chang}, C. and {Eifler}, T.~F. and {Friedrich}, O. and {Gruen}, D. and {MacCrann}, N. and {Chen}, A. and {Davis}, C. and {DeRose}, J. and {Dodelson}, S. and {Gatti}, M. and {Hoyle}, B. and {Huterer}, D. and {Jarvis}, M. and {Lacasa}, F. and {Lemos}, P. and {Peiris}, H.~V. and {Prat}, J. and {Samuroff}, S. and {S{\'a}nchez}, C. and {Sheldon}, E. and {Vielzeuf}, P. and {Wang}, M. and {Zuntz}, J. and {Lahav}, O. and {Abdalla}, F.~B. and {Allam}, S. and {Annis}, J. and {Avila}, S. and {Bertin}, E. and {Brooks}, D. and {Burke}, D.~L. and {Carnero Rosell}, A. and {Carrasco Kind}, M. and {Carretero}, J. and {Crocce}, M. and {Cunha}, C.~E. and {D'Andrea}, C.~B. and {da Costa}, L.~N. and {De Vicente}, J. and {Diehl}, H.~T. and {Doel}, P. and {Evrard}, A.~E. and {Flaugher}, B. and {Fosalba}, P. and {Frieman}, J. and {Garc{\'\i}a-Bellido}, J. and {Gaztanaga}, E. and {Gerdes}, D.~W. and {Gruendl}, R.~A. and {Gschwend}, J. and {Gutierrez}, G. and {Hartley}, W.~G. and {Hollowood}, D.~L. and {Honscheid}, K. and {James}, D.~J. and {Kirk}, D. and {Kuehn}, K. and {Kuropatkin}, N. and {Li}, T.~S. and {Lima}, M. and {March}, M. and {Menanteau}, F. and {Miquel}, R. and {Mohr}, J.~J. and {Ogando}, R.~L.~C. and {Plazas}, A.~A. and {Roodman}, A. and {Sanchez}, E. and {Scarpine}, V. and {Schindler}, R. and {Sevilla-Noarbe}, I. and {Smith}, M. and {Soares-Santos}, M. and {Sobreira}, F. and {Suchyta}, E. and {Swanson}, M.~E.~C. and {Thomas}, D. and {Walker}, A.~R. and {Wechsler}, R.~H.},
        title = "{Survey geometry and the internal consistency of recent cosmic shear measurements}",
      journal = {\mnras},
         year = 2018,
        month = oct,
       volume = {479},
       number = {4},
        pages = {4998-5004},
          doi = {10.1093/mnras/sty1889},
archivePrefix = {arXiv},
       eprint = {1804.10663},
 primaryClass = {astro-ph.CO},
       adsurl = {https://ui.adsabs.harvard.edu/abs/2018MNRAS.479.4998T}
}

@ARTICLE{2013PhRvD..87l3504T,
       author = {{Takada}, Masahiro and {Hu}, Wayne},
        title = "{Power spectrum super-sample covariance}",
      journal = {\prd},
         year = 2013,
        month = jun,
       volume = {87},
       number = {12},
          eid = {123504},
        pages = {123504},
          doi = {10.1103/PhysRevD.87.123504},
archivePrefix = {arXiv},
       eprint = {1302.6994},
 primaryClass = {astro-ph.CO},
       adsurl = {https://ui.adsabs.harvard.edu/abs/2013PhRvD..87l3504T}
}

@ARTICLE{2025JCAP...01..145R,
       author = {{Rashkovetskyi}, M. and {Forero-S{\'a}nchez}, D. and {de Mattia}, A. and {Eisenstein}, D.~J. and {Padmanabhan}, N. and {Seo}, H. and {Ross}, A.~J. and {Aguilar}, J. and {Ahlen}, S. and {Alves}, O. and {Andrade}, U. and {Brooks}, D. and {Burtin}, E. and {Chen}, X. and {Claybaugh}, T. and {Cole}, S. and {de la Macorra}, A. and {Ding}, Z. and {Doel}, P. and {Fanning}, K. and {Ferraro}, S. and {Font-Ribera}, A. and {Forero-Romero}, J.~E. and {Garcia-Quintero}, C. and {Gil-Mar{\'\i}n}, H. and {Gontcho A Gontcho}, S. and {Gonzalez-Morales}, A.~X. and {Gutierrez}, G. and {Honscheid}, K. and {Howlett}, C. and {Juneau}, S. and {Kremin}, A. and {Le Guillou}, L. and {Manera}, M. and {Medina-Varela}, L. and {Mena-Fern{\'a}ndez}, J. and {Miquel}, R. and {Mueller}, E. and {Mu{\~n}oz-Guti{\'e}rrez}, A. and {Myers}, A.~D. and {Nie}, J. and {Niz}, G. and {Paillas}, E. and {Percival}, W.~J. and {Poppett}, C. and {P{\'e}rez-Fern{\'a}ndez}, A. and {Rezaie}, M. and {Rosado-Marin}, A. and {Rossi}, G. and {Ruggeri}, R. and {Sanchez}, E. and {Saulder}, C. and {Schlegel}, D. and {Schubnell}, M. and {Sprayberry}, D. and {Tarl{\'e}}, G. and {Weaver}, B.~A. and {Yu}, J. and {Zhao}, C. and {Zou}, H.},
        title = "{Semi-analytical covariance matrices for two-point correlation function for DESI 2024 data}",
      journal = {\jcap},
         year = 2025,
        month = jan,
       volume = {2025},
       number = {1},
          eid = {145},
        pages = {145},
          doi = {10.1088/1475-7516/2025/01/145},
archivePrefix = {arXiv},
       eprint = {2404.03007},
 primaryClass = {astro-ph.CO},
       adsurl = {https://ui.adsabs.harvard.edu/abs/2025JCAP...01..145R}
}

@ARTICLE{2023MNRAS.525.4367H,
       author = {{Hadzhiyska}, Boryana and {Yuan}, S. and {Blake}, C. and {Eisenstein}, D.~J. and {Aguilar}, J. and {Ahlen}, S. and {Brooks}, D. and {Claybaugh}, T. and {de la Macorra}, A. and {Doel}, P. and {Emas}, N. and {Forero-Romero}, J.~E. and {Garcia-Quintero}, C. and {Ishak}, M. and {Joudaki}, S. and {Jullo}, E. and {Kehoe}, R. and {Kisner}, T. and {Kremin}, A. and {Krolewski}, A. and {Landriau}, M. and {Lange}, J.~U. and {Manera}, M. and {Miquel}, R. and {Nie}, Jundan and {Poppett}, C. and {Porredon}, A. and {Rossi}, G. and {Ruggeri}, R. and {Saulder}, C. and {Schubnell}, M. and {Tarl{\'e}}, G. and {Weaver}, B.~A. and {Xhakaj}, E. and {Zhou}, Zhimin},
        title = "{Synthetic light-cone catalogues of modern redshift and weak lensing surveys waith ABACUSSUMMIT}",
      journal = {\mnras},
         year = 2023,
        month = nov,
       volume = {525},
       number = {3},
        pages = {4367-4387},
          doi = {10.1093/mnras/stad2563},
archivePrefix = {arXiv},
       eprint = {2305.11935},
 primaryClass = {astro-ph.CO},
       adsurl = {https://ui.adsabs.harvard.edu/abs/2023MNRAS.525.4367H}
}

@ARTICLE{2021MNRAS.508.4017M,
       author = {{Maksimova}, Nina A. and {Garrison}, Lehman H. and {Eisenstein}, Daniel J. and {Hadzhiyska}, Boryana and {Bose}, Sownak and {Satterthwaite}, Thomas P.},
        title = "{ABACUSSUMMIT: a massive set of high-accuracy, high-resolution N-body simulations}",
      journal = {\mnras},
         year = 2021,
        month = dec,
       volume = {508},
       number = {3},
        pages = {4017-4037},
          doi = {10.1093/mnras/stab2484},
archivePrefix = {arXiv},
       eprint = {2110.11398},
 primaryClass = {astro-ph.CO},
       adsurl = {https://ui.adsabs.harvard.edu/abs/2021MNRAS.508.4017M}
}

@ARTICLE{2022MNRAS.509.2194H,
       author = {{Hadzhiyska}, Boryana and {Garrison}, Lehman H. and {Eisenstein}, Daniel and {Bose}, Sownak},
        title = "{The halo light-cone catalogues of ABACUSSUMMIT}",
      journal = {\mnras},
         year = 2022,
        month = jan,
       volume = {509},
       number = {2},
        pages = {2194-2208},
          doi = {10.1093/mnras/stab3066},
archivePrefix = {arXiv},
       eprint = {2110.11413},
 primaryClass = {astro-ph.CO},
       adsurl = {https://ui.adsabs.harvard.edu/abs/2022MNRAS.509.2194H}
}

@ARTICLE{2024MNRAS.530..947Y,
       author = {{Yuan}, Sihan and {Zhang}, Hanyu and {Ross}, Ashley J. and {Donald-McCann}, Jamie and {Hadzhiyska}, Boryana and {Wechsler}, Risa H. and {Zheng}, Zheng and {Alam}, Shadab and {Gonzalez-Perez}, Violeta and {Aguilar}, Jessica Nicole and {Ahlen}, Steven and {Bianchi}, Davide and {Brooks}, David and {de la Macorra}, Axel and {Fanning}, Kevin and {Forero-Romero}, Jaime E. and {Honscheid}, Klaus and {Ishak}, Mustapha and {Kehoe}, Robert and {Lasker}, James and {Landriau}, Martin and {Manera}, Marc and {Martini}, Paul and {Meisner}, Aaron and {Miquel}, Ramon and {Moustakas}, John and {Nadathur}, Seshadri and {Newman}, Jeffrey A. and {Nie}, Jundan and {Percival}, Will and {Poppett}, Claire and {Rocher}, Antoine and {Rossi}, Graziano and {Sanchez}, Eusebio and {Samushia}, Lado and {Schubnell}, Michael and {Seo}, Hee-Jong and {Tarl{\'e}}, Gregory and {Weaver}, Benjamin Alan and {Yu}, Jiaxi and {Zhou}, Zhimin and {Zou}, Hu},
        title = "{The DESI one-per cent survey: exploring the halo occupation distribution of luminous red galaxies and quasi-stellar objects with ABACUSSUMMIT}",
      journal = {\mnras},
         year = 2024,
        month = may,
       volume = {530},
       number = {1},
        pages = {947-965},
          doi = {10.1093/mnras/stae359},
archivePrefix = {arXiv},
       eprint = {2306.06314},
 primaryClass = {astro-ph.CO},
       adsurl = {https://ui.adsabs.harvard.edu/abs/2024MNRAS.530..947Y}
}

@ARTICLE{2021MNRAS.508..575G,
       author = {{Garrison}, Lehman H. and {Eisenstein}, Daniel J. and {Ferrer}, Douglas and {Maksimova}, Nina A. and {Pinto}, Philip A.},
        title = "{The ABACUS cosmological N-body code}",
      journal = {\mnras},
         year = 2021,
        month = nov,
       volume = {508},
       number = {1},
        pages = {575-596},
          doi = {10.1093/mnras/stab2482},
archivePrefix = {arXiv},
       eprint = {2110.11392},
 primaryClass = {astro-ph.CO},
       adsurl = {https://ui.adsabs.harvard.edu/abs/2021MNRAS.508..575G}
}

@ARTICLE{maus2025,
       author = {{Maus}, M. and {Chen}, S. and {White}, M. and {Aguilar}, J. and {Ahlen}, S. and {Aviles}, A. and {Brieden}, S. and {Brooks}, D. and {Claybaugh}, T. and {Cole}, S. and {de la Macorra}, A. and {Dey}, Arjun and {Doel}, P. and {Ferraro}, S. and {Findlay}, N. and {Forero-Romero}, J.~E. and {Gazta{\~n}aga}, E. and {Gil-Mar{\'\i}n}, H. and {Gontcho}, S. Gontcho A. and {Hahn}, C. and {Honscheid}, K. and {Howlett}, C. and {Ishak}, M. and {Juneau}, S. and {Kremin}, A. and {Lai}, Y. and {Landriau}, M. and {Levi}, M.~E. and {Manera}, M. and {Miquel}, R. and {Mueller}, E. and {Myers}, A.~D. and {Nadathur}, S. and {Nie}, J. and {Noriega}, H.~E. and {Palanque-Delabrouille}, N. and {Percival}, W.~J. and {Poppett}, C. and {Ramirez-Solano}, S. and {Rezaie}, M. and {Rocher}, A. and {Rossi}, G. and {Sanchez}, E. and {Schlegel}, D. and {Schubnell}, M. and {Seo}, H. and {Sprayberry}, D. and {Tarl{\'e}}, G. and {Vargas-Maga{\~n}a}, M. and {Weaver}, B.~A. and {Yuan}, S. and {Zarrouk}, P. and {Zhang}, H. and {Zhou}, R. and {Zou}, H.},
        title = "{An analysis of parameter compression and Full-Modeling techniques with Velocileptors for DESI 2024 and beyond}",
      journal = {\jcap},
         year = 2025,
        month = jan,
       volume = {2025},
       number = {1},
          eid = {138},
        pages = {138},
          doi = {10.1088/1475-7516/2025/01/138},
archivePrefix = {arXiv},
       eprint = {2404.07312},
 primaryClass = {astro-ph.CO},
       adsurl = {https://ui.adsabs.harvard.edu/abs/2025JCAP...01..138M}
}

@ARTICLE{pezzotta2021,
       author = {{Pezzotta}, Andrea and {Crocce}, Martin and {Eggemeier}, Alexander and {S{\'a}nchez}, Ariel G. and {Scoccimarro}, Rom{\'a}n},
        title = "{Testing one-loop galaxy bias: Cosmological constraints from the power spectrum}",
      journal = {\prd},
         year = 2021,
        month = aug,
       volume = {104},
       number = {4},
          eid = {043531},
        pages = {043531},
          doi = {10.1103/PhysRevD.104.043531},
archivePrefix = {arXiv},
       eprint = {2102.08315},
 primaryClass = {astro-ph.CO},
       adsurl = {https://ui.adsabs.harvard.edu/abs/2021PhRvD.104d3531P}
}

@ARTICLE{asgari2023,
       author = {{Asgari}, Marika and {Mead}, Alexander J. and {Heymans}, Catherine},
        title = "{The halo model for cosmology: a pedagogical review}",
      journal = {The Open Journal of Astrophysics},
         year = 2023,
        month = nov,
       volume = {6},
          eid = {39},
        pages = {39},
          doi = {10.21105/astro.2303.08752},
archivePrefix = {arXiv},
       eprint = {2303.08752},
 primaryClass = {astro-ph.CO},
       adsurl = {https://ui.adsabs.harvard.edu/abs/2023OJAp....6E..39A}
}

@ARTICLE{chavesmontero2023,
       author = {{Chaves-Montero}, Jon{\'a}s and {Angulo}, Raul E. and {Contreras}, Sergio},
        title = "{The galaxy formation origin of the lensing is low problem}",
      journal = {\mnras},
         year = 2023,
        month = may,
       volume = {521},
       number = {1},
        pages = {937-951},
          doi = {10.1093/mnras/stad243},
archivePrefix = {arXiv},
       eprint = {2211.01744},
 primaryClass = {astro-ph.CO},
       adsurl = {https://ui.adsabs.harvard.edu/abs/2023MNRAS.521..937C}
}

@ARTICLE{mead2015,
       author = {{Mead}, A.~J. and {Peacock}, J.~A. and {Heymans}, C. and {Joudaki}, S. and {Heavens}, A.~F.},
        title = "{An accurate halo model for fitting non-linear cosmological power spectra and baryonic feedback models}",
      journal = {\mnras},
         year = 2015,
        month = dec,
       volume = {454},
       number = {2},
        pages = {1958-1975},
          doi = {10.1093/mnras/stv2036},
archivePrefix = {arXiv},
       eprint = {1505.07833},
 primaryClass = {astro-ph.CO},
       adsurl = {https://ui.adsabs.harvard.edu/abs/2015MNRAS.454.1958M}
}

@ARTICLE{nishimichi2017,
       author = {{Nishimichi}, Takahiro and {Bernardeau}, Francis and {Taruya}, Atsushi},
        title = "{Moving around the cosmological parameter space: A nonlinear power spectrum reconstruction based on high-resolution cosmic responses}",
      journal = {\prd},
         year = 2017,
        month = dec,
       volume = {96},
       number = {12},
          eid = {123515},
        pages = {123515},
          doi = {10.1103/PhysRevD.96.123515},
archivePrefix = {arXiv},
       eprint = {1708.08946},
 primaryClass = {astro-ph.CO},
       adsurl = {https://ui.adsabs.harvard.edu/abs/2017PhRvD..96l3515N}
}

@ARTICLE{modi2020,
       author = {{Modi}, Chirag and {Chen}, Shi-Fan and {White}, Martin},
        title = "{Simulations and symmetries}",
      journal = {\mnras},
         year = 2020,
        month = mar,
       volume = {492},
       number = {4},
        pages = {5754-5763},
          doi = {10.1093/mnras/staa251},
archivePrefix = {arXiv},
       eprint = {1910.07097},
 primaryClass = {astro-ph.CO},
       adsurl = {https://ui.adsabs.harvard.edu/abs/2020MNRAS.492.5754M}
}

@ARTICLE{chen2024,
       author = {{Chen}, S. and {DeRose}, J. and {Zhou}, R. and {White}, M. and {Ferraro}, S. and {Blake}, C. and {Lange}, J.~U. and {Wechsler}, R.~H. and {Aguilar}, J. and {Ahlen}, S. and {Brooks}, D. and {Claybaugh}, T. and {Dawson}, K. and {de la Macorra}, A. and {Doel}, P. and {Font-Ribera}, A. and {Gazta{\~n}aga}, E. and {Gontcho A Gontcho}, S. and {Gutierrez}, G. and {Honscheid}, K. and {Howlett}, C. and {Kehoe}, R. and {Kirkby}, D. and {Kisner}, T. and {Kremin}, A. and {Landriau}, M. and {Le Guillou}, L. and {Manera}, M. and {Meisner}, A. and {Miquel}, R. and {Newman}, J.~A. and {Niz}, G. and {Palanque-Delabrouille}, N. and {Percival}, W.~J. and {Prada}, F. and {Rossi}, G. and {Sanchez}, E. and {Schlegel}, D. and {Schubnell}, M. and {Sprayberry}, D. and {Tarl{\'e}}, G. and {Weaver}, B.~A.},
        title = "{Analysis of DESI{\texttimes}DES using the Lagrangian effective theory of LSS}",
      journal = {\prd},
         year = 2024,
        month = nov,
       volume = {110},
       number = {10},
          eid = {103518},
        pages = {103518},
          doi = {10.1103/PhysRevD.110.103518},
archivePrefix = {arXiv},
       eprint = {2407.04795},
 primaryClass = {astro-ph.CO},
       adsurl = {https://ui.adsabs.harvard.edu/abs/2024PhRvD.110j3518C}
}

@ARTICLE{hadzhiyska2021,
       author = {{Hadzhiyska}, Boryana and {Garc{\'\i}a-Garc{\'\i}a}, Carlos and {Alonso}, David and {Nicola}, Andrina and {Slosar}, An{\v{z}}e},
        title = "{Hefty enhancement of cosmological constraints from the DES Y1 data using a hybrid effective field theory approach to galaxy bias}",
      journal = {\jcap},
         year = 2021,
        month = sep,
       volume = {2021},
       number = {9},
          eid = {020},
        pages = {020},
          doi = {10.1088/1475-7516/2021/09/020},
archivePrefix = {arXiv},
       eprint = {2103.09820},
 primaryClass = {astro-ph.CO},
       adsurl = {https://ui.adsabs.harvard.edu/abs/2021JCAP...09..020H}
}

@ARTICLE{eggemeier2023,
       author = {{Eggemeier}, Alexander and {Camacho-Quevedo}, Benjamin and {Pezzotta}, Andrea and {Crocce}, Martin and {Scoccimarro}, Rom{\'a}n and {S{\'a}nchez}, Ariel G.},
        title = "{COMET: Clustering observables modelled by emulated perturbation theory}",
      journal = {\mnras},
         year = 2023,
        month = feb,
       volume = {519},
       number = {2},
        pages = {2962-2980},
          doi = {10.1093/mnras/stac3667},
archivePrefix = {arXiv},
       eprint = {2208.01070},
 primaryClass = {astro-ph.CO},
       adsurl = {https://ui.adsabs.harvard.edu/abs/2023MNRAS.519.2962E}
}

@ARTICLE{eggemeier2025,
       author = {{Eggemeier}, Alexander and {Lee}, Nanoom and {Scoccimarro}, Rom{\'a}n and {Camacho-Quevedo}, Benjamin and {Pezzotta}, Andrea and {Crocce}, Martin and {S{\'a}nchez}, Ariel G.},
        title = "{Boosting galaxy clustering analyses with non-perturbative modelling of redshift-space distortions}",
      journal = {arXiv e-prints},
         year = 2025,
        month = jan,
          eid = {arXiv:2501.18597},
        pages = {arXiv:2501.18597},
          doi = {10.48550/arXiv.2501.18597},
archivePrefix = {arXiv},
       eprint = {2501.18597},
 primaryClass = {astro-ph.CO},
       adsurl = {https://ui.adsabs.harvard.edu/abs/2025arXiv250118597E}
}

@ARTICLE{damico2020,
       author = {{d'Amico}, Guido and {Gleyzes}, J{\'e}r{\^o}me and {Kokron}, Nickolas and {Markovic}, Katarina and {Senatore}, Leonardo and {Zhang}, Pierre and {Beutler}, Florian and {Gil-Mar{\'\i}n}, H{\'e}ctor},
        title = "{The cosmological analysis of the SDSS/BOSS data from the Effective Field Theory of Large-Scale Structure}",
      journal = {\jcap},
         year = 2020,
        month = may,
       volume = {2020},
       number = {5},
          eid = {005},
        pages = {005},
          doi = {10.1088/1475-7516/2020/05/005},
archivePrefix = {arXiv},
       eprint = {1909.05271},
 primaryClass = {astro-ph.CO},
       adsurl = {https://ui.adsabs.harvard.edu/abs/2020JCAP...05..005D}
}

@ARTICLE{ivanov2020,
       author = {{Ivanov}, Mikhail M. and {Simonovi{\'c}}, Marko and {Zaldarriaga}, Matias},
        title = "{Cosmological parameters from the BOSS galaxy power spectrum}",
      journal = {\jcap},
         year = 2020,
        month = may,
       volume = {2020},
       number = {5},
          eid = {042},
        pages = {042},
          doi = {10.1088/1475-7516/2020/05/042},
archivePrefix = {arXiv},
       eprint = {1909.05277},
 primaryClass = {astro-ph.CO},
       adsurl = {https://ui.adsabs.harvard.edu/abs/2020JCAP...05..042I}
}

@ARTICLE{baumann2012,
       author = {{Baumann}, Daniel and {Nicolis}, Alberto and {Senatore}, Leonardo and {Zaldarriaga}, Matias},
        title = "{Cosmological non-linearities as an effective fluid}",
      journal = {\jcap},
         year = 2012,
        month = jul,
       volume = {2012},
       number = {7},
          eid = {051},
        pages = {051},
          doi = {10.1088/1475-7516/2012/07/051},
archivePrefix = {arXiv},
       eprint = {1004.2488},
 primaryClass = {astro-ph.CO},
       adsurl = {https://ui.adsabs.harvard.edu/abs/2012JCAP...07..051B}
}

@ARTICLE{senatore2014,
       author = {{Senatore}, Leonardo and {Zaldarriaga}, Matias},
        title = "{Redshift Space Distortions in the Effective Field Theory of Large Scale Structures}",
      journal = {arXiv e-prints},
         year = 2014,
        month = sep,
          eid = {arXiv:1409.1225},
        pages = {arXiv:1409.1225},
          doi = {10.48550/arXiv.1409.1225},
archivePrefix = {arXiv},
       eprint = {1409.1225},
 primaryClass = {astro-ph.CO},
       adsurl = {https://ui.adsabs.harvard.edu/abs/2014arXiv1409.1225S}
}

@ARTICLE{desjacques18,
       author = {{Desjacques}, Vincent and {Jeong}, Donghui and {Schmidt}, Fabian},
        title = "{The galaxy power spectrum and bispectrum in redshift space}",
      journal = {J.\ Cosmology\ Astropart.\ Phys.},
         year = 2018,
        month = dec,
       volume = {2018},
       number = {12},
          eid = {035},
        pages = {035},
          doi = {10.1088/1475-7516/2018/12/035},
archivePrefix = {arXiv},
       eprint = {1806.04015},
 primaryClass = {astro-ph.CO},
       adsurl = {https://ui.adsabs.harvard.edu/abs/2018JCAP...12..035D}
}

@article{pueblas2009,
  title = {Generation of vorticity and velocity dispersion by orbit crossing},
  author = {Pueblas, Sebasti\'an and Scoccimarro, Rom\'an},
  journal = {Phys. Rev. D},
  volume = {80},
  issue = {4},
  pages = {043504},
  numpages = {21},
  year = {2009},
  month = {Aug},
  publisher = {American Physical Society},
  doi = {10.1103/PhysRevD.80.043504},
  url = {https://link.aps.org/doi/10.1103/PhysRevD.80.043504}
}

@ARTICLE{carrasco2012,
       author = {{Carrasco}, John Joseph M. and {Hertzberg}, Mark P. and {Senatore}, Leonardo},
        title = "{The effective field theory of cosmological large scale structures}",
      journal = {Journal of High Energy Physics},
         year = 2012,
        month = sep,
       volume = {2012},
          eid = {82},
        pages = {82},
          doi = {10.1007/JHEP09(2012)082},
archivePrefix = {arXiv},
       eprint = {1206.2926},
 primaryClass = {astro-ph.CO},
       adsurl = {https://ui.adsabs.harvard.edu/abs/2012JHEP...09..082C}
}

@ARTICLE{philcox2022,
       author = {{Philcox}, Oliver H.~E. and {Ivanov}, Mikhail M.},
        title = "{BOSS DR12 full-shape cosmology: {\ensuremath{\Lambda}} CDM constraints from the large-scale galaxy power spectrum and bispectrum monopole}",
      journal = {\prd},
         year = 2022,
        month = feb,
       volume = {105},
       number = {4},
          eid = {043517},
        pages = {043517},
          doi = {10.1103/PhysRevD.105.043517},
archivePrefix = {arXiv},
       eprint = {2112.04515},
 primaryClass = {astro-ph.CO},
       adsurl = {https://ui.adsabs.harvard.edu/abs/2022PhRvD.105d3517P}
}

@ARTICLE{sanchez2017,
       author = {{S{\'a}nchez}, Ariel G. and {Scoccimarro}, Rom{\'a}n and {Crocce}, Mart{\'\i}n and {Grieb}, Jan Niklas and {Salazar-Albornoz}, Salvador and {Dalla Vecchia}, Claudio and {Lippich}, Martha and {Beutler}, Florian and {Brownstein}, Joel R. and {Chuang}, Chia-Hsun and {Eisenstein}, Daniel J. and {Kitaura}, Francisco-Shu and {Olmstead}, Matthew D. and {Percival}, Will J. and {Prada}, Francisco and {Rodr{\'\i}guez-Torres}, Sergio and {Ross}, Ashley J. and {Samushia}, Lado and {Seo}, Hee-Jong and {Tinker}, Jeremy and {Tojeiro}, Rita and {Vargas-Maga{\~n}a}, Mariana and {Wang}, Yuting and {Zhao}, Gong-Bo},
        title = "{The clustering of galaxies in the completed SDSS-III Baryon Oscillation Spectroscopic Survey: Cosmological implications of the configuration-space clustering wedges}",
      journal = {\mnras},
         year = 2017,
        month = jan,
       volume = {464},
       number = {2},
        pages = {1640-1658},
          doi = {10.1093/mnras/stw2443},
archivePrefix = {arXiv},
       eprint = {1607.03147},
 primaryClass = {astro-ph.CO},
       adsurl = {https://ui.adsabs.harvard.edu/abs/2017MNRAS.464.1640S}
}

@ARTICLE{sheth1996,
       author = {{Sheth}, Ravi K.},
        title = "{The distribution of pairwise peculiar velocities in the non-linear regime}",
      journal = {\mnras},
         year = 1996,
        month = apr,
       volume = {279},
        pages = {1310},
          doi = {10.1093/mnras/279.4.1310},
archivePrefix = {arXiv},
       eprint = {astro-ph/9511068},
 primaryClass = {astro-ph},
       adsurl = {https://ui.adsabs.harvard.edu/abs/1996MNRAS.279.1310S}
}

@ARTICLE{juszkiewicz1998,
       author = {{Juszkiewicz}, Roman and {Fisher}, Karl B. and {Szapudi}, Istv{\'a}n},
        title = "{Skewed Exponential Pairwise Velocities from Gaussian Initial Conditions}",
      journal = {\apjl},
         year = 1998,
        month = sep,
       volume = {504},
       number = {1},
        pages = {L1-L4},
          doi = {10.1086/311558},
archivePrefix = {arXiv},
       eprint = {astro-ph/9804277},
 primaryClass = {astro-ph},
       adsurl = {https://ui.adsabs.harvard.edu/abs/1998ApJ...504L...1J}
}

@article{scoccimarro2004,
  title = {Redshift-space distortions, pairwise velocities, and nonlinearities},
  author = {Scoccimarro, Rom\'an},
  journal = {Phys. Rev. D},
  volume = {70},
  issue = {8},
  pages = {083007},
  numpages = {19},
  year = {2004},
  month = {Oct},
  publisher = {American Physical Society},
  doi = {10.1103/PhysRevD.70.083007},
  url = {https://link.aps.org/doi/10.1103/PhysRevD.70.083007}
}

@ARTICLE{cuestalazaro2020,
       author = {{Cuesta-Lazaro}, Carolina and {Li}, Baojiu and {Eggemeier}, Alexander and {Zarrouk}, Pauline and {Baugh}, Carlton M. and {Nishimichi}, Takahiro and {Takada}, Masahiro},
        title = "{Towards a non-Gaussian model of redshift space distortions}",
      journal = {\mnras},
         year = 2020,
        month = oct,
       volume = {498},
       number = {1},
        pages = {1175-1193},
          doi = {10.1093/mnras/staa2249},
archivePrefix = {arXiv},
       eprint = {2002.02683},
 primaryClass = {astro-ph.CO},
       adsurl = {https://ui.adsabs.harvard.edu/abs/2020MNRAS.498.1175C}
}

@ARTICLE{hou2021,
       author = {{Hou}, Jiamin and {S{\'a}nchez}, Ariel G. and {Ross}, Ashley J. and {Smith}, Alex and {Neveux}, Richard and {Bautista}, Julian and {Burtin}, Etienne and {Zhao}, Cheng and {Scoccimarro}, Rom{\'a}n and {Dawson}, Kyle S. and {de Mattia}, Arnaud and {de la Macorra}, Axel and {du Mas des Bourboux}, H{\'e}lion and {Eisenstein}, Daniel J. and {Gil-Mar{\'\i}n}, H{\'e}ctor and {Lyke}, Brad W. and {Mohammad}, Faizan G. and {Mueller}, Eva-Maria and {Percival}, Will J. and {Rossi}, Graziano and {Vargas Maga{\~n}a}, Mariana and {Zarrouk}, Pauline and {Zhao}, Gong-Bo and {Brinkmann}, Jonathan and {Brownstein}, Joel R. and {Chuang}, Chia-Hsun and {Myers}, Adam D. and {Newman}, Jeffrey A. and {Schneider}, Donald P. and {Vivek}, M.},
        title = "{The completed SDSS-IV extended Baryon Oscillation Spectroscopic Survey: BAO and RSD measurements from anisotropic clustering analysis of the quasar sample in configuration space between redshift 0.8 and 2.2}",
      journal = {\mnras},
         year = 2021,
        month = jan,
       volume = {500},
       number = {1},
        pages = {1201-1221},
          doi = {10.1093/mnras/staa3234},
archivePrefix = {arXiv},
       eprint = {2007.08998},
 primaryClass = {astro-ph.CO},
       adsurl = {https://ui.adsabs.harvard.edu/abs/2021MNRAS.500.1201H}
}

@ARTICLE{semenaite2023,
       author = {{Semenaite}, Agne and {S{\'a}nchez}, Ariel G. and {Pezzotta}, Andrea and {Hou}, Jiamin and {Eggemeier}, Alexander and {Crocce}, Martin and {Zhao}, Cheng and {Brownstein}, Joel R. and {Rossi}, Graziano and {Schneider}, Donald P.},
        title = "{Beyond - {\ensuremath{\Lambda}}CDM constraints from the full shape clustering measurements from BOSS and eBOSS}",
      journal = {\mnras},
         year = 2023,
        month = jun,
       volume = {521},
       number = {4},
        pages = {5013-5025},
          doi = {10.1093/mnras/stad849},
archivePrefix = {arXiv},
       eprint = {2210.07304},
 primaryClass = {astro-ph.CO},
       adsurl = {https://ui.adsabs.harvard.edu/abs/2023MNRAS.521.5013S}
}

@ARTICLE{semenaite2022,
       author = {{Semenaite}, Agne and {S{\'a}nchez}, Ariel G. and {Pezzotta}, Andrea and {Hou}, Jiamin and {Scoccimarro}, Roman and {Eggemeier}, Alexander and {Crocce}, Martin and {Chuang}, Chia-Hsun and {Smith}, Alexander and {Zhao}, Cheng and {Brownstein}, Joel R. and {Rossi}, Graziano and {Schneider}, Donald P.},
        title = "{Cosmological implications of the full shape of anisotropic clustering measurements in BOSS and eBOSS}",
      journal = {\mnras},
         year = 2022,
        month = jun,
       volume = {512},
       number = {4},
        pages = {5657-5670},
          doi = {10.1093/mnras/stac829},
archivePrefix = {arXiv},
       eprint = {2111.03156},
 primaryClass = {astro-ph.CO},
       adsurl = {https://ui.adsabs.harvard.edu/abs/2022MNRAS.512.5657S}
}

@article{eggemeier2019,
  title = {Bias loop corrections to the galaxy bispectrum},
  author = {Eggemeier, Alexander and Scoccimarro, Rom\'an and Smith, Robert E.},
  journal = {Phys. Rev. D},
  volume = {99},
  issue = {12},
  pages = {123514},
  numpages = {36},
  year = {2019},
  month = {Jun},
  publisher = {American Physical Society},
  doi = {10.1103/PhysRevD.99.123514},
  url = {https://link.aps.org/doi/10.1103/PhysRevD.99.123514}
}

@article{eggemeier2020,
  title = {Testing one-loop galaxy bias: Power spectrum},
  author = {Eggemeier, Alexander and Scoccimarro, Rom\'an and Crocce, Martin and Pezzotta, Andrea and S\'anchez, Ariel G.},
  journal = {Phys. Rev. D},
  volume = {102},
  issue = {10},
  pages = {103530},
  numpages = {29},
  year = {2020},
  month = {Nov},
  publisher = {American Physical Society},
  doi = {10.1103/PhysRevD.102.103530},
  url = {https://link.aps.org/doi/10.1103/PhysRevD.102.103530}
}

@article{fry1996,
	doi = {10.1086/310006},
	url = {https://doi.org/10.1086/310006},
	year = 1996,
	month = {apr},
	publisher = {American Astronomical Society},
	volume = {461},
	number = {2},
	author = {J. N. Fry},
	title = {The Evolution of Bias},
	journal = {The Astrophysical Journal}
}

@ARTICLE{chan2012,
       author = {{Chan}, Kwan Chuen and {Scoccimarro}, Rom{\'a}n and {Sheth}, Ravi K.},
        title = "{Gravity and large-scale nonlocal bias}",
      journal = {\prd},
         year = 2012,
        month = apr,
       volume = {85},
       number = {8},
          eid = {083509},
        pages = {083509},
          doi = {10.1103/PhysRevD.85.083509},
archivePrefix = {arXiv},
       eprint = {1201.3614},
 primaryClass = {astro-ph.CO},
       adsurl = {https://ui.adsabs.harvard.edu/abs/2012PhRvD..85h3509C}
}

@ARTICLE{catelan1998,
       author = {{Catelan}, Paolo and {Lucchin}, Francesco and {Matarrese}, Sabino and {Porciani}, Cristiano},
        title = "{The bias field of dark matter haloes}",
      journal = {\mnras},
         year = 1998,
        month = jul,
       volume = {297},
       number = {3},
        pages = {692-712},
          doi = {10.1046/j.1365-8711.1998.01455.x},
archivePrefix = {arXiv},
       eprint = {astro-ph/9708067},
 primaryClass = {astro-ph},
       adsurl = {https://ui.adsabs.harvard.edu/abs/1998MNRAS.297..692C}
}

@ARTICLE{eggemeier2021,
       author = {{Eggemeier}, Alexander and {Scoccimarro}, Rom{\'a}n and {Smith}, Robert E. and {Crocce}, Martin and {Pezzotta}, Andrea and {S{\'a}nchez}, Ariel G.},
        title = "{Testing one-loop galaxy bias: Joint analysis of power spectrum and bispectrum}",
      journal = {\prd},
         year = 2021,
        month = jun,
       volume = {103},
       number = {12},
          eid = {123550},
        pages = {123550},
          doi = {10.1103/PhysRevD.103.123550},
archivePrefix = {arXiv},
       eprint = {2102.06902},
 primaryClass = {astro-ph.CO},
       adsurl = {https://ui.adsabs.harvard.edu/abs/2021PhRvD.103l3550E}
}

@ARTICLE{assassi2014,
       author = {{Assassi}, Valentin and {Baumann}, Daniel and {Green}, Daniel and {Zaldarriaga}, Matias},
        title = "{Renormalized halo bias}",
      journal = {\jcap},
         year = 2014,
        month = aug,
       volume = {2014},
       number = {8},
        pages = {056-056},
          doi = {10.1088/1475-7516/2014/08/056},
archivePrefix = {arXiv},
       eprint = {1402.5916},
 primaryClass = {astro-ph.CO},
       adsurl = {https://ui.adsabs.harvard.edu/abs/2014JCAP...08..056A}
}

@ARTICLE{chen2020,
       author = {{Chen}, Shi-Fan and {Vlah}, Zvonimir and {White}, Martin},
        title = "{Consistent modeling of velocity statistics and redshift-space distortions in one-loop perturbation theory}",
      journal = {\jcap},
         year = 2020,
        month = jul,
       volume = {2020},
       number = {7},
          eid = {062},
        pages = {062},
          doi = {10.1088/1475-7516/2020/07/062},
archivePrefix = {arXiv},
       eprint = {2005.00523},
 primaryClass = {astro-ph.CO},
       adsurl = {https://ui.adsabs.harvard.edu/abs/2020JCAP...07..062C}
}

@Article{alcock1979,
author={Alcock, Charles
and Paczy{\'{n}}ski, Bohdan},
title={An evolution free test for non-zero cosmological constant},
journal={Nature},
year={1979},
month={Oct},
day={01},
volume={281},
number={5730},
pages={358-359},
issn={1476-4687},
doi={10.1038/281358a0},
url={https://doi.org/10.1038/281358a0}
}

@ARTICLE{sanchez2022,
       author = {{S{\'a}nchez}, Ariel G. and {Ruiz}, Andr{\'e}s N. and {Jara}, Jenny Gonzalez and {Padilla}, Nelson D.},
        title = "{Evolution mapping: a new approach to describe matter clustering in the non-linear regime}",
      journal = {\mnras},
         year = 2022,
        month = aug,
       volume = {514},
       number = {4},
        pages = {5673-5685},
          doi = {10.1093/mnras/stac1656},
archivePrefix = {arXiv},
       eprint = {2108.12710},
 primaryClass = {astro-ph.CO},
       adsurl = {https://ui.adsabs.harvard.edu/abs/2022MNRAS.514.5673S}
}

@ARTICLE{sanchez2020,
       author = {{S{\'a}nchez}, Ariel G.},
        title = "{Arguments against using h$^{-1}$ Mpc units in observational cosmology}",
      journal = {\prd},
         year = 2020,
        month = dec,
       volume = {102},
       number = {12},
          eid = {123511},
        pages = {123511},
          doi = {10.1103/PhysRevD.102.123511},
archivePrefix = {arXiv},
       eprint = {2002.07829},
 primaryClass = {astro-ph.CO},
       adsurl = {https://ui.adsabs.harvard.edu/abs/2020PhRvD.102l3511S}
}

@ARTICLE{esposito2024,
       author = {{Esposito}, Matteo and {S{\'a}nchez}, Ariel G. and {Bel}, Julien and {Ruiz}, Andr{\'e}s N.},
        title = "{Evolution mapping - II. Describing statistics of the non-linear cosmic velocity field}",
      journal = {\mnras},
         year = 2024,
        month = nov,
       volume = {534},
       number = {4},
        pages = {3906-3915},
          doi = {10.1093/mnras/stae2351},
archivePrefix = {arXiv},
       eprint = {2406.08539},
 primaryClass = {astro-ph.CO},
       adsurl = {https://ui.adsabs.harvard.edu/abs/2024MNRAS.534.3906E}
}

@ARTICLE{pezzotta2025,
       author = {{Pezzotta}, Andrea and {Eggemeier}, Alexander and {Gambardella}, Giosu{\`e} and {Finkbeiner}, Lukas and {S{\'a}nchez}, Ariel G. and {Camacho Quevedo}, Benjamin and {Crocce}, Martin and {Lee}, Nanoom and {Parimbelli}, Gabriele and {Scoccimarro}, Rom{\'a}n},
        title = "{Extending evolution mapping to massive neutrinos with COMET}",
      journal = {\prd},
         year = 2025,
        month = jul,
       volume = {112},
       number = {2},
          eid = {023520},
        pages = {023520},
          doi = {10.1103/vy3h-p92n},
archivePrefix = {arXiv},
       eprint = {2503.16160},
 primaryClass = {astro-ph.CO},
       adsurl = {https://ui.adsabs.harvard.edu/abs/2025PhRvD.112b3520P}
}

@article{karamanis2021,
title={hankl: A lightweight Python implementation of the FFTLog algorithm for Cosmology},
author={Karamanis, Minas and Beutler, Florian},
journal={arXiv preprint arXiv:2106.06331},
year={2021}
}

@ARTICLE{derose2019,
       author = {{DeRose}, Joseph and {Wechsler}, Risa H. and {Becker}, Matthew R. and {Busha}, Michael T. and {Rykoff}, Eli S. and {MacCrann}, Niall and {Erickson}, Brandon and {Evrard}, August E. and {Kravtsov}, Andrey and {Gruen}, Daniel and {Allam}, Sahar and {Avila}, Santiago and {Bridle}, Sarah and {Brooks}, David and {Buckley-Geer}, Elizabeth and {Carnero Rosell}, Aurelio and {Carrasco Kind}, Matias and {Carretero}, Jorge and {Castander}, Francisco J. and {Cawthon}, Ross and {Crocce}, Martin and {da Costa}, Luiz N. and {Davis}, Christopher and {De Vicente}, Juan and {Dietrich}, J{\"o}rg P. and {Doel}, Peter and {Drlica-Wagner}, Alex and {Fosalba}, Pablo and {Frieman}, Josh and {Garcia-Bellido}, Juan and {Gutierrez}, Gaston and {Hartley}, Will G. and {Hollowood}, Devon L. and {Hoyle}, Ben and {James}, David J. and {Krause}, Elisabeth and {Kuehn}, Kyler and {Kuropatkin}, Nikolay and {Lima}, Marcos and {Maia}, Marcio A.~G. and {Menanteau}, Felipe and {Miller}, Christopher J. and {Miquel}, Ramon and {Ogando}, Ricardo L.~C. and {Plazas Malag{\'o}n}, Andr{\'e}s and {Romer}, A. Kathy and {Sanchez}, Eusebio and {Schindler}, Rafe and {Serrano}, Santiago and {Sevilla-Noarbe}, Ignacio and {Smith}, Mathew and {Suchyta}, Eric and {Swanson}, Molly E.~C. and {Tarle}, Gregory and {Vikram}, Vinu},
        title = "{The Buzzard Flock: Dark Energy Survey Synthetic Sky Catalogs}",
      journal = {arXiv e-prints},
         year = 2019,
        month = jan,
          eid = {arXiv:1901.02401},
        pages = {arXiv:1901.02401},
          doi = {10.48550/arXiv.1901.02401},
archivePrefix = {arXiv},
       eprint = {1901.02401},
 primaryClass = {astro-ph.CO},
       adsurl = {https://ui.adsabs.harvard.edu/abs/2019arXiv190102401D}
}

@ARTICLE{guzik2001,
       author = {{Guzik}, Jacek and {Seljak}, Uro{\v{s}}},
        title = "{Galaxy-dark matter correlations applied to galaxy-galaxy lensing: predictions from the semi-analytic galaxy formation models}",
      journal = {\mnras},
         year = 2001,
        month = mar,
       volume = {321},
       number = {3},
        pages = {439-449},
          doi = {10.1046/j.1365-8711.2001.04081.x},
archivePrefix = {arXiv},
       eprint = {astro-ph/0007067},
 primaryClass = {astro-ph},
       adsurl = {https://ui.adsabs.harvard.edu/abs/2001MNRAS.321..439G}
}

@article{hu2004,
  title = {Joint galaxy-lensing observables and the dark energy},
  author = {Hu, Wayne and Jain, Bhuvnesh},
  journal = {Phys. Rev. D},
  volume = {70},
  issue = {4},
  pages = {043009},
  numpages = {16},
  year = {2004},
  month = {Aug},
  publisher = {American Physical Society},
  doi = {10.1103/PhysRevD.70.043009},
  url = {https://link.aps.org/doi/10.1103/PhysRevD.70.043009}
}

@ARTICLE{joachimi2010,
       author = {{Joachimi}, B. and {Bridle}, S.~L.},
        title = "{Simultaneous measurement of cosmology and intrinsic alignments using joint cosmic shear and galaxy number density correlations}",
      journal = {\aap},
         year = 2010,
        month = nov,
       volume = {523},
          eid = {A1},
        pages = {A1},
          doi = {10.1051/0004-6361/200913657},
archivePrefix = {arXiv},
       eprint = {0911.2454},
 primaryClass = {astro-ph.CO},
       adsurl = {https://ui.adsabs.harvard.edu/abs/2010A&A...523A...1J}
}

@ARTICLE{mead2021,
       author = {{Mead}, A.~J. and {Brieden}, S. and {Tr{\"o}ster}, T. and {Heymans}, C.},
        title = "{HMCODE-2020: improved modelling of non-linear cosmological power spectra with baryonic feedback}",
      journal = {\mnras},
         year = 2021,
        month = mar,
       volume = {502},
       number = {1},
        pages = {1401-1422},
          doi = {10.1093/mnras/stab082},
archivePrefix = {arXiv},
       eprint = {2009.01858},
 primaryClass = {astro-ph.CO},
       adsurl = {https://ui.adsabs.harvard.edu/abs/2021MNRAS.502.1401M}
}

@ARTICLE{lewis2000,
       author = {{Lewis}, Antony and {Challinor}, Anthony and {Lasenby}, Anthony},
        title = "{Efficient Computation of Cosmic Microwave Background Anisotropies in Closed Friedmann-Robertson-Walker Models}",
      journal = {\apj},
         year = 2000,
        month = aug,
       volume = {538},
       number = {2},
        pages = {473-476},
          doi = {10.1086/309179},
archivePrefix = {arXiv},
       eprint = {astro-ph/9911177},
 primaryClass = {astro-ph},
       adsurl = {https://ui.adsabs.harvard.edu/abs/2000ApJ...538..473L}
}

@ARTICLE{poole2023,
       author = {{Elvin-Poole}, J. and {MacCrann}, N. and {Everett}, S. and {Prat}, J. and {Rykoff}, E.~S. and {De Vicente}, J. and {Yanny}, B. and {Herner}, K. and {Fert{\'e}}, A. and {Di Valentino}, E. and {Choi}, A. and {Burke}, D.~L. and {Sevilla-Noarbe}, I. and {Alarcon}, A. and {Alves}, O. and {Amon}, A. and {Andrade-Oliveira}, F. and {Baxter}, E. and {Bechtol}, K. and {Becker}, M.~R. and {Bernstein}, G.~M. and {Blazek}, J. and {Camacho}, H. and {Campos}, A. and {Carnero Rosell}, A. and {Carrasco Kind}, M. and {Cawthon}, R. and {Chang}, C. and {Chen}, R. and {Cordero}, J. and {Crocce}, M. and {Davis}, C. and {DeRose}, J. and {Diehl}, H.~T. and {Dodelson}, S. and {Doux}, C. and {Drlica-Wagner}, A. and {Eckert}, K. and {Eifler}, T.~F. and {Elsner}, F. and {Fang}, X. and {Fosalba}, P. and {Friedrich}, O. and {Gatti}, M. and {Giannini}, G. and {Gruen}, D. and {Gruendl}, R.~A. and {Harrison}, I. and {Hartley}, W.~G. and {Huang}, H. and {Huff}, E.~M. and {Huterer}, D. and {Krause}, E. and {Kuropatkin}, N. and {Leget}, P. -F. and {Lemos}, P. and {Liddle}, A.~R. and {McCullough}, J. and {Muir}, J. and {Myles}, J. and {Navarro-Alsina}, A. and {Pandey}, S. and {Park}, Y. and {Porredon}, A. and {Raveri}, M. and {Rodriguez-Monroy}, M. and {Rollins}, R.~P. and {Roodman}, A. and {Rosenfeld}, R. and {Ross}, A.~J. and {S{\'a}nchez}, C. and {Sanchez}, J. and {Secco}, L.~F. and {Sheldon}, E. and {Shin}, T. and {Troxel}, M.~A. and {Tutusaus}, I. and {Varga}, T.~N. and {Weaverdyck}, N. and {Wechsler}, R.~H. and {Yin}, B. and {Zhang}, Y. and {Zuntz}, J. and {Aguena}, M. and {Avila}, S. and {Bacon}, D. and {Bertin}, E. and {Bocquet}, S. and {Brooks}, D. and {Garc{\'\i}a-Bellido}, J. and {Honscheid}, K. and {Jarvis}, M. and {Li}, T.~S. and {Mena-Fern{\'a}ndez}, J. and {To}, C. and {Wilkinson}, R.~D. and {DES Collaboration}},
        title = "{Dark Energy Survey Year 3 results: magnification modelling and impact on cosmological constraints from galaxy clustering and galaxy-galaxy lensing}",
      journal = {\mnras},
         year = 2023,
        month = aug,
       volume = {523},
       number = {3},
        pages = {3649-3670},
          doi = {10.1093/mnras/stad1594},
archivePrefix = {arXiv},
       eprint = {2209.09782},
 primaryClass = {astro-ph.CO},
       adsurl = {https://ui.adsabs.harvard.edu/abs/2023MNRAS.523.3649E}
}

@ARTICLE{lamman2024,
       author = {{Lamman}, Claire and {Tsaprazi}, Eleni and {Shi}, Jingjing and {{\v{S}}ar{\v{c}}evi{\'c}}, Nikolina Niko and {Pyne}, Susan and {Legnani}, Elisa and {Ferreira}, Tassia},
        title = "{The IA Guide: A Breakdown of Intrinsic Alignment Formalisms}",
      journal = {The Open Journal of Astrophysics},
         year = 2024,
        month = feb,
       volume = {7},
          eid = {14},
        pages = {14},
          doi = {10.21105/astro.2309.08605},
archivePrefix = {arXiv},
       eprint = {2309.08605},
 primaryClass = {astro-ph.CO},
       adsurl = {https://ui.adsabs.harvard.edu/abs/2024OJAp....7E..14L}
}

@ARTICLE{hirata2004,
       author = {{Hirata}, Christopher M. and {Seljak}, Uro{\v{s}}},
        title = "{Intrinsic alignment-lensing interference as a contaminant of cosmic shear}",
      journal = {\prd},
         year = 2004,
        month = sep,
       volume = {70},
       number = {6},
          eid = {063526},
        pages = {063526},
          doi = {10.1103/PhysRevD.70.063526},
archivePrefix = {arXiv},
       eprint = {astro-ph/0406275},
 primaryClass = {astro-ph},
       adsurl = {https://ui.adsabs.harvard.edu/abs/2004PhRvD..70f3526H}
}

@ARTICLE{bridle2007,
       author = {{Bridle}, Sarah and {King}, Lindsay},
        title = "{Dark energy constraints from cosmic shear power spectra: impact of intrinsic alignments on photometric redshift requirements}",
      journal = {New Journal of Physics},
         year = 2007,
        month = dec,
       volume = {9},
       number = {12},
        pages = {444},
          doi = {10.1088/1367-2630/9/12/444},
archivePrefix = {arXiv},
       eprint = {0705.0166},
 primaryClass = {astro-ph},
       adsurl = {https://ui.adsabs.harvard.edu/abs/2007NJPh....9..444B}
}

@ARTICLE{scoccimarro1999,
       author = {{Scoccimarro}, Rom{\'a}n and {Couchman}, H.~M.~P. and {Frieman}, Joshua A.},
        title = "{The Bispectrum as a Signature of Gravitational Instability in Redshift Space}",
      journal = {\apj},
         year = 1999,
        month = jun,
       volume = {517},
       number = {2},
        pages = {531-540},
          doi = {10.1086/307220},
archivePrefix = {arXiv},
       eprint = {astro-ph/9808305},
 primaryClass = {astro-ph},
       adsurl = {https://ui.adsabs.harvard.edu/abs/1999ApJ...517..531S}
}

@ARTICLE{zuntz2015,
       author = {{Zuntz}, J. and {Paterno}, M. and {Jennings}, E. and {Rudd}, D. and {Manzotti}, A. and {Dodelson}, S. and {Bridle}, S. and {Sehrish}, S. and {Kowalkowski}, J.},
        title = "{CosmoSIS: Modular cosmological parameter estimation}",
      journal = {Astronomy and Computing},
         year = 2015,
        month = sep,
       volume = {12},
        pages = {45-59},
          doi = {10.1016/j.ascom.2015.05.005},
archivePrefix = {arXiv},
       eprint = {1409.3409},
 primaryClass = {astro-ph.CO},
       adsurl = {https://ui.adsabs.harvard.edu/abs/2015A&C....12...45Z}
}

@ARTICLE{lange2023,
       author = {{Lange}, Johannes U.},
        title = "{NAUTILUS: boosting Bayesian importance nested sampling with deep learning}",
      journal = {\mnras},
         year = 2023,
        month = oct,
       volume = {525},
       number = {2},
        pages = {3181-3194},
          doi = {10.1093/mnras/stad2441},
archivePrefix = {arXiv},
       eprint = {2306.16923},
 primaryClass = {astro-ph.IM},
       adsurl = {https://ui.adsabs.harvard.edu/abs/2023MNRAS.525.3181L}
}

@ARTICLE{ramirez2025,
       author = {{Ramirez-Solano}, S. and {Icaza-Lizaola}, M. and {Noriega}, H.~E. and {Vargas-Maga{\~n}a}, M. and {Fromenteau}, S. and {Aviles}, A. and {Rodr{\'\i}guez-Mart{\'\i}nez}, F. and {Aguilar}, J. and {Ahlen}, S. and {Alves}, O. and {Brieden}, S. and {Brooks}, D. and {Claybaugh}, T. and {Cole}, S. and {de la Macorra}, A. and {Dey}, Arjun and {Dey}, B. and {Doel}, P. and {Fanning}, K. and {Forero-Romero}, J.~E. and {Gazta{\~n}aga}, E. and {Gil-Mar{\'\i}n}, H. and {Gontcho A Gontcho}, S. and {Honscheid}, K. and {Howlett}, C. and {Juneau}, S. and {Lai}, Y. and {Landriau}, M. and {Manera}, M. and {Maus}, M. and {Miquel}, R. and {Mueller}, E. and {Mu{\~n}oz-Guti{\'e}rrez}, A. and {Myers}, A.~D. and {Nadathur}, S. and {Nie}, J. and {Percival}, W.~J. and {Poppett}, C. and {Rezaie}, M. and {Rossi}, G. and {Sanchez}, E. and {Schlegel}, D. and {Schubnell}, M. and {Seo}, H. and {Sprayberry}, D. and {Tarl{\'e}}, G. and {Verde}, L. and {Weaver}, B.~A. and {Wechsler}, R.~H. and {Yuan}, S. and {Zarrouk}, P. and {Zou}, H.},
        title = "{Full Modeling and parameter compression methods in configuration space for DESI 2024 and beyond}",
      journal = {\jcap},
         year = 2025,
        month = jan,
       volume = {2025},
       number = {1},
          eid = {129},
        pages = {129},
          doi = {10.1088/1475-7516/2025/01/129},
archivePrefix = {arXiv},
       eprint = {2404.07268},
 primaryClass = {astro-ph.CO},
       adsurl = {https://ui.adsabs.harvard.edu/abs/2025JCAP...01..129R}
}

@ARTICLE{carrilho2023,
       author = {{Carrilho}, Pedro and {Moretti}, Chiara and {Pourtsidou}, Alkistis},
        title = "{Cosmology with the EFTofLSS and BOSS: dark energy constraints and a note on priors}",
      journal = {\jcap},
         year = 2023,
        month = jan,
       volume = {2023},
       number = {1},
          eid = {028},
        pages = {028},
          doi = {10.1088/1475-7516/2023/01/028},
archivePrefix = {arXiv},
       eprint = {2207.14784},
 primaryClass = {astro-ph.CO},
       adsurl = {https://ui.adsabs.harvard.edu/abs/2023JCAP...01..028C}
}

@ARTICLE{tsedrik2025b,
       author = {{Tsedrik}, Maria and {Carrilho}, Pedro and {Moretti}, Chiara},
        title = "{The simple way to measure evolving dark energy without prior-volume effects}",
      journal = {arXiv e-prints},
         year = 2025,
        month = sep,
          eid = {arXiv:2509.09562},
        pages = {arXiv:2509.09562},
          doi = {10.48550/arXiv.2509.09562},
archivePrefix = {arXiv},
       eprint = {2509.09562},
 primaryClass = {astro-ph.CO},
       adsurl = {https://ui.adsabs.harvard.edu/abs/2025arXiv250909562T}
}

@ARTICLE{simon2022,
       author = {{Simon}, Th{\'e}o and {Zhang}, Pierre and {Poulin}, Vivian and {Smith}, Tristan L.},
        title = "{On the consistency of effective field theory analyses of BOSS power spectrum}",
      journal = {arXiv e-prints},
         year = 2022,
        month = aug,
          eid = {arXiv:2208.05929},
        pages = {arXiv:2208.05929},
          doi = {10.48550/arXiv.2208.05929},
archivePrefix = {arXiv},
       eprint = {2208.05929},
 primaryClass = {astro-ph.CO},
       adsurl = {https://ui.adsabs.harvard.edu/abs/2022arXiv220805929S}
}

@ARTICLE{yuan2022,
       author = {{Yuan}, Sihan and {Garrison}, Lehman H. and {Hadzhiyska}, Boryana and {Bose}, Sownak and {Eisenstein}, Daniel J.},
        title = "{ABACUSHOD: a highly efficient extended multitracer HOD framework and its application to BOSS and eBOSS data}",
      journal = {\mnras},
         year = 2022,
        month = mar,
       volume = {510},
       number = {3},
        pages = {3301-3320},
          doi = {10.1093/mnras/stab3355},
archivePrefix = {arXiv},
       eprint = {2110.11412},
 primaryClass = {astro-ph.CO},
       adsurl = {https://ui.adsabs.harvard.edu/abs/2022MNRAS.510.3301Y}
}

@article{paradiso2025,
   title={Reducing nuisance prior sensitivity via non-linear reparameterization, with application to EFT analyses of large-scale structure},
   volume={2025},
   ISSN={1475-7516},
   url={http://dx.doi.org/10.1088/1475-7516/2025/07/005},
   DOI={10.1088/1475-7516/2025/07/005},
   number={07},
   journal={Journal of Cosmology and Astroparticle Physics},
   publisher={IOP Publishing},
   author={Paradiso, S. and Bonici, M. and Chen, M. and Percival, W.J. and D’Amico, G. and Zhang, H. and McGee, G.},
   year={2025},
   month=jul, pages={005} }

@article{donald2023,
   title={Analysis of unified galaxy power spectrum multipole measurements},
   volume={526},
   ISSN={1365-2966},
   url={http://dx.doi.org/10.1093/mnras/stad2957},
   DOI={10.1093/mnras/stad2957},
   number={3},
   journal={Monthly Notices of the Royal Astronomical Society},
   publisher={Oxford University Press (OUP)},
   author={Donald-McCann, Jamie and Gsponer, Rafaela and Zhao, Ruiyang and Koyama, Kazuya and Beutler, Florian},
   year={2023},
   month=sep, pages={3461–3481} }

@ARTICLE{desi_fs,
       author = {{DESI Collaboration} and {Adame}, A.~G. and {Aguilar}, J. and {Ahlen}, S. and {Alam}, S. and {Alexander}, D.~M. and {Allende Prieto}, C. and {Alvarez}, M. and {Alves}, O. and {Anand}, A. and {Andrade}, U. and {Armengaud}, E. and {Avila}, S. and {Aviles}, A. and {Awan}, H. and {Bahr-Kalus}, B. and {Bailey}, S. and {Baltay}, C. and {Bault}, A. and {Behera}, J. and {BenZvi}, S. and {Beutler}, F. and {Bianchi}, D. and {Blake}, C. and {Blum}, R. and {Bonici}, M. and {Brieden}, S. and {Brodzeller}, A. and {Brooks}, D. and {Buckley-Geer}, E. and {Burtin}, E. and {Calderon}, R. and {Canning}, R. and {Carnero Rosell}, A. and {Cereskaite}, R. and {Cervantes-Cota}, J.~L. and {Chabanier}, S. and {Chaussidon}, E. and {Chaves-Montero}, J. and {Chebat}, D. and {Chen}, S. and {Chen}, X. and {Claybaugh}, T. and {Cole}, S. and {Cuceu}, A. and {Davis}, T.~M. and {Dawson}, K. and {de la Macorra}, A. and {de Mattia}, A. and {Deiosso}, N. and {Dey}, A. and {Dey}, B. and {Ding}, Z. and {Doel}, P. and {Edelstein}, J. and {Eftekharzadeh}, S. and {Eisenstein}, D.~J. and {Elbers}, W. and {Elliott}, A. and {Fagrelius}, P. and {Fanning}, K. and {Ferraro}, S. and {Ereza}, J. and {Findlay}, N. and {Flaugher}, B. and {Font-Ribera}, A. and {Forero-S{\'a}nchez}, D. and {Forero-Romero}, J.~E. and {Frenk}, C.~S. and {Garcia-Quintero}, C. and {Garrison}, L.~H. and {Gazta{\~n}aga}, E. and {Gil-Mar{\'\i}n}, H. and {Gontcho}, S. Gontcho A. and {Gonzalez-Morales}, A.~X. and {Gonzalez-Perez}, V. and {Gordon}, C. and {Green}, D. and {Gruen}, D. and {Gsponer}, R. and {Gutierrez}, G. and {Guy}, J. and {Hadzhiyska}, B. and {Hahn}, C. and {Hanif}, M.~M.~S. and {Herrera-Alcantar}, H.~K. and {Honscheid}, K. and {Howlett}, C. and {Huterer}, D. and {Ir{\v{s}}i{\v{c}}}, V. and {Ishak}, M. and {Joyce}, R. and {Juneau}, S. and {Kara{\c{c}}ayl{\i}}, N.~G. and {Kehoe}, R. and {Kent}, S. and {Kirkby}, D. and {Kong}, H. and {Koposov}, S.~E. and {Kremin}, A. and {Krolewski}, A. and {Lahav}, O. and {Lai}, Y. and {Lan}, T.-W. and {Landriau}, M. and {Lang}, D. and {Lasker}, J. and {Le Goff}, J.~M. and {Le Guillou}, L. and {Leauthaud}, A. and {Levi}, M.~E. and {Li}, T.~S. and {Lodha}, K. and {Magneville}, C. and {Manera}, M. and {Margala}, D. and {Martini}, P. and {Matthewson}, W. and {Maus}, M. and {McDonald}, P. and {Medina-Varela}, L. and {Meisner}, A. and {Mena-Fern{\'a}ndez}, J. and {Miquel}, R. and {Moon}, J. and {Moore}, S. and {Moustakas}, J. and {Mudur}, N. and {Mueller}, E. and {Mu{\~n}oz-Guti{\'e}rrez}, A. and {Myers}, A.~D. and {Nadathur}, S. and {Napolitano}, L. and {Neveux}, R. and {Newman}, J.~A. and {Nguyen}, N.~M. and {Nie}, J. and {Niz}, G. and {Noriega}, H.~E. and {Padmanabhan}, N. and {Paillas}, E. and {Palanque-Delabrouille}, N. and {Pan}, J. and {Penmetsa}, S. and {Percival}, W.~J. and {Pieri}, M.~M. and {Pinon}, M. and {Poppett}, C. and {Porredon}, A. and {Prada}, F. and {P{\'e}rez-Fern{\'a}ndez}, A. and {P{\'e}rez-R{\`a}fols}, I. and {Rabinowitz}, D. and {Raichoor}, A. and {Ram{\'\i}rez-P{\'e}rez}, C. and {Ramirez-Solano}, S. and {Rashkovetskyi}, M. and {Ravoux}, C. and {Rezaie}, M. and {Rich}, J. and {Rocher}, A. and {Rockosi}, C. and {Roe}, N.~A. and {Rosado-Marin}, A. and {Ross}, A.~J. and {Rossi}, G. and {Ruggeri}, R. and {Ruhlmann-Kleider}, V. and {Samushia}, L. and {Sanchez}, E. and {Saulder}, C. and {Schlafly}, E.~F. and {Schlegel}, D. and {Schubnell}, M. and {Seo}, H. and {Shafieloo}, A. and {Sharples}, R. and {Silber}, J. and {Slosar}, A. and {Smith}, A. and {Sprayberry}, D. and {Tan}, T. and {Tarl{\'e}}, G. and {Taylor}, P. and {Trusov}, S. and {Vaisakh}, R. and {Valcin}, D. and {Valdes}, F. and {Valogiannis}, G. and {Vargas-Maga{\~n}a}, M. and {Verde}, L. and {Walther}, M. and {Wang}, B. and {Wang}, M.~S. and {Weaver}, B.~A. and {Weaverdyck}, N. and {Wechsler}, R.~H. and {Weinberg}, D.~H. and {White}, M. and {Wilson}, M.~J. and {Yi}, L.},
        title = "{DESI 2024 VII: cosmological constraints from the full-shape modeling of clustering measurements}",
      journal = {\jcap},
         year = 2025,
        month = jul,
       volume = {2025},
       number = {7},
          eid = {028},
        pages = {028},
          doi = {10.1088/1475-7516/2025/07/028},
archivePrefix = {arXiv},
       eprint = {2411.12022},
 primaryClass = {astro-ph.CO},
       adsurl = {https://ui.adsabs.harvard.edu/abs/2025JCAP...07..028A}
}

@ARTICLE{derose2025,
       author = {{DeRose}, Joseph and {Chen}, Shi-Fan},
        title = "{The Lensing Counter Narrative: An Effective Description of Small-Scale Clustering in Weak Lensing Power Spectra}",
      journal = {arXiv e-prints},
         year = 2025,
        month = oct,
          eid = {arXiv:2510.18981},
        pages = {arXiv:2510.18981},
          doi = {10.48550/arXiv.2510.18981},
archivePrefix = {arXiv},
       eprint = {2510.18981},
 primaryClass = {astro-ph.CO},
       adsurl = {https://ui.adsabs.harvard.edu/abs/2025arXiv251018981D}
}

@ARTICLE{tsedrik2025a,
       author = {{Tsedrik}, M. and {Lee}, S. and {Markovic}, K. and {Carrilho}, P. and {Pourtsidou}, A. and {Moretti}, C. and {Bose}, B. and {Huff}, E. and {Robertson}, A. and {Taylor}, P.~L. and {Zuntz}, J.},
        title = "{Interacting dark energy constraints from the full-shape analyses of BOSS DR12 and DES Year 3 measurements}",
      journal = {\mnras},
         year = 2025,
        month = jul,
       volume = {541},
       number = {1},
        pages = {L65-L70},
          doi = {10.1093/mnrasl/slaf055},
archivePrefix = {arXiv},
       eprint = {2502.03390},
 primaryClass = {astro-ph.CO},
       adsurl = {https://ui.adsabs.harvard.edu/abs/2025MNRAS.541L..65T}
}

@ARTICLE{perlmutter1999,
       author = {{Perlmutter}, S. and {Aldering}, G. and {Goldhaber}, G. and {Knop}, R.~A. and {Nugent}, P. and {Castro}, P.~G. and {Deustua}, S. and {Fabbro}, S. and {Goobar}, A. and {Groom}, D.~E. and {Hook}, I.~M. and {Kim}, A.~G. and {Kim}, M.~Y. and {Lee}, J.~C. and {Nunes}, N.~J. and {Pain}, R. and {Pennypacker}, C.~R. and {Quimby}, R. and {Lidman}, C. and {Ellis}, R.~S. and {Irwin}, M. and {McMahon}, R.~G. and {Ruiz-Lapuente}, P. and {Walton}, N. and {Schaefer}, B. and {Boyle}, B.~J. and {Filippenko}, A.~V. and {Matheson}, T. and {Fruchter}, A.~S. and {Panagia}, N. and {Newberg}, H.~J.~M. and {Couch}, W.~J. and {Project}, The Supernova Cosmology},
        title = "{Measurements of {\ensuremath{\Omega}} and {\ensuremath{\Lambda}} from 42 High-Redshift Supernovae}",
      journal = {\apj},
         year = 1999,
        month = jun,
       volume = {517},
       number = {2},
        pages = {565-586},
          doi = {10.1086/307221},
archivePrefix = {arXiv},
       eprint = {astro-ph/9812133},
 primaryClass = {astro-ph},
       adsurl = {https://ui.adsabs.harvard.edu/abs/1999ApJ...517..565P}
}

@ARTICLE{riess1998,
       author = {{Riess}, Adam G. and {Filippenko}, Alexei V. and {Challis}, Peter and {Clocchiatti}, Alejandro and {Diercks}, Alan and {Garnavich}, Peter M. and {Gilliland}, Ron L. and {Hogan}, Craig J. and {Jha}, Saurabh and {Kirshner}, Robert P. and {Leibundgut}, B. and {Phillips}, M.~M. and {Reiss}, David and {Schmidt}, Brian P. and {Schommer}, Robert A. and {Smith}, R. Chris and {Spyromilio}, J. and {Stubbs}, Christopher and {Suntzeff}, Nicholas B. and {Tonry}, John},
        title = "{Observational Evidence from Supernovae for an Accelerating Universe and a Cosmological Constant}",
      journal = {\aj},
         year = 1998,
        month = sep,
       volume = {116},
       number = {3},
        pages = {1009-1038},
          doi = {10.1086/300499},
archivePrefix = {arXiv},
       eprint = {astro-ph/9805201},
 primaryClass = {astro-ph},
       adsurl = {https://ui.adsabs.harvard.edu/abs/1998AJ....116.1009R}
}

@article{DiValentino2021,
   title={In the realm of the Hubble tension—a review of solutions
						            *},
   volume={38},
   ISSN={1361-6382},
   url={http://dx.doi.org/10.1088/1361-6382/ac086d},
   DOI={10.1088/1361-6382/ac086d},
   number={15},
   journal={Classical and Quantum Gravity},
   publisher={IOP Publishing},
   author={Di Valentino, Eleonora and Mena, Olga and Pan, Supriya and Visinelli, Luca and Yang, Weiqiang and Melchiorri, Alessandro and Mota, David F and Riess, Adam G and Silk, Joseph},
   year={2021},
   month=jul, pages={153001} }

\section*{Affiliations}
\scriptsize
\noindent
$^{1}$ Centre for Astrophysics \& Supercomputing, Swinburne University of Technology, P.O. Box 218, Hawthorn, VIC 3122, Australia\\
$^{2}$ CIEMAT, Avenida Complutense 40, E-28040 Madrid, Spain\\
$^{3}$ Institute for Astronomy, University of Edinburgh, Royal Observatory, Blackford Hill, Edinburgh EH9 3HJ, UK\\
$^{4}$ Ruhr University Bochum, Faculty of Physics and Astronomy, Astronomical Institute (AIRUB), German Centre for Cosmological Lensing, 44780 Bochum, Germany\\
$^{5}$ The Ohio State University, Columbus, 43210 OH, USA\\
$^{6}$ Lawrence Berkeley National Laboratory, 1 Cyclotron Road, Berkeley, CA 94720, USA\\
$^{7}$ Department of Physics, Boston University, 590 Commonwealth Avenue, Boston, MA 02215 USA\\
$^{8}$ Dipartimento di Fisica ``Aldo Pontremoli'', Universit\`a degli Studi di Milano, Via Celoria 16, I-20133 Milano, Italy\\
$^{9}$ INAF-Osservatorio Astronomico di Brera, Via Brera 28, 20122 Milano, Italy\\
$^{10}$ Department of Physics \& Astronomy, University College London, Gower Street, London, WC1E 6BT, UK\\
$^{11}$ Institut d'Estudis Espacials de Catalunya (IEEC), c/ Esteve Terradas 1, Edifici RDIT, Campus PMT-UPC, 08860 Castelldefels, Spain\\
$^{12}$ Institute of Space Sciences, ICE-CSIC, Campus UAB, Carrer de Can Magrans s/n, 08913 Bellaterra, Barcelona, Spain\\
$^{13}$ Department of Physics and Astronomy, The University of Utah, 115 South 1400 East, Salt Lake City, UT 84112, USA\\
$^{14}$ Instituto de F\'{\i}sica, Universidad Nacional Aut\'{o}noma de M\'{e}xico,  Circuito de la Investigaci\'{o}n Cient\'{\i}fica, Ciudad Universitaria, Cd. de M\'{e}xico  C.~P.~04510,  M\'{e}xico\\
$^{15}$ Department of Astronomy \& Astrophysics, University of Toronto, Toronto, ON M5S 3H4, Canada\\
$^{16}$ Department of Physics \& Astronomy and Pittsburgh Particle Physics, Astrophysics, and Cosmology Center (PITT PACC), University of Pittsburgh, 3941 O'Hara Street, Pittsburgh, PA 15260, USA\\
$^{17}$ \\
$^{18}$ Department of Physics, The Ohio State University, 191 West Woodruff Avenue, Columbus, OH 43210, USA\\
$^{19}$ University of California, Berkeley, 110 Sproul Hall \#5800 Berkeley, CA 94720, USA\\
$^{20}$ Institut de F\'{i}sica d’Altes Energies (IFAE), The Barcelona Institute of Science and Technology, Edifici Cn, Campus UAB, 08193, Bellaterra (Barcelona), Spain\\
$^{21}$ Departamento de F\'isica, Universidad de los Andes, Cra. 1 No. 18A-10, Edificio Ip, CP 111711, Bogot\'a, Colombia\\
$^{22}$ Observatorio Astron\'omico, Universidad de los Andes, Cra. 1 No. 18A-10, Edificio H, CP 111711 Bogot\'a, Colombia\\
$^{23}$ Center for Astrophysics $|$ Harvard \& Smithsonian, 60 Garden Street, Cambridge, MA 02138, USA\\
$^{24}$ Institute of Cosmology and Gravitation, University of Portsmouth, Dennis Sciama Building, Portsmouth, PO1 3FX, UK\\
$^{25}$ University of Virginia, Department of Astronomy, Charlottesville, VA 22904, USA\\
$^{26}$ Fermi National Accelerator Laboratory, PO Box 500, Batavia, IL 60510, USA\\
$^{27}$ Institute of Astronomy, University of Cambridge, Madingley Road, Cambridge CB3 0HA, UK\\
$^{28}$ Institut d'Astrophysique de Paris. 98 bis boulevard Arago. 75014 Paris, France\\
$^{29}$ IRFU, CEA, Universit\'{e} Paris-Saclay, F-91191 Gif-sur-Yvette, France\\
$^{30}$ Department of Astronomy and Astrophysics, UCO/Lick Observatory, University of California, 1156 High Street, Santa Cruz, CA 95064, USA\\
$^{31}$ Center for Cosmology and AstroParticle Physics, The Ohio State University, 191 West Woodruff Avenue, Columbus, OH 43210, USA\\
$^{32}$ School of Mathematics and Physics, University of Queensland, Brisbane, QLD 4072, Australia\\
$^{33}$ Department of Physics, University of Michigan, 450 Church Street, Ann Arbor, MI 48109, USA\\
$^{34}$ University of Michigan, 500 S. State Street, Ann Arbor, MI 48109, USA\\
$^{35}$ Department of Physics, The University of Texas at Dallas, 800 W. Campbell Rd., Richardson, TX 75080, USA\\
$^{36}$ NSF NOIRLab, 950 N. Cherry Ave., Tucson, AZ 85719, USA\\
$^{37}$ Aix Marseille Univ, CNRS, CNES, LAM, Marseille, France\\
$^{38}$ Department of Physics and Astronomy, University of California, Irvine, 92697, USA\\
$^{39}$ Department of Physics and Astronomy, University of Waterloo, 200 University Ave W, Waterloo, ON N2L 3G1, Canada\\
$^{40}$ Perimeter Institute for Theoretical Physics, 31 Caroline St. North, Waterloo, ON N2L 2Y5, Canada\\
$^{41}$ Waterloo Centre for Astrophysics, University of Waterloo, 200 University Ave W, Waterloo, ON N2L 3G1, Canada\\
$^{42}$ Department of Physics, American University, 4400 Massachusetts Avenue NW, Washington, DC 20016, USA\\
$^{43}$ Sorbonne Universit\'{e}, CNRS/IN2P3, Laboratoire de Physique Nucl\'{e}aire et de Hautes Energies (LPNHE), FR-75005 Paris, France\\
$^{44}$ Department of Astronomy and Astrophysics, University of California, Santa Cruz, 1156 High Street, Santa Cruz, CA 95065, USA\\
$^{45}$ Departament de F\'{i}sica, Serra H\'{u}nter, Universitat Aut\`{o}noma de Barcelona, 08193 Bellaterra (Barcelona), Spain\\
$^{46}$ Instituci\'{o} Catalana de Recerca i Estudis Avan\c{c}ats, Passeig de Llu\'{\i}s Companys, 23, 08010 Barcelona, Spain\\
$^{47}$ Department of Physics and Astronomy, Siena University, 515 Loudon Road, Loudonville, NY 12211, USA\\
$^{48}$ Space Sciences Laboratory, University of California, Berkeley, 7 Gauss Way, Berkeley, CA  94720, USA\\
$^{49}$ Instituto de Astrof\'{i}sica de Andaluc\'{i}a (CSIC), Glorieta de la Astronom\'{i}a, s/n, E-18008 Granada, Spain\\
$^{50}$ Departament de F\'isica, EEBE, Universitat Polit\`ecnica de Catalunya, c/Eduard Maristany 10, 08930 Barcelona, Spain\\
$^{51}$ Department of Physics and Astronomy, Sejong University, 209 Neungdong-ro, Gwangjin-gu, Seoul 05006, Republic of Korea\\
$^{52}$ Queensland University of Technology,  School of Chemistry \& Physics, George St, Brisbane 4001, Australia\\
$^{53}$ Max Planck Institute for Extraterrestrial Physics, Gie\ss enbachstra\ss e 1, 85748 Garching, Germany\\
$^{54}$ Department of Physics \& Astronomy, Ohio University, 139 University Terrace, Athens, OH 45701, USA\\
$^{55}$ National Astronomical Observatories, Chinese Academy of Sciences, A20 Datun Road, Chaoyang District, Beijing, 100101, P.~R.~China\\
$^{56}$ NASA Einstein Fellow\\
\normalsize

\appendix

\section{Gaussian analytical cross-covariance}
\label{sec:covxil}

In this section we present relations for the Gaussian analytical cross-covariance between the galaxy correlation function multipoles and the galaxy-galaxy lensing and cosmic shear correlations in configuration space.  We note that the analytical covariances involving cosmic shear, galaxy-galaxy lensing and projected clustering in configuration space can be found in \cite{2024MNRAS.533..589Y} and \cite{2025OJAp....8E..24B}.  \cite{2022PhRvD.106f3536T} present related cross-covariances for Fourier-space statistics.

\subsection{Relation of correlation function multipoles to power spectra}

We start with the standard relation between the galaxy correlation function multipoles $\xi^\ell_{gg}(s)$ as a function of separation $s$, and the galaxy power spectrum $P_{gg}(\boldsymbol{k})$ as a function of wavenumber $\boldsymbol{k}$:
\begin{equation}
  \xi^\ell_{gg}(s) = \frac{2\ell+1}{2} \int_{-1}^{+1} d\mu_s \, L_\ell(\mu_s) \, \xi_{gg}(s,\mu_s) = \frac{2\ell+1}{2} \int_{-1}^{+1} d\mu_s \, L_\ell(\mu_s) \int \frac{d^3\boldsymbol{k}}{(2\pi)^3} \, P_{gg}(\boldsymbol{k}) \, e^{-i\boldsymbol{k} \cdot \boldsymbol{s}} ,
\label{eq:xilpk}
\end{equation}
where $\mu_s$ is the cosine of the angle with respect to the line-of-sight, and $L_\ell(\mu_s)$ are the Legendre polynomials.  This expression may be simplified to produce the usual transformation between the correlation function and power spectrum multipoles, by substituting $P_{gg}(\boldsymbol{k}) = \sum_\ell P_{\ell}(k) \, L_\ell(\mu_k)$, which yields,
\begin{equation}
  \xi^\ell_{gg}(s) = i^\ell \int_0^\infty \frac{dk \, k^2}{2\pi^2} \, P_\ell(k) \, j_\ell(ks) ,
\end{equation}
using $\int_{-1}^{+1} d\mu_s \, L_\ell(\mu_s) \, e^{-i\boldsymbol{k} \cdot \boldsymbol{s}} = 2 i^\ell \, j_\ell(ks) \, L_\ell(\mu_k)$, where $j_\ell$ is a spherical Bessel function.

\subsection{Covariance of correlation function multipoles}

The covariance of $\xi^\ell_{gg}(s)$ in the Gaussian approximation follows from Eq.\ref{eq:xilpk} \citep[see also,][]{2015MNRAS.447..234W, 2016MNRAS.457.1577G},
\begin{equation}
\begin{split}
    {\rm Cov} \left[ \xi^\ell_{gg}(s) , \xi^{\ell'}_{gg}(s') \right] &=
    \frac{(2\ell+1)(2\ell'+1)}{4} \int d\mu_s L_\ell(\mu_s)
    \int d\mu'_s L_{\ell'}(\mu'_s) \int \frac{d^3\boldsymbol{k}}{(2\pi)^3} \int \frac{d^3\boldsymbol{k}'}{(2\pi)^3} {\rm Cov} \left[ P_{gg}(\boldsymbol{k}) , P_{gg}(\boldsymbol{k}') \right] e^{-i\boldsymbol{k} \cdot \boldsymbol{s}} e^{i\boldsymbol{k}'.\boldsymbol{s}'} \\
    &= \frac{(2\ell+1)(2\ell'+1)}{4} \int_{-1}^{+1} d\mu_s \, L_\ell(\mu_s) \int_{-1}^{+1} d\mu'_s \, L_{\ell'}(\mu'_s) \, \frac{2}{V_s} \int \frac{d^3\boldsymbol{k}}{(2\pi)^3} \left( P_{gg}(k, \mu_k) + \frac{1}{n_g} \right)^2 \, e^{-i\boldsymbol{k} \cdot \boldsymbol{s}} \, e^{i\boldsymbol{k}' \cdot \boldsymbol{s}'} \\
    &= \frac{i^{\ell+\ell'}(2\ell+1)(2\ell'+1)}{V_s} \int_0^\infty \frac{dk \, k^2}{2\pi^2} \, j_\ell(ks) \, j_{\ell'}(ks') \int_{-1}^{+1} d\mu_k \left( P_{gg}(k, \mu_k) + \frac{1}{n_g} \right)^2 \, L_\ell(\mu_k) \, L_{\ell'}(\mu_k) ,
\end{split}
\label{eq:covxil}
\end{equation}
where $V_s$ is the survey volume.   In the first line of Eq.\ref{eq:covxil} we have used ${\rm Cov} \left[ P_{gg}(\boldsymbol{k}), P_{gg}(\boldsymbol{k}') \right] = 2 \, \left[ P_{gg}(\boldsymbol{k}) + \frac{1}{n_g} \right]^2 \, \tilde{\delta}_D(\boldsymbol{k}-\boldsymbol{k}')$, where the factor of 2 accounts for the fact that the modes $-\boldsymbol{k}$ and $\boldsymbol{k}$ are not independent.

\subsection{Cross-covariance between correlation function multipoles and galaxy-galaxy lensing}

To determine the cross-covariance between the galaxy-galaxy lensing and clustering correlation functions, we first consider the relation between the differential projected mass density $\Delta \Sigma(R)$ as a function of projected separation $R$, and the galaxy-matter cross-power spectrum $P_{gm}(k)$ \citep[see for example,][]{2024MNRAS.533..589Y},
\begin{equation}
    \Delta \Sigma(R) = \overline{\rho}_m \int \frac{d^2\boldsymbol{k}_\perp}{(2\pi)^2} \, P_{gm}(\boldsymbol{k}_\perp) \, J_2(k_\perp R) ,
\label{eq:dsig}
\end{equation}
where $\overline{\rho}_m$ is the mean cosmic matter density, $\boldsymbol{k}_\perp$ is the 2D Fourier wavevector in the plane of the sky, $k_\parallel$ is the wavenumber in the line-of-sight direction, and $J_n$ indicates a Bessel function of the first kind.  Eqs.\ref{eq:dsig} and \ref{eq:xilpk} may be used to deduce the cross-covariance between $\Delta\Sigma(R)$ and $\xi^\ell_{gg}(s)$,
\begin{equation}
\begin{split}
    {\rm Cov} \left[ \Delta\Sigma(R) , \xi^\ell_{gg}(s) \right] &= \frac{\overline{\rho}_m \, (2\ell+1)}{2} \int \frac{d^3\boldsymbol{k}}{(2\pi)^3} \int \frac{d^3\boldsymbol{k}'}{(2\pi)^3} \, {\rm Cov} \left[ P_{gm}(\boldsymbol{k}) , P_{gg}(\boldsymbol{k}') \right] \left[ L_\parallel \, \tilde{\delta}_D(k_\parallel) \right] J_2(k_\perp R) \int_{-1}^{+1} d\mu_s \, L_\ell(\mu_s) \, e^{-i\boldsymbol{k}' \cdot \boldsymbol{s}} \\
    &= \frac{2 \, \overline{\rho}_m \, (2\ell+1) \, i^\ell}{V_s} \int \frac{d^3\boldsymbol{k}}{(2\pi)^3} \, P_{gm}(\boldsymbol{k}) \left[ P_{gg}(\boldsymbol{k}) + \frac{1}{n_g} \right] \left[ L_\parallel \, \tilde{\delta}(k_\parallel) \right] J_2(k_\perp R) \, j_\ell(ks) \, L_\ell(\mu_k) \\
    &= \frac{2 \, \overline{\rho}_m \, (2\ell+1) \, i^\ell \, L_\ell(0)}{V_s} \int \frac{dk_\perp \, k_\perp}{2\pi} P_{gm}(k_\perp) \left[ P_{gg}(k_\perp) + \frac{1}{n_g} \right] \, J_2(k_\perp R) \, j_\ell(k_\perp s) .
\end{split}
\label{eq:covdsxil}
\end{equation}
In the first line of Eq.\ref{eq:covdsxil} we have used ${\rm Cov} \left[ P_{gm}(\boldsymbol{k}) , P_{gg}(\boldsymbol{k}') \right] = 2 P_{gm}(\boldsymbol{k}) \left[ P_{gg}(\boldsymbol{k}') + \frac{1}{n_g} \right] \, \tilde{\delta}_D(\boldsymbol{k}-\boldsymbol{k}')$, and we also note that $i^\ell \, L_\ell(0) = (1, 1/2, 3/8)$ for $\ell = (0,2,4)$.  To derive the cross-covariance involving the average tangential shear $\gamma_t(\theta)$, we use the approximation $\gamma^{\rm Nar}_t(\theta) = \overline{\Sigma_c^{-1}}(\chi_{\rm eff}) \, \Delta\Sigma(R)$ \citep[see for example,][]{2024MNRAS.533..589Y}, where $\gamma^{\rm Nar}_t$ denotes the approximation that the galaxy sample forms a uniform narrow redshift slice at effective distance $\chi_{\rm eff}$, and the average inverse critical surface mass density is
\begin{equation}
  \overline{\Sigma_c^{-1}}(\chi) = \int_\chi^\infty d\chi_s \, p_s(\chi_s) \, \Sigma_c^{-1}(\chi,\chi_s) ,
\end{equation}
in terms of the critical surface mass density $\Sigma_c$ and the source distance distribution $p_s(\chi)$.  The cross-covariance between $\gamma_t(\theta)$ and $\xi_{gg}^\ell(s)$ may then be written as,
\begin{equation}
{\rm Cov} \left[ \gamma^{\rm Nar}_t(\theta) , \xi^\ell_{gg}(s) \right] = \overline{\Sigma_c^{-1}}(\chi_{\rm eff}) \, {\rm Cov} \left[ \Delta\Sigma(R) , \xi^\ell_{gg}(s) \right] .
\end{equation}

\subsection{Cross-covariance between correlation function multipoles and cosmic shear}

To determine the cross-covariance between the cosmic shear and clustering correlation functions, we first consider the relation between the cosmic shear correlation functions and the matter power spectrum,
\begin{equation}
    \xi_\pm(\theta) = \int \frac{d^2\boldsymbol{\ell}}{(2\pi)^2} \, C_{\kappa\kappa}(\boldsymbol{\ell}) \, J_{0/4}(\ell \theta) ,
\label{eq:xipm}
\end{equation}
where for a cosmic shear signal measured using two tomographic bins with average inverse critical surface mass densities $\overline{\Sigma_{c,1}^{-1}}$ and $\overline{\Sigma_{c,2}^{-1}}$ \citep[see for example,][]{2025OJAp....8E..24B},
\begin{equation}
  C_{\kappa\kappa}(\boldsymbol{\ell}) = \overline{\rho}_m^2 \int_0^\infty d\chi \, \frac{\overline{\Sigma_{c,1}^{-1}}(\chi) \, \overline{\Sigma_{c,2}^{-1}}(\chi)}{\chi^2} \, P_{mm} \left( \frac{\boldsymbol{\ell}}{\chi} , \chi \right) .
\end{equation}
The cross-covariance of the cosmic shear correlation functions $\xi_\pm(\theta)$ and galaxy multipole correlation functions $\xi^\ell_{gg}(s)$ may be written using Eq.\ref{eq:xipm} and Eq.\ref{eq:xilpk} as,
\begin{equation}
  {\rm Cov} \left[ \xi_\pm(\theta) , \xi^\ell_{gg}(s) \right] = (2\ell+1) \, i^\ell \int \frac{d^2\boldsymbol{\ell}}{(2\pi)^2} \int \frac{d^3\boldsymbol{k}}{(2\pi)^3} \, {\rm Cov} \left[ C_{\kappa\kappa}(\boldsymbol{\ell}) , P_{gg}(\boldsymbol{k}) \right] \, J_{0/4}(\ell\theta) \, j_\ell(ks) \, L_\ell(\mu_k) .
\end{equation}
Now we consider,
\begin{equation}
\begin{split}
  {\rm Cov} \left[ C_{\kappa\kappa}(\boldsymbol{\ell}) , P_{gg}(\boldsymbol{k}) \right] &= \overline{\rho}_m^2 \int_0^\infty d\chi \, \frac{\overline{\Sigma_{c,1}^{-1}}(\chi) \, \overline{\Sigma_{c,2}^{-1}}(\chi)}{\chi^2} \, {\rm Cov} \left[ P_{mm} \left( \frac{\boldsymbol{\ell}}{\chi} , \chi \right) , P_{gg}(\boldsymbol{k}) \right] \\
  &= \overline{\rho}_m^2 \, L_\parallel \, \frac{\overline{\Sigma_{c,1}^{-1}}(\chi_{\rm eff}) \, \overline{\Sigma_{c,2}^{-1}}(\chi_{\rm eff}) }{\chi_{\rm eff}^2} \, {\rm Cov} \left[ P_{mm}(\boldsymbol{k}'_\perp) , P_{gg}(\boldsymbol{k}_\perp) \right] \, \tilde{\delta}_D(k_\parallel) \\
  &= \overline{\rho}_m^2 \, L_\parallel \, \frac{ \overline{\Sigma_{c,1}^{-1}}(\chi_{\rm eff}) \, \overline{\Sigma_{c,2}^{-1}}(\chi_{\rm eff})}{\chi_{\rm eff}^2} \, 2 \, P^2_{gm}(\boldsymbol{k}_\perp) \, \tilde{\delta}_D(k_\parallel) \, \tilde{\delta}_D \left( \frac{\boldsymbol{\ell}}{\chi_{\rm eff}} - \boldsymbol{k}_\perp \right) .
\end{split}
\label{eq:covclkkpgg}
\end{equation}
Hence using Eq.\ref{eq:covclkkpgg},
\begin{equation}
\begin{split}
  {\rm Cov} \left[ \xi_\pm(\theta) , \xi^\ell_{gg}(s) \right] &= 2 \, \overline{\rho}_m^2 \, (2\ell+1) \, i^\ell \, \overline{\Sigma_{c,1}^{-1}}(\chi_{\rm eff}) \, \overline{\Sigma_{c,2}^{-1}}(\chi_{\rm eff}) \int \frac{1}{\chi_{\rm eff}^2} \int \frac{d^2\boldsymbol{\ell}}{(2\pi)^2} \int \frac{d^2\boldsymbol{k}_\perp}{(2\pi)^2} \, \tilde{\delta}_D \left( \frac{\boldsymbol{\ell}}{\chi_{\rm eff}} - \boldsymbol{k}_\perp \right) P^2_{gm}(\boldsymbol{k}_\perp) \\
  &\times \int_{-\infty}^\infty \frac{dk_\parallel}{2\pi} \, L_\parallel \, \tilde{\delta}_D(k_\parallel) \, J_{0/4}(\ell\theta) \, j_\ell(ks) \, L_\ell(\mu_k) \\
  &= \frac{2 \, \overline{\rho}_m^2 \, (2\ell+1) \, i^\ell \, L_\ell(0) \, \overline{\Sigma_{c,1}^{-1}}(\chi_{\rm eff}) \, \overline{\Sigma_{c,2}^{-1}}(\chi_{\rm eff})}{A_s} \int \frac{dk_\perp \, k_\perp}{2\pi} \, P^2_{gm}(k_\perp) \, J_{0/4}(k_\perp \chi_{\rm eff} \theta) \, j_\ell(k_\perp s) ,
\end{split}
\end{equation}
where $A_s = \Omega_s \, \chi_{\rm eff}^2$ is the projected survey area at the effective lens distance $\chi_{\rm eff}$.

\section{Effect of cross-covariance on results}
\label{sec:crosscovariance}

We perform an explicit test to assess the effect that the  cross-covariance terms between the projected 2D lensing quantities and the 3D galaxy clustering might have on the joint constraints for DESI-Y1 like analysis. We present the posteriors obtained by fitting the $3\times2$pt measurements of {\sc Abacus} mocks, corresponding to a DESI$\times$DES-like setup for two cases: with and without including the cross-covariance terms in the covariance.  As for previous tests, we do not scale the covariance when fitting to the mock mean.

\begin{figure}[h]
\centering
\includegraphics[width=0.7\columnwidth]{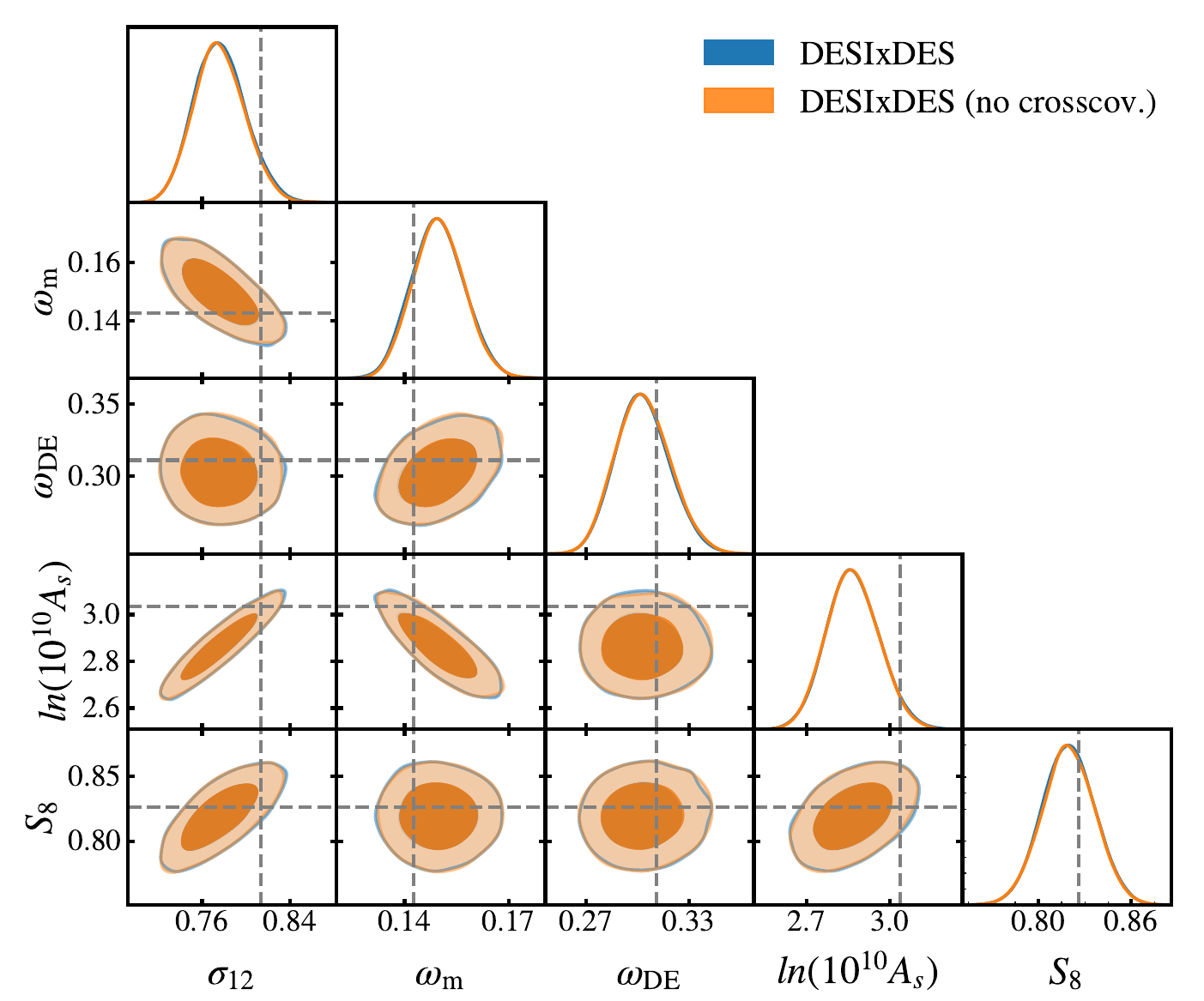}
\caption{Constraints from {\sc Abacus} mock measurement fits corresponding to DESI$\times$DES-like joint analysis. Here in orange the posteriors obtained when assuming no cross-correlation between galaxy clustering and shear and in blue when including shear-galaxy clustering cross terms.}
\label{fig:nocrosscov}
\end{figure}

Our results are presented in Figure \ref{fig:nocrosscov} and show that omitting the cross-covariance terms has no effect on the final posteriors. This confirms earlier work \citep[see, for example,][]{taylor2022} that forecasts negligible cross-covariance efects for Stage-IV surveys.

\section{Individual tracer results}
\label{sec:tracers}

In this Appendix we provide a comparison between the individual full shape fits derived in this work and DESI key cosmology full shape analysis of clustering in Fourier space \citep{2025JCAP...09..008A}. For this comparison, we additionally impose Gaussian priors on $n_{\rm{s}}$ and $\omega_{\rm{b}}$ to match those used in the key analysis. 

\begin{figure}[htbp]
    \centering
    \begin{subfigure}[b]{0.45\textwidth}
        \centering
        \includegraphics[width=\textwidth]{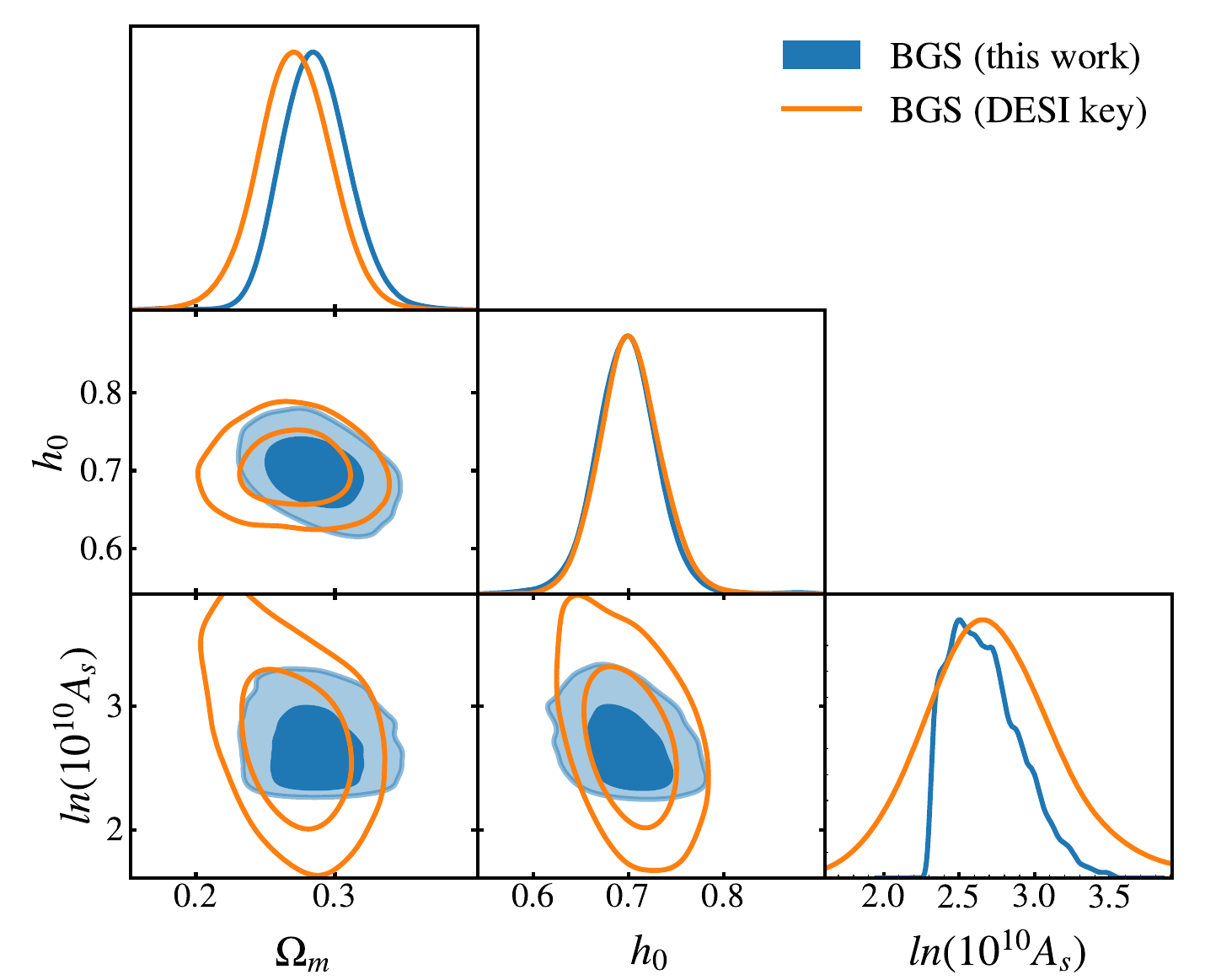}
        \caption{BGS}
        \label{fig:panel_a}
    \end{subfigure}
    \hfill
    \begin{subfigure}[b]{0.45\textwidth}
        \centering
        \includegraphics[width=\textwidth]{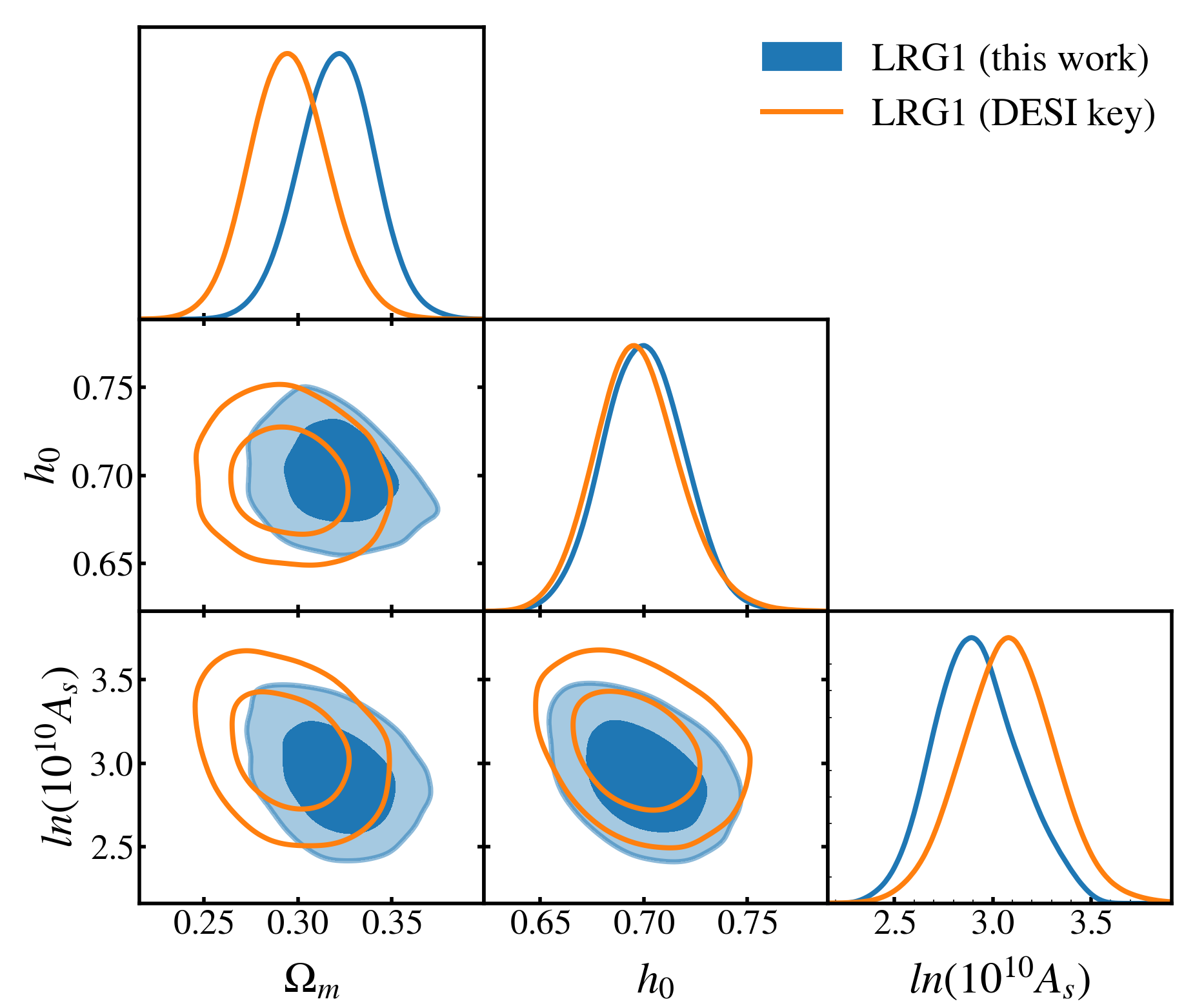}
        \caption{LRG1}
        \label{fig:panel_b}
    \end{subfigure}
    
    \vspace{1em}
    
    \begin{subfigure}[b]{0.45\textwidth}
        \centering
        \includegraphics[width=\textwidth]{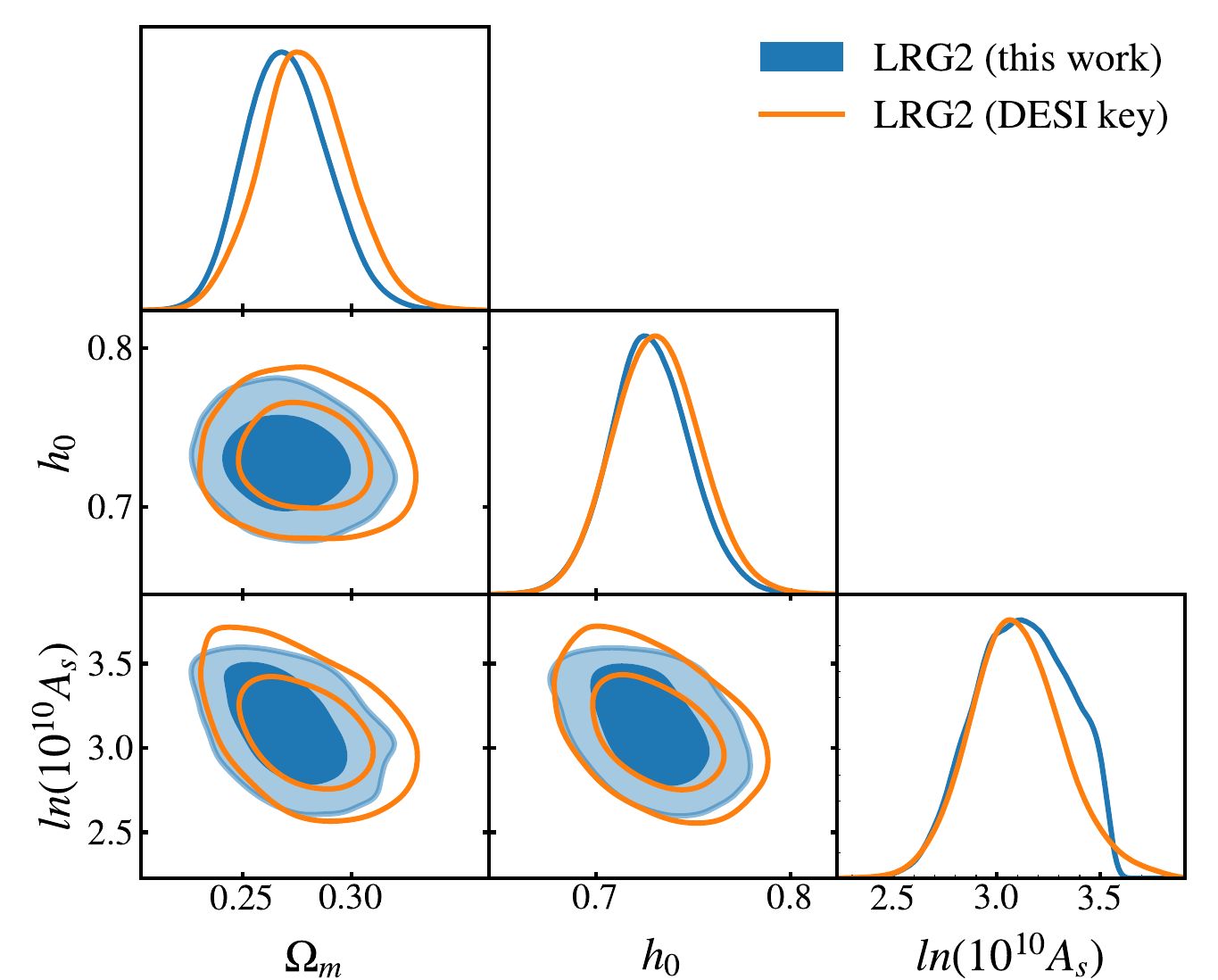}
        \caption{LRG2}
        \label{fig:panel_c}
    \end{subfigure}
    \hfill
    \begin{subfigure}[b]{0.45\textwidth}
        \centering
        \includegraphics[width=\textwidth]{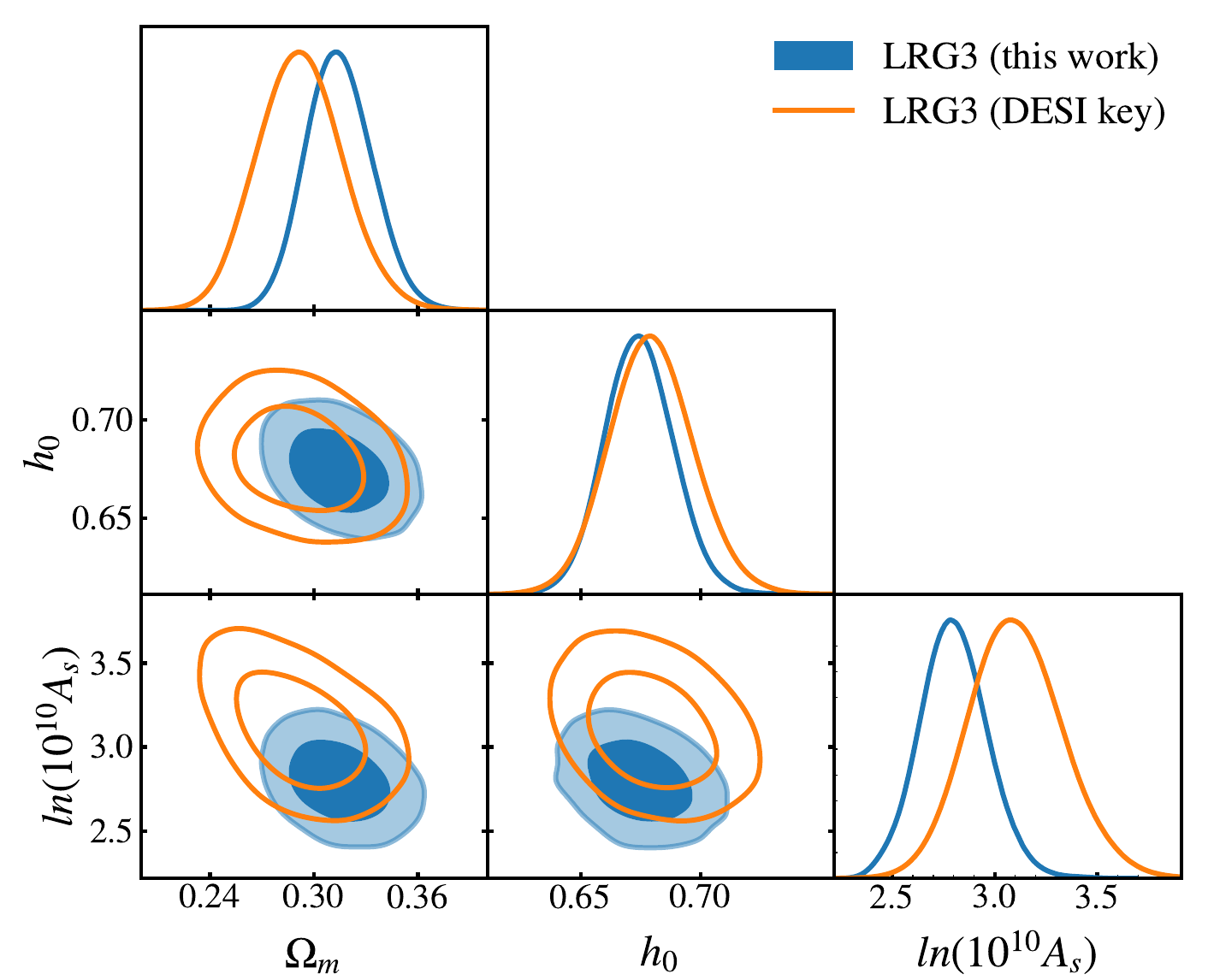}
        \caption{LRG3}
        \label{fig:panel_d}
    \end{subfigure}
    
    \caption{Tracer-by-tracer comparison of the configuration space clustering multipole fits presented in this work (blue) and the fits from DESI DR1 key cosmology full shape analysis in Fourier space \citep[orange lines,][]{2025JCAP...09..008A} }
    \label{fig:tracers_herevsdesi}
\end{figure}

As can be seen in Figure \ref{fig:tracers_herevsdesi}, we find good agreement between the two sets of analyses, although the configuration space fits do show lower values for $\ln(10^{10}A_{\rm{s}})$ and higher values for $\Omega_{\mathrm{m}}$ than the Fourier space analysis. We find an excellent agreement in the posterior of $h_0$, which is expected, as this parameter is well constrained by the BAO feature. We do not necessarily expect the remaining contours to match exactly, as the two analyses employ different sampling schemes and full-shape models and the configuration space measurements will have different noise properties to the Fourier space measurement. We also note that the individual tracer results are affected by the limited range of $\ln(10^{10}A_{\rm{s}})$ supported by our emulator. We leave a more detailed comparison for future analyses.

\section{Robustness tests}
\label{sec:validation_extra}

In this section we present the corner plots further testing the effects of our analysis setup by performing fits to {\sc Abacus} BGS and LRG clustering two-point correlation function multipoles. 

\begin{figure}
    \begin{subfigure}[b]{0.45\textwidth}
        \centering
        \includegraphics[width=\textwidth]{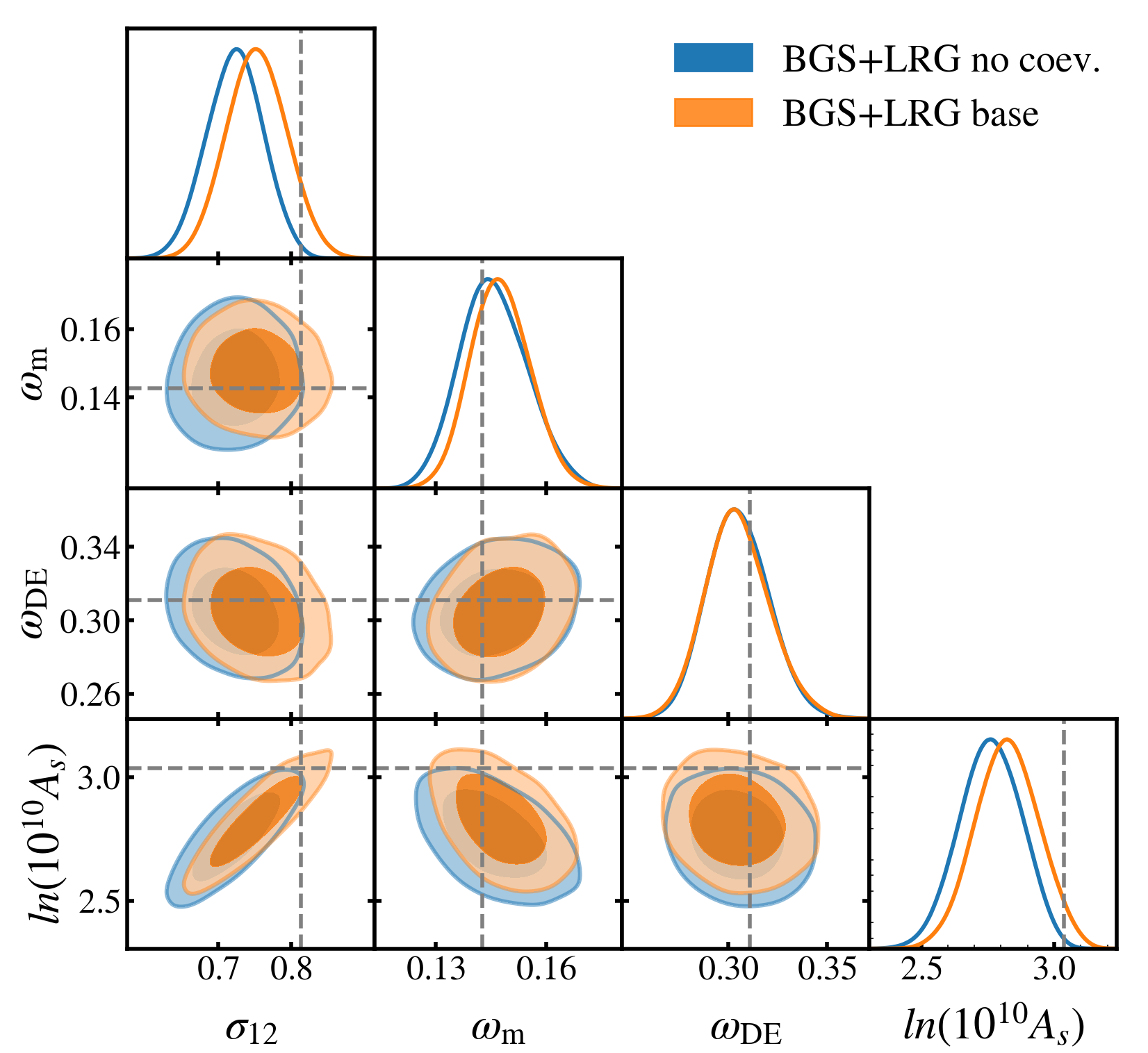}
        \caption{Coevolution test}
        \label{fig:coev}
    \end{subfigure}
    \hfill
    \begin{subfigure}[b]{0.45\textwidth}
        \centering
        \includegraphics[width=\textwidth]{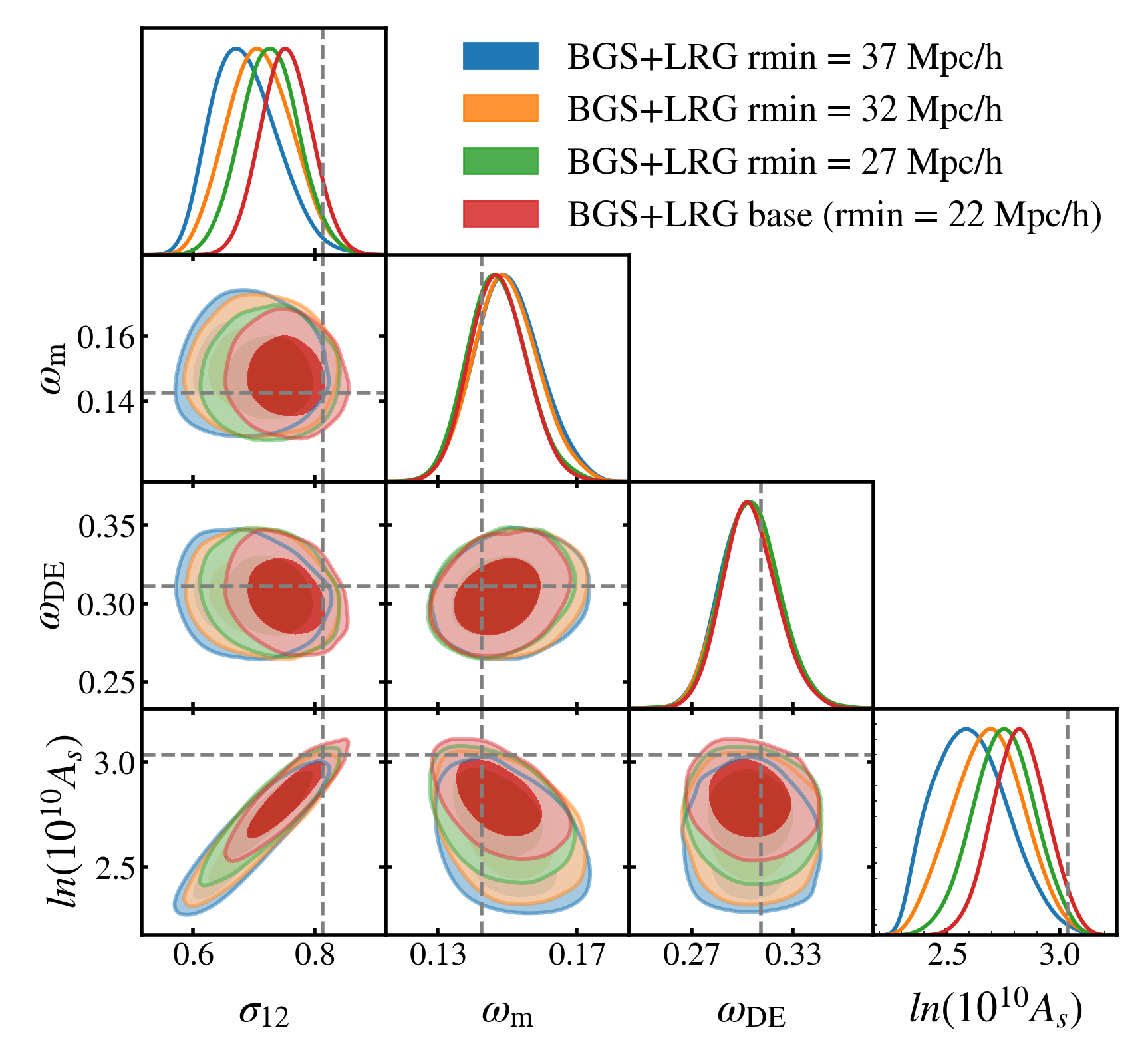}
        \caption{Scale cut test}
        \label{fig:scale}
    \end{subfigure}
\caption{Robustness tests for (a) the tidal bias coevolution assumption: we show the comparison of the fits to {\sc Abacus} BGS and LRG clustering multipoles when allowing $\gamma_{2}$ and $\gamma_{21}$ to vary freely with flat priors (in blue) and fixing to the coevolution relations (orange); and (b) the scale cuts - each set of contours represents fits to {\sc Abacus} BGS and LRG clustering multipoles after removing the measurement corresponding to three, two or one smallest separation bin(s), compared to the base case used in this analysis (red).}
\label{fig:robust}
\end{figure}

First of all, we test allowing the tidal bias parameters $\gamma_2$ and $\gamma_{21}$ to vary freely, instead of fixing them to coevolution relations given in Eq. \ref{eq:gamma2} and Eq. \ref{eq:gamma21}. We set flat priors of $\gamma_{2} \in U(-2,2)$ and $\gamma_{21} \in U(-7,7)$. The resulting posteriors are presented in panel (a) of Figure \ref{fig:robust} and show that, as expected, introducing additional free parameters worsens the projection effects for amplitude parameters, whilst having a more limited effect on $\omega_{\rm{m}}$ and $\omega_{\rm{DE}}$. 

Our scale cut tests are presented in panel (b) of Figure \ref{fig:robust} and show the result of performing fits to two-point correlation multipoles after removing the measurements at one, two or three separation bins corresponding to the smallest scales. Once again, we see a very similar effect as in our coevolution test - reducing the constraining power significantly affects our ability to recover the amplitude parameters, whilst having little effect on the rest of the parameter space. This is expected, as the measurements at the smallest separations carry the most significant constraining power and removing them results in a significant loss in information. This test once again illustrates the significance of projection effects in configuration space.  

\end{document}